\documentclass[11pt,a4paper]{article}
\usepackage{jheppub}
\pdfoutput=1

\usepackage{amsfonts}
\usepackage{amscd}
\usepackage{amssymb}
\usepackage{amsmath,bbm}
\usepackage{graphicx}
\usepackage{epsfig}
\usepackage{latexsym}
\usepackage{mathtools}
\usepackage{hyperref}
\usepackage{tikz-cd}
\usepackage[vcentermath]{youngtab}

\def \be  {\begin{equation}}
\def \ee  {\end{equation}}
\def \bea {\begin{equation}\begin{aligned}}
\def \eea {\end{aligned}\end{equation}}
\def \ba  {\begin{eqnarray}}
\def \ea  {\end{eqnarray}}
\def \bb  {\begin{thebibliography}}
\def \eb  {\end{thebibliography}}
\def \lab #1 {\label{#1}}

\newcommand\cond{\mathrel{\,\hspace{.75ex}\joinrel\rhook\joinrel\hspace{-.75ex}\joinrel\rightarrow}}

\newcommand\cA{\mathcal{A}}

\newcommand\cF{\mathcal{F}}
\newcommand\cG{\mathcal{G}}
\newcommand\cH{\mathcal{H}}

\newcommand\cO{\mathcal{O}}

\newcommand\cT{\mathcal{T}}

\newcommand\cZ{\mathcal{Z}}

\newcommand\C{\mathsf{C}}
\newcommand\B{\mathsf{B}}
\newcommand\T{\mathsf{T}}

\newcommand\bC{\mathbb{C }}

\newcommand\bR{\mathbb{R}}

\newcommand\bZ{\mathbb{Z}}
\newcommand\R{\mathbb{R}}

\newcommand\vect{\mathsf{Vec}}
\newcommand\rep{\mathsf{Rep}}

\definecolor{cardinal}{rgb}{0.6,0,0}
\definecolor{darkgreen}{rgb}{0,0.5,0}
\definecolor{golden}{rgb}{0.92, 0.7, 0}
\definecolor{midnight}{rgb}{0, 0, 0.5}
\definecolor{darkblue}{rgb}{0.2, 0, 0.8}

\usepackage{braket}
\usepackage{stackengine}
\usepackage{pict2e}

\newcommand\yleftarrow[2][]{\xleftarrow[{\raisebox{1.25ex-\heightof{$\scriptstyle#1$}}[0pt]{$\scriptstyle#1$}}]{#2}}

\newcommand{\id}{\text{\usefont{U}{bbold}{m}{n}1}}
\MakeRobust{\id}

\newcommand\genone[2][]{\big\langle \unitlength=1pt\thicklines
 \begin{picture}(20,1)
 \put(12,-3){\line(0,1){14}}
 \put(0,4){\line(1,0){24}}
 \put(12,4){\circle*{3}}
 \put(1,5.5){$\scriptstyle #1$}
 \put(13.5,-4.5){$\scriptstyle #2$}
  \end{picture} \hspace{3.5pt} \big\rangle}

\newcommand\gentwo[2][]{\big\langle \unitlength=1pt\thicklines
 \begin{picture}(20,1)
 \put(12,-3){\line(0,1){14}}
 \put(0,4){\line(1,0){24}}
 \put(12,4){\circle*{3}}
 \put(0.5,5.5){$\scriptstyle m$}
 \put(13.5,-4.5){$\scriptstyle #1$}
 \put(13.5,8){$\scriptstyle #2$}
  \end{picture} \hspace{3.5pt} \big\rangle}

\newcommand\genthree[1][]{\big\langle \unitlength=1pt\thicklines
 \begin{picture}(20,1)
 \put(12,-3){\line(0,1){14}}
 \put(0,4){\line(1,0){24}}
 \put(12,4){\circle*{3}}
 \put(0.5,5.5){$\scriptstyle m$}
 \put(13.5,-4.5){$\scriptstyle m$}
 \put(13.4,5.4){$\scriptstyle #1$}
  \end{picture} \hspace{3.5pt} \big\rangle}
  
\newcommand\genfour[3][]{\big\langle \hspace{6pt} \unitlength=1pt\thicklines
 \begin{picture}(20,15)
 \put(0,1.5){\line(1,0){20}}
 \put(0,1.5){\line(4,5){4}}
 \put(20,1.5){\line(4,5){4}}
 \put(4,6.5){\line(1,0){6}}
 \put(14,6.5){\line(1,0){10}}
  \put(12,-3){\line(0,1){3}}
 \put(12,4){\line(0,1){7}}
 \put(12,4){\circle*{3}}
 \put(-6,3){$\scriptstyle #1$}
 \put(13.5,-5.5){$\scriptstyle #2$}
 \put(13.5,9.5){$\scriptstyle #3$}
\end{picture} \hspace{4.5pt} \big\rangle}

\title{Representation theory for categorical symmetries}
\author{Thomas Bartsch, Mathew Bullimore, Andrea Grigoletto}
\affiliation{Department of Mathematical Sciences, Durham University, \\
Upper Mountjoy, Stockton Road, Durham, DH1 3LE, United Kingdom}
\emailAdd{thomas.d.bartsch@durham.ac.uk}
\emailAdd{mathew.r.bullimore@durham.ac.uk}
\emailAdd{andrea.grigoletto@durham.ac.uk}
\date{}

\abstract{This paper addresses the question of how categorical symmetries act on extended operators in quantum field theory. Building on recent results in two dimensions, we introduce higher tube categories and algebras associated to higher fusion category symmetries. We show that twisted sector extended operators transform in higher representations of higher tube algebras and interpret this result from the perspective of the sandwich construction of finite symmetries via the Drinfeld center. Focusing on three dimensions, we discuss a variety of examples to illustrate the general constructions. In the case of invertible symmetries, we show that higher tube algebras are higher analogues of twisted Drinfeld doubles of finite groups, generalising known constructions in two dimensions. Building on this foundation, we discuss non-invertible Ising-like symmetry categories obtained by gauging finite subgroups. We also consider non-invertible topological symmetry lines described by braided fusion categories and discuss connections to the M\"uger center and braided module categories.}

\begin{document}
\maketitle

\section{Introduction}
\label{sec:intro}

It has recently been understood that quantum field theories admit finite symmetries captured by the mathematical structure of higher fusion categories~\cite{2018arXiv181211933D,Gaiotto:2019xmp,Johnson-Freyd:2022wm}. This structure encapsulates the relevant properties of topological extended operators and includes both finite invertible (or group-like) and non-invertible symmetries. The latter have been studied extensively in dimension $D=2$ \cite{Verlinde:1988sn,Petkova:2000ip,Frohlich:2004ef,Frohlich:2006ch,Frohlich:2009gb,Carqueville:2012dk,Brunner:2013ota,Brunner:2013xna,Bhardwaj:2017xup,Chang:2018iay,Thorngren:2019iar,Ji:2019ugf,Lin:2019hks,Komargodski:2020mxz,Chang:2020imq,Nguyen:2021naa,Burbano:2021loy,Huang:2021nvb,Thorngren:2021yso,Sharpe:2021srf,Lin:2022dhv,Chang:2022hud,Lin:2023uvm} and recent work has demonstrated their existence and utility in dimension $D>2$ \cite{Rudelius:2020orz,Heidenreich:2021xpr,Koide:2021zxj,Choi:2021kmx,Kaidi:2021xfk,Wang:2021vki,Chen:2021xuc,Cordova:2022rer,Benini:2022hzx,Roumpedakis:2022aik,DelZotto:2022ras,Bhardwaj:2022yxj,Hayashi:2022fkw,Arias-Tamargo:2022nlf,Choi:2022zal,Kaidi:2022uux,Choi:2022jqy,Cordova:2022ieu,Antinucci:2022eat,Bashmakov:2022jtl,Damia:2022rxw,Damia:2022bcd,Moradi:2022lqp,Choi:2022rfe,Bhardwaj:2022lsg,Bartsch:2022mpm,Lin:2022xod,Lu:2022ver,GarciaEtxebarria:2022vzq,Heckman:2022muc,Kaidi:2022cpf,Niro:2022ctq,Mekareeya:2022spm,Antinucci:2022vyk,Chen:2022cyw,Bashmakov:2022uek,Karasik:2022kkq,Cordova:2022fhg,Decoppet:2022dnz,GarciaEtxebarria:2022jky,Choi:2022fgx,Bhardwaj:2022kot,Bartsch:2022ytj,Bhardwaj:2022maz,Antinucci:2022cdi,Apte:2022xtu,Delcamp:2023kew,Kaidi:2023maf,Brennan:2023kpw,Radhakrishnan:2023zcq,Putrov:2023jqi,Carta:2023bqn,Bhardwaj:2023wzd,Bartsch:2023pzl,Cao:2023doz,Inamura:2023qzl}.

An important question is how categorical symmetries act on and thereby organise the spectrum of (not necessarily topological) extended operators in a quantum field theory. In other words, a comprehensive representation theory for categorical symmetries in quantum field theory is needed. This paper seeks to address this question in $D\geq 2$.

\subsection{Background}

\subsubsection{Two dimensions}

In two dimensions, one may consider quantum field theories with a spherical fusion category symmetry $\C$. Objects in the symmetry category $\C$ correspond to topological symmetry lines and morphisms to junctions between them.

\begin{figure}[h]
	\centering
	\includegraphics[height=3.7cm]{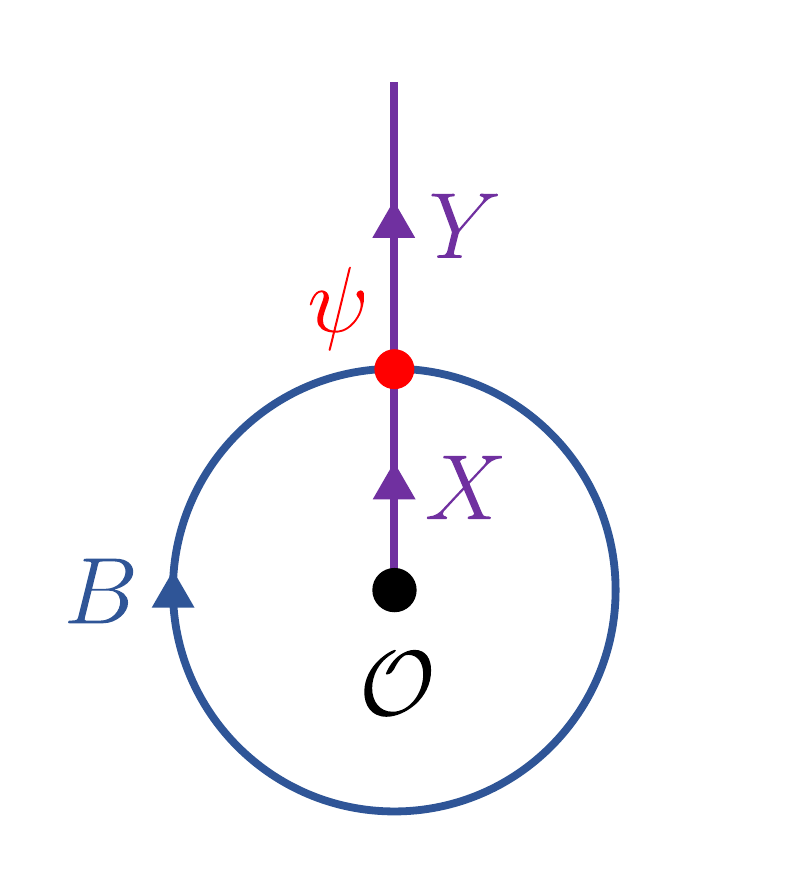}
	\vspace{-5pt}
	\caption{}
	\label{fig:intro-2d-1-twisted-action}
\end{figure}

The topological lines in $\C$ act on twisted sector local operators, which are local operators that are themselves attached to topological symmetry lines. Crucially, this action may transform one twisted sector to another~\cite{Chang:2018iay}. This action is captured by the type of configuration illustrated in figure~\ref{fig:intro-2d-1-twisted-action}.

In more detail, figure \ref{fig:intro-2d-1-twisted-action} captures the linking action of a topological symmetry line $B \in \C$ on a twisted sector local operator $\mathcal{O}$ that is attached to another topological symmetry line $X \in \mathsf{C}$. This requires a point-like intersection $\psi$ connecting $X$ to another topological symmetry line $Y \in \mathsf{C}$. Shrinking the circle then implements the action of $B$ on $\mathcal{O}$, transforming one twisted sector to another.

The collection of data associated to configurations of the type illustrated in figure~\ref{fig:intro-2d-1-twisted-action} is conveniently encoded in the \textit{tube category} $\mathsf{TC}$ of $\mathsf{C}$. Schematically, this is a finite semi-simple category whose
\begin{itemize}
\item objects are twisted sectors $X, Y, ... \in \mathsf{C}$,
\item morphisms $X \to Y$ are certain equivalence classes of pairs $(B,\psi)$ consisting of a topological symmetry line $B \in \C$ and a morphism
\be
\psi : \; B \otimes X \; \to \; Y \otimes B \, .
\ee
\end{itemize}
A more detailed description will be given in the main text. 

The action of the symmetry category $\mathsf{C}$ on twisted sector local operators may then be described as a functor 
\be
\mathcal{F}: \mathsf{TC} \to \mathsf{Vec}
\ee
into the category of finite-dimensional vector spaces. This functor assigns a vector space to each object $X \in \T\C$, corresponding to a collection of twisted sector operators attached to the line $X$. Furthermore, to each morphism $X \to Y$ it assigns a linear map between vector spaces arising from configurations as in figure~\ref{fig:intro-2d-1-twisted-action}. We call this a \emph{tube representation} of $\C$. The collection of tube representations forms a category $[\mathsf{TC},\mathsf{Vec}]$.

There is an equivalent formulation in terms of the associated \textit{tube algebra} $\mathcal{A}(\mathsf{C})$, as discussed in~\cite{Lin:2022dhv}, where the above remarks have their genesis. Briefly, the tube algebra is generated by morphisms in the tube category with the algebra product given by composition of morphisms. There is an equivalence
\begin{equation}
[\mathsf{TC}, \mathsf{Vec}] \; \cong \; \mathsf{Rep}( \mathcal{A}(\mathsf{C})) \, , 
\end{equation}
meaning that tube representations may be regarded as ordinary representations of the tube algebra.

\begin{figure}[h]
	\centering
	\includegraphics[height=4.6cm]{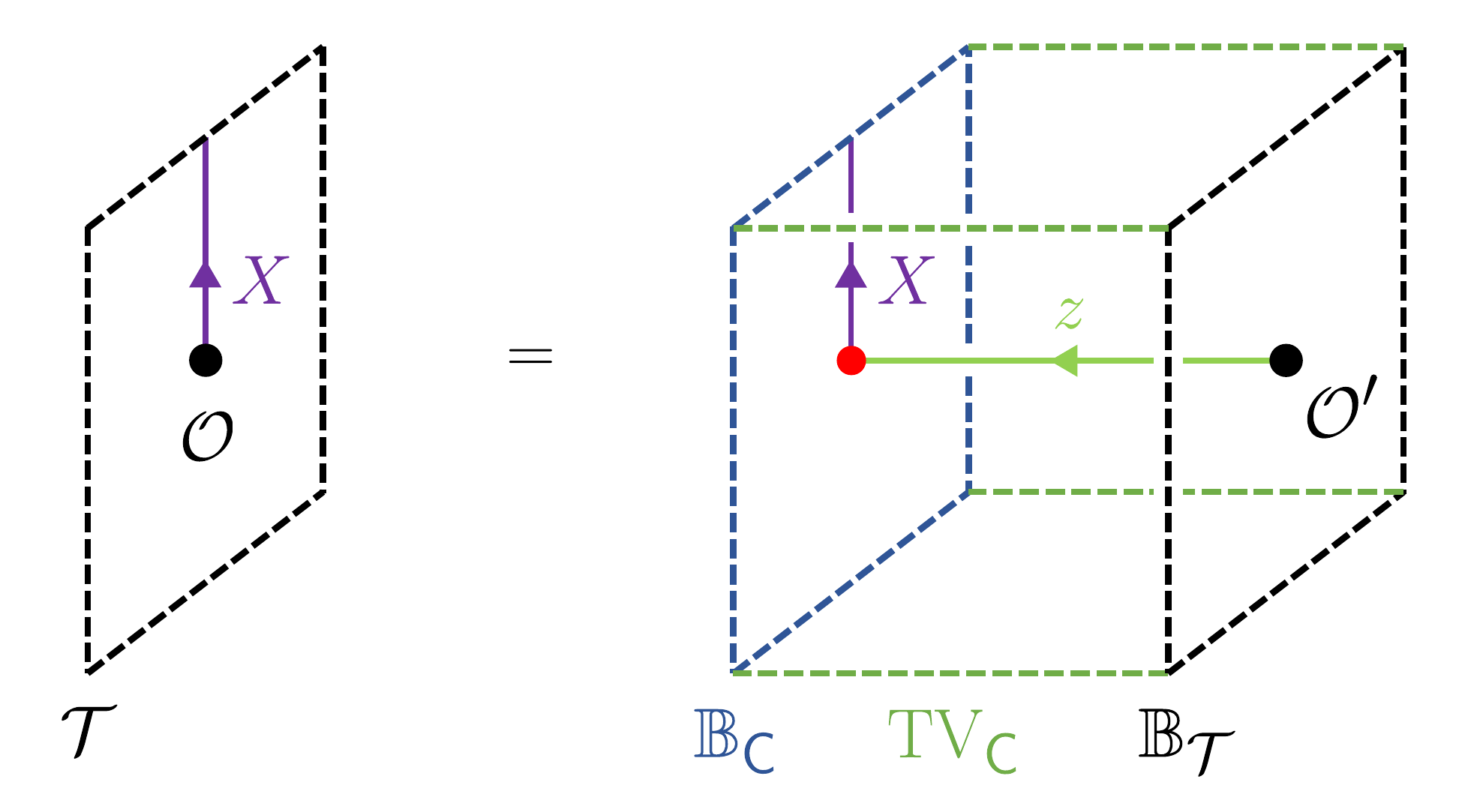}
	\vspace{-5pt}
	\caption{}
	\label{fig:intro-2d-sandwich-construction}
\end{figure}

As emphasised in~\cite{Lin:2022dhv}, this has deep connections to the sandwich construction of finite symmetries~\cite{Freed:2012bs,Gaiotto:2020iye,Ji:2019jhk,Kong:2020cie,Freed:2022qnc,Freed:2022iao}\footnote{See also~\cite{Apruzzi:2021nmk,Apruzzi:2022dlm,Apruzzi:2022rei,vanBeest:2022fss,Chen:2023qnv} for realisations of the sandwich construction in string theory.}. In this framework, a two-dimensional theory $\mathcal{T}$ with fusion category symmetry $\C$ is viewed as an interval compactification of the associated three-dimensional Turaev-Viro theory $\text{TV}_{\C}$~\cite{TURAEV1992865,Barrett:1993ab}, as illustrated in figure~\ref{fig:intro-2d-sandwich-construction}. 

The left boundary condition $\mathbb{B}_{\mathsf{C}}$ is a canonical gapped boundary supporting the symmetry $\C$, while the right boundary condition $\mathbb{B}_{\mathcal{T}}$ contains information about the specific theory $\mathcal{T}$. This has the conceptual advantage that universal information about the categorical symmetry $\mathsf{C}$ is abstracted onto the left boundary, while information about its action on a specific two-dimensional theory is separated onto the right boundary.

Concretely, the relationship between tube representations and the sandwich construction takes the form of a braided equivalence~\cite{Evans1995ONOT,Izumi:2000aa,MUGER2003159}
\begin{equation}
[\mathsf{TC},\mathsf{Vec}] \,\; \cong \;\, \text{TV}_{\mathsf{C}}(S^1) \, ,
\end{equation}
where the right-hand side is the category of topological lines in $\text{TV}_\C$ that are linked by $S^1$. A tube representation corresponding to a bulk topological line is realised by local operators at the junction (shown in red in figure~\ref{fig:intro-2d-sandwich-construction}) between the bulk topological line $z \in \mathcal{Z}(\mathsf{C})$ and boundary topological lines $X \in \mathsf{C}$ on the canonical gapped boundary condition.

The topological line defects in the bulk topological theory $\text{TV}_\C$ are given by the factorisation homology of $\C$ on $S^1$, 
\begin{equation}
 \text{TV}_{\mathsf{C}}(S^1) \,\; := \,\; \int_{S^1} \!\mathsf{C} \,\; = \,\; \mathcal{Z}(\mathsf{C}) \, ,
\end{equation}
where $\cZ(\C)$ denotes the \textit{Drinfeld center} of $\C$\footnote{We work with spherical fusion categories and do not distinguish between center and co-center.}. In summary, tube representations of $\C$ are captured by the Drinfeld center $\cZ(\C)$.

\subsubsection{Extended operators}

A natural question is how this construction generalizes to dimensions $D \geq 2$. A first step was initiated in~\cite{Bartsch:2023pzl} (see also~\cite{Bhardwaj:2023wzd}), which studied the action of invertible symmetries on genuine (untwisted) extended operators in any number of dimensions. The proposal is that \emph{higher representation theory} is the natural language: genuine $(n-1)$-dimensional extended operators transform in $n$-representations of an invertible symmetry. 

An \textit{$n$-representation} of an invertible symmetry $\mathcal{G}$ is an $n$-functor 
\be
\cF : \cG \to \mathsf{nVec} \, ,
\ee
where $\cG$ is understood as $n$-groupoid and $\mathsf{nVec}$ is the $n$-category of finite-dimensional $n$-vector spaces. The collection such $n$-representations forms as $n$-category 
\begin{equation}
\mathsf{nRep}(\mathcal{G}) \; = \; [\mathcal{G},\mathsf{nVec}] \, .
\end{equation}

\begin{figure}[h]
	\centering
	\includegraphics[height=4.6cm]{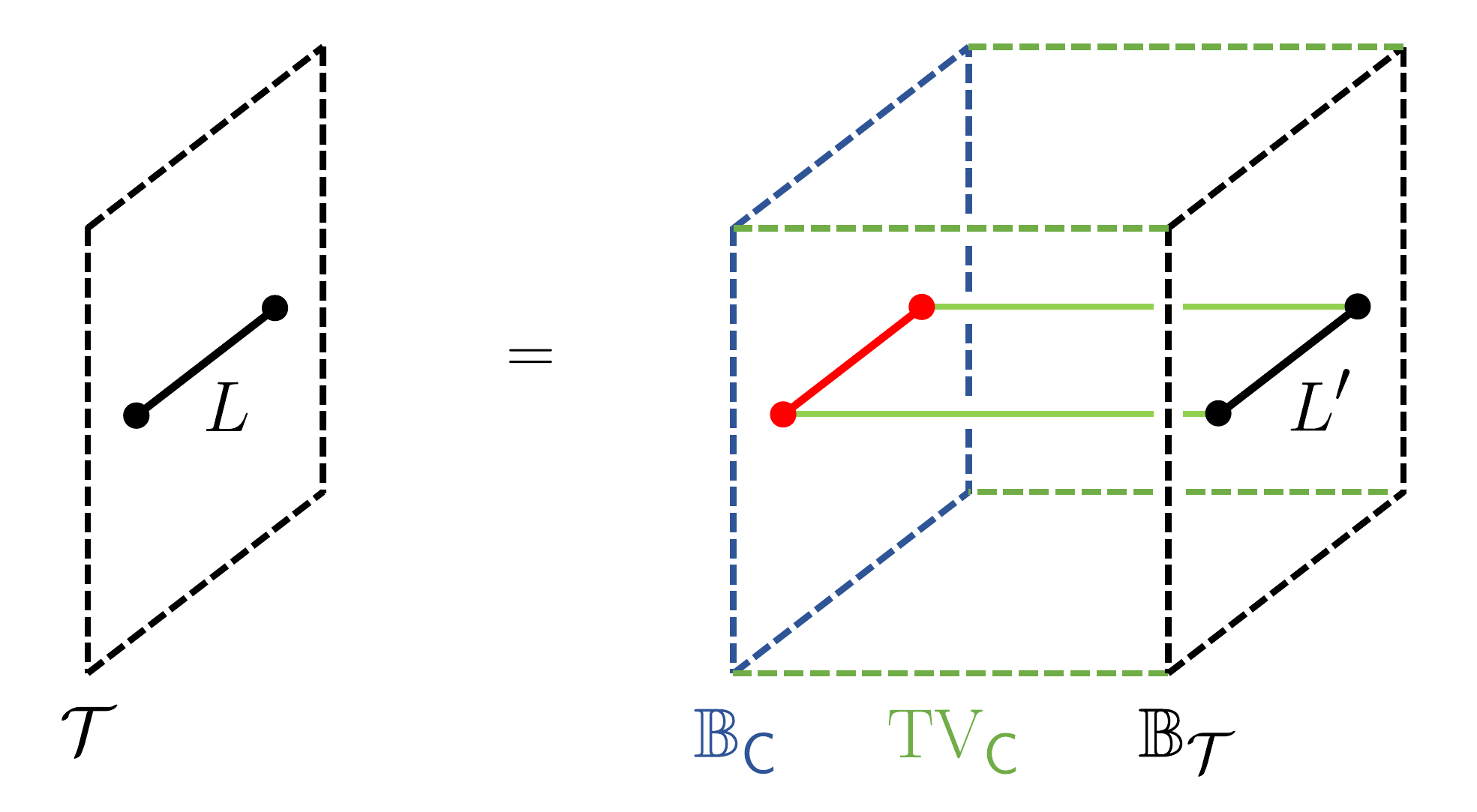}
	\vspace{-5pt}
	\caption{}
	\label{fig:intro-higher-group-action}
\end{figure}

This again has deep connections to the sandwich construction of finite symmetries. The $(D+1)$-dimensional bulk topological theory $\text{TV}_\C$ is now a generalised Dijkgraaf-Witten theory constructed from $\cG$. The $(n-1)$-dimensional extended operators $L$ transforming in $n$-representations of $\cG$ now arise from junctions with a particular class of $n$-dimensional bulk topological defects with the property that they are trivial when brought to the canonical boundary condition $\mathbb{B}_\C$. This is illustrated in figure~\ref{fig:intro-higher-group-action}.

More precisely, such an $n$-dimensional bulk topological defect becomes trivial on $\mathbb{B}_\C$ in the absence of background fields for the invertible symmetry. Otherwise, it becomes a $n$-dimensional background Wilson line, or extended $n$-dimensional TQFT equipped with a $\mathcal{G}$-equivariant structure, which is another description of an $n$-representation of $\cG$. This reproduces the categorical perspective in~\cite{Bartsch:2023pzl}.

\subsubsection{Synthesis}

The idea of this paper is to combine the above constructions to develop the higher representation theory for non-invertible symmetries described by a spherical higher fusion category $\C$. This will capture the action of $\C$ on all local and extended twisted sector operators.

\subsection{Summary}

The most general setup we contemplate is a quantum field theory in $D$ dimensions with a spherical fusion $(D-1)$-category symmetry $\mathsf{C}$\footnote{Ultimately, in this paper we focus on $D = 3$, where such structures are entirely well-defined.}. This captures the properties of topological symmetry defects of various dimensions, including those arising from condensations of lower dimensional topological symmetry defects.

\begin{figure}[h]
	\centering
	\includegraphics[height=3.1cm]{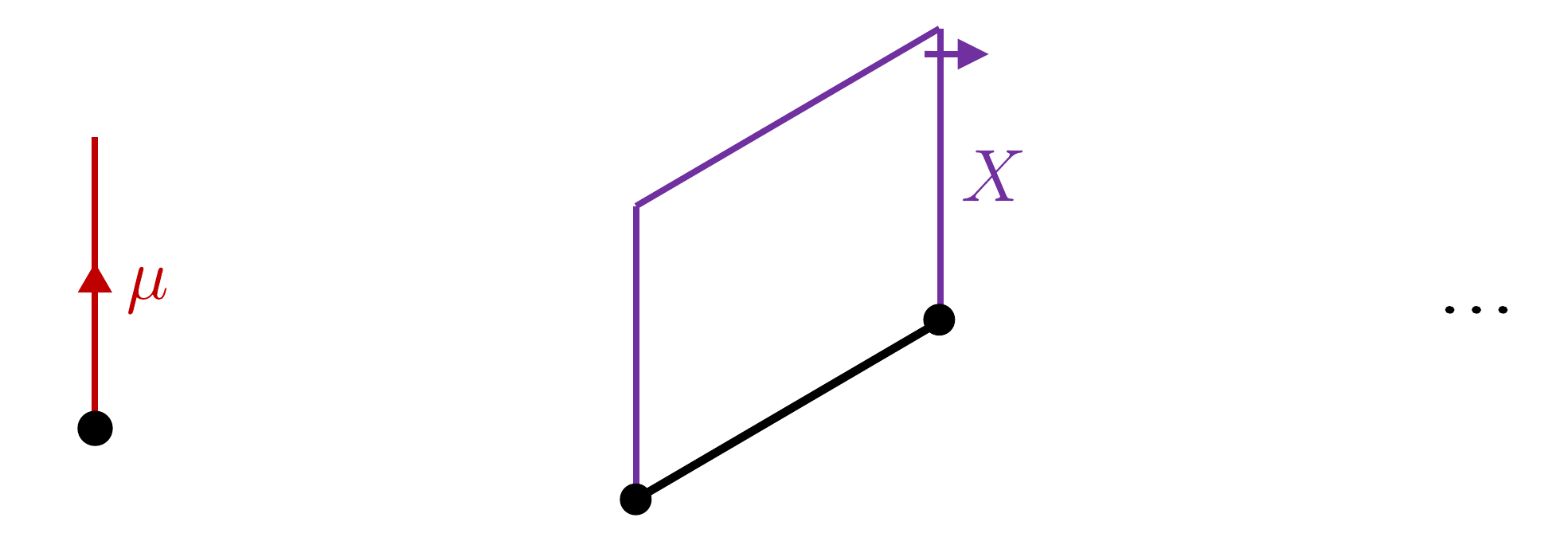}
	\vspace{-5pt}
	\caption{}
	\label{fig:intro-general-twisted-sectors}
\end{figure}

We want to consider the action of $\C$ on (not necessarily topological) twisted sector extended operators attached to topological symmetry defects. Concretely, we can consider $(n-1)$-dimensional extended operators attached to $n$-dimensional topological symmetry defects, as illustrated schematically in figure \ref{fig:intro-general-twisted-sectors}. We refer to such extended operators as \textit{$n$-twisted sector} operators.

\subsubsection{Higher tube categories}

Expanding on the results from two dimensions, we expect the data associated to configurations of topological symmetry defects acting on $n$-twisted sector operators to be encoded in a \textit{tube $n$-category}
\begin{equation}
\mathsf{T}_{S^{d}}(\mathsf{C}) \, ,
\end{equation}
where $d:= D-n$ and $S^{d}$ denotes a sphere that links $n$-twisted sector operators in $D$ dimensions. The intuition is that the tube $n$-category is obtained from $\C$ by compactification on $S^d$. Schematically, its objects are $n$-twisted sectors and morphisms capture actions of all possible topological symmetry defects supported on $S^d \times \R^{n-1}$.

\begin{figure}[h]
	\centering
	\includegraphics[height=4.8cm]{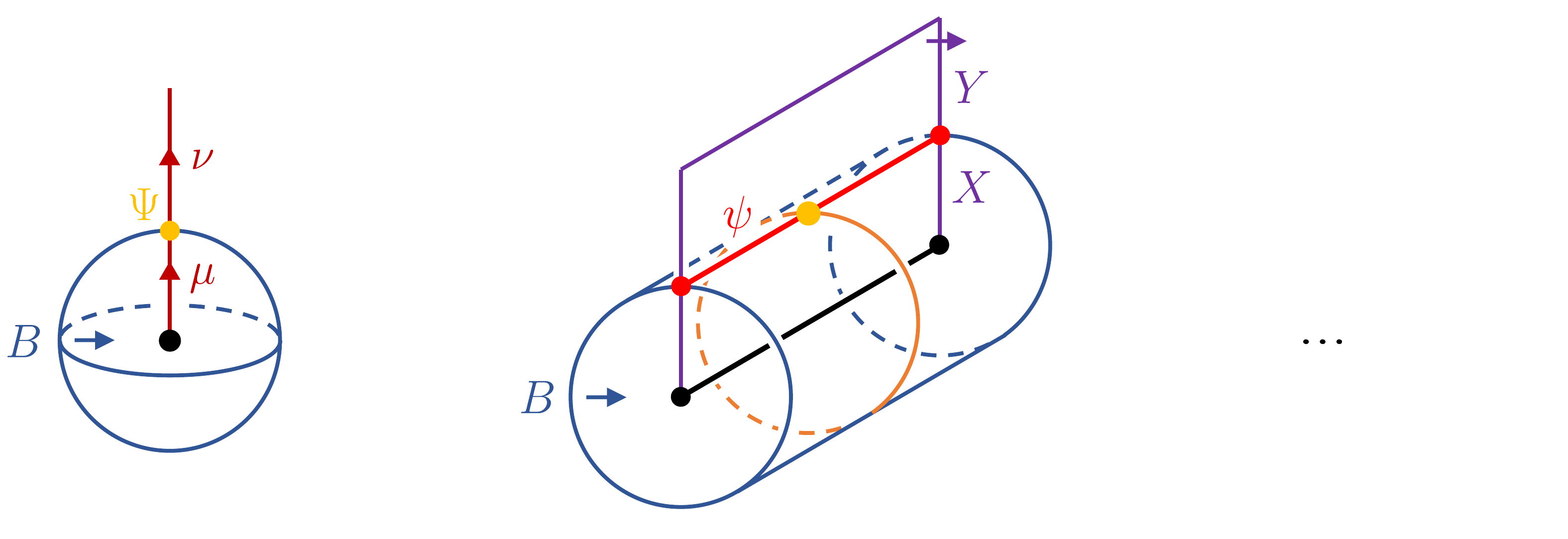}
	\vspace{-5pt}
	\caption{}
	\label{fig:intro-general-twisted-sector-action}
\end{figure}

In more detail, $\mathsf{T}_{S^{d}}(\mathsf{C})$ is a finite semi-simple $n$-category with
\begin{itemize}
\item objects $\mu, \nu, ... \in \Omega^{d-1}(\mathsf{C})$\footnote{We define $\Omega\mathsf{C} := \text{1-End}_{\mathsf{C}}(\mathbf{1})$, where $\mathbf{1}$ denotes the monoidal unit of $\mathsf{C}$. The operation $\Omega^m$ for $m \in \mathbb{N}$ is then defined by applying $\Omega$ iteratively $m$ times.} given by $n$-twisted sectors or genuine $n$-dimensional topological symmetry defects,
\item 1-morphisms $\mu \to \nu$ given by certain equivalence classes of pairs $(B,\Psi)$ consisting of an object $B \in \mathsf{C}$ and a $d$-morphism
\begin{equation}
\Psi: \;\; \text{Id}^{d-1}_B \otimes \, \mu \;\; \to \;\; \nu \, \otimes \text{Id}^{d-1}_B
\end{equation}
that describes the wrapping of $n$-twisted sector operators by codimension-one topological symmetry defects $B$ on $S^{d} \times \bR^{n-1}$.

\item higher morphisms given by the wrapping / linking of $n$-twisted sector operators by topological symmetry defects of higher codimension.

\end{itemize}
This is illustrated schematically in figure~\ref{fig:intro-general-twisted-sector-action}. A complete description in dimensions $D = 2,3$ will be given in the main text.

The action of the symmetry category $\mathsf{C}$ on $n$-twisted sector operators is then captured by an $n$-functor
\be
\mathcal{F}: \mathsf{T}_{S^d}(\mathsf{C}) \to \mathsf{nVec}
\ee
which we call a \textit{tube $n$-representation}. The collection of tube $n$-representations forms a fusion $n$-category 
\begin{equation}
[\mathsf{T}_{S^d}(\mathsf{C}), \mathsf{nVec}] \, ,
\end{equation}
which we expect to be the Karoubi completion of $\mathsf{T}_{S^d}(\mathsf{C})$. 

There is some redundancy is this description as tube $m$-representations can be recovered from tube $n$-representations for $m < n$. In principle, it is therefore enough to consider the top tube $(D-1)$-category $\T_{S^1}(\C)$. Nevertheless, there is some efficiency and utility in working with the intermediate tube categories.

\subsubsection{Higher tube algebras}

There is again an alternative description in terms of \textit{higher tube algebras $\mathcal{A}_{S^d}(\mathsf{C})$}, which are finite semi-simple $n$-algebras generated by 1-morphisms in the tube category $\mathsf{T}_{S^d}(\mathsf{C})$. They have the property that their $n$-representations or module $(n-1)$-categories are the same as tube $n$-representations,
\begin{equation}
[\mathsf{T}_{S^d}(\mathsf{C}), \mathsf{nVec}] \; \cong \; \mathsf{nRep}( \mathcal{A}_{S^d}(\mathsf{C})) \, .
\end{equation}
Higher tube algebras therefore provide an equivalent formulation of the representation theory of categorical symmetries. 

The tube algebras perspective is often useful for explicit computations and to exhibit some structural features. We describe the relation between the two perspectives in more detail in the main text.

\subsubsection{Sandwich construction}

The formulation of tube $n$-representations again has deep connections to the sandwich construction of finite symmetries, which generalises the correspondence between tube representations of a fusion category $\C$ and its Drinfeld center $\mathcal{Z}(\C)$. 

\begin{figure}[h]
	\centering
	\includegraphics[height=4.6cm]{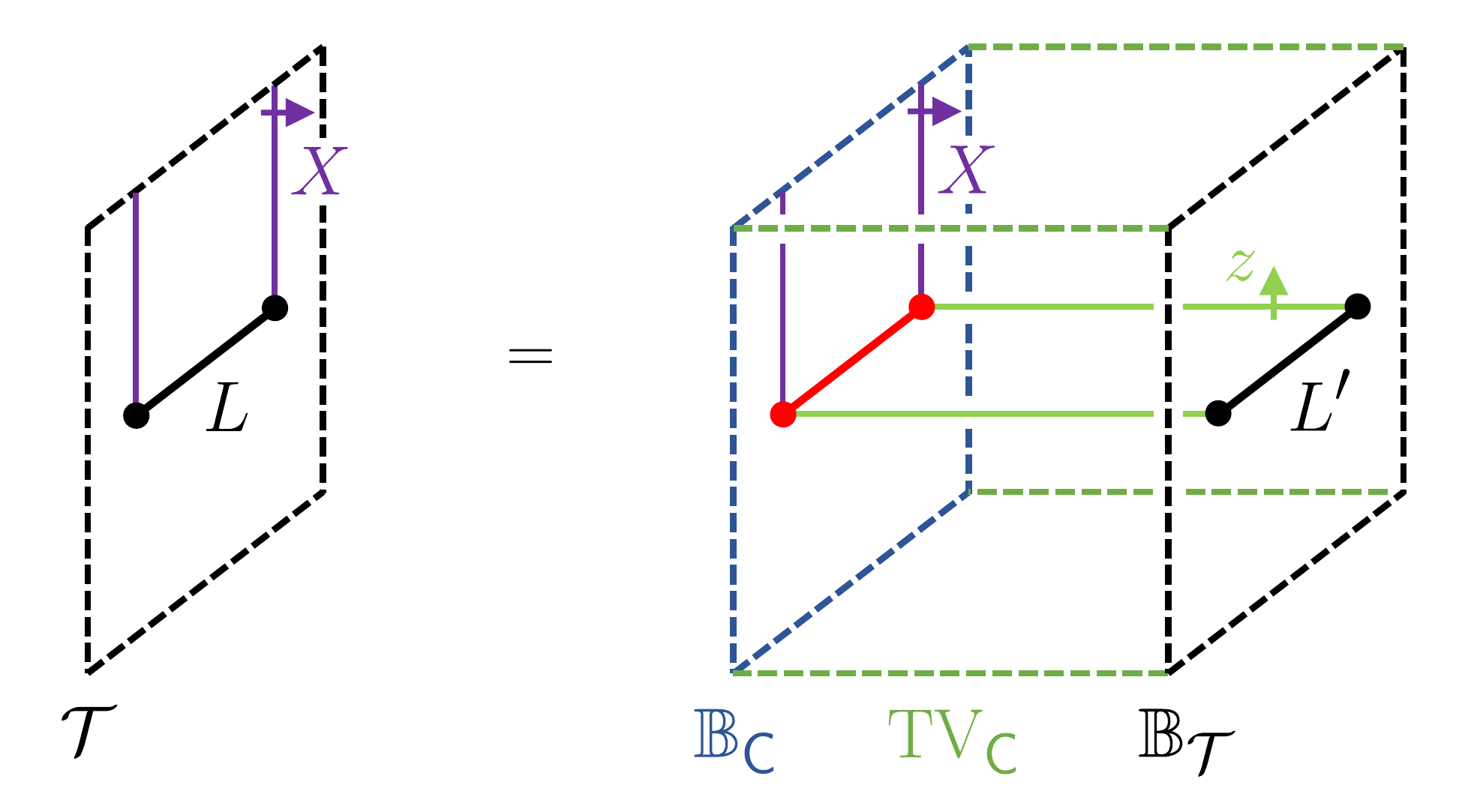}
	\vspace{-5pt}
	\caption{}
	\label{fig:intro-general-sandwich-construction}
\end{figure}

Concretely, a $D$-dimensional theory $\mathcal{T}$ with higher fusion category symmetry $\C$ is viewed as an interval compactification of an associated $(D+1)$-dimensional topological theory $\text{TV}_{\mathsf{C}}$. The left boundary condition is a canonical gapped boundary supporting the symmetry $\C$, while the right boundary condition contains information about the specific $D$-dimensional theory $\mathcal{T}$. This is illustrated in figure~\ref{fig:intro-general-sandwich-construction}.

The relationship between higher tube representations and the sandwich construction now takes the form of an equivalence
\begin{equation}
[\mathsf{T}_{S^d}(\mathsf{C}),\mathsf{nVec}] \,\; \cong \,\; \text{TV}_{\mathsf{C}}(S^d) \, ,
\end{equation}
where the right-hand side is the category of $n$-dimensional topological defects in $\text{TV}_\C$ that are linked by $S^d$. From a physical perspective, tube $n$-representations are realised by $(n-1)$-dimensional extended operators at the junction (shown in red in figure \ref{fig:intro-general-sandwich-construction}) between $n$-dimensional defects in the bulk and topological symmetry defects on the canonical gapped boundary condition.

The $n$-dimensional topological defects in the bulk topological theory $\text{TV}_\C$ are given by factorisation homology of $\C$ on $S^d$,
\begin{equation}
 \text{TV}_{\mathsf{C}}(S^d) \,\; := \;\, \int_{S^d} \!\mathsf{C} \,\; = \;\, \Omega^{d-1}\mathcal{Z}(\mathsf{C}) \, .
\end{equation}
In summary, higher tube representations of $\C$ are captured by factorisation homology of $\C$, or iterated loopings of the Drinfeld center of $\C$. In particular, representations of the top tube category $\T_{S^1}(\C)$ are equivalent to the full Drinfeld center $\mathcal{Z}(\C)$.

\subsubsection{Group-theoretical symmetries}

Higher tube representations are interesting even for invertible symmetries $\cG$ in $D \geq 2$. Indeed, the $(D+1)$-dimensional bulk topological theory $\text{TV}_\C$ is a generalised Dijkgraaf-Witten theory constructed from $\cG$ and the tube algebras $\cA_{S^d}(\C)$ provide higher analogues of the twisted Drinfeld double construction.

A simple example is an ordinary finite group symmetry $G$ in $D$ dimensions with an 't Hooft anomaly specified by a group cocycle $\alpha \in Z^{D+1}(G,U(1))$. The associated symmetry category is
\begin{equation}
\mathsf{C} \; = \; \mathsf{(D-1)Vec}^{\alpha}_G \, .
\end{equation}
In this case, the higher tube algebras $\mathcal{A}_{S^d}(\mathsf{C})$ provide higher categorical versions of the twisted Drinfeld double of finite groups, to which they reduce when $D=2$ and $d=1$. The higher tube representations in this case can be determined systematically. For example, the irreducible $(D-1)$-representations of $\T_{S^1}\C$ are labelled by
\begin{enumerate}
\item a conjugacy class $[x] \in G/G$ with representative $x \in G$,
\item an irreducible projective $(D-1)$-representation of the centraliser $C_x(G)$ with projective $D$-cocycle 
\be
\tau_x(\alpha) \in Z^D(C_x(G),U(1)) \, ,
\ee
where $\tau(\alpha)$ is the \textit{transgression} of the 't Hooft anomaly $\alpha$.
\end{enumerate}
This provides a description of simple objects in $\mathcal{Z}(\C)$, or topological defects in the $(D+1)$-dimensional Dijkgraaf-Witten theory associated to $G$ and $\alpha$. In particular, note that focusing on tube representations supported on the trivial conjugacy class recovers the fact that genuine $(D-2)$-dimensional extended operators transform in $(D-1)$-representations of the symmetry group $G$. This general class of examples was studied for $D =3$ in~\cite{Kong:2019brm,Bullivant:2020xhy,Bullivant:2021pkd}.

The above construction has extensions to finite higher group symmetries $\cG$ and provides a launch pad to study a large class of categorical symmetries in $D >2$ obtained by gauging finite higher subgroups $\cH \subset \cG$ with topological phases. This typically leads to non-invertible symmetries whose structure is nevertheless determined by group-theoretical data. The resulting group-theoretical higher fusion categories were studied in~\cite{Bartsch:2022ytj}. In particular, they all correspond to gapped boundary conditions of the same $(D+1)$-dimensional Dijkgraaf-Witten theory associated to $\mathcal{G}$ and $\alpha$. Thus, while the associated higher tube categories and algebras may be vastly different, they share the same higher tube representations and Drinfeld center. A number of Ising-like non-invertible examples of this type are studied in the main text.

\subsection{Outline}

The paper is organised by increasing spacetime and extended operator dimension, as summarised in table~\ref{tab:outline-table}.

\vspace{6pt}
\begin{table}[h]
\begin{center}
\renewcommand{\arraystretch}{1.8}
	\setlength{\tabcolsep}{10pt}

\begin{tabular}{ c | c | c | c | c } 
Section & $\;\; D \;\;$ & \shortstack{Twisted \\[2pt] sector action} & \shortstack{Tube \\[2pt] category} & \shortstack{Tube \\[2pt] representations}  \\ 

\hline
 
\ref{sec:2d} & 2 & \begin{minipage}{0.9in} \centering \vspace{6pt} \includegraphics[height=1.8cm]{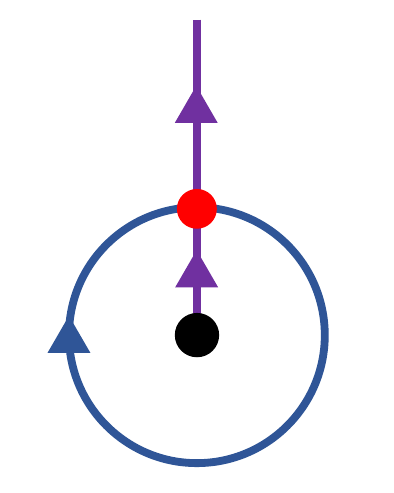} \vspace{6pt} \end{minipage} & $\mathsf{TC}$ & \begin{minipage}{1in} \centering \shortstack{$[\mathsf{TC},\mathsf{Vec}]$ \\[3pt] $= \mathcal{Z}(\mathsf{C})$} \end{minipage}\\ 

\ref{sec:3d-ops} & 3 & \begin{minipage}{0.9in} \centering \vspace{6pt} \includegraphics[height=1.8cm]{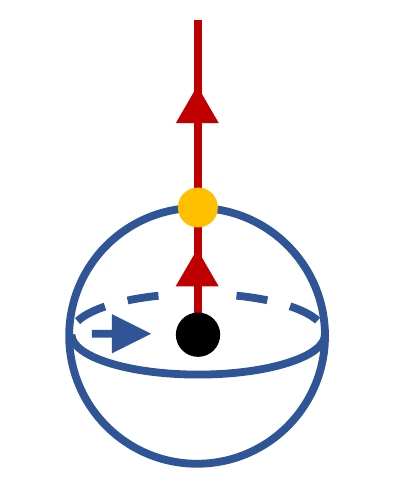} \vspace{6pt} \end{minipage}  & $\mathsf{T}_{S^2}(\mathsf{C})$ & \begin{minipage}{1in} \centering \shortstack{$[\mathsf{T}_{S^2}(\mathsf{C}),\mathsf{Vec}]$ \\[3pt] $= \, \Omega\mathcal{Z}(\mathsf{C})$} \end{minipage} \\ 

\ref{sec:3d-lines} & 3 & \begin{minipage}{0.9in} \centering \vspace{6pt} \includegraphics[height=2.6cm]{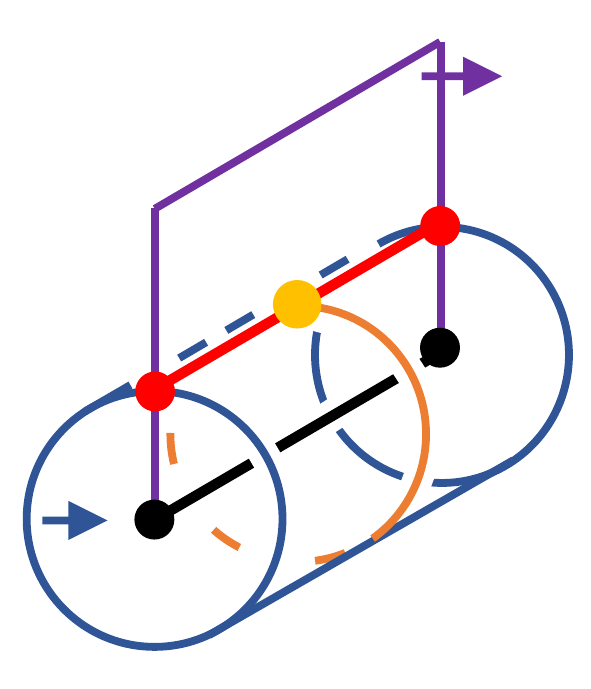}  \vspace{-6pt} \end{minipage} & $\mathsf{T}_{S^1}(\mathsf{C})$ & \begin{minipage}{1in} \centering \shortstack{$[\mathsf{T}_{S^1}(\mathsf{C}),\mathsf{2Vec}]$ \\[3pt] $= \, \mathcal{Z}(\mathsf{C})$} \end{minipage} 
\end{tabular}

\end{center}
\vspace{-6pt}
\caption{}
\label{tab:outline-table}
\end{table}

\begin{itemize}
\item In section \ref{sec:2d} we review the action of a fusion category symmetry $\C$ on twisted sector local operators attached to topological symmetry lines in two dimensions, following~\cite{Lin:2022dhv}. We review the tube category $\T\C = \mathsf{T}_{S^1}(\mathsf{C})$ associated to $\mathsf{C}$ and show that its category of representations reproduces the Drinfeld center $\mathcal{Z}(\C)$. As an example, we consider an anomalous finite group symmetry and illustrate how local operators in the twisted sector transform in representations of its twisted Drinfeld double. Furthermore, we study tube representations of Tambara-Yamagami categories that capture the non-invertible symmetries of the two-dimensional critical Ising model.

\item In section \ref{sec:3d-ops} we study the action of a fusion 2-category symmetry $\C$ on 1-twisted sector local operators attached to topological line defect in three dimensions. We introduce the tube category $\T_{S^2}(\C)$ and show that its category of representations reproduces the looping of the Drinfeld center $\Omega \mathcal{Z}(\C)$. As an example, we consider an anomalous finite 2-group symmetry and provide a complete description of the associated tube category and tube representations. Furthermore, we study a non-invertible Ising-like example that is obtained by gauging a finite subgroup. These examples are realised concretely in dynamical gauge theories with gauge algebra $\mathfrak{spin}(4N)$. Finally, we study fusion 2-categories of the form $\C = \text{Mod}(\B)$ where $\B$ is a braided fusion category and formulate tube representations in terms of the M\"uger center of $\B$.

\item In section \ref{sec:3d-lines} we study the action of a fusion 2-category symmetry on 2-twisted sector line operators attached to topological surface defects in three dimensions. We introduce the tube 2-category $\mathsf{T}_{S^1}(\mathsf{C})$ and show that its 2-category of 2-representations reproduces the Drinfeld center $\mathcal{Z}(\C)$. As an example, we consider an anomalous finite 2-group symmetry and illustrate how 2-twisted sector line operators transform in 2-representations of a higher categorical analogue of its twisted Drinfeld double.  Furthermore, we discus compatibility with gauging finite groups and twisted sectors for condensation defects. These examples are again realised concretely in dynamical gauge theories with gauge algebra $\mathfrak{spin}(4N)$. Finally, we study fusion 2-categories of the form $\C = \text{Mod}(\B)$ and formulate their tube 2-representations as braided module categories over $\B$.
\end{itemize}

\emph{Note added: The authors are grateful to Lakshya Bhardwaj and Sakura Sch\"afer-Nameki for coordination on potentially overlapping work.}


\section{Two dimensions}
\label{sec:2d}

We consider a two-dimensional theory with a spherical fusion category symmetry $\C$. We review how twisted sector local operators transform in irreducible representations of the tube category or the tube algebra associated to $\C$ and explain the relation to the sandwich construction via the Drinfeld center~\cite{Lin:2022dhv}. We illustrate these ideas for a finite group symmetry with 't Hooft anomaly and Tambara-Yamagami fusion categories of Ising type.

\subsection{Fusion category symmetry} 

We consider a quantum field theory in two dimensions whose symmetries are described by a spherical fusion category $\mathsf{C}$. We do not provide a complete definition of the latter here but review some salient features and notation.

\begin{figure}[h]
	\centering
	\includegraphics[height=2.1cm]{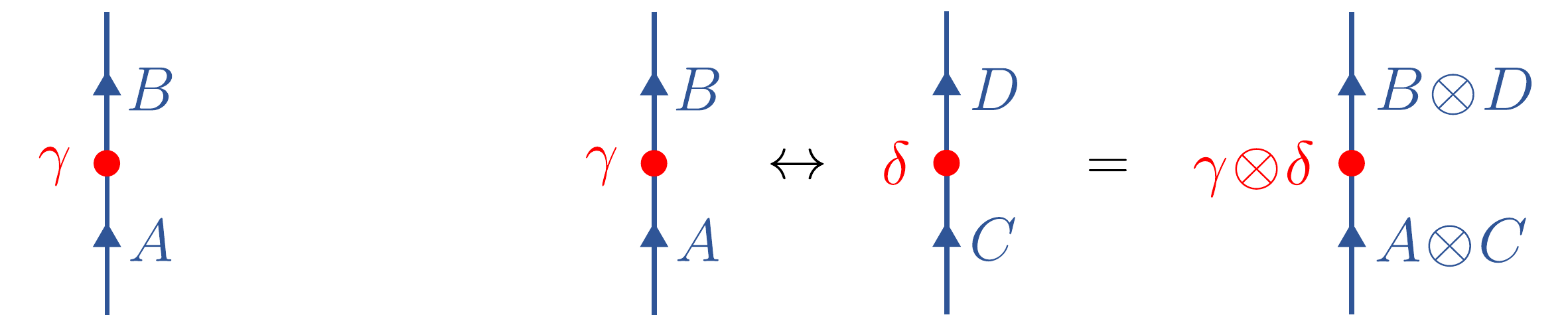}
	\vspace{-5pt}
	\caption{}
	\label{fig:2d-defects+morphisms+fusion}
\end{figure}

Objects $A \in \mathsf{C}$ are topological line defects and morphisms $\gamma: A \to B$ are point-like topological junctions between lines. The tensor structure $\otimes: \; \mathsf{C} \times \mathsf{C} \; \to \; \mathsf{C}$ captures the fusion of topological lines and their junctions. This is illustrated in figure \ref{fig:2d-defects+morphisms+fusion}. 

There is a simple tensor unit $\mathbf{1} \in \mathsf{C}$ and associativity is controlled by a natural isomorphism $\alpha: (. \otimes .) \otimes . \Rightarrow . \otimes (. \otimes .)$ with component morphisms
\vspace{-6pt}
\begin{equation}
\begin{gathered}
\includegraphics[height=1.18cm]{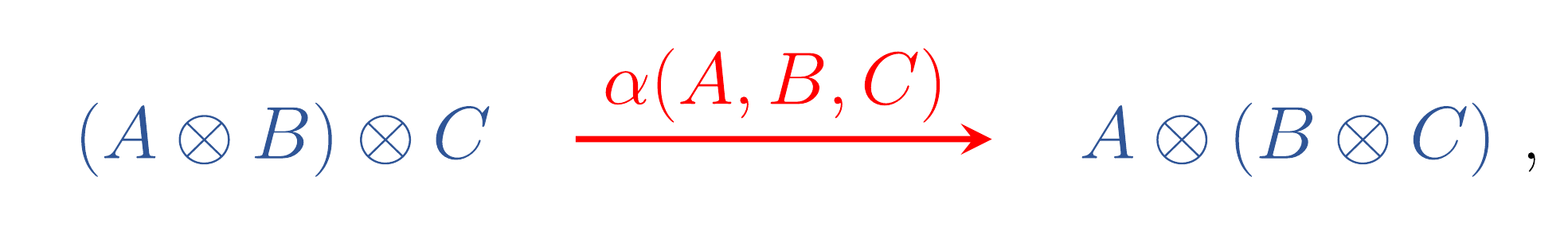}
\end{gathered}
\vspace{-6pt}
\end{equation}
which we call the \textit{associator}.

\begin{figure}[h]
	\centering
	\includegraphics[height=2.35cm]{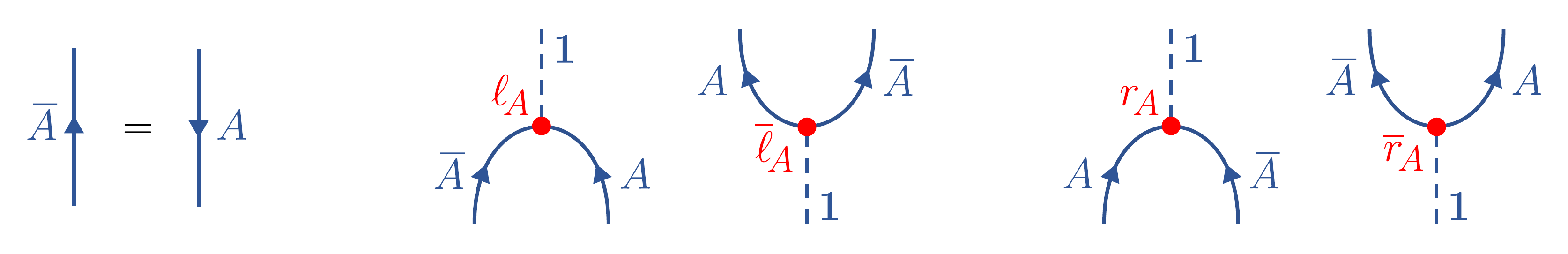}
	\vspace{-5pt}
	\caption{}
	\label{fig:2d-orientation}
\end{figure}

As a fusion category, $\mathsf{C}$ is equipped with additional structures that capture properties of topological symmetry lines and their junctions. For instance, for each object $A \in \mathsf{C}$ there is a dual object $\overline{\!A} \in \mathsf{C}$ that corresponds to the orientation reversal of $A$. There are left and right evaluation morphisms $\ell_A$, $r_A$ and co-evaluation morphisms $\bar\ell_A$, $\bar r_A$ that ensure that the topological line $A$ can be bent to the left and to the right consistently. These structures are illustrated in figure \ref{fig:2d-orientation}.

\begin{figure}[h]
	\centering
	\includegraphics[height=2.35cm]{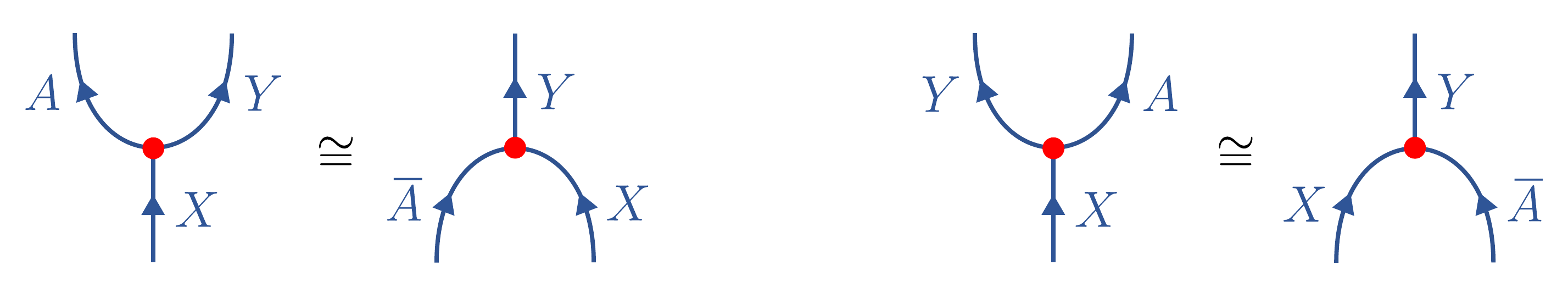}
	\vspace{-5pt}
	\caption{}
	\label{fig:2d-junction-bending}
\end{figure}

In particular, the left and right evaluation and co-evaluation morphisms allow us to canonically identify morphism spaces between tensor products of objects, such as
\begin{equation}\label{eq:coevaluation-isomorphisms}
\begin{aligned}
\text{Hom}_{\mathsf{C}}(X, A \otimes Y) \; &\cong \; \text{Hom}_{\mathsf{C}}(\overline{A} \otimes X,Y) \, , \\
\text{Hom}_{\mathsf{C}}(X, Y \otimes A) \; &\cong \; \text{Hom}_{\mathsf{C}}(X \otimes \overline{A}, Y) \, .
\end{aligned}
\end{equation}
Physically, this means that topological lines can be freely bent around junctions with other lines as illustrated in figure \ref{fig:2d-junction-bending}.

\begin{figure}[h]
	\centering
	\includegraphics[height=3.5cm]{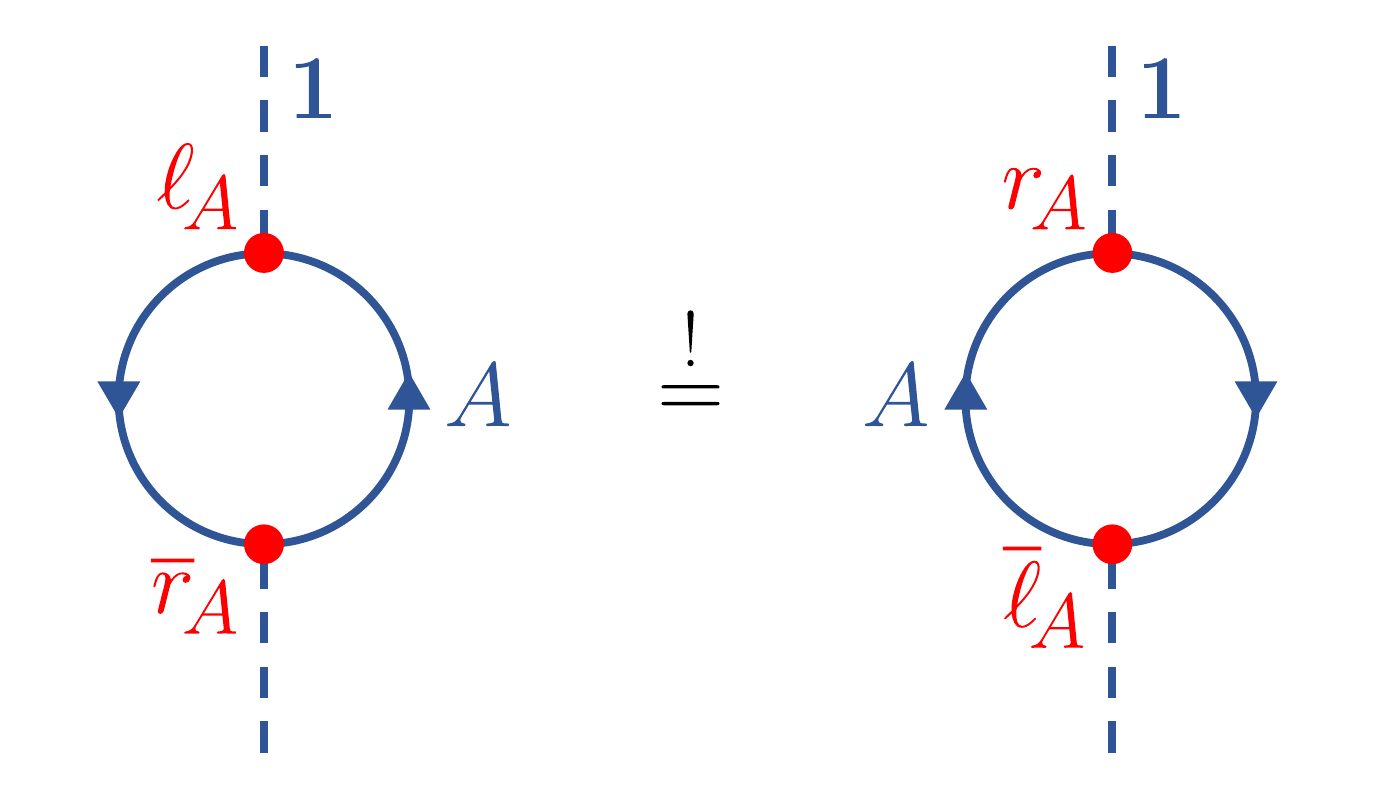}
	\vspace{-5pt}
	\caption{}
	\label{fig:2d-spherical-structure}
\end{figure}

Finally, the spherical structure ensures that there is an unambiguous notion of dimension for each object $A \in \C$, which is obtained by placing the corresponding topological symmetry defect on a circle, as illustrated in figure \ref{fig:2d-spherical-structure}. The corresponding dimension is then given by $\text{dim}(A) = \ell_A \circ \overline{r}_A = r_A \circ \overline{\ell}_A \in \text{End}_{\mathsf{C}}(\mathbf{1}) = \mathbb{C}$.

\subsubsection{Connected components}

An object $S \in \mathsf{C}$ is called \textit{simple} if it cannot be decomposed as a direct sum of non-trivial sub-objects. As a consequence of the semi-simplicity of $\mathsf{C}$, the relation
\begin{equation}
S \, \sim \, T \quad \Leftrightarrow \quad \text{Hom}_{\mathsf{C}}(S,T) \, \neq \, 0
\end{equation}
then defines an equivalence relation on the set of simple objects, whose equivalence classes are the \textit{connected components} of $\mathsf{C}$, i.e.
\begin{equation}
\pi_0(\mathsf{C}) \; := \; \lbrace \text{simple objects} \; S \in \mathsf{C} \rbrace \,/ \sim \, .
\end{equation}
In particular, since there are no non-zero morphisms between non-isomorphic simple objects, $\pi_0(\mathsf{C})$ enumerates isomorphism classes of simple objects in $\mathsf{C}$.

\subsection{Twisted sector operators}

\begin{figure}[h]
	\centering
	\includegraphics[height=2cm]{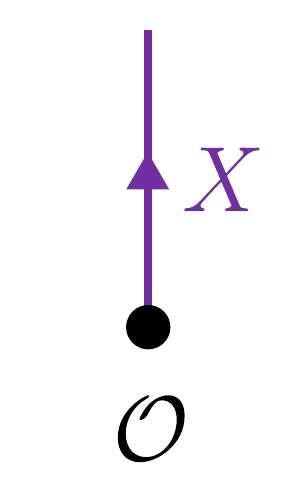}
	\vspace{-5pt}
	\caption{}
	\label{fig:2d-twisted-sectors}
\end{figure}

We now consider twisted sector local operators $\mathcal{O}$ attached to outwardly oriented topological lines $X \in \mathsf{C}$, as illustrated in figure \ref{fig:2d-twisted-sectors}. Local operators of this type are said to be in the \textit{$X$-twisted sector}. They need not be topological themselves and correlations functions may depend on  where they are inserted. 

In the following, we will denote by   
\be
\cF(X) \, \in \, \mathsf{Vec}
\label{eq:2d-F-objects}
\ee
a finite-dimensional vector space spanned by local operators in the $X$-twisted sector. The reason for this notation will be explained momentarily.

\begin{figure}[h]
	\centering
	\includegraphics[height=3.6cm]{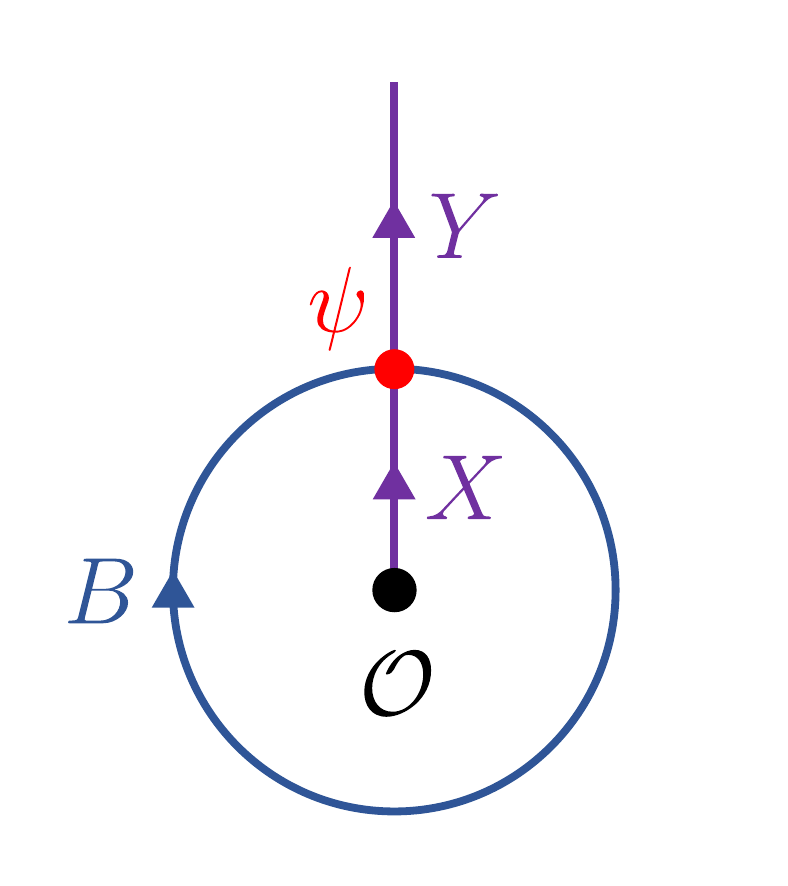}
	\vspace{-5pt}
	\caption{}
	\label{fig:2d-local-action}
\end{figure}

A topological symmetry line $B \in \mathsf{C}$ may act on $X$-twisted sector operators $\mathcal{O} \in \mathcal{F}(X)$ by linking with $S^1$. Due to the topological symmetry line $X$ attached to $\mathcal{O}$, this requires a choice of morphism
\vspace{-8pt}
\begin{equation}\label{eq:2d-tube-morphisms}
\begin{gathered}
\includegraphics[height=1.19cm]{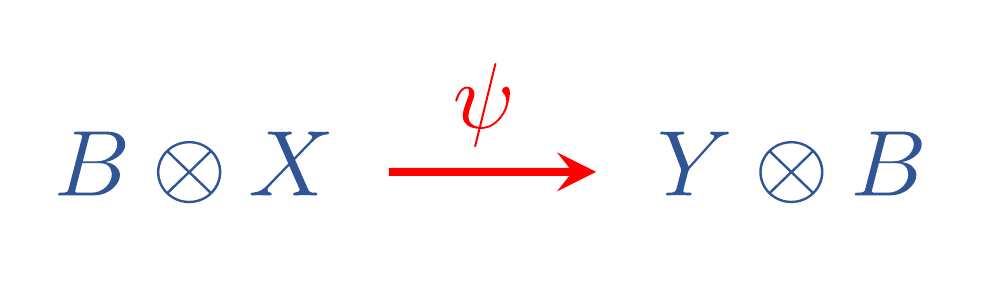}
\end{gathered}
\vspace{-8pt}
\end{equation}
that describes how $B$ intersects $X$ and transforms it into a new line $Y$. This is illustrated in figure~\ref{fig:2d-local-action}. Upon shrinking the $S^1$ down to $\mathcal{O}$, the pair $(B,\psi)$ then induces a linear map
\begin{equation}\label{eq:2d-F-morphisms}
\begin{gathered}
\includegraphics[height=1.19cm]{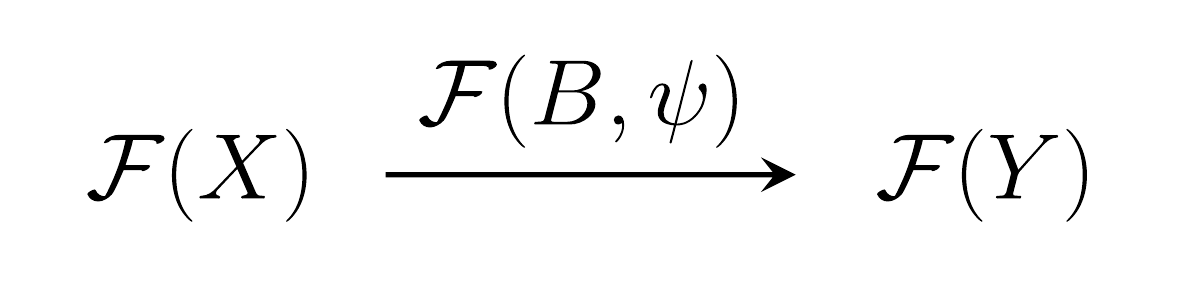}
\end{gathered}
\vspace{-4pt}
\end{equation}
between twisted sectors.

\begin{figure}[h]
	\centering
	\includegraphics[height=5cm]{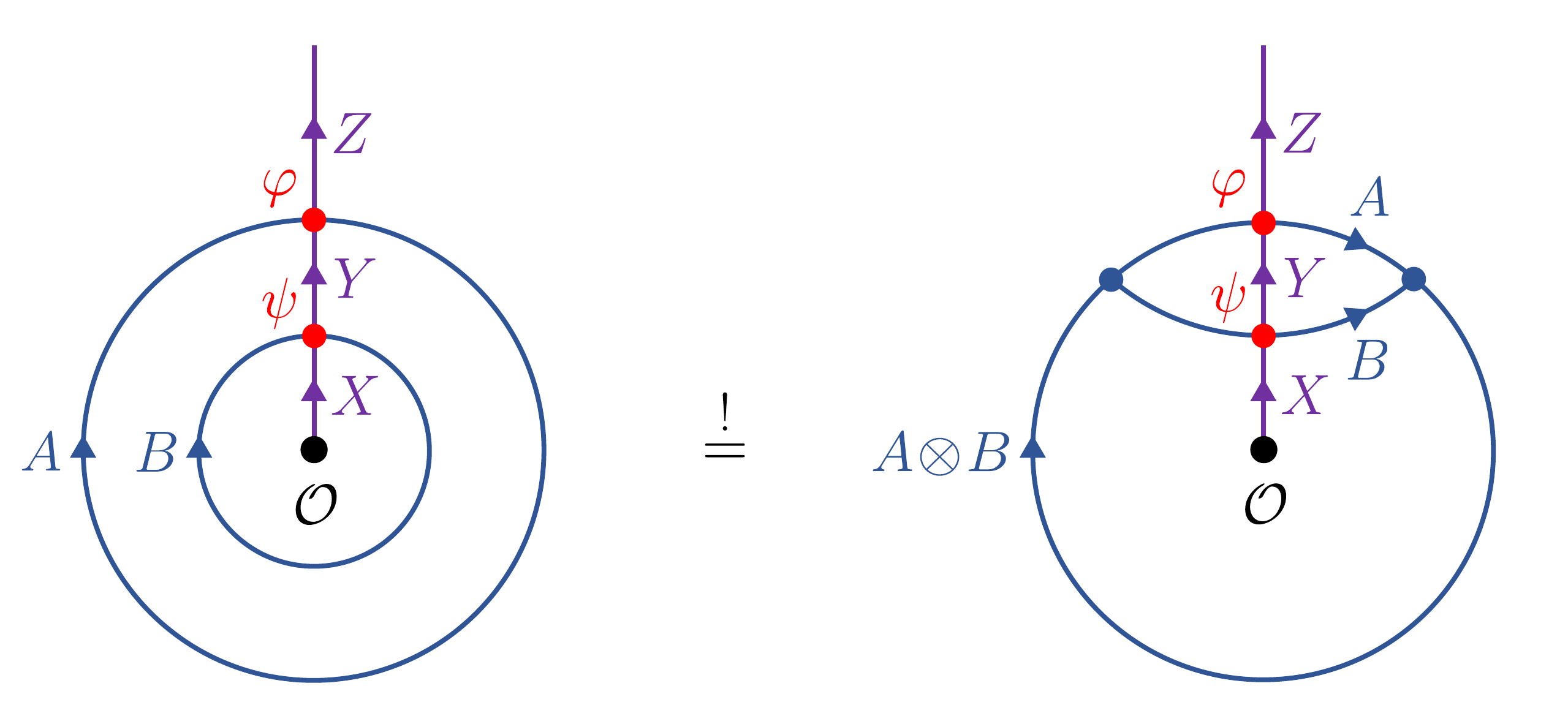}
	\vspace{-5pt}
	\caption{}
	\label{fig:2d-composition}
\end{figure}

These linear maps must be compatible with the consecutive action of two topological symmetry lines $A$ and $B$ in the sense that the diagram
\begin{equation}\label{eq:2d-F-composition}
\begin{gathered}
\includegraphics[height=2.85cm]{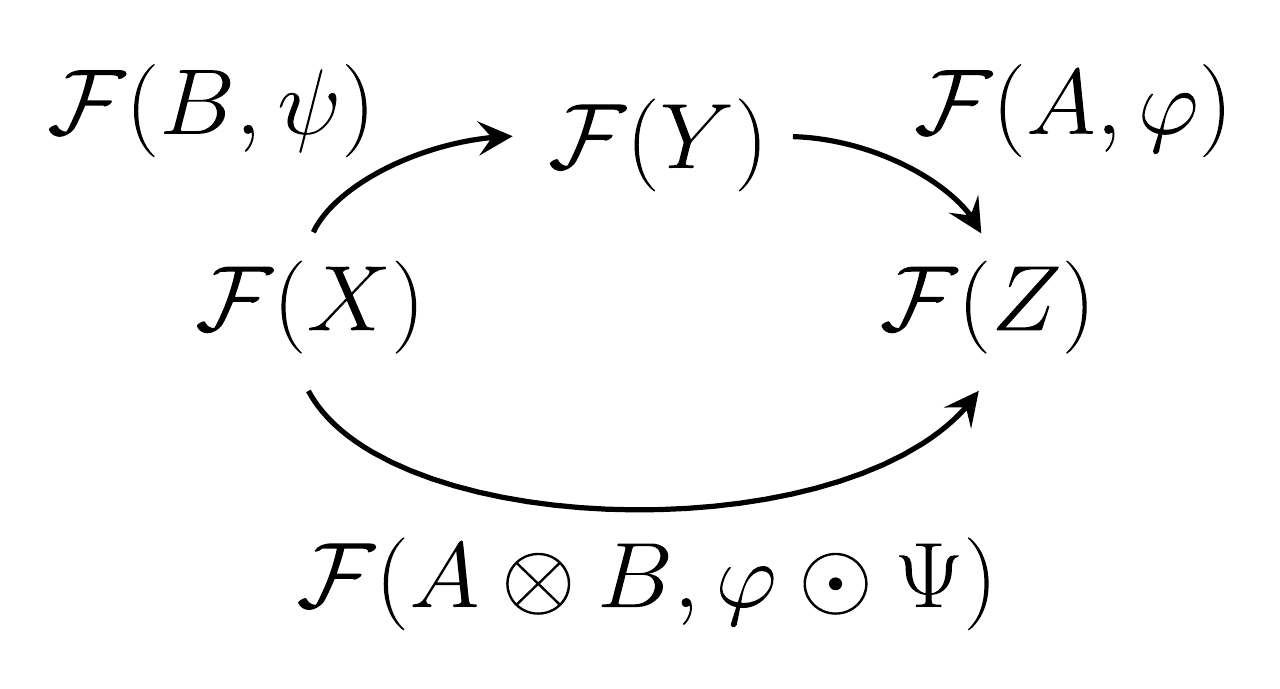}
\end{gathered}
\vspace{-4pt}
\end{equation}
commutes, where the morphism $\varphi \odot \psi$ is defined by the diagram 
\begin{equation}\label{eq:tube-composition}
\begin{gathered}
\includegraphics[height=7.2cm]{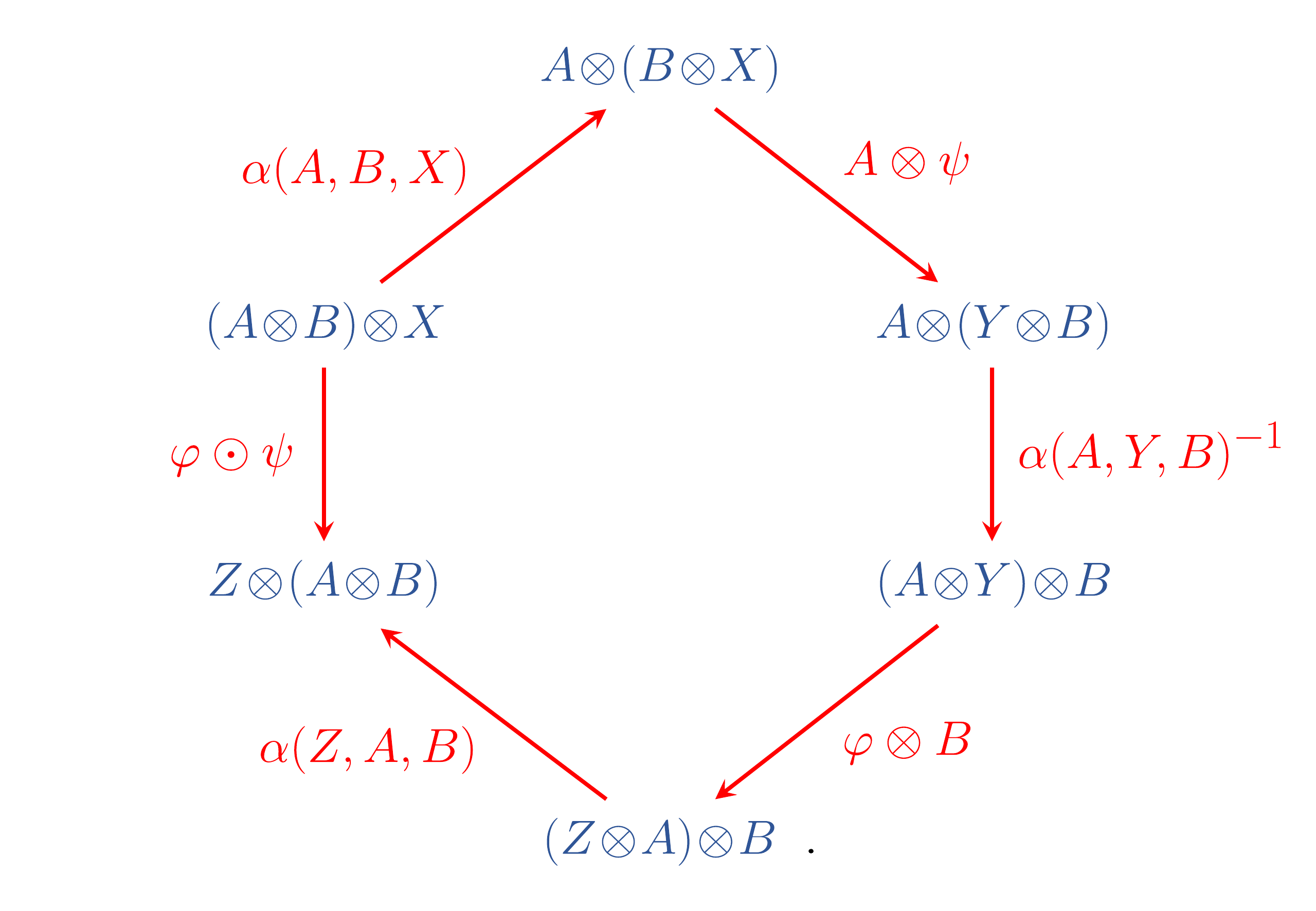}
\end{gathered}
\end{equation}
This condition simply states that acting on $\mathcal{O}$ with $A$ and $B$ consecutively is equivalent to acting on $\mathcal{O}$ with their fusion $A \otimes B$, as illustrated in figure \ref{fig:2d-composition}.

\begin{figure}[h]
	\centering
	\includegraphics[height=3.6cm]{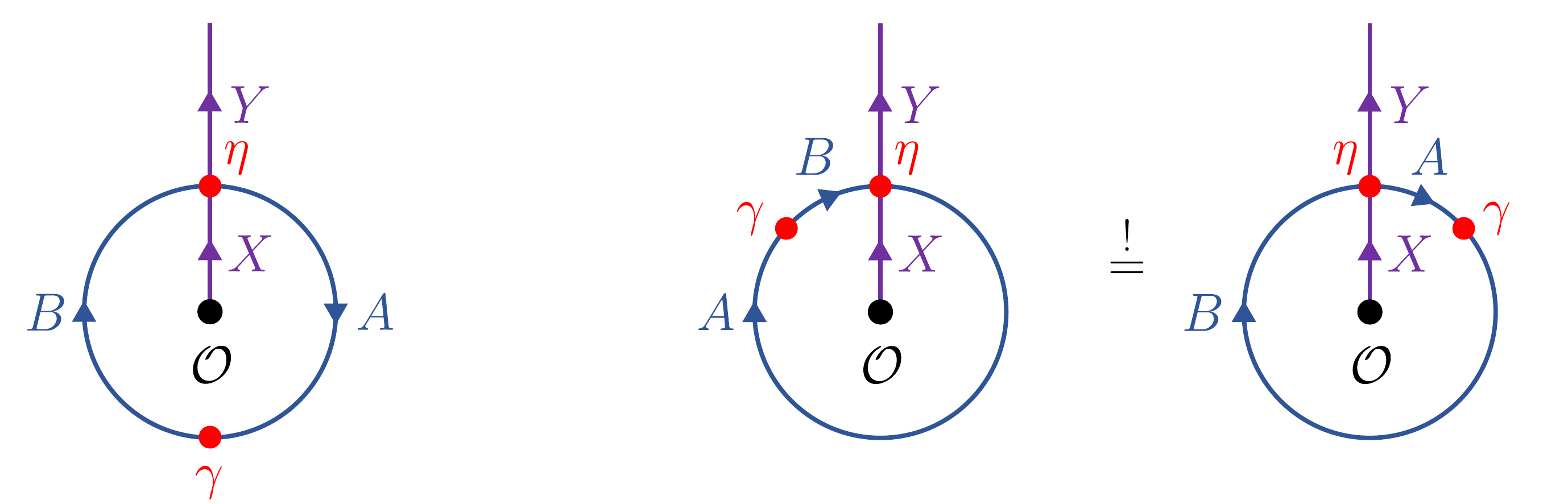}
	\vspace{-5pt}
	\caption{}
	\label{fig:2d-tube-equivalence}
\end{figure}

Lastly, the action of topological lines on twisted sector local operators must be compatible with the possibility to move topological junctions around the linking $S^1$. Concretely, consider a configuration as on the left-hand side of figure \ref{fig:2d-tube-equivalence}, where two line defects $A$ and $B$ that are connected by a morphism $\gamma: A \to B$ link a local operator $\mathcal{O}$ in the $X$-twisted sector via a specified morphism
\vspace{-8pt}
\begin{equation}
\begin{gathered}
\includegraphics[height=1.19cm]{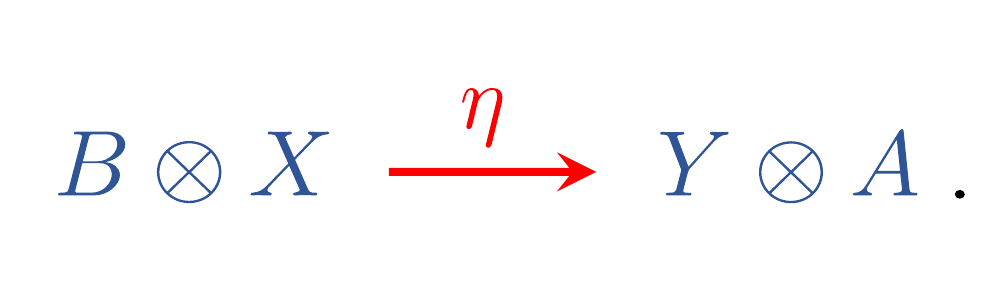}
\end{gathered}
\vspace{-10pt}
\end{equation}
By moving the topological junction $\gamma$ around the 1-sphere towards $\eta$ from the left and from the right, we can either regard this configuration as 
\begin{enumerate}
\item the defect $A$ acting on $\mathcal{O}$ via the intersection morphism $\varphi = \eta \circ (\gamma \otimes X)$,
\item the defect $B$ acting on $\mathcal{O}$ via the intersection morphism $\psi = (Y \otimes \gamma) \circ \eta$.
\end{enumerate}
Since both configurations are physically equivalent, the corresponding linear actions on $\mathcal{O}$ must coincide in the sense that
\begin{equation}\label{eq:2d-F-invariance}
\mathcal{F}(A,\varphi) \; \stackrel{!}{=} \; \mathcal{F}(B,\psi) \, .
\end{equation}
This is illustrated on the right-hand side of figure \ref{fig:2d-tube-equivalence}.

For fixed topological lines $X$ and $Y$, the above relations generate an equivalence relation $\sim$ on the vector space
\begin{equation}
\bigoplus_{B \, \in \, \mathsf{C}} \, \text{Hom}_{\mathsf{C}}(B \otimes X, Y \otimes B)
\end{equation}
capturing configurations that act identically on twisted sector local operators. We will denote the equivalence class of a pair $(B,\psi)$ under $\sim$ by $[B,\psi]$ in what follows.

\subsection{Tube representations}

The above action of the symmetry category $\mathsf{C}$ on twisted sector local operators may be formulated in terms of the representation theory of the \textit{tube category} or \emph{tube algebra} of $\C$. This is a vast generalisation of the fact the local operators transform in representations of a finite symmetry group.

\subsubsection{Tube category}

We first introduce the \textit{tube category} $\T\C = \T_{S^1}\C$\footnote{In two dimensions, there is only one tube category and so we drop the subscript $S^1$}. This is a finite semi-simple category whose objects are twisted sectors $X \in \C$ and morphisms are actions of topological symmetry defects on twisted sector operators via linking with $S^1$. This has appeared in many guises in the mathematical literature, such as for instance~\cite{kirillov2011stringnet,hardiman2020graphical,hardiman_king_2020,Mousaaid_2021,Kirillov_Jr__2022}.

It has the following explicit description:
\begin{itemize}
\item Its objects are objects of $\mathsf{C}$, i.e. $\text{Ob}(\T\C) = \text{Ob}(\mathsf{C})$.

\item Its morphism spaces between objects $X,Y \in \T\C$ are given by
\begin{equation}
\text{Hom}_{\mathsf{TC}}(X,Y) \; = \; \bigg( \!\bigoplus_{B \, \in \, \mathsf{C}} \, \text{Hom}_{\mathsf{C}}(B \otimes X, Y \otimes B) \bigg) \, \bigg/ \, \sim
\end{equation}
with the equivalence relation $\sim$ as above. More concretely, morphisms are linear combinations of equivalence classes of pairs 
\vspace{-4pt}
\begin{equation}
\begin{gathered}
\includegraphics[height=1.19cm]{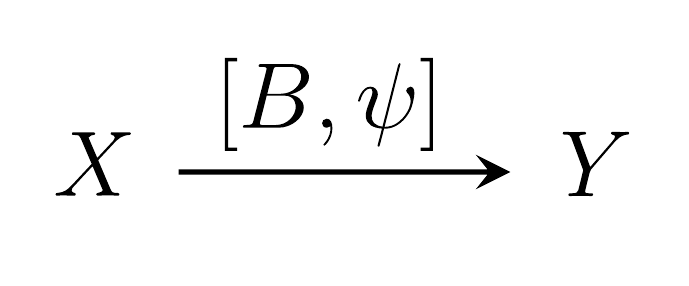}
\end{gathered}
\vspace{-8pt}
\end{equation}
consisting of an object $B \in \mathsf{C}$ and a morphism
\begin{equation}
\psi \; \in \; \text{Hom}_{\mathsf{C}}(B \otimes X, Y \otimes B) \, .
\end{equation}

\item The composition of two morphisms 
\vspace{-4pt}
\begin{equation}
\begin{gathered}
\includegraphics[height=1.19cm]{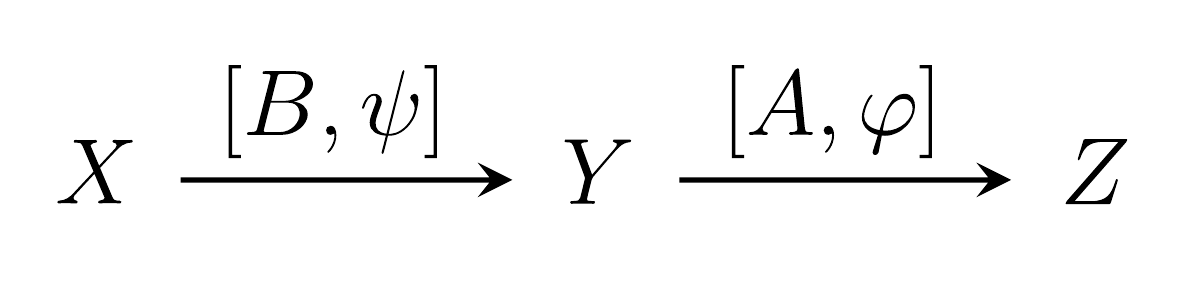}
\end{gathered}
\vspace{-8pt}
\end{equation}
is defined by the composition law
\begin{equation}\label{eq:2d-tube-composition}
[A,\varphi] \, \circ \, [B,\psi] \; := \; [A \otimes B, \, \varphi \odot \psi]
\end{equation}
with the morphism $\varphi \odot \psi$ defined as in (\ref{eq:tube-composition}).
\end{itemize}
There is an inclusion functor $\C \hookrightarrow \T\C$ whose image corresponds to morphisms with $B = \mathbf{1}$.  The tube category $\T\C$ is obtained by horizontal compactification on $S^1$: it has the same objects as $\C$ but additional morphisms arising from intersections with horizontal topological defects wrapping $S^1$.

The above definition of morphism spaces in $\mathsf{TC}$ permits a clean physical interpretation in terms of symmetry defects acting on twisted sectors, but may be rather abstract in view of concrete computations. Fortunately, there exists a more concrete but less canonical description of morphism spaces in $\mathsf{TC}$. Concretely, there is an isomorphism
\begin{equation}\label{eq:2d-tube-hom-isomorphism}
\begin{aligned}
\text{Hom}_{\T\C}(X,Y) \;\; \cong \bigoplus_{[S] \, \in \, \pi_0(\C)} \; \text{Hom}_{\C}(S \otimes X, Y \otimes S) \, ,
\end{aligned}
\end{equation}
where the sum ranges over a set of representatives $S$ of isomorphism classes $[S] \in \pi_0(\mathsf{C})$ of simple objects in $\C$. For an explicit proof of this isomorphism we refer the reader to~\cite{bhowmick2018tube}; a schematic proof is presented in appendix \ref{app:1-condensations}. From a physical perspective, this captures the fact the action of a generic symmetry defect is completely determined by the action of its simple constituents.

With the definition of $\mathsf{TC}$ in hand, the collection of vector spaces (\ref{eq:2d-F-objects}) and linear maps (\ref{eq:2d-F-morphisms}) together with the compatibility conditions (\ref{eq:2d-F-composition}) and (\ref{eq:2d-F-invariance}) can now be summarised conveniently as a functor
\begin{equation}
\mathcal{F}: \; \T\C \, \to \, \mathsf{Vec} \, ,
\end{equation}
which can be regarded as a linear representation of $\T\C$. This explains our choice of notation. In summary, twisted sector local operators transform in linear representations of the tube category $\mathsf{TC}$. 

We denote the category of linear representations by 
\be
[\T\C,\mathsf{Vec}] \, .
\ee
Although we do not discuss it here, this is the idempotent completion or Karoubi completion of $\T\C$ and has the structure of a non-degenerate braided fusion category. The braided fusion structure captures the action of the symmetry category on products of twisted sector local operators supported at separated spacetime points, whose configuration space in two dimensions is homotopic to $S^1$.

\subsubsection{Tube algebra}

There is an equivalent formulation of the above representation theory in terms of the \emph{tube algebra} $\cA(\C) = \cA_{S^1}(\mathsf{C})$\footnote{Again, we drop the subscript $S^1$ in two dimensions.}. This is the finite semi-simple associative algebra
\begin{equation}
\mathcal{A}(\mathsf{C}) \; := \; \text{End}_{\T\C}\Big( \bigoplus_{[S]  \, \in \, \pi_0(\C)} \, S \Big) \, ,
\end{equation} 
where $S$ again runs over a complete set of representatives of isomorphism classes of simple objects in $\mathsf{C}$ and the algebra product is given by composition of endomorphisms in $\T\C$. As a finite semi-simple associative algebra, it has a decomposition as a sum of simple algebras indexed by primitive central idempotents.  

Any representation $\cF : \T\C \to \vect$ of the tube category then determines a representation of the tube algebra on the vector space 
\be
V \; := \; \mathcal{F}\Big(\bigoplus_{[S]} \, S \Big)
\ee 
by sending algebra elements
\begin{equation}
a \in \mathcal{A}(\C) \;\; \mapsto \;\; \mathcal{F}(a) \in \text{End}(V) \, .
\end{equation}
The irreducible representations arise from simple objects in $[\T\C,\vect]$ and correspond to primitive central idempotents in $\cA(\C)$. This sets up an equivalence of non-degenerate braided fusion categories
\begin{equation}\label{eq:2d-morita-equivalent-tube-reps}
[ \T \C , \vect ]  \; \cong \; \rep(\cA(\C)) \, .
\end{equation}
From either perspective, this provides a complete description of the representation theory of a fusion category symmetry $\C$ on twisted sector local operators in dimension $D = 2$. We will uniformly speak of representations of the tube category and the tube algebra as \textit{tube representations} in what follows.

\subsection{Sandwich construction}

We now rephrase the construction of tube representations in the context of the sandwich construction of categorical symmetries~\cite{Freed:2012bs,Gaiotto:2020iye,Ji:2019jhk,Kong:2020cie,Freed:2022qnc,Freed:2022iao}. This is realised via the braided equivalence between category of tube representations and the Drinfeld center 
discovered in~\cite{Evans1995ONOT,Izumi:2000aa,MUGER2003159}. We now provide a quick summary of this equivalence.

An object of the Drinfeld center $\cZ(\C)$ is a pair $z = (U,\tau)$ consisting of 
\begin{enumerate}
\item an object $U \in \mathsf{C}$,
\item a half-braiding $(.) \otimes U \xRightarrow{\, \tau \,} U \otimes (.)$ with component morphisms
\vspace{-8pt}
\begin{equation}\label{eq:2d-half-braiding}
\begin{gathered}
\includegraphics[height=1.1cm]{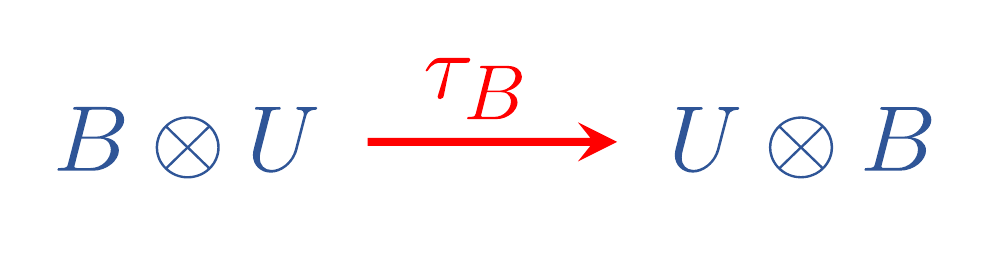}
\end{gathered}
\vspace{-10pt}
\end{equation}
for each object $B \in \mathsf{C}$. 
\end{enumerate}
Note that there is a forgetful functor $F : \mathcal{Z}(\C) \to \C$ that discards the information of the half-braiding and sends $z \mapsto F(z) = U$.

To each object $z \in \mathcal{Z}(\mathsf{C})$ of the Drinfeld center, we can associate a tube representation $\mathcal{F}_{z} \in [\T\C,\vect]$ by defining the functor $\mathcal{F}_z$ as follows:
\begin{itemize}
\item To an object $X \in \T\C$ it assigns the vector space
\be
\mathcal{F}_z(X) \; :=  \; \text{Hom}_{\mathsf{C}}(U,X) \, .
\ee 

\item To a morphism $[B,\psi] \in \text{Hom}_{\T\C}(X,Y)$ it assigns the linear map
\begin{equation}
\begin{gathered}
\includegraphics[height=1.08cm]{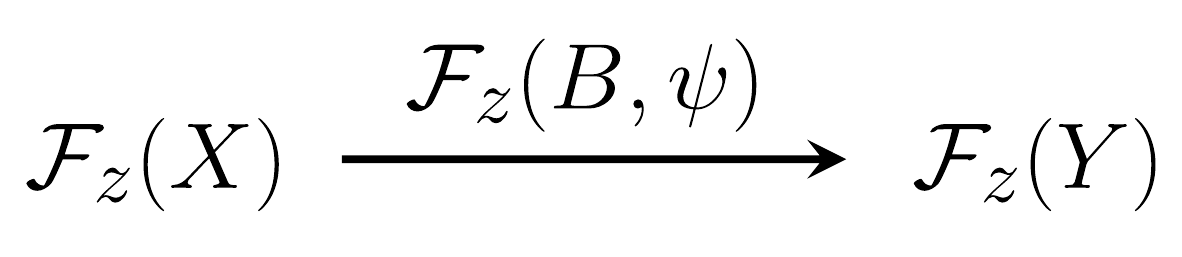}
\end{gathered}
\end{equation}
that sends $\mathcal{O} \in \text{Hom}_{\mathsf{C}}(U,X)$ to $\mathcal{O}' \in \text{Hom}_{\mathsf{C}}(U,Y)$ defined by the diagram\footnote{Here, we use the canonical isomorphisms
\begin{equation}
\begin{aligned}
\text{Hom}_{\mathsf{C}}(\overline{B} \otimes U, U \otimes \overline{B}) \; &\cong \; \text{Hom}_{\mathsf{C}}(U, B \otimes (U \otimes \overline{B})) \\
\text{Hom}_{\mathsf{C}}(B \otimes X, Y \otimes B) \; &\cong \; \text{Hom}_{\mathsf{C}}((B \otimes X) \otimes \overline{B},Y)
\end{aligned}
\end{equation}
induced by the evaluation and co-evaluation morphisms as in (\ref{eq:coevaluation-isomorphisms}). Their use is implicit when applied to the morphisms $\tau_{\Bar{B}}$ and $\psi$ in (\ref{eq:2d-drinfeld-induced-linear-map}).}
\begin{equation}\label{eq:2d-drinfeld-induced-linear-map}
\begin{gathered}
\includegraphics[height=5.1cm]{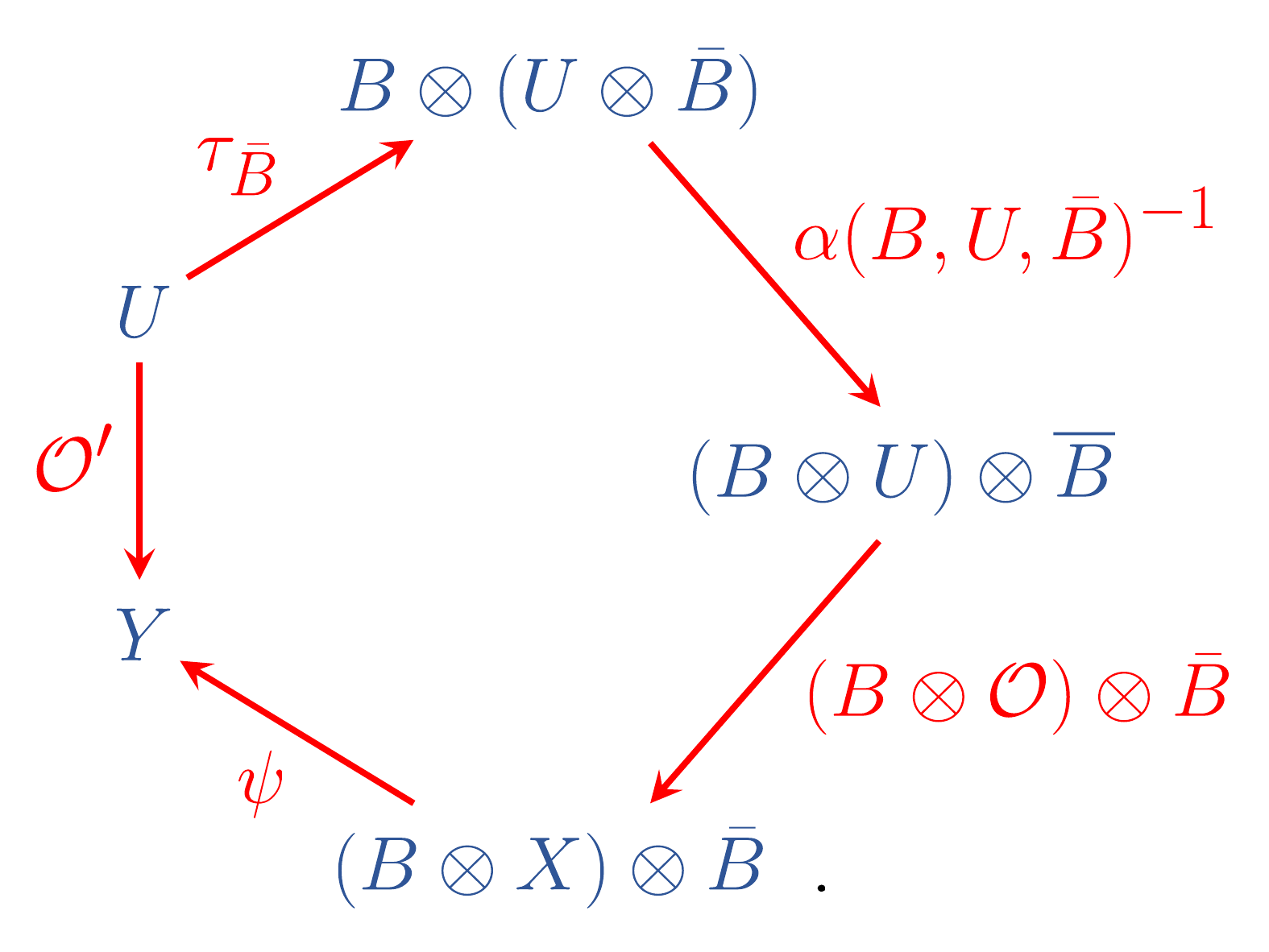}
\end{gathered}
\end{equation}
\end{itemize}
It is known that, up to equivalence, every tube representation takes the form $\mathcal{F} = \mathcal{F}_z$ for some $z \in \mathcal{Z}(\mathsf{C})$. The mapping $z \mapsto \mathcal{F}_z$ then extends to an equivalence
\begin{equation}\label{eq:2d-tube-reps-and-drinfeld}
\mathcal{Z}(\mathsf{C}) \; \cong \; [\mathsf{TC},\mathsf{Vec}] 
\end{equation}
of braided fusion categories.

\begin{figure}[h]
	\centering
	\includegraphics[height=4.7cm]{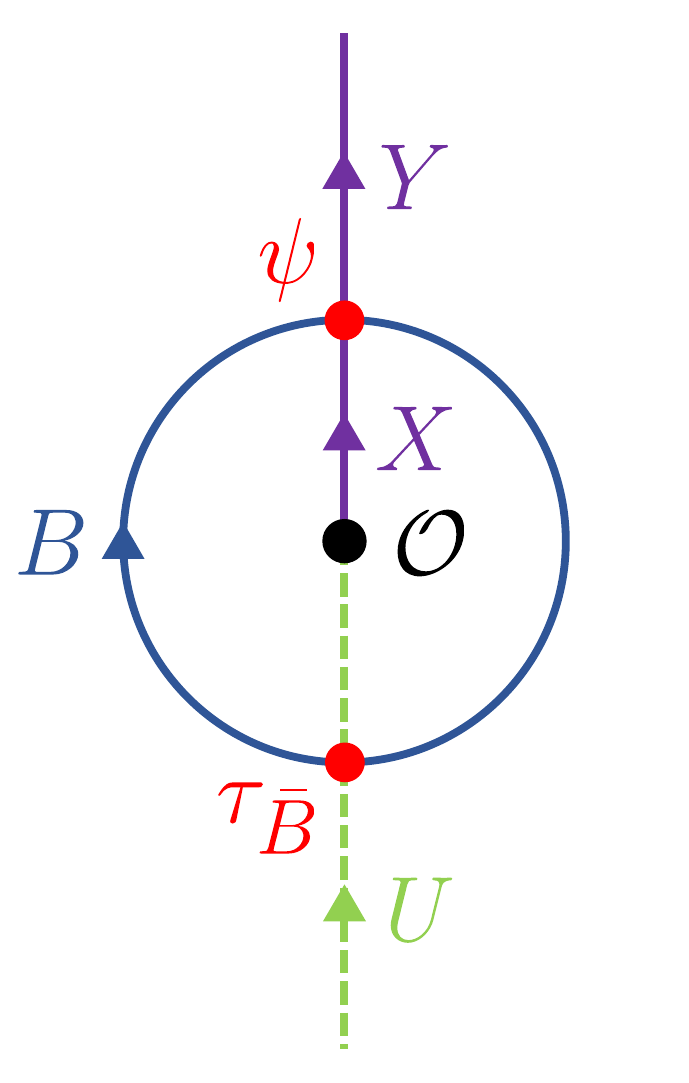}
	\vspace{-5pt}
	\caption{}
	\label{fig:2d-drinfeld-isomorphism}
\end{figure}

Let us now explain the physical picture underpinning this equivalence. We first give an intrinsically two-dimensional perspective that extends our previous work~\cite{Bartsch:2023pzl}. Here, local operators $\mathcal{O}$ in the $X$-twisted sector are viewed as junctions between $X$ and an auxiliary topological line $U$, as illustrated in figure \ref{fig:2d-drinfeld-isomorphism}. This identifies $\cF_z(X) = \text{Hom}_\C(U,X)$. The action of morphisms $[B,\psi]$ is then obtained by surrounding $\mathcal{O}$ with a circular line $B$ that intersects the auxiliary line $U$ via the half-braiding $\tau$. Shrinking the circle down to $\mathcal{O}$ then induces the linear map $\cF_z(B,\psi): \, \cF_z(X) \to \cF_z(Y)$.

\begin{figure}[h]
	\centering
	\includegraphics[height=4.7cm]{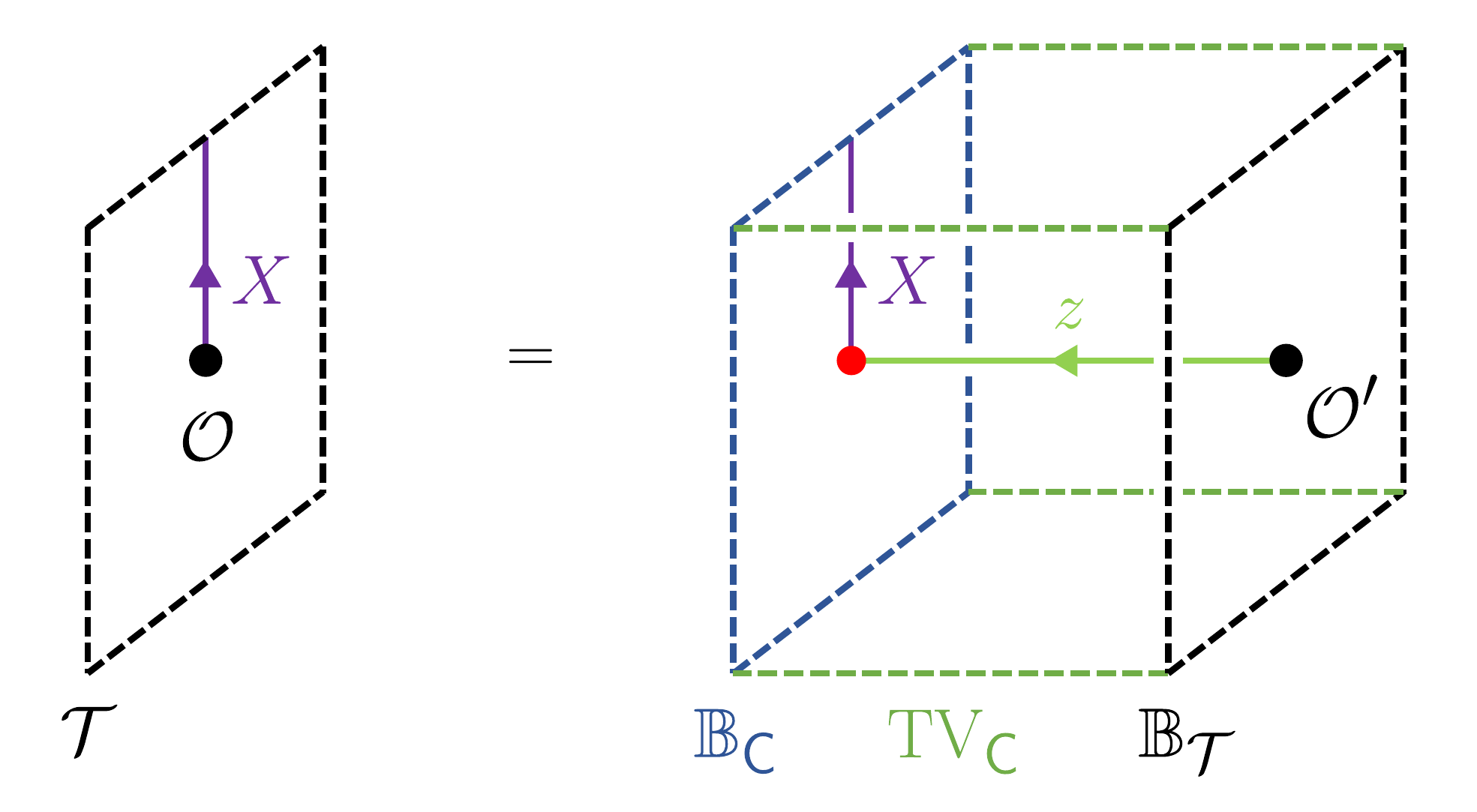}
	\vspace{-5pt}
	\caption{}
	\label{fig:2d-sandwich-construction}
\end{figure}

In the sandwich construction, we view a two-dimensional theory $\cT$ with fusion category symmetry $\C$ as an interval compactification of the associated three-dimensional topological theory $\text{TV}_{\mathsf{C}}$. The Drinfeld centre is identified with the braided fusion category of topological lines
\be
\text{TV}_\C(S^1) \;:= \; \int_{S^1}\C \; = \; \cZ(\C) \, ,
\ee
which is what the three-dimensional topological theory associates to $S^1$. The left boundary condition is a canonical gapped boundary condition $\mathbb{B}_\C$ with bulk-to-boundary functor $F : \mathcal{Z}(\C) \to \C$, while the right boundary condition $\mathbb{B}_\cT$ contains informations about the specific theory $\cT$. This is illustrated in figure~\ref{fig:2d-sandwich-construction}.

A tube representation $\cF_z \in [\T\C,\vect]$ with underlying vector space $\cF_z(X)$ given by $\text{Hom}_\C(U,X)$ now corresponds to local operators on the gapped boundary condition $\mathbb{B}_\C$ that sit at the junction between a bulk topological line $z = (U,\tau)$ and a boundary line $X \in \C$, as described in~\cite{Lin:2022dhv}.

\subsection{Example: group symmetry}

Let us consider a finite symmetry group $G$ with 't Hooft anomaly specified by a normalised 3-cocycle $\alpha \in Z^3(G,U(1))$. This corresponds to the spherical fusion category $\mathsf{C} = \mathsf{Vec}_G^{\alpha}$ of finite-dimensiuonal $G$-graded vector spaces with associator twisted by $\alpha$. The associated tube algebra is the twisted Drinfeld double
\be
\cA(\C) \; = \;{}^{\tau(\alpha)}\bC[G/\!/\! \,G] \, ,
\ee
where $\tau(\alpha)$ is the transgression of the 't Hooft anomaly~\cite{DIJKGRAAF199160,Freed_1994,Willerton_2008}. Its representation category
is one description of the three-dimensional topological theory $\text{TV}_\C$, which is Dijkgraaf-Witten theory constructed from $G$ and $\alpha$.

\subsubsection{Symmetry category}

The symmetry category $\mathsf{C} = \mathsf{Vec}_G^{\alpha}$ is the spherical fusion category of finite-dimensional $G$-graded vector spaces with associator twisted by the 3-cocycle $\alpha \in Z^3(G,U(1))$. It has the following explicit description:
\begin{itemize}
\item The simple objects are one-dimensional vector spaces $\mathbb{C}_g$ with a single graded component $(\mathbb{C}_g)_h \; = \; \delta_{g,h} \cdot \mathbb{C}$. 

\item The morphism spaces between simple objects are given by
\begin{equation}
\text{Hom}_\C(\mathbb{C}_g, \mathbb{C}_h) \; = \; \delta_{g,h} \cdot \mathbb{C} \; .
\end{equation}

\item The fusion of simple objects is given by $\mathbb{C}_g \otimes \mathbb{C}_h = \mathbb{C}_{gh}$
with associator
\vspace{-4pt}
\begin{equation}
\begin{gathered}
\includegraphics[height=1.16cm]{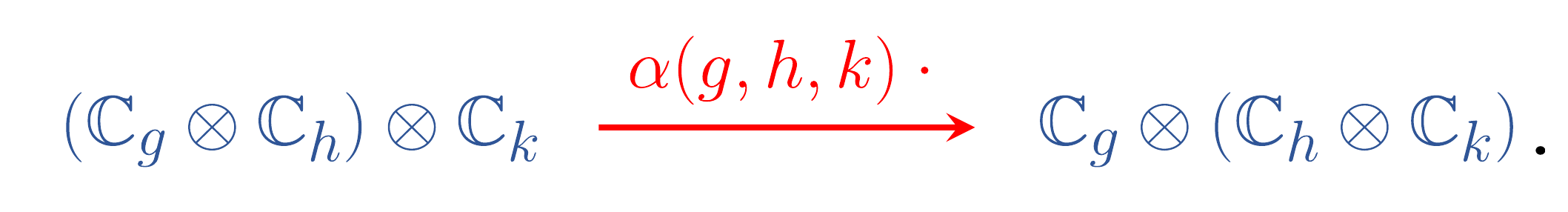}
\end{gathered}
\vspace{-4pt}
\end{equation}
\end{itemize}

\begin{figure}[h]
	\centering
	\includegraphics[height=3.2cm]{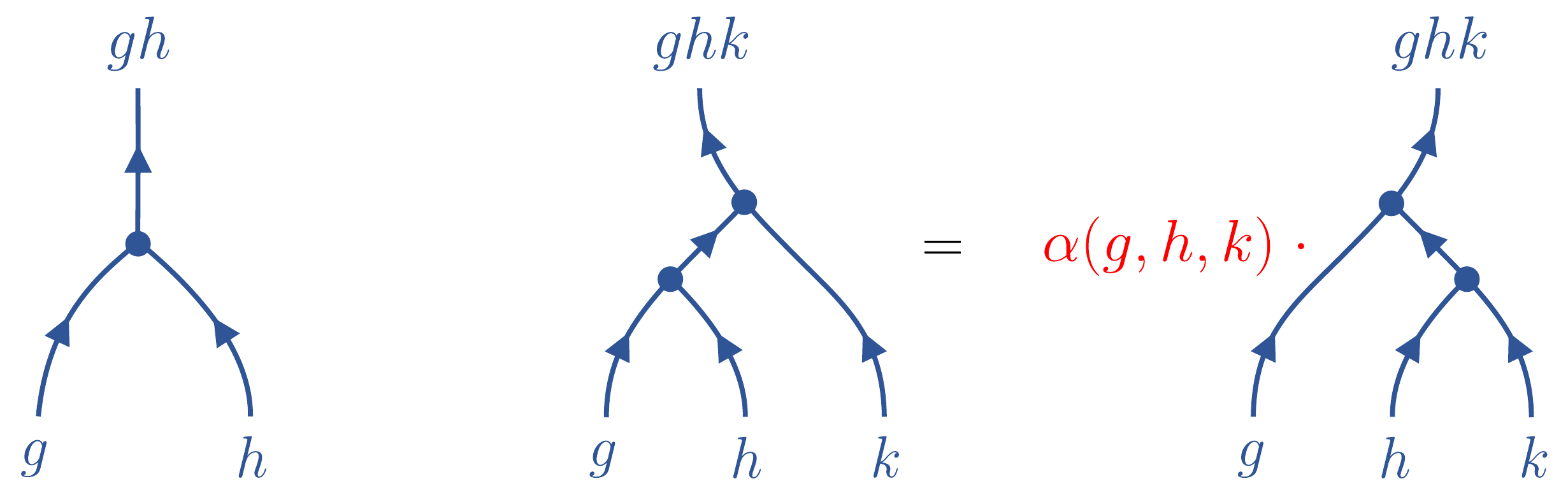}
	\vspace{-5pt}
	\caption{}
	\label{fig:2d-grpex-fusion+associator}
\end{figure}

From a physical perspective, the simple objects are topological symmetry lines labelled by group elements $g \in G$ whose fusion is determined by the group law of $G$ and associative up to multiplicative phases parameterised by $\alpha$. This is illustrated in figure \ref{fig:2d-grpex-fusion+associator}.

\subsubsection{Tube algebra}

The associated tube algebra is generated by equivalence classes of pairs $[B,\psi]$ where $B = \mathbb{C}_g$ and $\psi$ has a single graded component
\vspace{-8pt}
\begin{equation}
\begin{gathered}
\includegraphics[height=1.18cm]{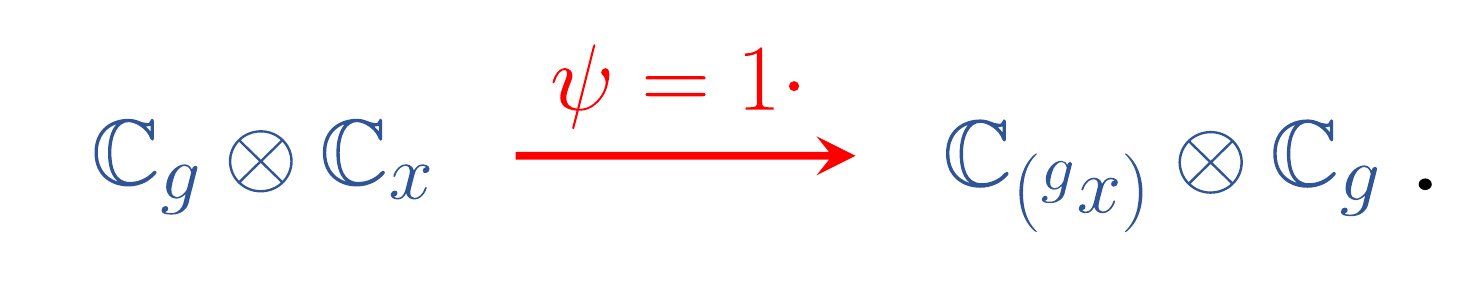}
\end{gathered}
\vspace{-8pt}
\end{equation}
They correspond to intersections of a vertical topological line $x$ with a horizontal topological line $g \in G$, whereupon the former is transformed into ${}^gx = g x g^{-1}$. 

Following standard notation, we denote the above generators by $\braket{ \yleftarrow{\;g\,}\! x }$. Using the definition (\ref{eq:2d-tube-composition}), their tube algebra product can be determined to be
\vspace{-4pt}
\begin{equation} \label{eq:2d-tube-alg-product}
\braket{ \yleftarrow{\;g\,}\! x } \, \circ \, \braket{ \yleftarrow{\;h\,}\! y } \;\; = \;\; \delta_{x, {}^{h}y} \, \cdot \, \tau_y(\alpha)(g,h) \, \cdot \, \braket{ \yleftarrow{\;g \, \cdot \, h\,}\! y } \, ,
\end{equation}
where the multiplicative phase is given by
\begin{equation}\label{eq:2d-transgression}
\tau_y(\alpha)(g,h) \; = \; \frac{\alpha(g,h,y) \cdot \alpha({}^{gh}y,g,h)}{\alpha(g,{}^hy,h)} \, .
\end{equation}
As a consequence of the 3-cocycle condition obeyed by the associator, it satisfies
\begin{equation}
\frac{\tau_y(\alpha)(h,k) \cdot \tau_y(\alpha)(g,h k)}{\tau_y(\alpha)(g h,k) \cdot \tau_{({}^ky)}(\alpha)(g,h)} \; = \; 1 \, ,
\label{eq:groupoid-2cocycle}
\end{equation}
which ensures that the algebra product in (\ref{eq:2d-tube-alg-product}) is associative. Furthermore, equation (\ref{eq:groupoid-2cocycle}) defines a 2-cocycle
\be
\tau(\alpha) \, \in \, Z^2\big(G/\! /_{\!\,} G,U(1)\big)
\ee
on the action groupoid $G/\! /_{\!\,} G$ for the conjugation action of $G$ on itself. This is known as the \textit{transgression} of the 't Hooft anomaly $\alpha \in Z^3(G,U(1))$.

In summary, the tube algebra in this case can be identified with the twisted Drinfeld double or twisted groupoid algebra
\be
\cA(\C) \; = \; {}^{\tau(\alpha)}\bC[G/\! /_{\!\,} G] \, .
\ee

\subsubsection{Tube representations}

A general tube representation $\mathcal{F}$ is a collection of finite-dimensional complex vector spaces $V_x := \mathcal{F}(\mathbb{C}_x)$ together with linear maps 
\begin{equation}
\begin{gathered}
\includegraphics[height=1.1cm]{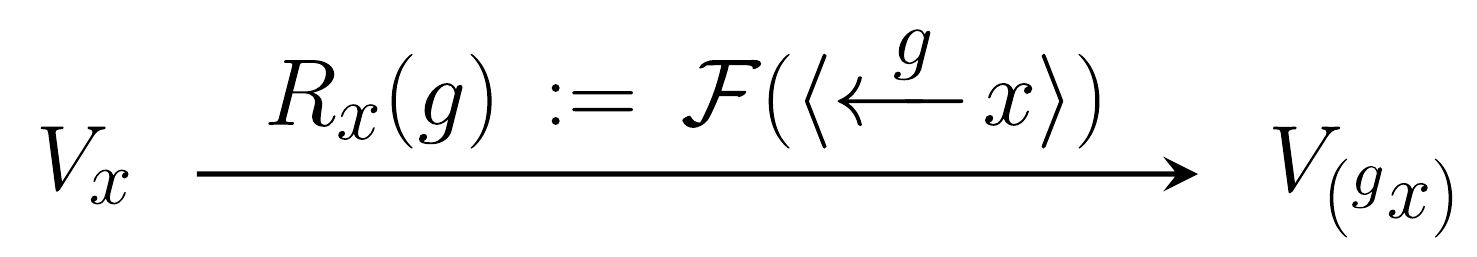}
\end{gathered}
\vspace{-3pt}
\end{equation}
that satisfy the composition rule
\begin{equation}\label{eq:2d-grpex-tube-rep-composition-rule}
R_x(g) \, \circ \, R_y(h) \,\; = \,\; \delta_{x, {}^{h}y} \, \cdot \, \tau_y(\alpha)(g,h) \, \cdot \, R_y(gh) \, .
\end{equation}
This collection of data can be viewed as a vector bundle over $G$ equipped with a projective $G$-action that acts on the base by conjugation. 

From the composition rule (\ref{eq:2d-grpex-tube-rep-composition-rule}), it is clear that $\mathcal{F}$ will decompose as a direct sum of representations supported on conjugacy classes of $G$. Let us therefore fix a conjugacy class $[x] \in G/G$ with representative $x \in G$. If we restrict to group elements $g,h\in C_x(G)$ in the centraliser of $x$ in $G$, the composition rule becomes
\begin{equation}
R_x(g) \, \circ \, R_x(h) \;\; = \;\; \tau_x(\alpha)(g,h) \, \cdot \, R_x(gh) \, .
\end{equation}
Consequently, the map 
\begin{equation}
R_x : \; C_x(G) \, \to \, \text{End}(V_x) 
\end{equation}
defines an ordinary projective representation of $C_x(G)$ with 2-cocycle 
\begin{equation}
\tau_x(\alpha) \, \in \, Z^2(C_x(G),U(1)) \, .
\end{equation}
A tube representation supported on $[x] \in G/G$ is then irreducible if the associated projective representation $R_x$ of $C_x(G)$ is irreducible. Conversely, it is known that, up to equivalence, any irreducible projective representation of $C_x(G)$ determines an irreducible tube representation by induction.

In summary, the irreducible tube representations are determined by
\begin{enumerate}
\item a group element $x \in G$,
\item a projective representation of $C_x(G)$ with 2-cocycle 
\begin{equation}
\tau_x(\alpha)(g,h) \; = \; \frac{\alpha(g,h,x) \cdot \alpha(x,g,h)}{\alpha(g,x,h)} \; .
\end{equation}
\end{enumerate}
Up to equivalence, this data depends only on the conjugacy class $[x] \in G/G$ and the group cohomology class $[\tau_x(\alpha)] \in H^2(C_x(G),U(1))$. Note that tube representations on the trivial conjugacy class $[e] \in G/G$ reproduce the fact that genuine local operators transform in representations of $G$.

The above then reproduces the known equivalence 
\be
\mathsf{Rep}\big( {}^{\tau(\alpha)}\bC[G/\! /_{\!\,} G] \big) \; \cong \; \cZ(\vect_G^\alpha)
\ee
between representations of the twisted Drinfeld double and topological lines in the three-dimensional Dijkgraaf-Witten theory constructed from $G$ and $\alpha$. The latter is the three-dimensional topological theory $\text{TV}_\C$ associated to the symmetry category $\C = \vect_G^\alpha$.

\subsection{Example: Ising symmetry}

We now consider an intrinsically non-invertible example realised in the critical Ising model. General Tambara-Yamagami fusion categories $\C = \mathsf{TY}(A,\chi,s)$ \cite{tambara1998tensor} are specified by
\begin{enumerate}
\item a finite abelian group $A$,
\item a non-degenerate bicharacter $\chi: A \times A \to U(1)$,
\item a square-root $s$ of $|A|^{-1}$.
\end{enumerate}
We focus here on $A= \mathbb{Z}_2 =: \braket{x}$ with the unique non-degenerate bicharacter $\chi(x,x) = -1$ and $s = 1 / \sqrt{2}$. This symmetry is realised in the critical Ising model~\cite{Chang:2018iay,Thorngren:2019iar}.

\subsubsection{Symmetry category}

The symmetry category $\mathsf{C} = \mathsf{TY}(\mathbb{Z}_2,\chi,s)$ has the following explicit description:
\begin{itemize}
\item There are three simple objects $1$, $x$ and $m$ of dimension $1$, $1$ and $\sqrt{2}$, respectively. 

\item The morphism spaces between simple objects $S,S' \in \lbrace 1,x,m \rbrace$ are given by
\begin{equation}
\text{Hom}_{\mathsf{C}}(S,S') \; = \; \delta_{S,S'} \cdot \mathbb{C} \, .
\end{equation}

\item The fusion rules for simple objects are given by
\begin{equation}
x \otimes x \; = \; 1 \, , \qquad x \otimes m \; = \; m \otimes x \; = \; m \, , \qquad m \otimes m \; = \; 1 \oplus x \, .
\end{equation}

\item The non-trivial components of the associator are given by
\begin{equation}
\begin{gathered}
\includegraphics[height=1.26cm]{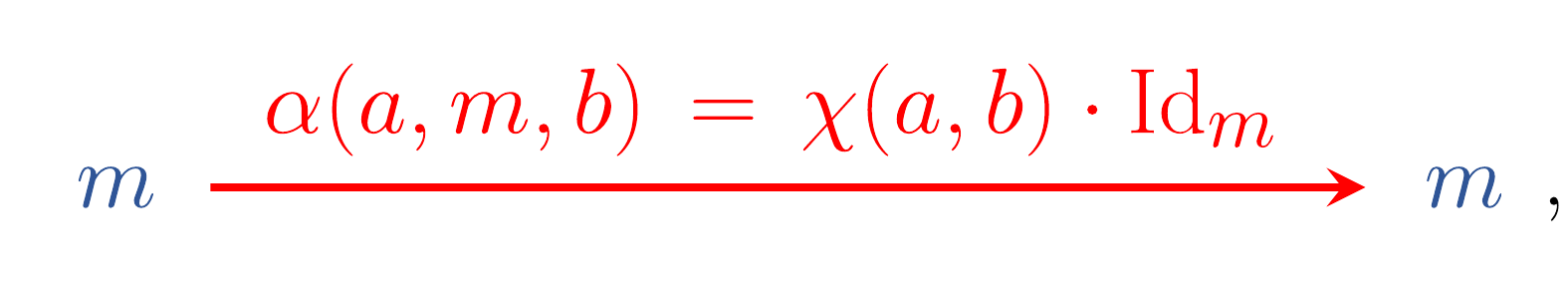} \\[-6pt]
\includegraphics[height=1.65cm]{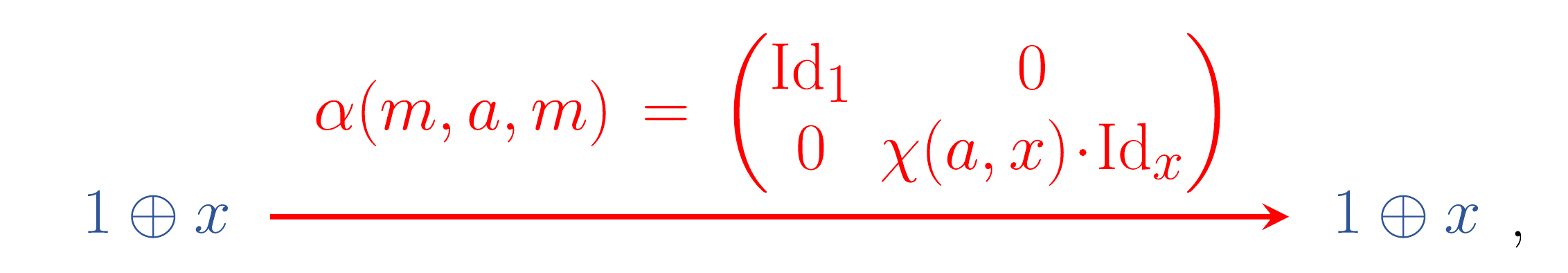} \\
\includegraphics[height=1.65cm]{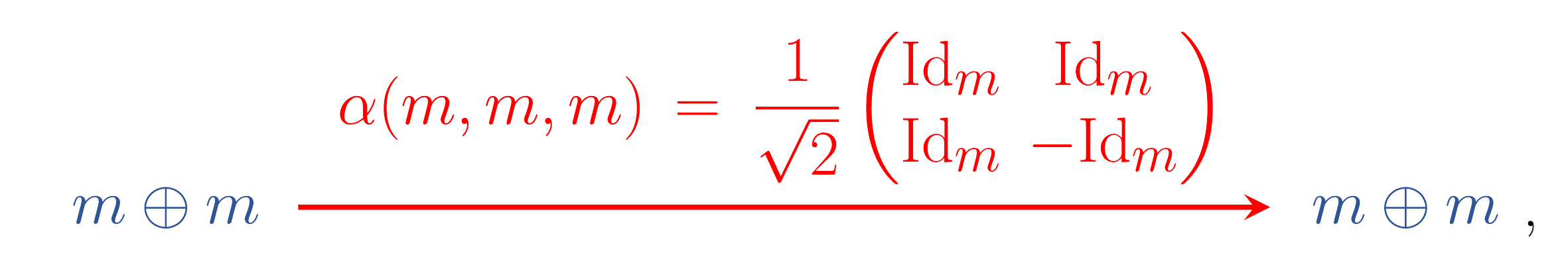}
\end{gathered}
\vspace{-4pt}
\end{equation}
where $a,b \in \mathbb{Z}_2$ denote generic elements.
\end{itemize}
From a physical perspective, the fact that the associator becomes trivial when restricted to simple objects $a,b,c \in \lbrace 1,x \rbrace$ captures the absence of an anomaly for the $\mathbb{Z}_2$ sub-symmetry generated by $x$.

\subsubsection{Tube algebra}

The associated tube algebra is generated by equivalence classes of pairs $[B,\psi]$ corresponding to the following combinations of an object $B$ and an intersection morphism $\psi$:
\begin{itemize}
\item For $a,b \in \mathbb{Z}_2$, we denote by $\genone[a]{b}$ those generators for which $B = a$ and
\vspace{-6pt}
\begin{equation}
\begin{gathered}
\includegraphics[height=1.1cm]{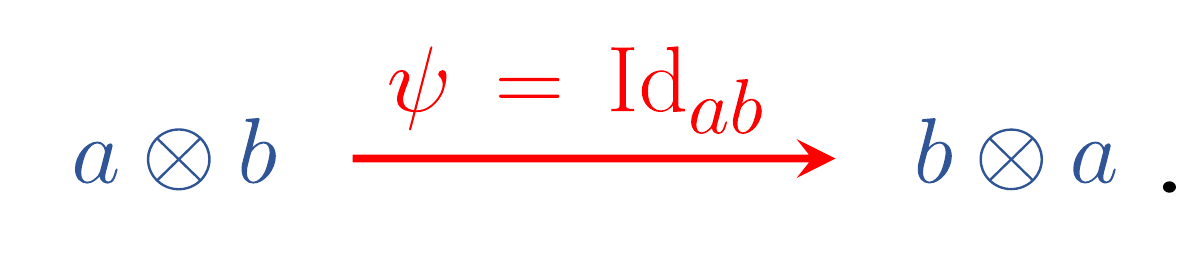}
\end{gathered}
\vspace{-8pt}
\end{equation}

\item For $b,c \in \mathbb{Z}_2$, we denote by $\gentwo[b]{c}$ those generators for which $B = m$ and
\vspace{-6pt}
\begin{equation}
\begin{gathered}
\includegraphics[height=1.1cm]{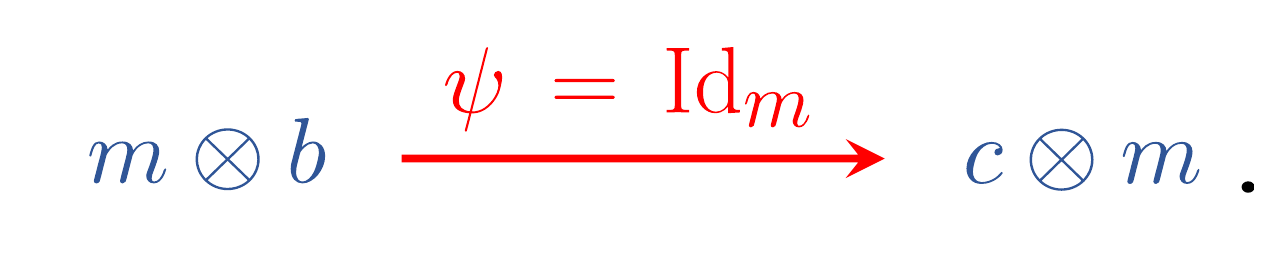}
\end{gathered}
\vspace{-8pt}
\end{equation}

\item For $a \in \mathbb{Z}_2$, we denote by $\genone[a]{m}$ those generators for which $B = a$ and
\vspace{-6pt}
\begin{equation}
\begin{gathered}
\includegraphics[height=1.1cm]{2d-dia-isingex-gen-2.pdf}
\end{gathered}
\vspace{-8pt}
\end{equation}

\item For $c \in \mathbb{Z}_2$, we denote by $\genthree[c]$ those generators for which $B=m$ and
\vspace{-6pt}
\begin{equation}
\begin{gathered}
\includegraphics[height=1.1cm]{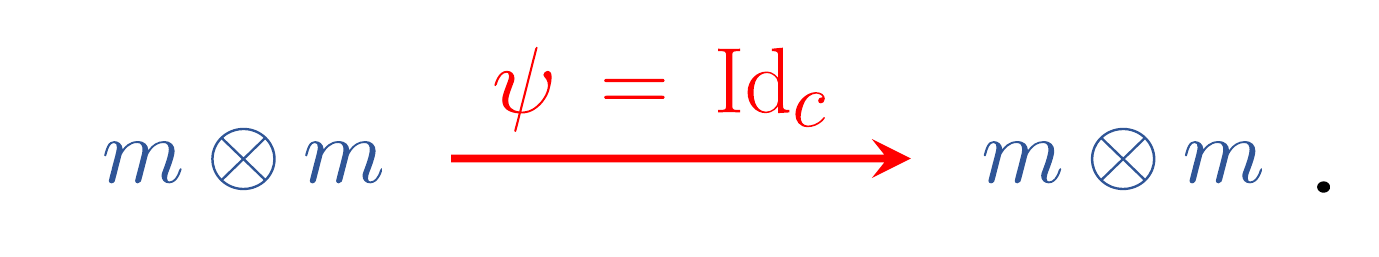}
\end{gathered}
\vspace{-8pt}
\end{equation}
\end{itemize}
The algebra product of these generators is straightforward to compute using~\eqref{eq:tube-composition}. For example, we find
\begin{equation}\label{eq:ising-composition}
\begin{aligned}
\gentwo[1]{1} \, \circ \, \gentwo[1]{1} \;\, &= \;\, \genone[1]{1} \, + \, \genone[x]{1} \, , \\[8pt]
\gentwo[x]{x} \, \circ \, \gentwo[x]{x} \;\, &= \;\, \genone[x]{x} \, - \, \genone[1]{x} \, , \\[4pt]
\genthree[1] \, \circ \, \genthree[1] \;\, &= \;\, \frac{1}{2 \sqrt{2}} \, \Big[ \, \genone[1]{m} + \genone[x]{m} \, \Big] \, , \\[3pt]
\genthree[1] \, \circ \, \genthree[x] \;\, &= \;\, \frac{1}{2 \sqrt{2}} \, \Big[ \, \genone[1]{m} - \genone[x]{m} \, \Big] \, , \\[3pt]
\genthree[x] \, \circ \, \genthree[x] \;\, &= \;\, - \frac{1}{2 \sqrt{2}} \, \Big[ \, \genone[1]{m} + \genone[x]{m} \, \Big] \, , \\[4pt]
\gentwo[1]{x} \, \circ \, \gentwo[x]{1} \;\, &= \;\, \genone[1]{x} \, + \, \genone[x]{x} \, , \\[8pt]
\gentwo[x]{1} \, \circ \, \gentwo[1]{x} \;\, &= \;\, \genone[1]{1} \, - \, \genone[x]{1} \, ,
\end{aligned}
\vspace{4pt}
\end{equation}
which are notably compatible with the fusion rule $m \otimes m = 1 \oplus x$.

\subsubsection{Tube representations}

\begin{figure}[h]
	\centering
	\includegraphics[height=6.4cm]{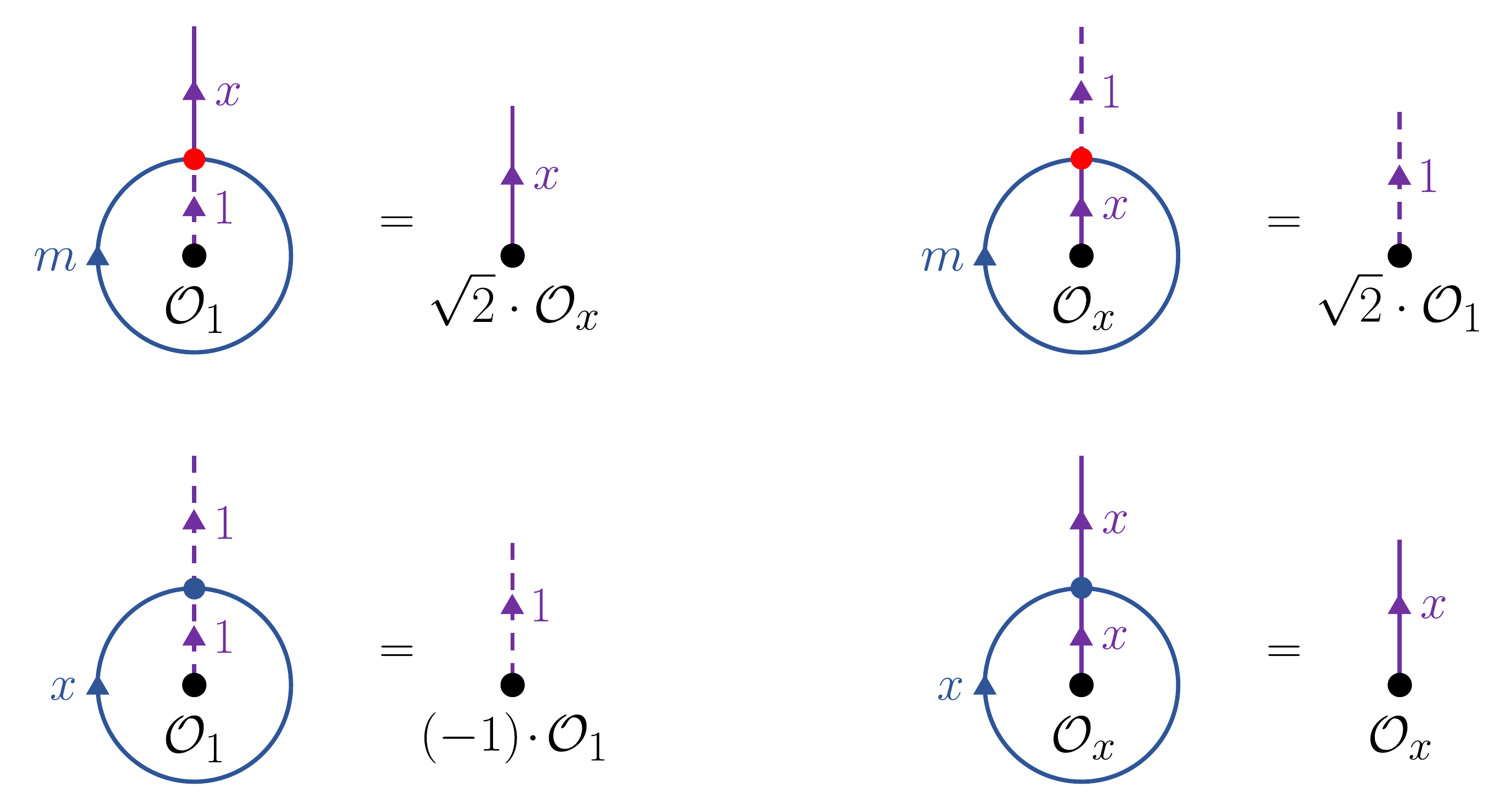}
	\vspace{-5pt}
	\caption{}
	\label{fig:2d-isingex-1}
\end{figure}

Let us now consider irreducible representations of the tube algebra. As the Drinfeld center of general Tambara-Yamagami fusion categories $\mathsf{TY}(A,\chi,s)$ was computed in~\cite{gelaki2009centers}, it is convenient to take this as the starting point to construct tube representations. The Drinfeld center has
\begin{equation}
\frac{n(n+7)}{2}
\end{equation}
simple objects, with $n = |A|$. When $A = \mathbb{Z}_2$, there are therefore $9$ simple objects, or equivalently $9$ irreducible tube representations, which can be described as follows:
\begin{itemize}

\item There are two one-dimensional irreducible tube representations $\mathcal{F}^{\pm}_1$ with untwisted sector $\mathcal{F}^{\pm}_1(1) = \mathbb{C}$ and non-trivial generator actions
\begin{equation}
\mathcal{F}^{\pm}_1\big( \genone[x]{1} \big) \; = \; 1 \, \qquad \text{and} \qquad \mathcal{F}^{\pm}_1\big( \gentwo[1]{1} \big) \; = \; \pm \sqrt{2} \, .
\end{equation}
These representations correspond to genuine local operators that are not charged under the $\bZ_2$ sub-symmetry generated by $x$ and that remain genuine local operators after the action of $m$. 

\item There are two one-dimensional irreducible tube representations $\mathcal{F}^{\pm}_x$ with twisted sector $\mathcal{F}^{\pm}_x(x) = \mathbb{C}$ and non-trivial generator actions
\begin{equation}
\mathcal{F}^{\pm}_x\big( \genone[x]{x} \big) \; = \; -1 \, \qquad \text{and} \qquad \mathcal{F}^{\pm}_x\big( \gentwo[x]{x} \big) \; = \; (\pm i) \cdot \sqrt{2} \, .
\end{equation}
These representations are now associated to $x$-twisted operators that are charged under the $\bZ_2$ sub-symmetry generated by $x$ and remain twisted sector operators after the action of $m$.

\item There are four one-dimensional irreducible tube representations $\mathcal{F}^{\pm}_m$ and $\widetilde{\mathcal{F}}^{\pm}_m \equiv \big(\mathcal{F}^{\pm}_m \big)^{\ast}$ with twisted sector $\mathcal{F}^{\pm}_m(m) = \widetilde{\mathcal{F}}^{\pm}_m(m) = \mathbb{C}$ and non-trivial generator actions
\begin{equation}
\begin{aligned}
\mathcal{F}^{\pm}_m\big(\genone[x]{m}\big) \; & = \; i \, , \\[4pt]
\mathcal{F}^{\pm}_m\big(\genthree[1]\big) \; & = \; \pm \, \frac{1}{\sqrt{2}} \, e^{\frac{i\pi}{8}}, \\[4pt]
\mathcal{F}^{\pm}_m\big(\genthree[x]\big) \; & = \; \pm \, \frac{1}{\sqrt{2}} \, e^{-\frac{i3\pi}{8}}  \, .
\end{aligned}
\end{equation}

\item There is one two-dimensional irreducible tube representation $\mathcal{F}$ with twisted sectors $\mathcal{F}(1) = \mathcal{F}(x) = \mathbb{C}$ and non-trivial generator actions
\begin{equation}
\begin{aligned}
&\mathcal{F}\big( \genone[x]{1} \big) \; = \; \begin{pmatrix} -1 & 0 \\ 0 & 0 \end{pmatrix} \, ,  &&\mathcal{F}\big( \genone[x]{x} \big) \; = \; \begin{pmatrix} 0 & 0 \\ 0 & +1 \end{pmatrix} \, ,  \\[6pt]
&\mathcal{F}\big( \gentwo[1]{x} \big) \; = \; \begin{pmatrix} 0 & 0 \\ \sqrt{2} & 0 \end{pmatrix} \, , \qquad  &&\mathcal{F}\big( \gentwo[x]{1} \big) \; = \; \begin{pmatrix} 0 & \sqrt{2} \\ 0 & 0 \end{pmatrix} \, . 
\end{aligned}
\end{equation}
This tube representations corresponds to a pair consisting of a genuine local operator $\mathcal{O}_1$ and an $x$-twisted sector local operator $\mathcal{O}_x$ that are exchanged by the non-invertible line $m$. The local operator $\mathcal{O}_1$ is charged under the $\mathbb{Z}_2$ sub-symmetry generated by $x$, whereas the twisted sector operator local operator $\mathcal{O}_x$ is not. This is illustrated in figure~\ref{fig:2d-isingex-1}. 

 \end{itemize}
Note that the two-dimensional irreducible tube representation is the only one that permits genuine local operators charged under the $\bZ_2$ symmetry generated by $x$: such charge is inevitably tied with the behaviour under the action of the non-invertible line $m$.

In the critical Ising model, the set of twisted sector primary fields and their transformation properties are such that each of these tube representations appears exactly once. To understand the correspondence between tube representations and primaries, it is useful to note that $\genthree[1]$ corresponds to the lassoing action of $m$ line, which in a conformal field theory is proportional to the exponentiated spin 
\be
\exp(2\pi i (h-\bar{h}))
\ee
for a primary operator $\mathcal{O}_m$ of weight $(h,\bar h)$ in the $m$-twisted sector. Following~\cite{Chang:2018iay,Thorngren:2019iar}, the tube representations generated by primaries are summarised as follows:
\begin{itemize}
\item The primaries $1$, $\varepsilon$ with conformal weights $(h,\bar h) = (0,0),(\frac12,\frac12)$ transform in the irreducible tube representations $\mathcal{F}^+_1$, $\mathcal{F}^-_1$.
\item The primaries $\psi$, $\tilde{\psi}$ in the $x$-twisted sector with conformal weights $(h,\bar{h})=(\frac{1}{2},0),(0,\frac{1}{2})$ transform in the irreducible tube representations $\mathcal{F}^+_x, \mathcal{F}^-_x$. This is compatible with the spin selection rules induced by them being charged under the non-anomalous $\bZ_2$ symmetry. 
\item The primaries  $s$, $\tilde{s}$, $\Lambda$, $\tilde{\Lambda}$ in the $m$-twisted sector with conformal weights 
\begin{equation}
(h,\bar{h}) \; = \; \left(\tfrac{1}{16},0\right), \, \left(0,\tfrac{1}{16}\right), \, \left(\tfrac{1}{16},\tfrac{1}{2}\right), \, \left(\tfrac{1}{2},\tfrac{1}{16}\right)
\end{equation}
transform in the irreducible tube representations $\mathcal{F}^{+}_m, \widetilde{\mathcal{F}}^{+}_m, \mathcal{F}^{-}_m, \widetilde{\mathcal{F}}^{-}_m$.  

\item The primaries $\sigma$, $\mu$ in the $1,x$-twisted sectors with conformal weights $(h,\bar h) = (\frac{1}{16},\frac{1}{16})$ transform in the two-dimensional irreducible tube representation $\mathcal{F}$.

\end{itemize}


\section{Three dimensions: local operators}
\label{sec:3d-ops}

We consider a three-dimensional theory with spherical fusion 2-category symmetry $\C$. This section considers twisted sector local operators attached to topological symmetry lines, while section~\ref{sec:3d-lines} will consider twisted sector line operators attached to topological symmetry surfaces. While the latter determines the former, this intermediate step nevertheless has some utility and will not require higher algebraic methods. 

We will motivate and introduce the tube category $\T_{S^2}\C$ and tube algebra $\cA_{S^2}(\C)$ associated to the manifold $S^2$ linking a point in three dimensions. We show that twisted sector local operators transform in linear representations thereof, which are in 1:1-correspondence with bulk topological lines in the sandwich construction. This extends to an equivalence
\be
[ \T_{S^2} \C , \vect  ] \; \cong \; \text{TV}_\C(S^2) \; := \; \int_{S^2} \C \; = \; \Omega\mathcal{Z}(\mathsf{C})
\ee
of symmetric fusion categories.

We illustrate these ideas in three examples: a finite invertible symmetry with 't Hooft anomaly, a non-invertible Ising-like symmetry obtained by gauging a finite 2-group, and general non-invertible topological lines described by a braided fusion category.


\subsection{Fusion 2-category symmetry} 

We consider a quantum field theory in three dimensions with a spherical fusion 2-category symmetry $\mathsf{C}$~\cite{2018arXiv181211933D,Gaiotto:2019xmp,Johnson-Freyd:2022wm}. This framework captures both finite 2-group symmetries with 't Hooft anomalies and finite non-invertible symmetries. We do not provide a full definition here but summarise key features that will appear prominently in what follows.

\begin{figure}[h]
	\centering
	\includegraphics[height=4.1cm]{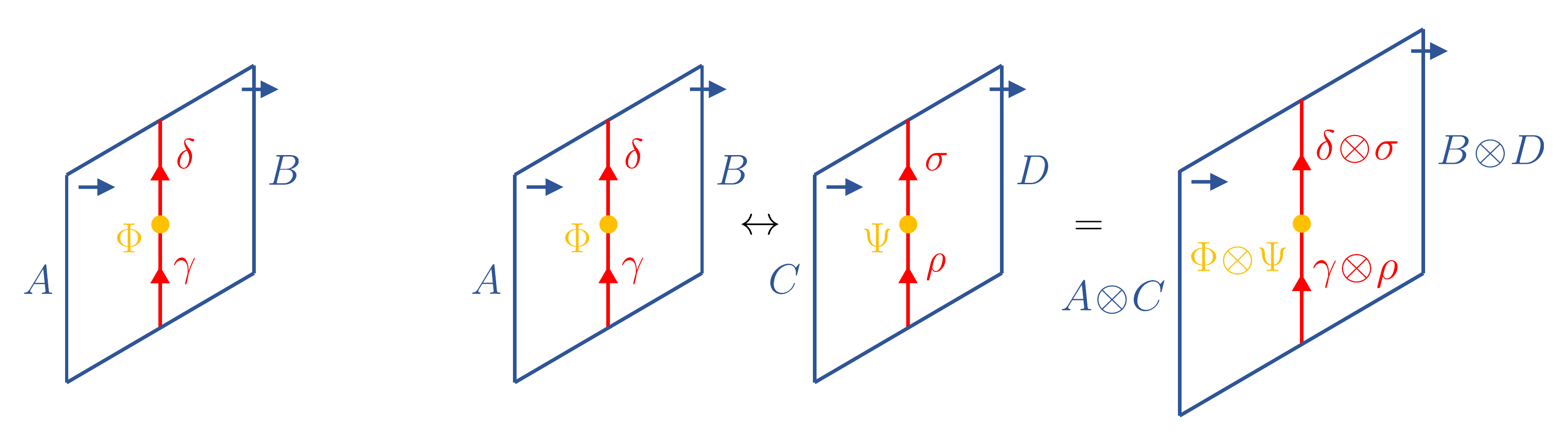}
	\vspace{-5pt}
	\caption{}
	\label{fig:3d-fusion}
\end{figure}

Objects $A \in \mathsf{C}$ correspond to topological surfaces defects, 1-morphisms $\gamma: A \to B$ to topological lines between surface defects, and 2-morphisms $\Phi: \gamma \Rightarrow \delta$ to point-like junctions between topological lines. This is illustrated on the left-hand side of figure \ref{fig:3d-fusion}.
The tensor structure $\otimes: \mathsf{C} \otimes \mathsf{C} \to \mathsf{C}$ captures the fusion of topological surfaces, their line interfaces and point-like junctions, as illustrated on the right-hand side of figure \ref{fig:3d-fusion}. The associativity of $\otimes$ is again controlled by an associator $\alpha$. The compatibility of the associator $\alpha$ with the fusion of four surface defects is implemented by an invertible modification $\Pi$ with component 2-isomorphisms
\begin{equation}
\begin{gathered}
\includegraphics[height=7.2cm]{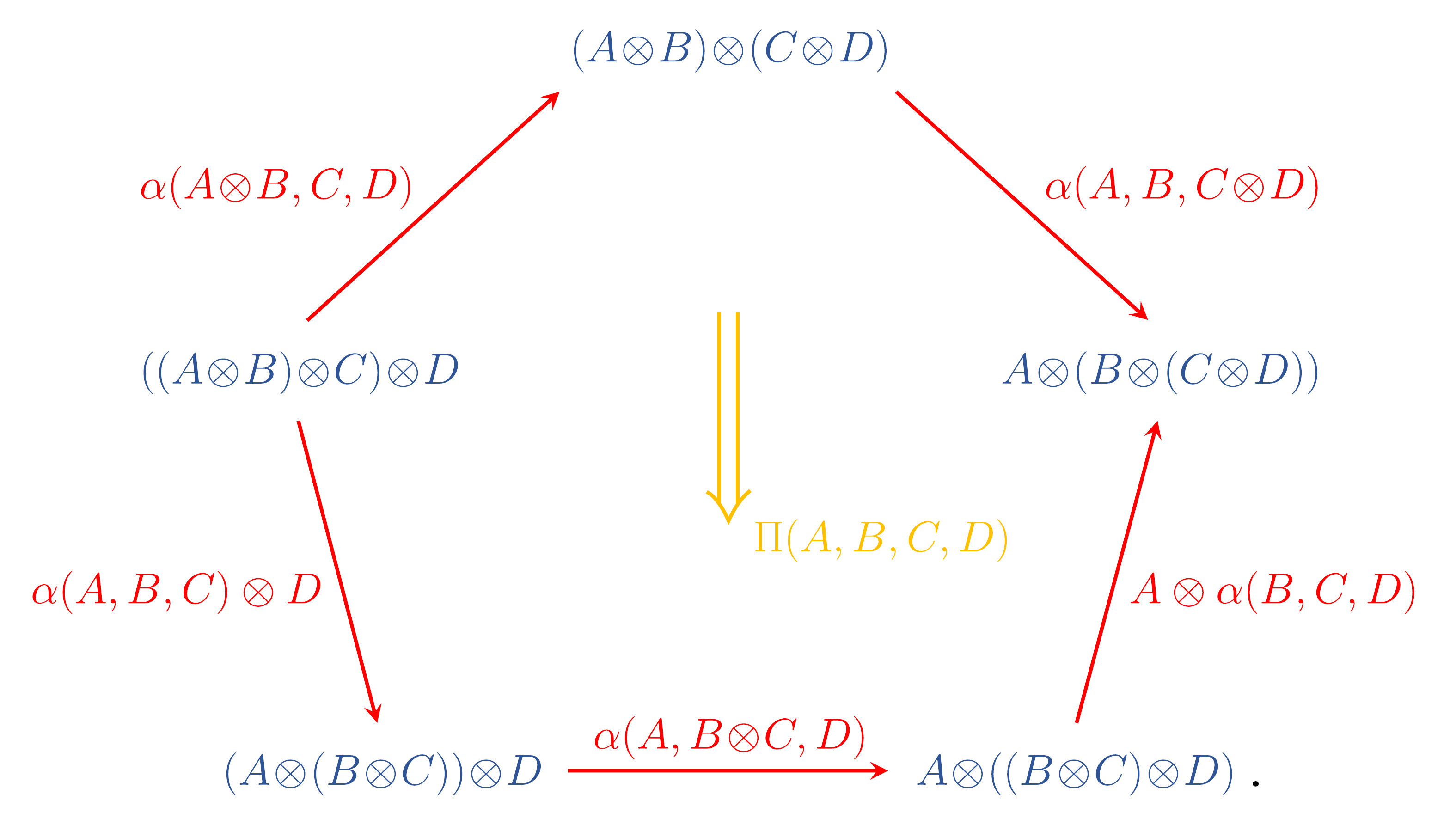}
\end{gathered}
\end{equation}
We call $\Pi$ the \textit{pentagonator} in what follows.

\begin{figure}[h]
	\vspace{-8pt}
	\centering
	\includegraphics[height=3.4cm]{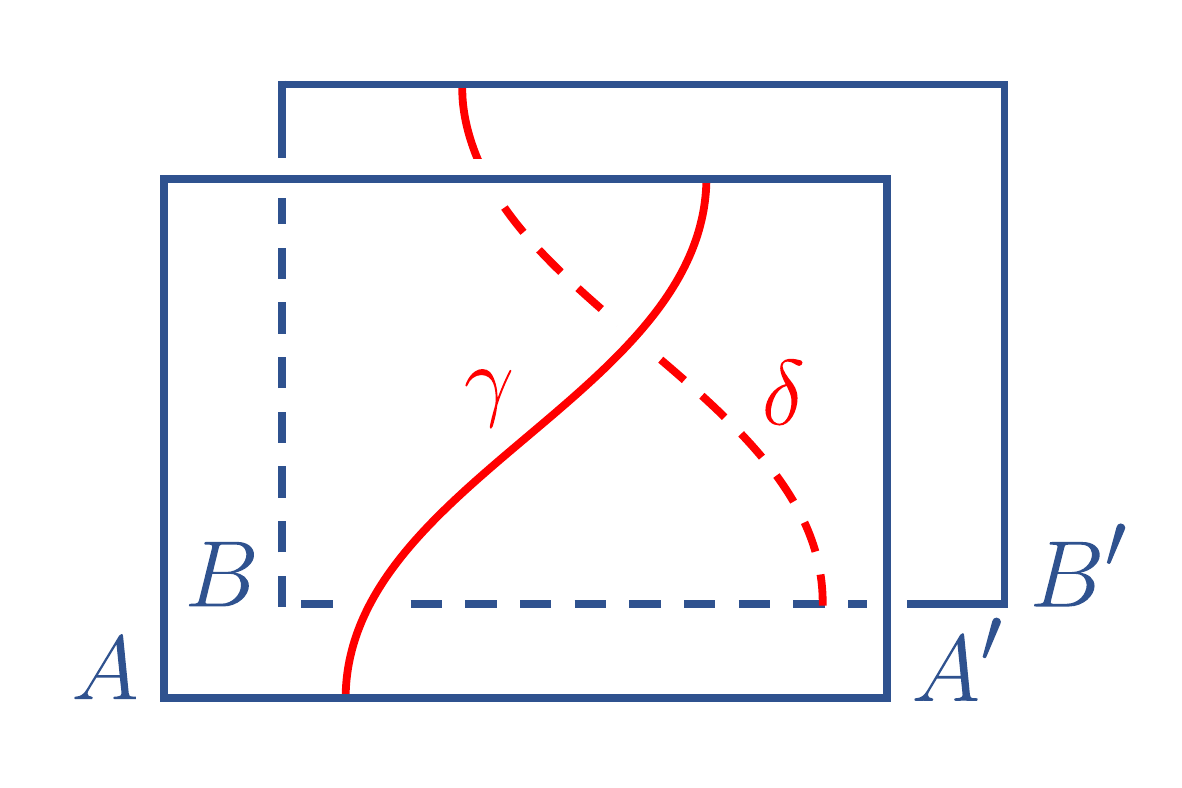}
	\vspace{-8pt}
	\caption{}
	\label{fig:3d-interchanger}
\end{figure}

In addition, there is now a \textit{2-associator} $\Lambda$ with component 2-isomorphisms 
\begin{equation}
\begin{gathered}
\includegraphics[height=3.5cm]{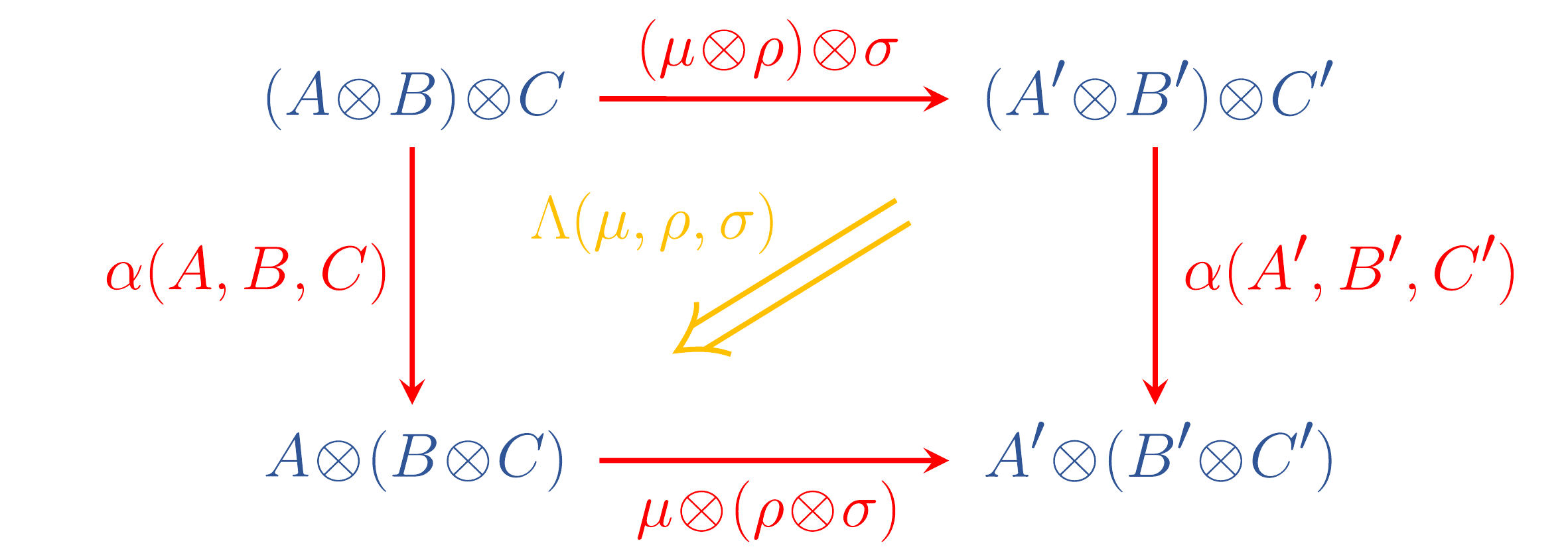}
\end{gathered}
\vspace{-3pt}
\end{equation}
that control the associativity of the fusion of three 1-morphisms $\mu: A \to A'$, $\rho: B \to B'$ and $\sigma: C \to C'$. Furthermore, there are \textit{interchanger} 2-morphisms
\begin{equation}
\begin{gathered}
\includegraphics[height=4.1cm]{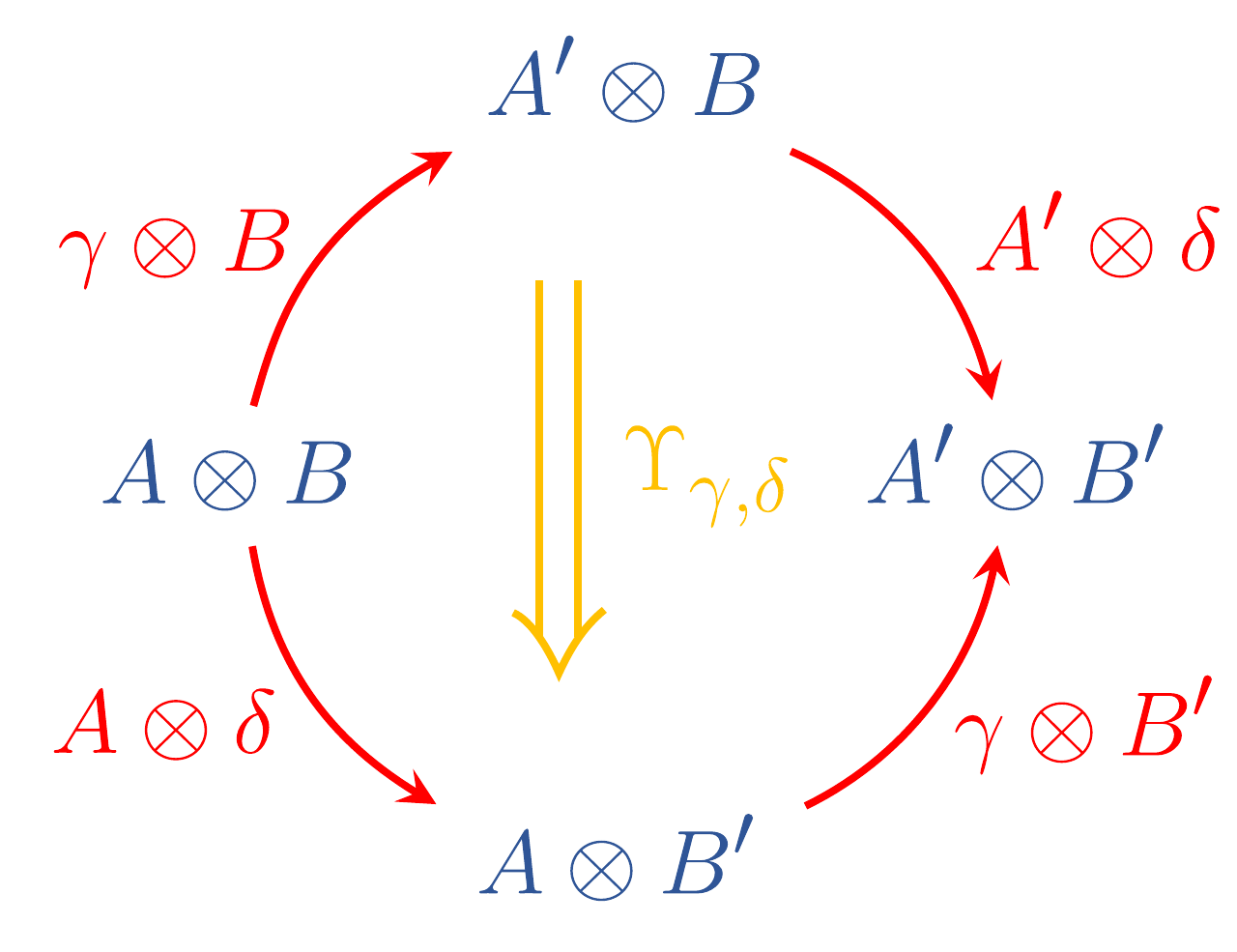}
\end{gathered}
\vspace{-4pt}
\end{equation}
that control the braiding of topological lines $\gamma: A \to A'$ and $\delta: B \to B'$ as illustrated in figure \ref{fig:3d-interchanger}.

\begin{figure}[h]
	\vspace{-8pt}
	\centering
	\includegraphics[height=4.3cm]{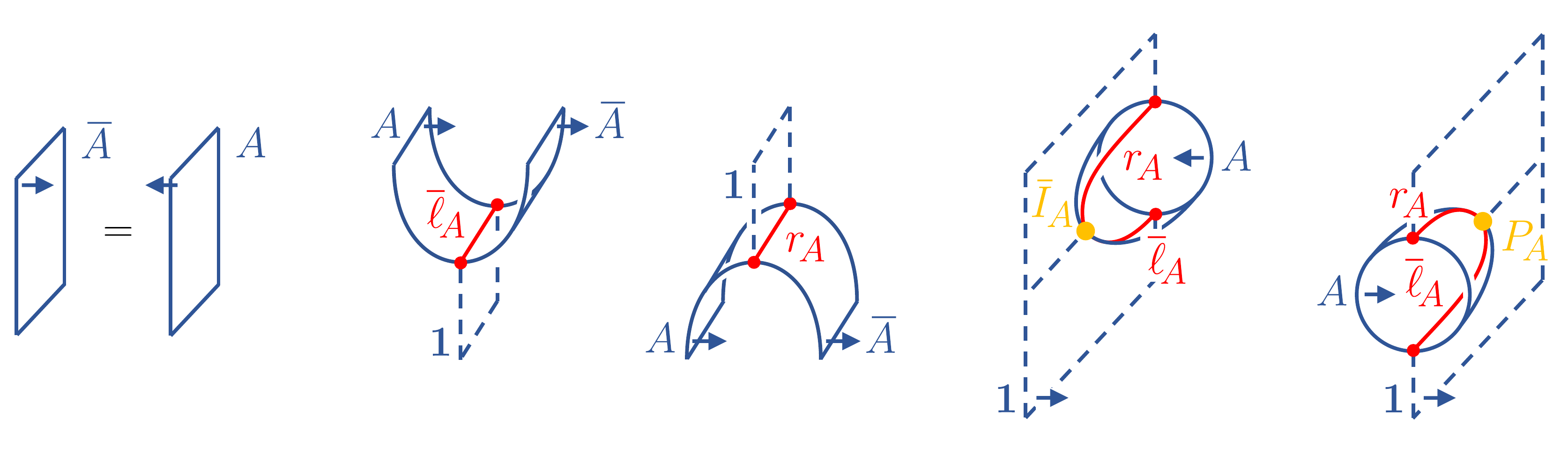}
	\vspace{-8pt}
	\caption{}
	\label{fig:3d-1-orientation}
\end{figure}

A spherical fusion 2-category $\mathsf{C}$ comes equipped with many additional structures~\cite{2018arXiv181211933D}. Each object $A \in \mathsf{C}$ has a dual object $\overline{\!A} \in \mathsf{C}$ defined by orientation reversal of the surface. There are left and right evaluation 1-morphisms $\ell_A$, $r_A$ and co-evaluation morphisms $\overline{\!\ell}_A$, $\overline{r}_A$ as well as front and back closure 2-morphisms $I_A$, $P_A$ and co-closure 2-morphisms $\overline{\!I}_A$, $\overline{\!P}_A$. These ensure topological surfaces can be bent to the left and to the right and closed up forwards and backwards consistently. This is illustrated in figure \ref{fig:3d-1-orientation}.

\begin{figure}[h]
	\vspace{-2pt}
	\centering
	\includegraphics[height=5.6cm]{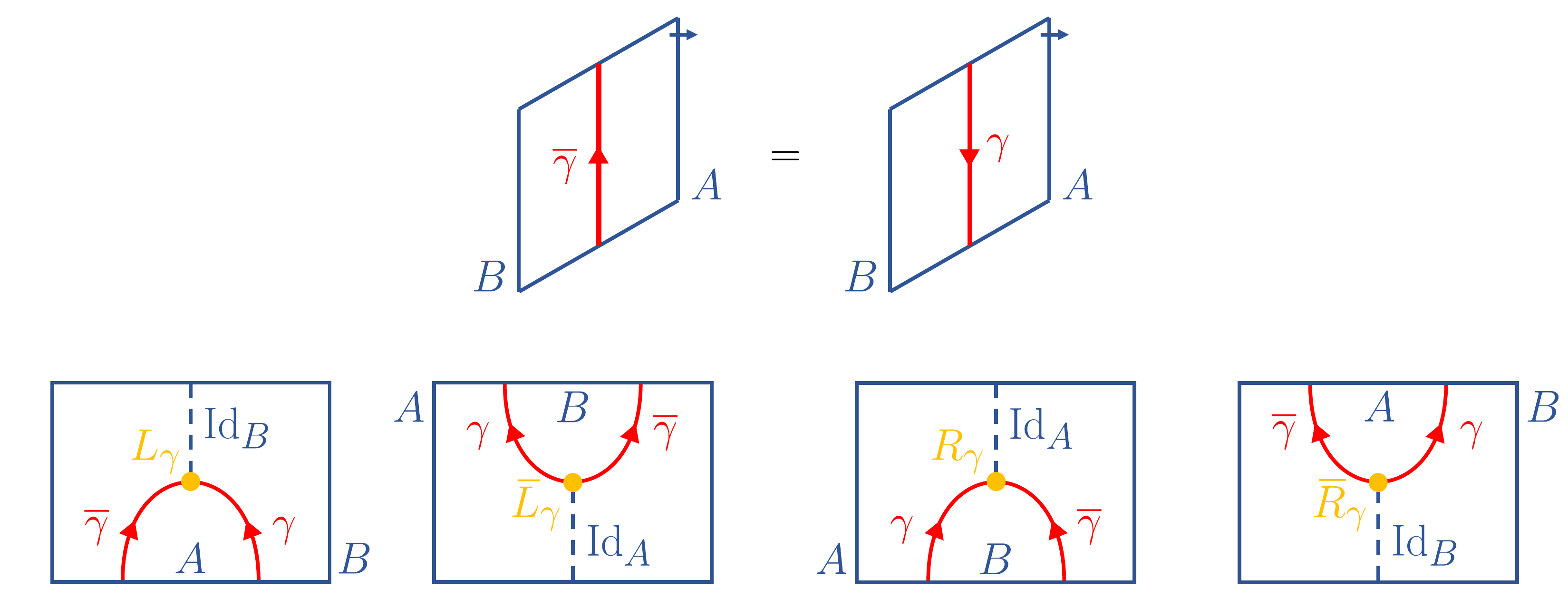}
	\vspace{-2pt}
	\caption{}
	\label{fig:3d-1-orientation-2}
\end{figure}

Similarly, each 1-morphism $\gamma: A \to B$ has a dual $\overline{\gamma}: B \to A$ defined by the orientation reversal of the line. There are left and right evaluation 2-morphisms $L_{\gamma}$, $R_{\gamma}$ and co-evaluation 2-morphisms $\overline{L}_{\gamma}$ and $\overline{R}_{\gamma}$ which ensure that line interface can be bent to the left and to the right consistently. This is illustrated in figure \ref{fig:3d-1-orientation-2}.

\begin{figure}[h]
	\centering
	\includegraphics[height=4.7cm]{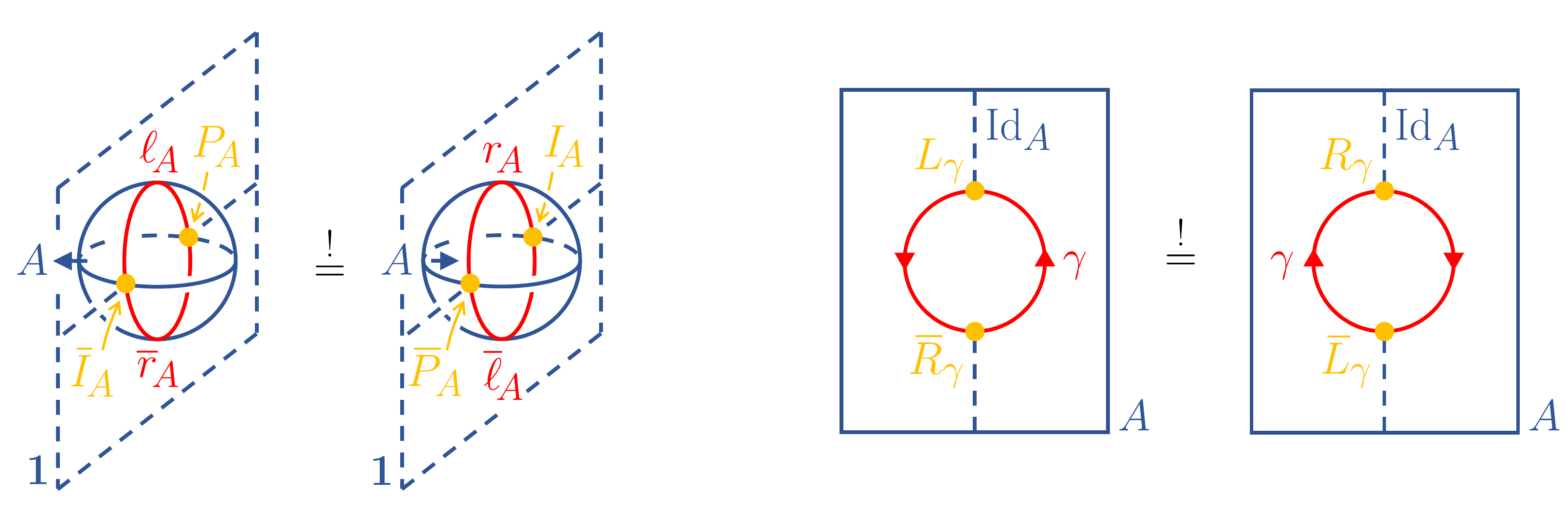}
	\vspace{-5pt}
	\caption{}
	\label{fig:3d-1-spherical-structure}
\end{figure}

Finally, the spherical structure ensures that there is a unambiguous notion of dimension for each object $A \in \C$, which is obtained by placing the corresponding topological symmetry defect on a sphere. Similarly, there is an unambiguous notion of dimension for each 1-endomorphisms $\gamma \in \text{1-End}_{\mathsf{C}}(A)$  by placing the corresponding line defect on a circle as before. This is illustrated in figure \ref{fig:3d-1-spherical-structure}.

\subsubsection{Homotopy groups and condensations}

An object $S \in \mathsf{C}$ is said to be \textit{simple} if it cannot be decomposed as a direct sum of non-trivial sub-objects. Similarly to an ordinary fusion 1-category, the semi-simplicity of $\mathsf{C}$ implies that
\begin{equation}
S \, \sim \, T \quad :\Leftrightarrow \quad \text{1-Hom}_{\mathsf{C}}(S,T) \, \neq \, 0
\end{equation}
defines an equivalence relation on the set of simple objects, whose equivalence classes are the \textit{connected components} of $\mathsf{C}$, i.e.
\begin{equation}
\pi_0(\mathsf{C}) \; := \; \lbrace \text{simple objects} \; S \in \mathsf{C} \rbrace \,/ \sim \, .
\end{equation}
However, unlike an ordinary fusion 1-category, simple objects $S$ and $T$ that lie in the same equivalence class $[S] \in \pi_0(\mathsf{C})$ need not be isomorphic but are instead related by a \textit{condensation} \cite{Gaiotto:2019xmp}. The set of connected components $\pi_0(\C)$ is therefore generally smaller than the set of isomorphism classes of simple objects in $\mathsf{C}$.

Abstractly, a \textit{2-condensation} from an object $A \in \mathsf{C}$ onto an object $B \in \mathsf{C}$ (denoted by $A \cond B$) consists of two pairs of 1- and 2-morphisms
\vspace{-4pt}
\begin{equation}\label{eq:2-condensation-1-2-morphisms}
\begin{gathered}
\includegraphics[height=2.75cm]{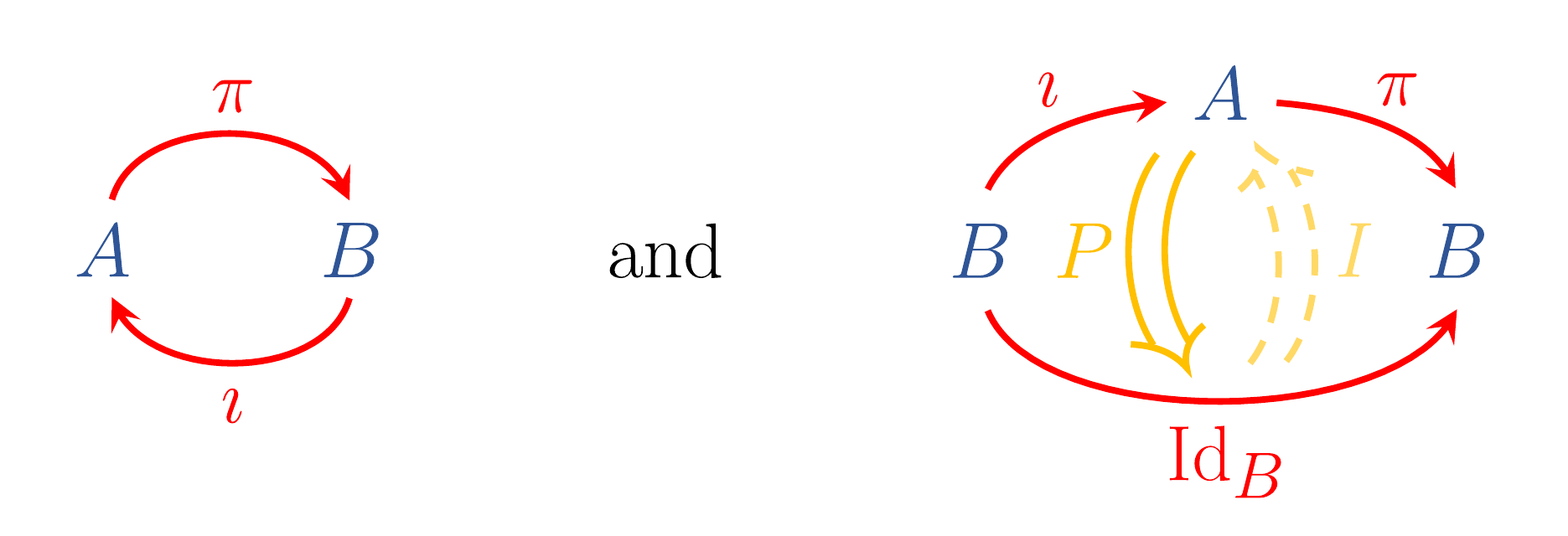}
\end{gathered}
\vspace{-4pt}
\end{equation}
such that $P \circ I = \text{Id}^2_B$. As a consequence, the 1-endomorphism $\varepsilon := \imath \circ \pi \in \text{1-End}_{\mathsf{C}}(A)$ defines a 1-condensation monad ($\varepsilon^2 \cond \varepsilon$) in the sense that there exist 2-morphisms
\vspace{-4pt}
\begin{equation}
\begin{gathered}
\includegraphics[height=3.5cm]{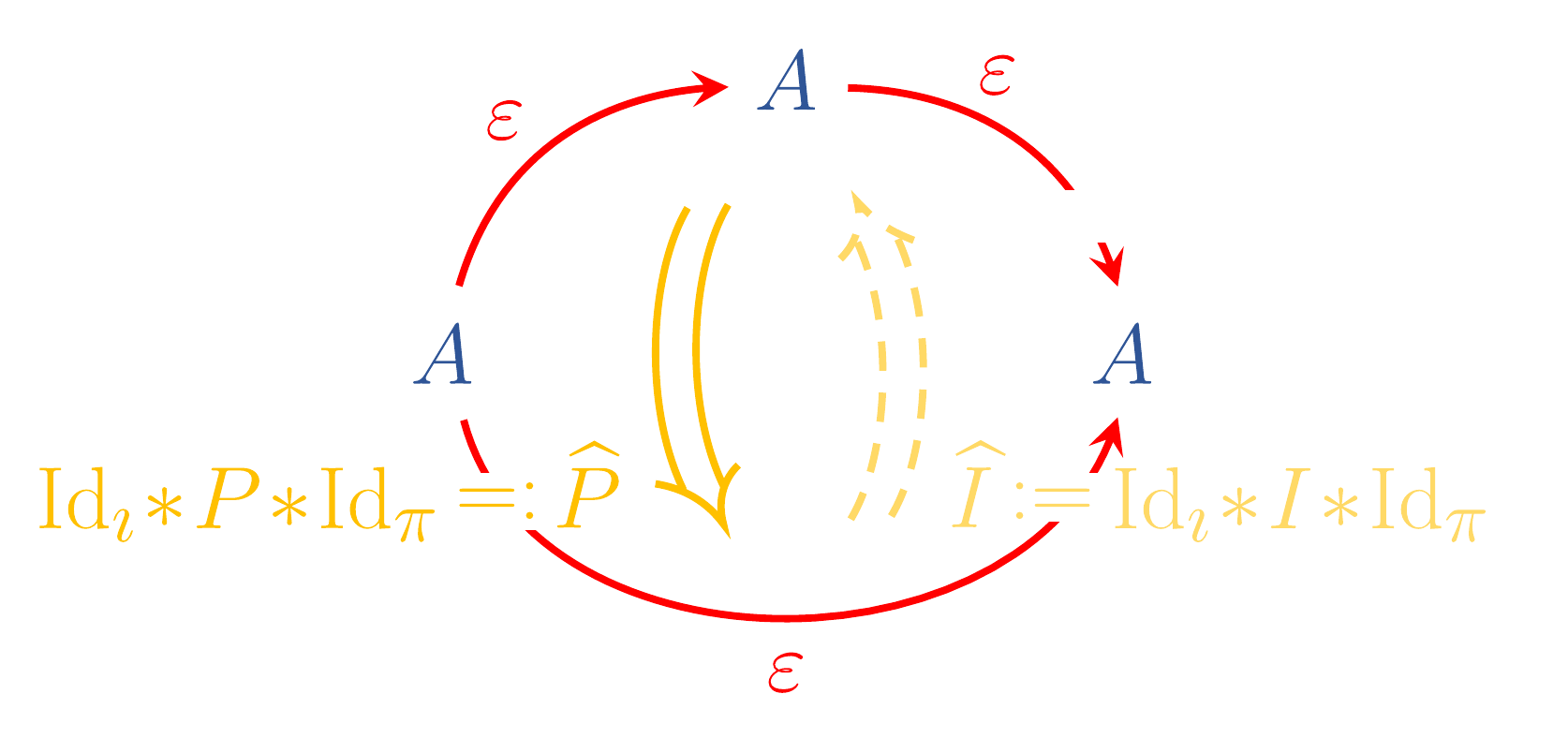}
\end{gathered}
\vspace{-4pt}
\end{equation}
satisfying $\widehat{P} \circ \widehat{I} = \text{Id}_{\varepsilon}$. 

\begin{figure}[h]
	\centering
	\includegraphics[height=3.45cm]{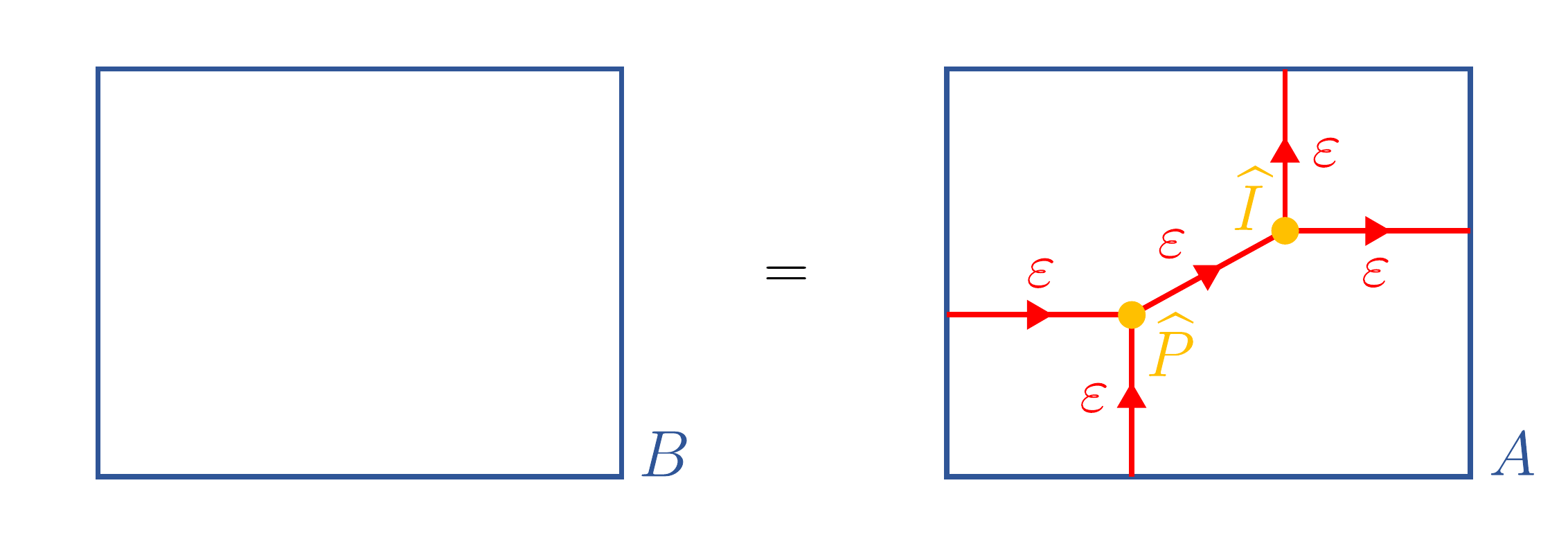}
	\vspace{-5pt}
	\caption{}
	\label{fig:3d-1-condensation}
\end{figure}

From a physical perspective, the topological defect $B$ is obtained by condensing the topological line $\varepsilon$ on $A$. In other words, $B$ is defined by inserting a sufficiently fine mesh of topological lines $\varepsilon$ on $A$ whose junctions are controlled by the 2-morphisms $\widehat{P}$ and $\widehat{I}$. This is illustrated schematically in figure \ref{fig:3d-1-condensation}. 

Upon decomposing the topological line $\varepsilon$ into a direct sum of simple lines, the insertion of this network may be viewed as the gauging of the sub-symmetry on $A$ that is generated by the simple constituents of $\varepsilon$. This physical perspective was introduced in the context of an ordinary 1-form symmetry $A$, corresponding to the fusion 2-category $\C = \mathsf{2Vec}_{A[1]}$, in~\cite{Roumpedakis:2022aik}. The interpretation of condensation defects when gauging a finite non-abelian group $G$ in three dimensions, corresponding to the fusion 2-category $\C = 2\rep(G)$, was discussed further in~\cite{Bartsch:2022mpm,Lin:2022xod,Bhardwaj:2022lsg}.

Since there are no non-trivial networks of topological lines on $S^2$, objects related by condensation correspond to topological surface defects that become equivalent when wrapped on $S^2$. In particular, objects related by condensation have the same dimension. We may thus equivalently think of $\pi_0(\C)$ as enumerating classes of simple topological surfaces that become indistinguishable on $S^2$.

Lastly, we consider the 1-endomorphism category
\begin{equation}
\Omega\mathsf{C} \; := \; \text{1-End}_{\mathsf{C}}(\mathbf{1}) \, ,
\end{equation}
which is an ordinary fusion 1-category due to the tensor unit $\mathbf{1} \in \mathsf{C}$ being a simple object of $\mathsf{C}$. We then define the \textit{first homotopy group} of $\mathsf{C}$ to be 
\begin{equation}
\pi_1(\mathsf{C}) \; := \; \pi_0(\Omega\mathsf{C}) \, ,
\end{equation}
which captures isomorphism classes of simple genuine topological line defects.


\subsection{Twisted sector operators}

We now consider \textit{1-twisted sector local operators}, which are local operators $\mathcal{O}$ attached to an outwardly oriented topological line defect $\mu \in \Omega\mathsf{C}$. Local operators of this type are said to be in the \textit{$\mu$-twisted sector}. The local operators need not be topological and correlation functions may depend on where they are inserted. 

By analogy with section~\ref{sec:2d}, we denote by 
\begin{equation}\label{eq:3d-1-F-objects}
\mathcal{F}(\mu) \; \in \; \mathsf{Vec}
\end{equation}
a finite-dimensional vector space spanned by local operators in the $\mu$-twisted sector. The reason for this notation will again be explained momentarily.

\begin{figure}[h]
	\centering
	\includegraphics[height=3.9cm]{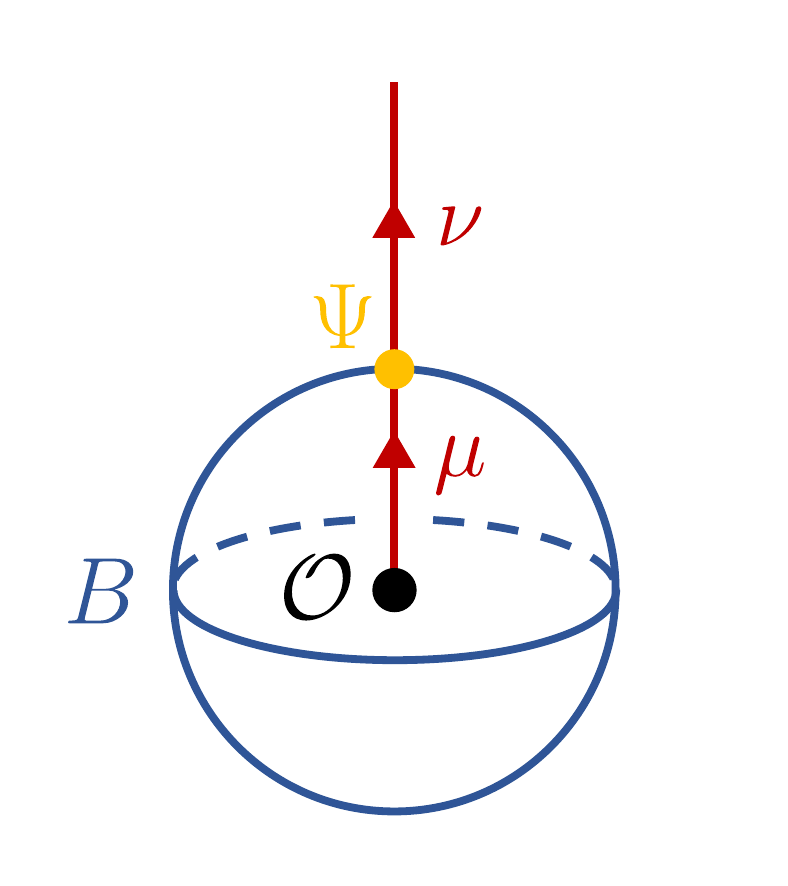}
	\vspace{-5pt}
	\caption{}
	\label{fig:3d-1-local-action}
\end{figure}

A topological symmetry defect $B \in \mathsf{C}$ may act on 1-twisted sector local operators $\cO \in \cF(\mu)$ by linking with a small $S^2$, as shown in figure \ref{fig:3d-1-local-action}. Due to the attached topological line $\mu$, this requires specifying a 2-morphism
\vspace{-8pt}
\begin{equation}\label{eq:3d-1-tube-morphisms}
\begin{gathered}
\includegraphics[height=2.7cm]{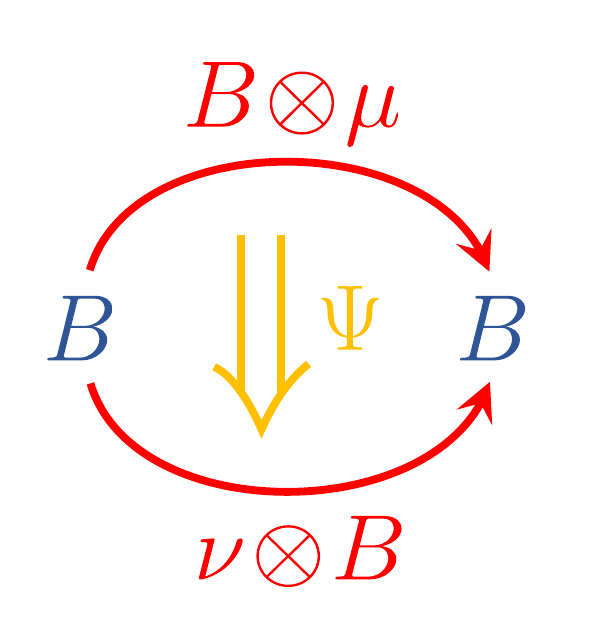}
\end{gathered}
\vspace{-8pt}
\end{equation}
that determines how the surface $B$ intersects the line $\mu$ and transforms it into a new line $\nu$. Upon shrinking the 2-sphere, the pair $(B,\Psi)$ then induces a linear map
\begin{equation}\label{eq:3d-1-F-morphisms}
\begin{gathered}
\includegraphics[height=1.19cm]{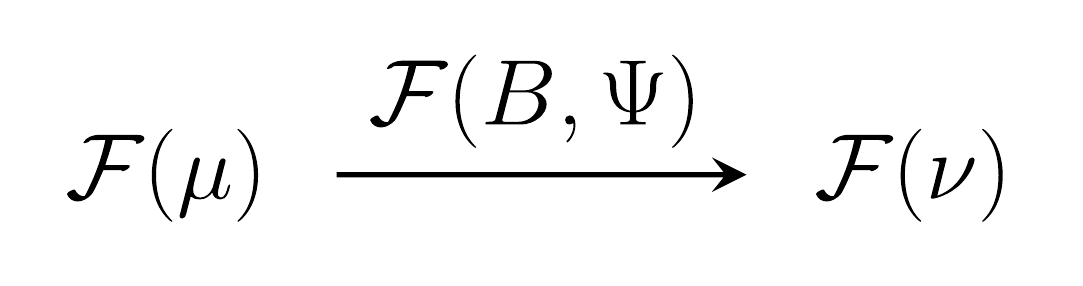}
\end{gathered}
\vspace{-4pt}
\end{equation}
from the $\mu$-twisted to the $\nu$-twisted sector.

\begin{figure}[h]
	\centering
	\includegraphics[height=5cm]{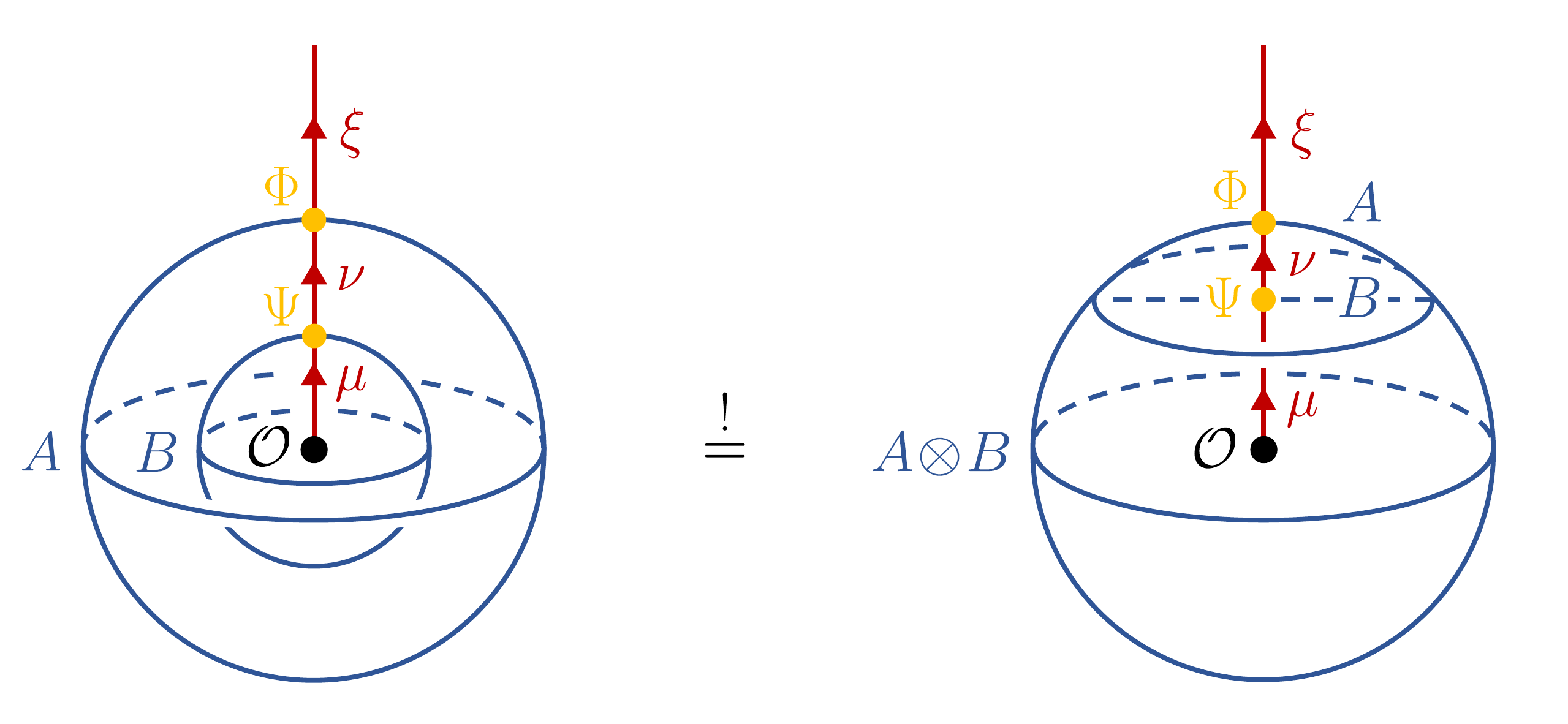}
	\vspace{-5pt}
	\caption{}
	\label{fig:3d-1-composition}
\end{figure}

These linear maps must be compatible with the consecutive action of two topological symmetry defects $A$ and $B$ in the sense that the diagram
\begin{equation}\label{eq:3d-1-F-composition}
\begin{gathered}
\includegraphics[height=2.85cm]{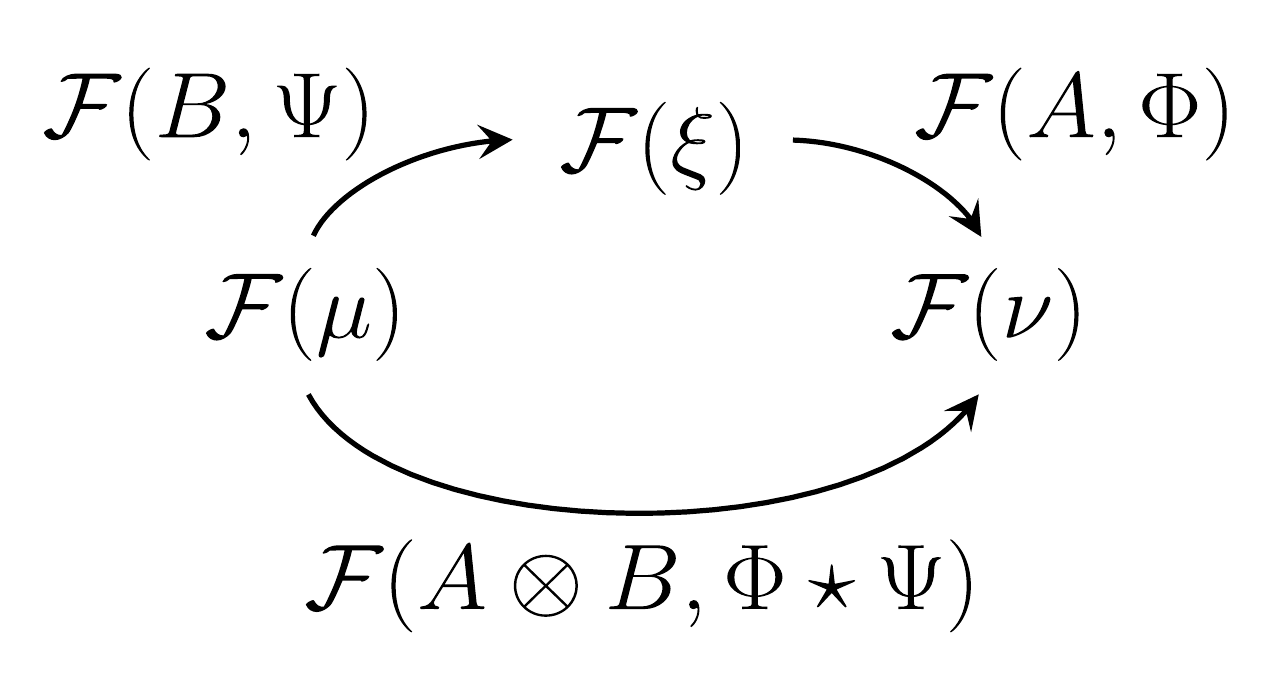}
\end{gathered}
\vspace{-4pt}
\end{equation}
commutes, where the 2-morphism $\Phi \star \Psi$ is defined by the diagram
\begin{equation}\label{eq:1-tube-composition}
\begin{gathered}
\includegraphics[height=5.8cm]{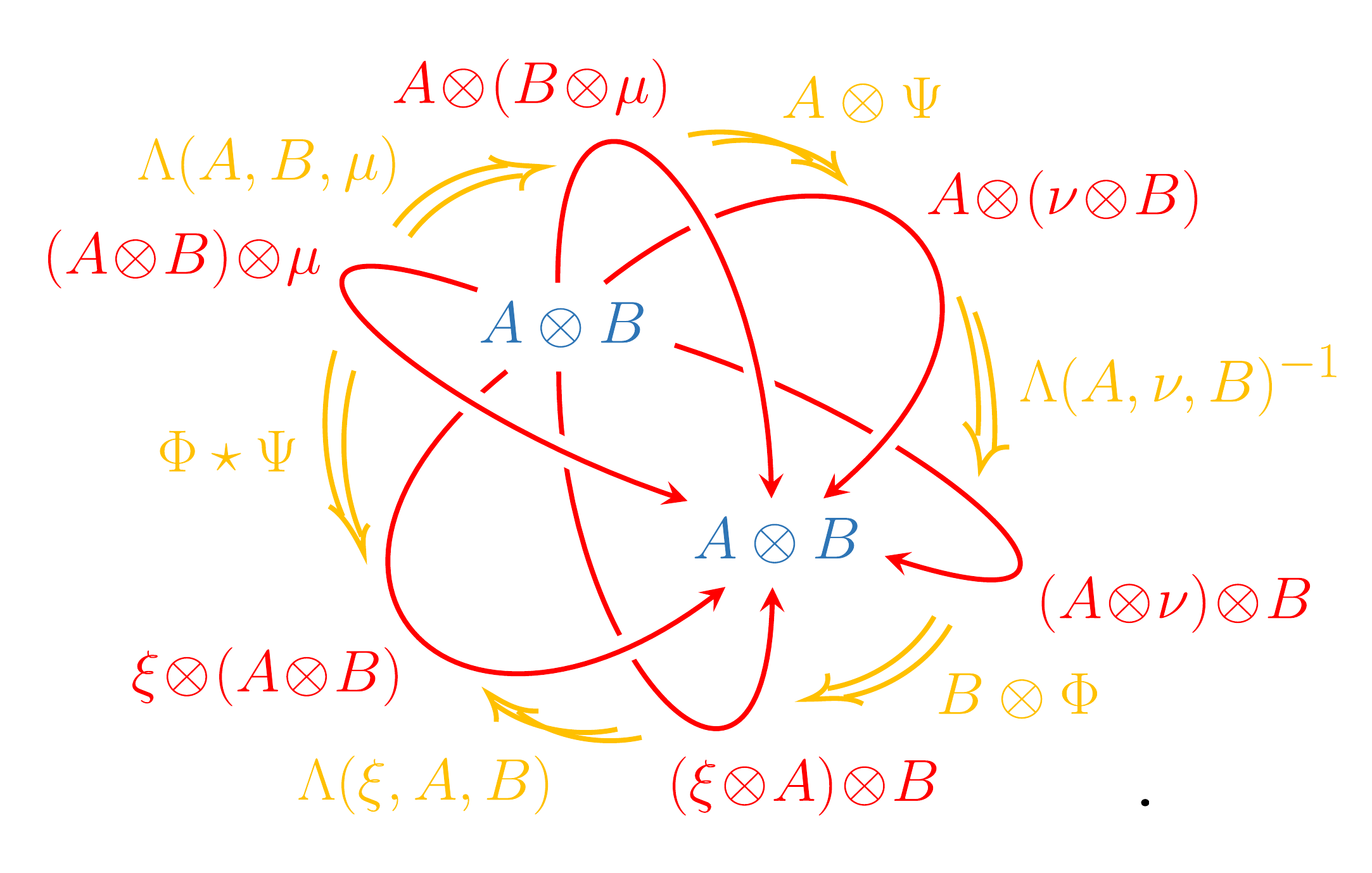}
\end{gathered}
\end{equation}
Despite its fearsome appearance, this condition simply states that acting with $A$ and $B$ consecutively is equivalent to acting with their fusion $A \otimes B$, as illustrated in figure \ref{fig:3d-1-composition}.

\begin{figure}[h]
	\centering
	\includegraphics[height=3.9cm]{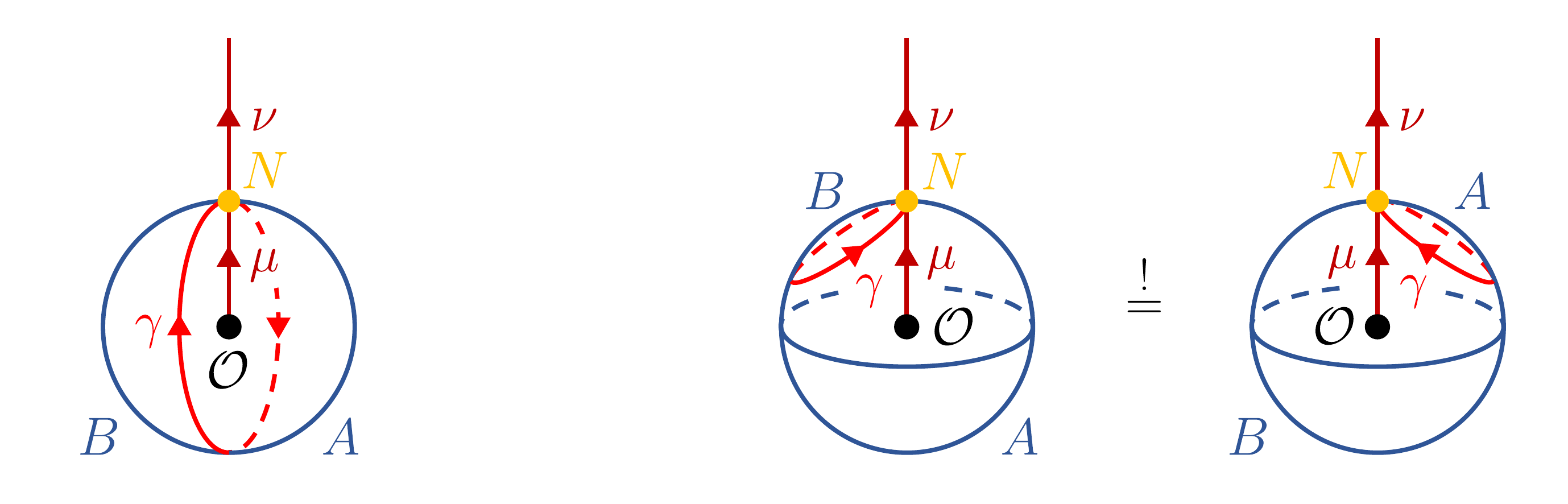}
	\vspace{-5pt}
	\caption{}
	\label{fig:3d-1-tube-equivalence}
\end{figure}

Finally, the action of topological symmetry defects on 1-twisted sector local operators must be compatible with the possibility to move topological lines around the encircling $S^2$. Concretely, consider a configuration as on the left-hand side of figure~\ref{fig:3d-1-tube-equivalence}, where two surface defects $A$ and $B$ that are connected by a 1-morphism $\gamma: A \to B$ link a local operator $\mathcal{O}$ in the $\mu$-twisted sector via a specified 2-morphism
\begin{equation}
\begin{gathered}
\includegraphics[height=2.7cm]{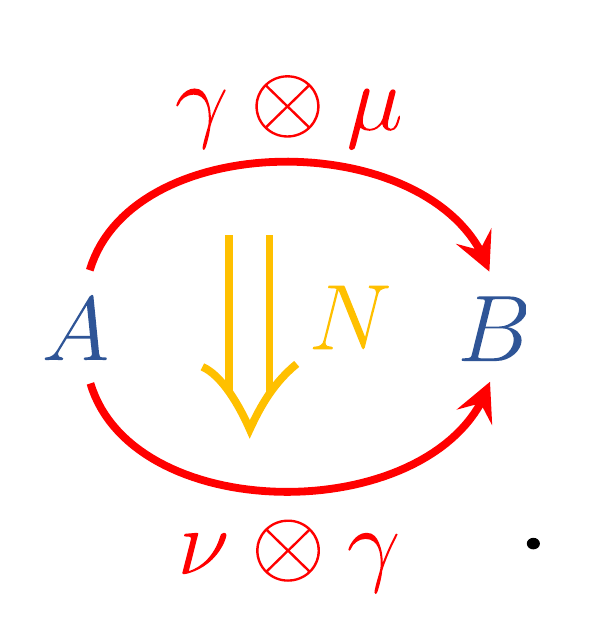}
\end{gathered}
\vspace{-4pt}
\end{equation}
By moving the topological line interface $\gamma$ around the 2-sphere towards $N$ from the left and from the right, we can either regard this configuration as
\begin{enumerate}
\item the defect $A$ acting on $\mathcal{O}$ via the intersection 2-morphism $\Phi$ defined by
\vspace{-2pt}
\begin{equation}
\begin{gathered}
\includegraphics[height=4.4cm]{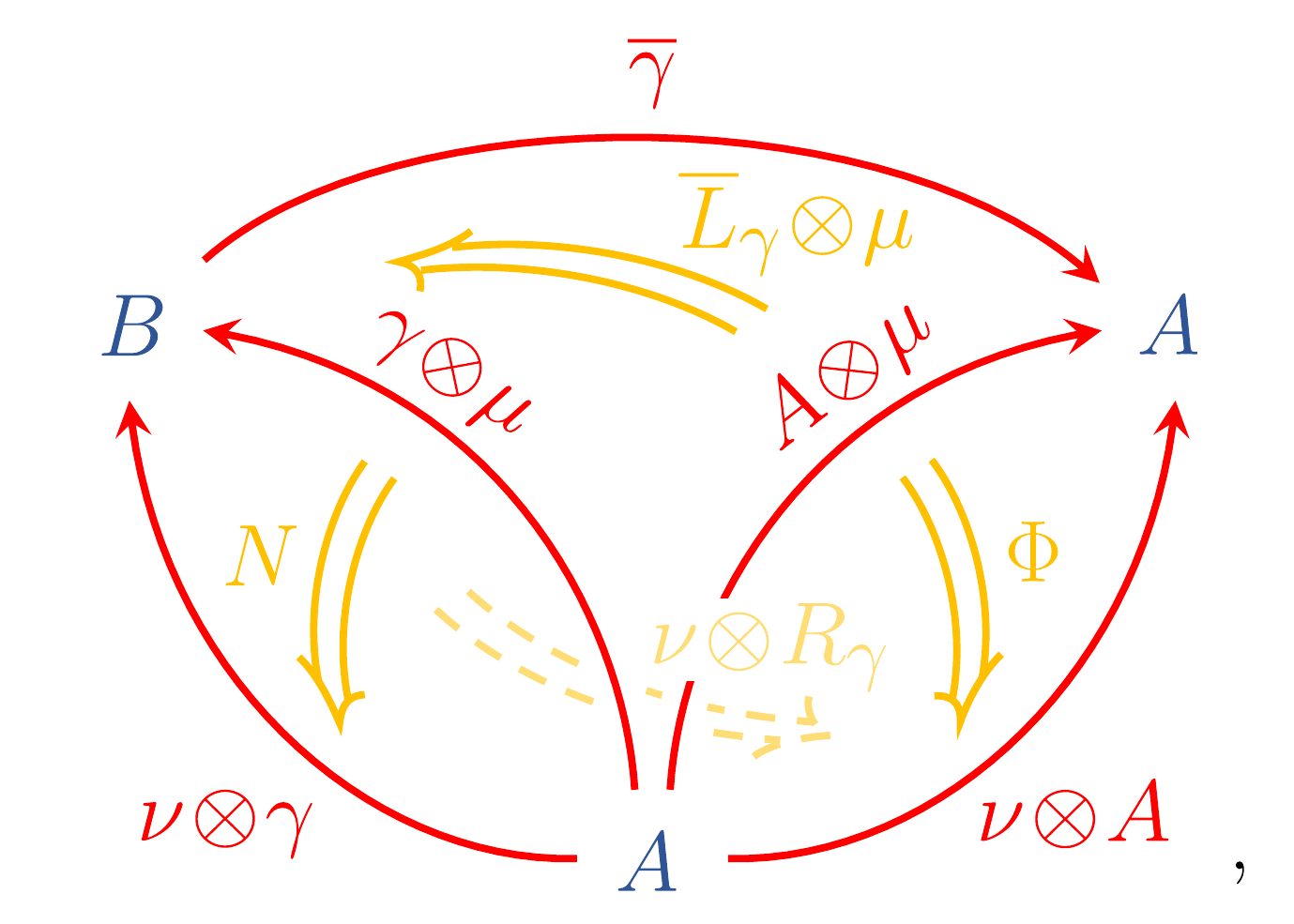}
\end{gathered}
\vspace{-4pt}
\end{equation}

\item the defect $B$ acting on $\mathcal{O}$ via the intersection 2-morphism $\Psi$ defined by
\vspace{-2pt}
\begin{equation}
\begin{gathered}
\includegraphics[height=4.4cm]{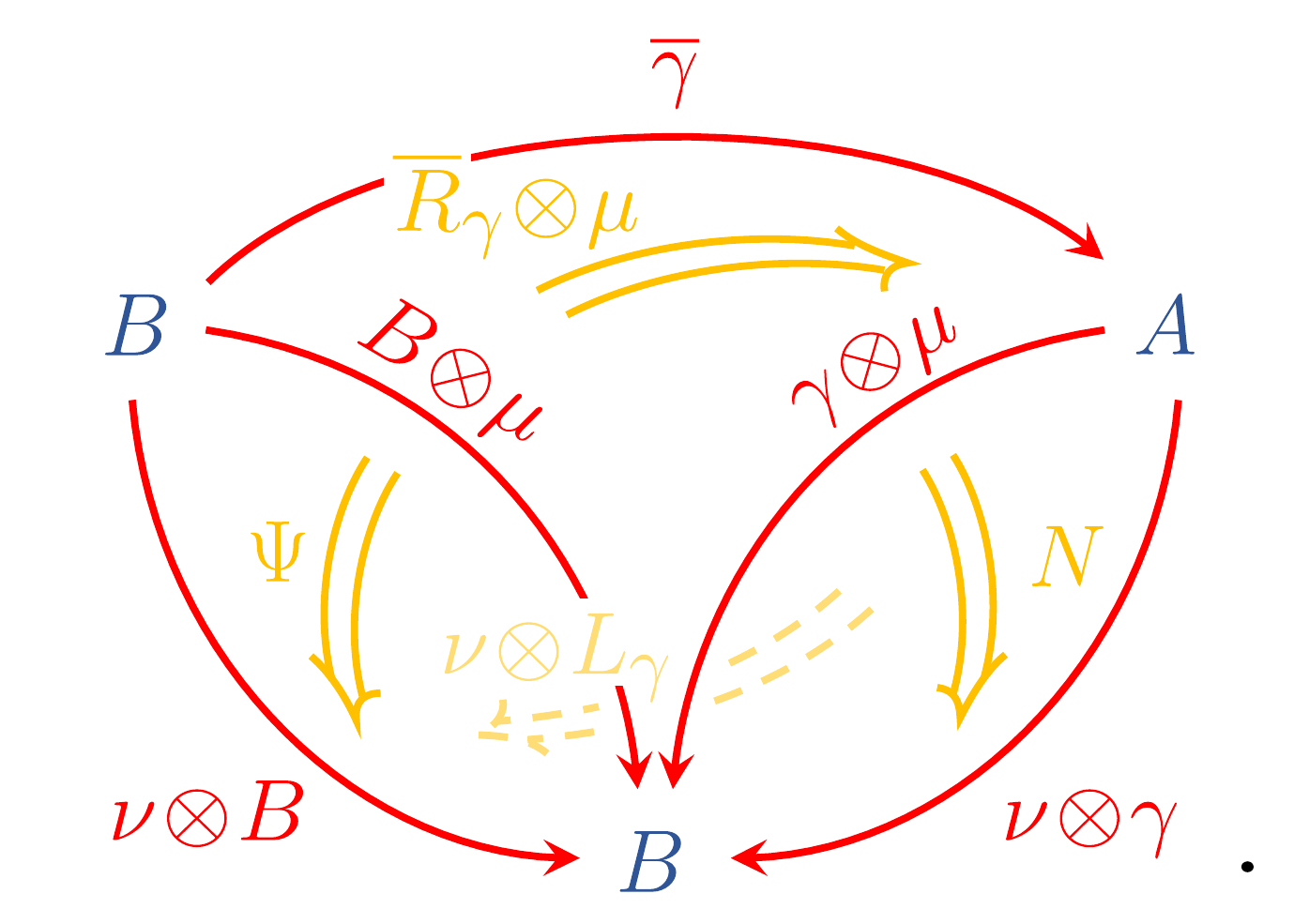}
\end{gathered}
\vspace{-4pt}
\end{equation}
\end{enumerate}
Since both configurations are physically equivalent, the corresponding linear actions on $\mathcal{O}$ must coincide in the sense that
\begin{equation}\label{eq:3d-1-F-invariance}
\mathcal{F}(A, \Phi) \; \stackrel{!}{=} \; \mathcal{F}(B,\Psi) \, .
\end{equation}
This is illustrated on the right-hand side of figure \ref{fig:3d-1-tube-equivalence}.

For fixed topological lines $\mu$ and $\nu$, the above relations generate an equivalence relation $\sim$ on the vector space
\begin{equation}
\bigoplus_{B \, \in \, \mathsf{C}} \, \text{2-Hom}_{\mathsf{C}}(\text{Id}_B \otimes \mu, \nu \otimes \text{Id}_B) \, ,
\end{equation}
which relates configurations acting identically on 1-twisted sector local operators as a consequence of placing the topological defects on $S^2$. We will denote the equivalence class of a pair $(B,\Psi)$ under $\sim$ by $[B,\Psi]$ in what follows.

\subsection{Tube representations}

We now formulate the above action of the symmetry category $\mathsf{C}$ on 1-twisted sector local operators  in terms of representation theory. We introduce the \textit{tube category} $\T_{S^2}\C$ and \textit{tube algebra} $\cA_{S^2}(\C)$ and show that their linear representation theory captures the structures presented above. This is a natural generalisation of the fact the local operators transform in representations of a finite symmetry group in three dimensions.

\subsubsection{Tube category}

We first introduce the tube category $\T_{S^2}\C$, which is a finite semi-simple category whose objects are 1-twisted sectors and morphisms are actions of topological symmetry defects on 1-twisted sector operators via linking with $S^2$.

This has the following explicit description:
\begin{itemize}
\item Its objects are those of $\C$, $\text{Ob}(\T_{S^2}\C) = \text{Ob}(\Omega\C)$.

\item Its morphism spaces between objects $\mu, \nu \in \T_{S^2}\C$ are given by
\begin{equation}
\text{Hom}_{\T_{S^2}\C}(\mu,\nu) \; = \; \bigg( \!\bigoplus_{B \, \in \, \mathsf{C}} \, \text{2-Hom}_{\mathsf{C}}(\text{Id}_B \otimes \mu, \nu \otimes \text{Id}_B) \bigg) \, \bigg/ \, \sim
\end{equation}
with $\sim$ as described in the previous sub-section. More concretely, morphisms between $\mu$ and $\nu$ in $\T_{S^2}\C$ are linear combinations of equivalence classes of pairs
\vspace{-4pt}
\begin{equation}
\begin{gathered}
\includegraphics[height=1.2cm]{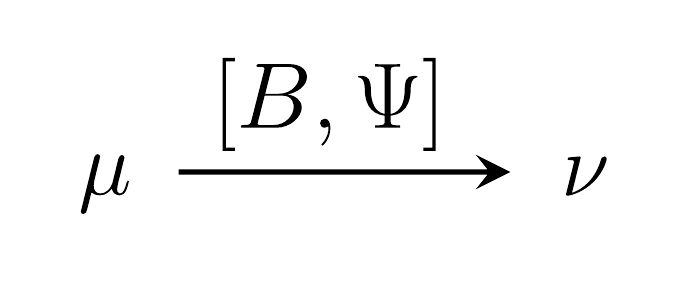}
\end{gathered}
\vspace{-8pt}
\end{equation}
consisting of an object $B \in \mathsf{C}$ and a 2-morphism 
\begin{equation}
\Psi \; \in \; \text{2-Hom}_{\mathsf{C}}(\text{Id}_B \otimes \mu, \nu \otimes \text{Id}_B) \, .
\end{equation}

\item The composition of morphisms
\vspace{-4pt}
\begin{equation}
\begin{gathered}
\includegraphics[height=1.2cm]{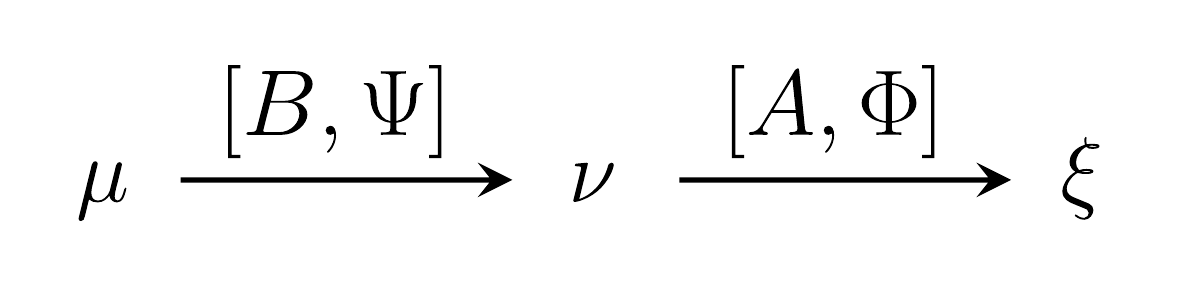}
\end{gathered}
\vspace{-8pt}
\end{equation}
is defined by the composition law
\begin{equation}\label{eq:3d-1-tube-composition}
[A,\Phi] \, \circ \, [B,\Psi] \; := \; [A \otimes B, \, \Phi \star \Psi]
\end{equation}
with the 2-morphism $\Phi \star \Psi$ defined as in (\ref{eq:1-tube-composition}).
\end{itemize}
There is again an inclusion functor $\Omega\C \hookrightarrow \T_{S^2}\C$ whose image corresponds to morphisms with $B = \mathbf{1}$.  The tube category $\T_{S^2}\C$ is obtained by horizontal compactification of $\Omega \C$ on $S^2$: it has the same objects as $\Omega\C$ but additional morphisms from intersections with topological surfaces wrapping $S^2$. 

The above definition of morphism spaces as a quotient allows for a clean physical interpretation in terms of symmetry defects acting on 1-twisted sectors, but may be abstract in view of concrete computations. Similarly to two dimensions, there exists a more concrete but less canonical description in terms of simple objects of $\mathsf{C}$, although this is now slightly more intricate. Concretely, there is an isomorphism
\begin{equation}\label{eq:3d-1-tube-hom-isomorphism}
\begin{aligned}
\text{Hom}_{\T_1\C}(\mu,\nu) \;\; \cong \bigoplus_{[S] \, \in \, \pi_0(\C)} \; \text{2-Hom}_{\mathsf{C}}(\text{Id}_S \otimes \mu, \nu \otimes \text{Id}_S) \; ,
\end{aligned}
\end{equation}
where the sum is over a collection of representatives $S$ of connected components $[S] \in \pi_0(\C)$ of $\C$. A schematic proof of this isomorphism is presented in appendix \ref{app:2-condensations-sphere}. From a physical perspective, this captures the intuitive fact that the action of a generic symmetry defect on 1-twisted sector operators is completely determined by the action of its simple constituents. Furthermore, the latter only need to be considered up to condensation, since condensations become trivial on $S^2$.

With the above definitions in hand, the collection of vector spaces (\ref{eq:3d-1-F-objects}) and linear maps (\ref{eq:3d-1-F-morphisms}) together with the compatibility conditions (\ref{eq:3d-1-F-composition}) and (\ref{eq:3d-1-F-invariance}) can now be summarised conveniently as a functor
\begin{equation}
\mathcal{F} \, : \; \T_{S^2}\C \; \to \; \mathsf{Vec} \, .
\end{equation}
In other words, 1-twisted sector local operators transform in linear representations of the tube category $\T_{S^2}\C$. 

We again denote the category of all such linear representations by 
\be
[\T_{S^2}\C,\mathsf{Vec}] \, .
\ee
This is the idempotent or Karoubi completion of $\T_{S^2}\C$ and has the structure of a symmetric braided fusion category. The fusion structure captures the action of $\C$ on products of 1-twisted sector local operators supported at separated spacetime points. The reason for a symmetric fusion structure is that the configuration space in three dimensions is now homotopic to $S^2$.

\subsubsection{Tube algebra}

There is an equivalent formulation of the above representation theory in terms of the \textit{tube algebra} $\cA_{S^2}(\C)$. This is the finite semi-simple associative algebra
\begin{equation}
\mathcal{A}_{S^2}(\mathsf{C}) \; := \; \text{End}_{\T_{S^2}\C}\Big( \bigoplus_{[\sigma] \, \in \, \pi_1(\C) } \sigma \, \Big) \, ,
\end{equation} 
where the sum ranges over a collection of representatives $\sigma$ of isomorphism classes $[\sigma] \in \pi_1(\mathsf{C})$ of simple topological lines in $\mathsf{C}$. The algebra product is again determined by the composition of morphisms in $\T_{S^2}\C$. 

Any representation $\mathcal{F} : \T_{S^2}\C \to \mathsf{Vec}$ of the tube category then determines a representation of the tube algebra $\mathcal{A}_{S^2}(\C)$ on the vector space 
\be
V \; := \; \mathcal{F}\Big(\bigoplus_{[\sigma]} \, \sigma\Big)
\ee 
by sending algebra elements
\begin{equation}
a \in \mathcal{A}_1(\C) \;\; \mapsto \;\; \mathcal{F}(a) \in \text{End}(V) \, .
\end{equation}
This sets up an equivalence of symmetric fusion categories
\begin{equation}\label{eq:3d-1-morita-equivalent-tube-reps}
[ \T_{S^2}\C , \vect ]  \; \cong \; \rep(\cA_{S^2}(\C)) \, .
\end{equation}
From either perspective, this provides a complete description of the representation theory of a spherical fusion 2-category symmetry $\C$ on 1-twisted local operators in three dimensions. We will again uniformly speak of \textit{tube representations} in what follows.

\subsection{Sandwich construction}

We now rephrase the above construction of tube representations in the context of the sandwich construction. This is realised by a symmetric equivalence of the category of tube representations and the category of topological lines in the associated four-dimensional Turaev-Viro theory $\text{TV}_\C$.

The latter is the looping
\begin{equation}
\text{TV}_\C(S^2) \; := \; \int_{S^2} \mathsf{C}  \; = \; \Omega \mathcal{Z}(\mathsf{C})
\end{equation}
of the Drinfeld center of the fusion 2-category $\mathsf{C}$. From the explicit description of 1-morphisms in $\cZ(\C)$~\cite{BAEZ1996196} (see also references~\cite{davydov2021braided,Kong:2019brm} for additional presentations), an object in $\Omega\mathcal{Z}(\mathsf{C})$ is given by a pair $p = (\omega,T)$ consisting of 
\begin{enumerate}
\item an object $\omega \in \Omega\mathsf{C}$,
\item a collection $T$ of 2-morphisms
\vspace{-4pt}
\begin{equation}
\begin{gathered}
\includegraphics[height=2.65cm]{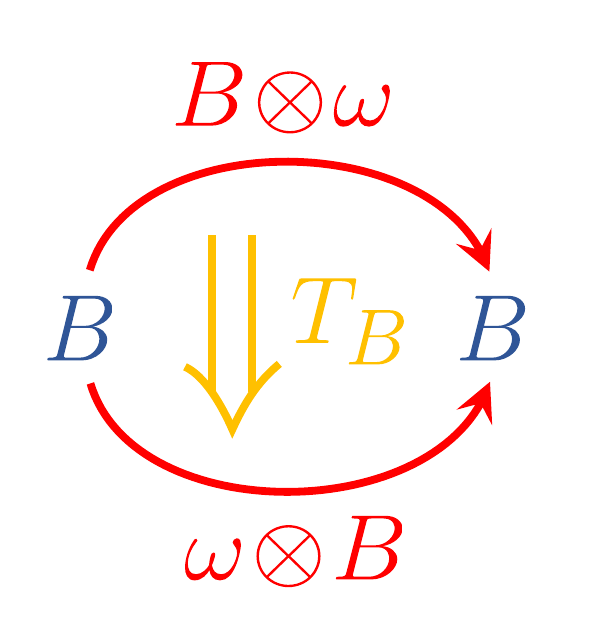}
\end{gathered}
\vspace{-6pt}
\end{equation}
for each object $B \in \mathsf{C}$, subject to suitable coherence conditions \cite{Kong:2019brm}.
\end{enumerate}
Note that there is a forgetful functor $F : \Omega \mathcal{Z}(\C) \to \Omega \C$ that discards the information of the additional 2-morphisms $T$ and sends $p \mapsto  F(p) = \omega$. 

To each object $p \in \Omega \mathcal{Z}(\mathsf{C})$, we can associate a tube representation $\mathcal{F}_p \in [\T_{S^2}\C,\mathsf{Vec}]$ by defining the functor $\mathcal{F}_p$ as follows:
\begin{itemize}
\item To an object $\mu \in \Omega\mathsf{C}$ it assigns the vector space
\begin{equation}
\mathcal{F}_p(\mu) \; := \; \text{2-Hom}_{\mathsf{C}}(\omega,\mu) \, .
\end{equation}

\item To a morphism $[B,\Psi] \in \text{Hom}_{\mathsf{T_1C}}(\mu,\nu)$ it assigns the linear map
\begin{equation}
\begin{gathered}
\includegraphics[height=1.08cm]{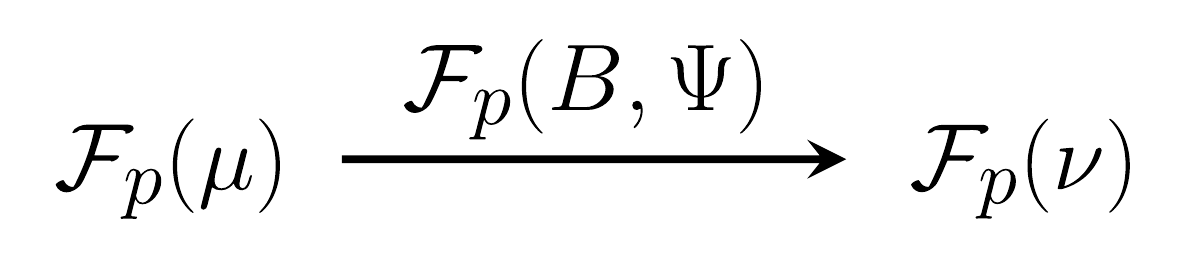}
\end{gathered}
\end{equation}
that sends $\mathcal{O} \in \text{2-Hom}_{\mathsf{C}}(\omega,\mu)$ to $\mathcal{O}' \in \text{2-Hom}_{\mathsf{C}}(\omega,\nu)$ defined by the diagram\footnote{Here, we use the existence of canonical isomorphisms
\begin{equation}
\begin{aligned}
\text{2-Hom}_{\mathsf{C}}(\text{Id}_{\overline{B}} \otimes \omega, \, \omega \otimes \text{Id}_{\overline{B}}) \; &\cong \; \text{2-Hom}_{\mathsf{C}}(\overline{\ell}_B \circ \omega, \, [\text{Id}_B \otimes (\omega \otimes \text{Id}_{\overline{B}})] \circ \overline{\ell}_B) \\
\text{2-Hom}_{\mathsf{C}}(\text{Id}_B \otimes \mu, \, \nu \otimes \text{Id}_B) \; &\cong \; \text{2-Hom}_{\mathsf{C}}(r_B \circ [\text{Id}_B \otimes (\mu \otimes \text{Id}_{\overline{B}})], \, \nu \circ r_B)
\end{aligned}
\end{equation}
that are induced by the (co)evaluation 2-morphisms of $\mathsf{C}$. Their usage is left implicit when applied to the morphisms $T_{\Bar{B}}$ and $\Psi$ in (\ref{eq:3d-1-dia-drinfeld-morphism-2}).}
\vspace{-4pt}
\begin{equation}\label{eq:3d-1-dia-drinfeld-morphism-2}
\begin{gathered}
\includegraphics[height=9.7cm]{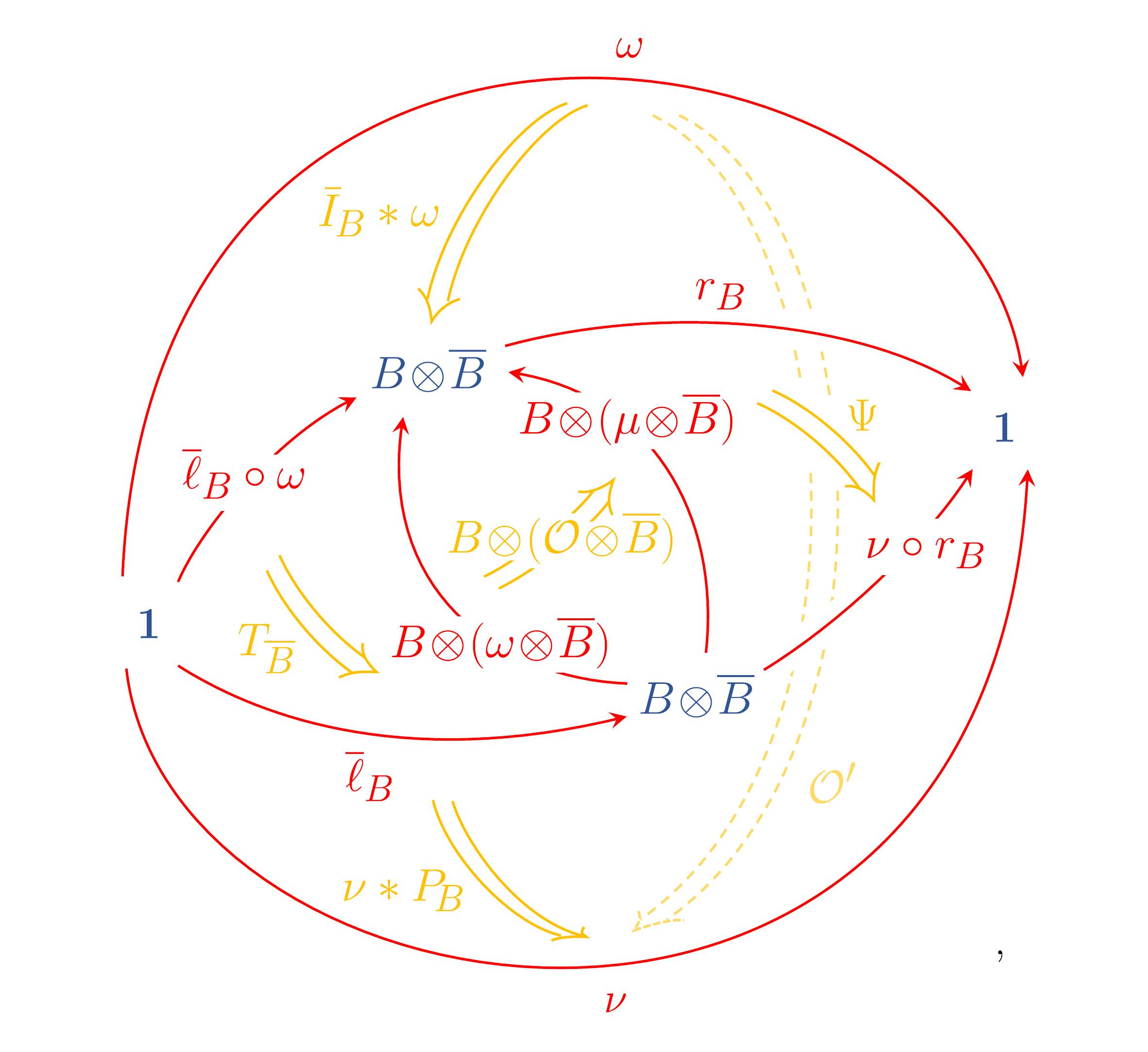}
\end{gathered}
\end{equation}
\end{itemize}
We expect that, up to equivalence, every tube representation is of the form $\mathcal{F} = \mathcal{F}_p$ for some $p \in \Omega \mathcal{Z}(\mathsf{C})$ and that the mapping $p \mapsto \mathcal{F}_p$ extends to an equivalence 
\begin{equation}\label{eq:3d-1-tube-reps-and-drinfeld}
\Omega \mathcal{Z}(\mathsf{C}) \; \cong \; [\T_{S^2}\C,\mathsf{Vec}] \, .
\end{equation}
of symmetric fusion categories.

Let us now explain the picture underpinning this equivalence. We first put forward an intrinsically three-dimensional perspective generalising the categorical perspective in our previous work~\cite{Bartsch:2023pzl}, before explaining the interpretation in the four-dimensional sandwich construction.

\begin{figure}[h]
	\centering
	\includegraphics[height=4.7cm]{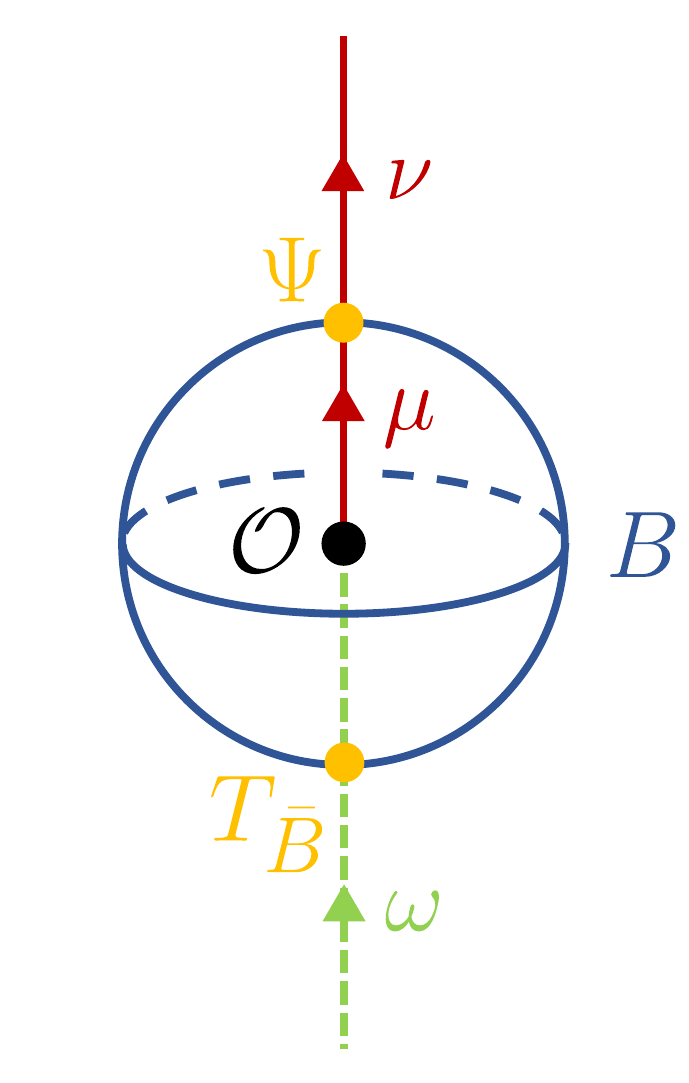}
	\vspace{-5pt}
	\caption{}
	\label{fig:3d-1-drinfeld-isomorphism}
\end{figure}

In the intrinsically three-dimensional perspective, local operators $\mathcal{O}$ in the $\mu$-twisted sector are viewed as junctions between an auxiliary topological line defect $\omega$ and $\mu$, as illustrated in figure \ref{fig:3d-1-drinfeld-isomorphism}. We then have the following interpretation:
\begin{itemize}
\item $\mathcal{F}_p(\mu)$ is identified with $\text{2-Hom}_{\mathsf{C}}(\omega,\mu)$.
\item  The action of a morphism $[B,\Psi]$ is obtained by surrounding $\cO$ with a spherical surface $B$ intersecting the auxiliary line $\omega$ via the associated component 2-morphism of $T$. Shrinking the sphere induces the linear map $\mathcal{F}_p(B,\Psi): \mathcal{F}_p(\mu) \to \mathcal{F}_p(\nu)$.
\end{itemize}

\begin{figure}[h]
	\centering
	\includegraphics[height=4.7cm]{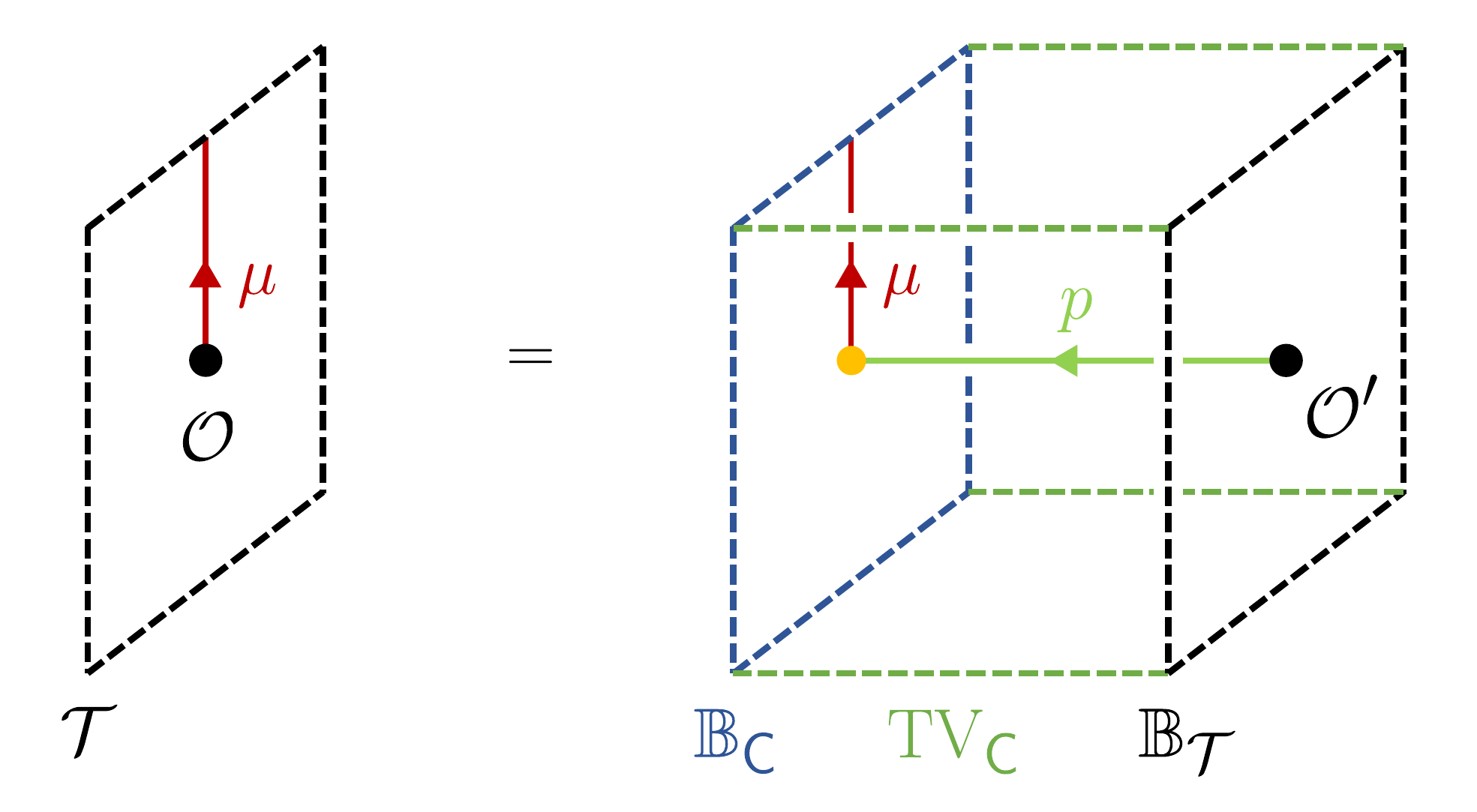}
	\vspace{-5pt}
	\caption{}
	\label{fig:3d-1-sandwich-construction}
\end{figure}

In the sandwich construction, we instead view a three-dimensional theory $\cT$ with spherical fusion 2-category symmetry $\C$ as an interval compactification of the four-dimensional topological theory $\text{TV}_{\mathsf{C}}$. Endomorphisms of the identity in the Drinfeld center are identified with the symmetric fusion category of topologigical lines
\begin{equation}
\text{TV}_\C(S^2) \; := \; \int_{S^2} \mathsf{C}  \; = \; \Omega \mathcal{Z}(\mathsf{C}) \, ,
\end{equation}
which is what the four-dimensional topological theory associates to $S^2$. The left boundary condition is the canonical gapped boundary condition $\mathbb{B}_\C$ with bulk-to-boundary functor $F : \cZ(\C) \to \C$, while the right boundary condition $\mathbb{B}_\cT$ contains informations about the theory $\cT$. This is illustrated in figure \ref{fig:3d-1-sandwich-construction}.

A tube representation $\cF_p \in [\T_{S^2}\C,\vect]$ with underlying vector space $\mathcal{F}_p(\mu)$ given by $\text{2-Hom}_{\mathsf{C}}(\omega,\mu)$ now corresponds to local operators on the gapped boundary condition $\mathbb{B}_\C$ that sit at the junction between a bulk topological line $p = (\omega,T) \in \Omega \mathcal{Z}(\C)$ and a boundary line $\mu \in \Omega \C$.

\subsection{Example: 2-group symmetry}

Let us now consider finite 2-group symmetry $\cG$ with 't Hooft anomaly $\lambda$\footnote{A discussion of 't Hooft anomalies for finite 2-group symmetries can be found in~\cite{Kapustin:2013uxa,Benini:2018reh}.}. This corresponds to the spherical fusion 2-category symmetry $\C = 2\vect^\lambda_\cG$. For concreteness, we specialise here to a split 2-group symmetry\footnote{This is sometimes also written as $\cG = A[1] \rtimes_{\rho} G$.} $\mathcal{G}=(G,A,\rho)$ consisting of
\begin{enumerate}
\item a finite 0-form symmetry group $G$,
\item a finite abelian 1-form symmetry group $A$,
\item a group action $\rho: G \to \text{Aut}(A)$ of $G$ on $A$.
\end{enumerate}
Furthermore, we restrict attention to a mixed 't Hooft anomaly between the 0-form and the 1-form component specified by a normalised 2-cocycle
\begin{equation}
\lambda \, \in \, Z^2_{\widehat{\rho}\,}(G,\widehat{A}) \, ,
\end{equation}
where $\widehat{A} := \text{Hom}(A,U(1))$ denotes the \textit{Pontryagin dual} of $A$.

We show that the associated tube algebra  is the twisted groupoid algebra
\begin{equation}
\mathcal{A}_{S^2}(\C)  \; = \; {}^{\theta(\lambda)}\bC[A /\! /_{\!\rho\, } G] \, .
\end{equation}
where $\theta(\lambda)$ denotes a groupoid 2-cocycle constructed from the 't Hooft anomaly $\lambda$. The linear representation theory of this tube algebra provides a concrete description of topological lines in the associated four-dimensional theory $\text{TV}_\C$, which is the four-dimensional Dijkgraaf-Witten theory constructed from $\cG$, $\lambda$.

\subsubsection{Symmetry 2-category}

The symmetry category $\C = 2\vect^\lambda_\cG$ is the spherical fusion 2-category of $\cG$-graded 2-vector spaces with 2-associator twisted by the presence of the 't Hooft anomaly $\lambda \, \in \, Z^2_{\widehat{\rho}\,}(G,\widehat{A})$. It has the following explicit description:
\begin{itemize}
\item The simple objects up to condensation\footnote{These are condensation defects for the 1-form symmetry $A$.} are one-dimensional $G$-graded 2-vector spaces $1_g$ with graded components $(1_g)_h = \delta_{g,h}$. Their fusion is given by
\begin{equation}
1_g \otimes 1_h \; = \; 1_{gh}
\end{equation}
with trivial associator and pentagonator.

\item The 1-morphism spaces between simple objects are fusion categories
\begin{equation}
\text{1-Hom}(1_g,1_h) \; = \; \delta_{g,h} \cdot \mathsf{Vec}_A \; ,
\end{equation}
whose simple objects are one-dimensional vector spaces $\mathbb{C}_a$ with graded components $(\mathbb{C}_a)_b = \delta_{a,b} \cdot \mathbb{C}$. 
The fusion of 1-morphisms $\mathbb{C}_a \in \text{1-End}(1_g)$ and $\mathbb{C}_b \in \text{1-End}(1_h)$ is given by
\begin{equation}
\mathbb{C}_a \, \otimes \, \mathbb{C}_b \; = \; \mathbb{C}_{a \, \cdot \, {}^gb} \; \in \; \text{1-End}(1_{gh}) \, ,
\end{equation}
where ${}^gb := \rho_g(b)$ denotes the action of $g \in G$ on $b \in A$. 

\item The 2-associator for the fusion of 1-morphisms is given by
\begin{equation}
\begin{gathered}
\includegraphics[height=3.7cm]{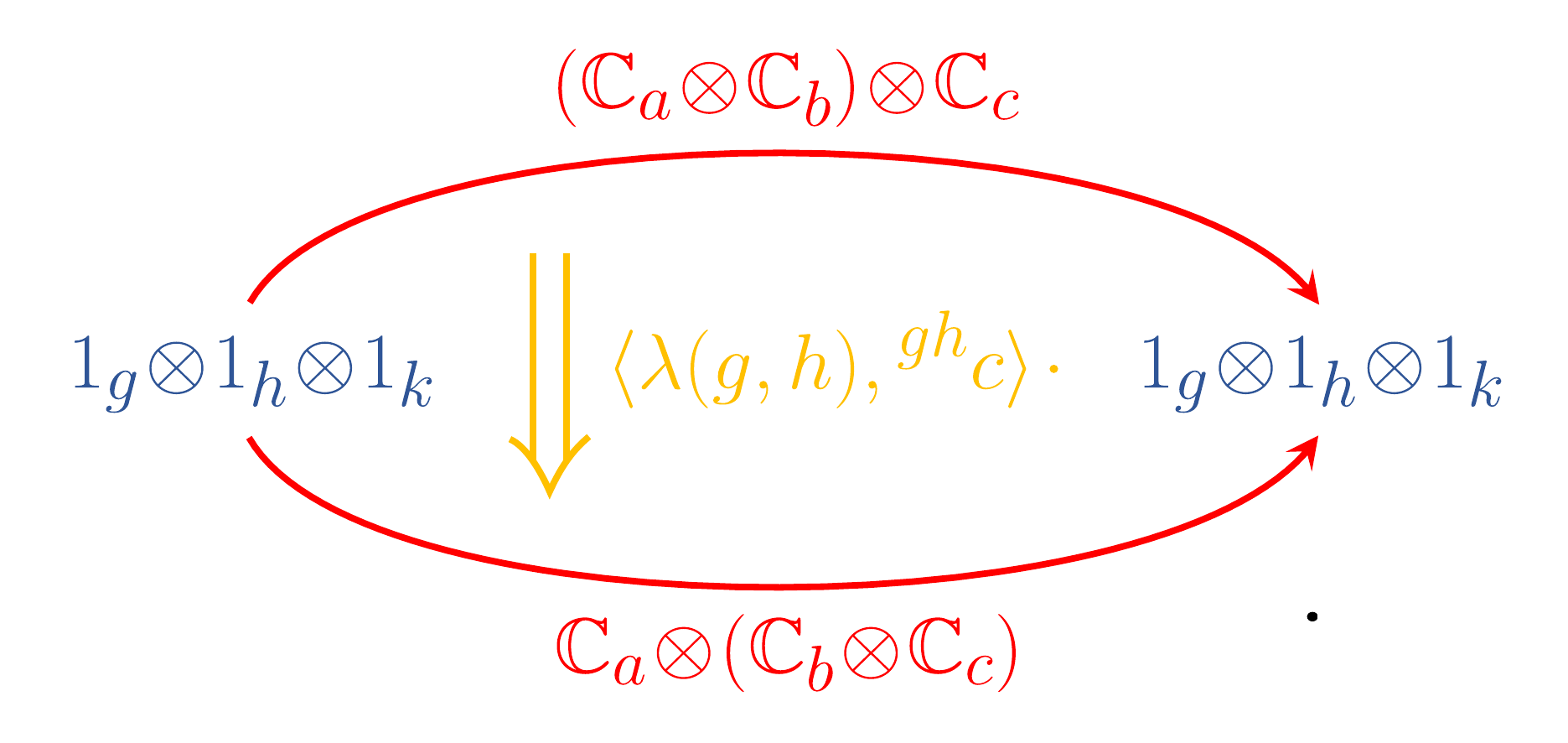}
\end{gathered}
\vspace{-6pt}
\end{equation}
\end{itemize}

In physical terms, simple objects up to condensations are topological surfaces generating the 0-form symmetry $G$, while simple 1-endomorphisms are topological lines generating the 1-form symmetry $A$. They correspond to the homotopy groups $\pi_0(\C)$ and $\pi_1(\C)$, respectively. 

\begin{figure}[h]
	\centering
	\includegraphics[height=7.8cm]{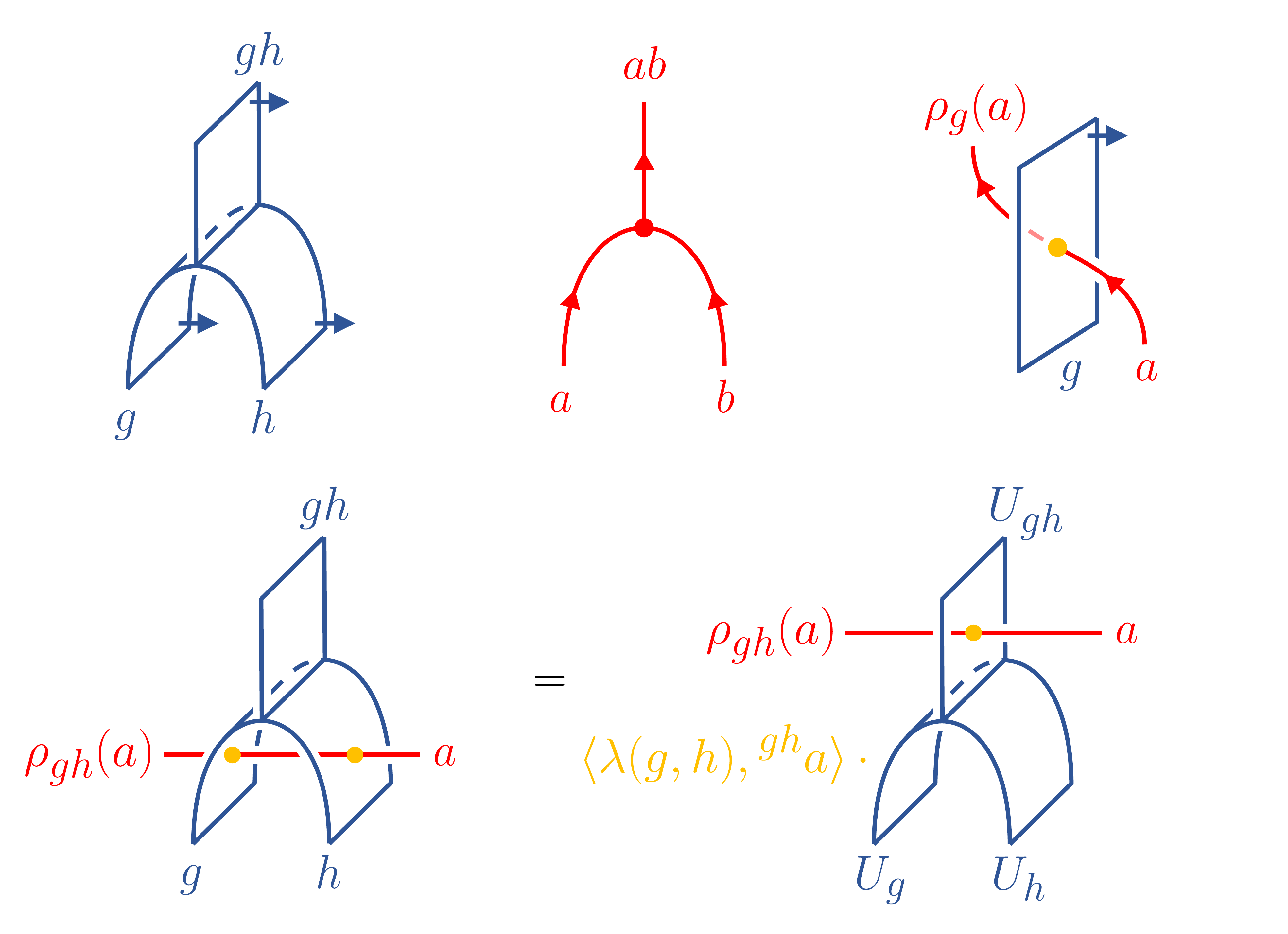}
	\vspace{-5pt}
	\caption{}
	\label{fig:3d-1-2-grpex-fusion+2-associator}
\end{figure}

Furthermore, passing a line $a \in A$ through a surface $g \in G$ transforms it into a new line $\rho_g(a) \in A$ determined by the homomorphism $\rho : G \to \text{Aut}(A)$. Passing a line $a \in A$ through the fusion of two surfaces labelled by $g,h \in G$ generates a multiplicative phase determined by the 't Hooft anomaly $\lambda$. This is illustrated in figure \ref{fig:3d-1-2-grpex-fusion+2-associator}.

\subsubsection{Tube algebra}

The associated tube algebra is generated by equivalences classes of pairs $[B,\Psi]$ where $B=1_g$ and $\Psi$ has a single graded component
\begin{equation}
\begin{gathered}
\includegraphics[height=2.7cm]{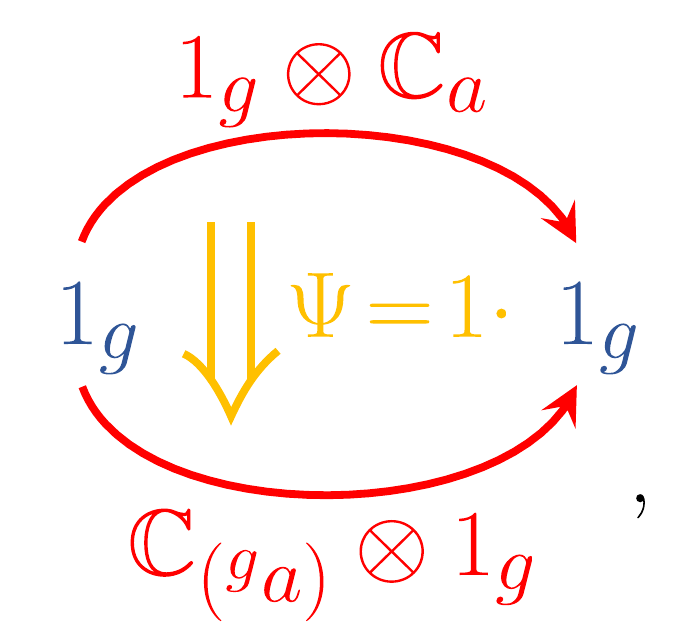}
\end{gathered}
\vspace{-2pt}
\end{equation}
From a physical perspective, this corresponds to intersecting a topological line $a \in A$ with a surface defect $g \in G$, whereupon the former is transformed into the line ${}^ga := \rho_g(a)$. Note that the generators do not include contributions from condensations for the 1-form symmetry $A$, which become trivial on $S^2$.

Continuing our notation from two dimensions, we denote the above generators by $\langle \yleftarrow{\;g\,}\! a \rangle$. Using the definition~\eqref{eq:3d-1-tube-composition}, their algebra product can be determined to be
\begin{equation} \label{eq:3d-1-tube-alg-product}
\braket{ \yleftarrow{\;g\,}\! a } \, \circ \, \braket{ \yleftarrow{\;h\,}\! b } \;\; = \;\;  \, \delta_{a, {}^{h}b}  \;\cdot\;\theta_b(\lambda)(g,h) \;\cdot\; \braket{ \yleftarrow{\;g \, \cdot \, h\,}\! b } \, ,
\end{equation}
where the multiplicative phase is given by
\be
\theta_b(\lambda)(g,h) \; = \; \braket{\lambda(g,h),{}^{gh}b} \, .
\ee
As a consequence of the 2-cocycle condition obeyed by $\lambda$, it satisfies
\begin{equation}
\frac{\theta_b(\lambda)(h,k) \cdot \theta_b(\lambda)(g,h k)}{\theta_b(\lambda)(g h,k) \cdot \theta_{({}^kb)}(\lambda)(g,h)} \; = \; 1 \, ,
\label{eq:3d-groupoid-2-cocycle}
\end{equation}
which ensures that the algebra product in (\ref{eq:3d-1-tube-alg-product}) is associative. Furthermore, equation (\ref{eq:3d-groupoid-2-cocycle}) defines a 2-cocycle 
\begin{equation}
\theta(\lambda) \, \in \, Z^2\big(A /\! /_{\!\rho\, } G, U(1)\big)
\end{equation}
on the action groupoid $A /\! /_{\!\rho\, } G$ associated to the group action $\rho : G \to \text{Aut}(A)$.

In summary, the tube algebra can be identified with the twisted groupoid algebra
\begin{equation}
\mathcal{A}_{S^2}(\C)  \; = \; {}^{\theta(\lambda)}\bC[A /\! /_{\!\rho\, } G] \, .
\end{equation}

\subsubsection{Tube representations}

A general tube representation $\mathcal{F}$ is a collection of finite-dimensional complex vector spaces $V_a := \mathcal{F}(\mathbb{C}_a)$ together with linear maps
\begin{equation}
\begin{gathered}
\includegraphics[height=1.1cm]{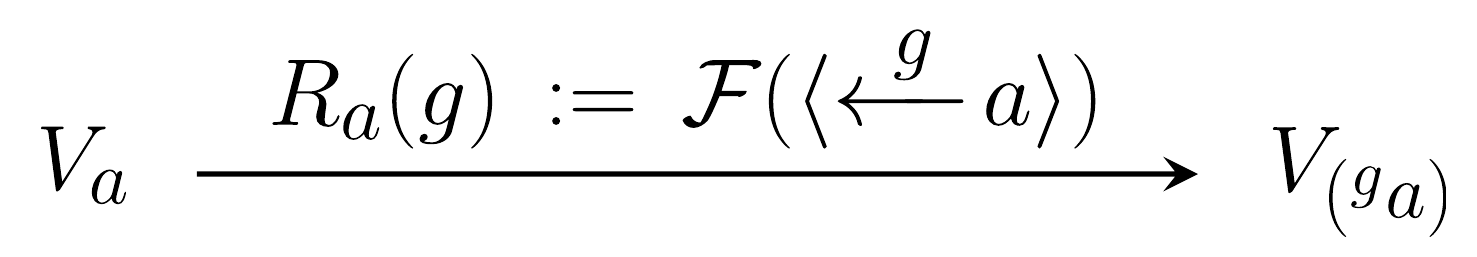}
\end{gathered}
\vspace{-3pt}
\end{equation}
satisfying the composition rule
\begin{equation}\label{eq:3d-2-grpex-tube-rep-composition-rule}
R_a(g) \, \circ \, R_b(h) \,\; = \,\; \delta_{a, {}^{h}b} \, \cdot \, \theta_b(\lambda)(g,h) \, \cdot \, R_b(gh) \, .
\end{equation}
This collection of data can be viewed as a vector bundle over the group $A$ equipped with a projective $G$-action that acts on the base via the group action $\rho$. 

From the composition rule (\ref{eq:3d-2-grpex-tube-rep-composition-rule}), it is clear that any tube representation $\mathcal{F}$ will decompose as a direct sum of representations supported on $G$-orbits in $A$. Let us therefore fix an orbit $[a] \in A/G$ with representative $a \in A$. If we restrict to group elements $g,h\in C_a(G)$ in the stabiliser of $a$, the composition rule becomes
\begin{equation}
R_a(g) \, \circ \, R_a(h) \;\; = \;\; \theta_a(\lambda)(g,h) \, \cdot \, R_a(gh) \, .
\end{equation}
Consequently, the map 
\begin{equation}
R_a : \; C_a(G) \, \to \, \text{End}(V_a) 
\end{equation}
defines an ordinary projective representation of $C_a(G)$ with 2-cocycle 
\begin{equation}
\theta_a(\lambda) \, \in \, Z^2(C_a(G),U(1)) \, .
\end{equation}
A tube representation supported on $[a] \in A/G$ is then irreducible if the associated projective representation $R_a$ of $C_a(G)$ is irreducible. Conversely, any irreducible projective representation of $C_a(G)$ determines an irreducible tube representation by induction.

In summary, the irreducible tube representations are determined by
\begin{enumerate}
\item a group element $a \in A$,
\item a projective representation of $C_a(G)$ with 2-cocycle 
\begin{equation}
\theta_a(\lambda)(g,h) \; = \; \braket{\lambda(g,h),a} \, .
\end{equation}
\end{enumerate}
Up to equivalence, this data depends only on the orbit $[a] \in A/G$ and the group cohomology class $[\theta_a(\lambda)] \in H^2(C_a(G),U(1))$. Note that tube representations supported in the identity orbit $[e] \in A / G$ reproduce the fact that genuine local operators transform in representations of the 0-form symmetry group $G$.

The above then sets up an equivalence
\be
 \rep\big( {}^{\theta(\lambda)}\bC[A /\! /_{\!\rho\, } G] \big) \; \cong \; \Omega\cZ(2\vect_\cG^\lambda)
\ee
between representations of the twisted groupoid algebra and topological lines in the four-dimensional Dijkgraaf-Witten theory constructed from $\cG$, $\lambda$, which is the four-dimensional topological theory $\text{TV}_\C$ associated to the symmetry category $\C = 2\vect_\cG^\lambda$.

\subsubsection{Gauge Theory}

Consider a pure gauge theory in three dimensions with a compact connected, simple gauge group $\mathbb{G}$. This has an anomaly-free split 2-group symmetry given by the automorphism 2-group of $\mathbb{G}$:
\begin{itemize}
\item Its 0-form component is charge conjugation symmetry $G = \text{Out}(\mathbb{G})$. Group elements are equivalence classes $[f] \in \text{Out}(\mathbb{G})$ of automorphisms $f \in \text{Aut}(\mathbb{G})$ modulo inner automorphisms. Thus $[f] = [f']$ iff $f' = f \circ \text{conj}_g$ for some $g \in \mathbb{G}$. 
\item Its 1-form component is the electric 1-form symmetry $A = Z(\mathbb{G})$.
\item The action $\rho: G \to \text{Aut}(A)$ of outer automorphisms $[f] \in \text{Out}(\mathbb{G})$ on central group elements $z \in Z(\mathbb{G})$ defined by $\rho_{[f]}(z) = f(z)$.
\end{itemize}
The 1-twisted sector local operators are fractional monopole operators on which topological Gukov-Witten line defects end. They will transform in irreducible tube representations of the automorphism 2-group.

Let us focus on $\mathbb{G} = \text{Spin}(4N)$ gauge theory with electric 1-form symmetry $A = \bZ_2 \times \bZ_2$, whose two factors are interchanged by the action of charge conjugation $G = \bZ_2$. The automorphism 2-group is therefore $\cG = (\bZ_2 \times \bZ_2)[1] \rtimes \bZ_2$. Let us denote the generator of charge conjugation symmetry by $c$ and the generators of the 1-form center symmetry by $a$ and $b$, so that ${}^ca = b$. There are five irreducible tube representations:
\begin{itemize}
\item There are four one-dimensional tube representations corresponding to a choice of 1-dimensional orbit $\{1\}$, $\{ab\}$ and an irreducible representation of the stabiliser $\mathbb{Z}_2$.
\item There is one two-dimensional tube representation corresponding to the 2-dimensional orbit $\{a,b\}$ with trivial stabiliser.
\end{itemize}

The one-dimensional tube representations simply capture the charge of fractional monopole operators in the untwisted and $ab$-twisted sectors that are invariant under charge conjugation symmetry. The two-dimensional tube representation captures a pair of fractional monopole operators in the $a$-twisted and $b$-twisted sectors that are exchanged under charge conjugation. An example of the latter is given by fractional monopole operators labelled by the spinor representations $S^+$ and $S^-$ of $\mathbb{G} = \text{Spin}(4N)$.

These irreducible tube representations are in 1:1-correspondence with irreducible representations of the finite group $D_8$. This is not a coincidence. Upon gauging the 1-form symmetry $A = \bZ_2 \times \bZ_2$, we obtain a $\text{PSO}(4N)$ gauge theory with symmetry group $D_8$ combining charge conjugation and the dual $\bZ_2 \times \bZ_2$ magnetic symmetry. The fractional monopole operators become genuine monopole operators after gauging and therefore transform in irreducible representations of $D_8$.

\subsection{Example: Ising-like symmetry}\label{sec:ising-sphere}

The analysis of tube representations for a 2-group symmetry in three dimensions provides a stepping stone to a vast number of non-invertible symmetries. Beginning with a three-dimensional theory with 2-group symmetry $\cG$ and 't Hooft anomaly $\alpha$, we may gauge an anomaly-free 2-subgroup $\cH \subset \cG$ with trivialisation $\alpha|_{\cH}= d\phi$, which generates the class of group-theoretical fusion 2-categories
\begin{equation}
\C \; = \; \C(\cG,\alpha \, | \, \cH , \phi) \, ,
\label{eq:grp-th-2fusion}
\end{equation}
introduced in~\cite{Bartsch:2022mpm}. This provides many non-invertible examples whose structure is simple to determine using group-theoretical techniques.

The above symmetry categories all arise as gapped boundary conditions of a common four-dimensional Dijkgraaf-Witten theory associated to $\cG$ and $\alpha$. In other words, $\text{TV}_\C$ is the same four-dimensional Dijkgraaf-Witten theory for all group theoretical fusion 2-categories with fixed $\cG$, $\alpha$. In particular, they share a common Drinfeld center. Thus, while the associated tube categories $\T_{S^2}\C$ and tube algebras $\cA_{S^2}(\C)$ may be wildly different, their categories of tube representations are equivalent. In practice, this highly non-trivial equivalence may be determined explicitly by tracking tube representations under the gauging of 2-subgroups $\cH \subset \cG$.

\subsubsection{Ising-like symmetry category}

In the following, we will content ourselves with a simple example of a group-theoretical fusion 2-category that shares some similarities with the Ising fusion category in two dimensions. Let us start from a theory with ordinary group symmetry $G = D_8 \cong D_4 \rtimes \bZ_2$ and symmetry category $\C = 2\vect_{D_8}$. There are no 1-twisted sectors and tube representations are simply representations of $D_8$ describing the transformation behaviour of genuine local operators. 

We now gauge the non-normal subgroup $H = \mathbb{Z}_2$. This results in the non-invertible group-theoretical fusion 2-category 
\begin{equation}
\C \; = \; 2\rep(D_4[1] \rtimes \bZ_2) \, .
\end{equation}
A summary of this 2-category can be found in~\cite{Bartsch:2022mpm}. Here, we content ourselves with the following brief description:
\begin{itemize}
\item There are three connected components
$\pi_0(\C) = \{1,V,D\}$, whose fusion rules on $S^2$ (obtained by setting condensations to be trivial) are given by
\begin{equation}
V \otimes V \, = \, 1 \, , \qquad V \otimes D \, = \, D \, , \qquad D \otimes D \, = \, 1 \oplus V \, . 
\label{eq:3d-ising-like-fusion}
\end{equation}
Note that $V$ generates an unbroken $\bZ_2$ 0-form symmetry.

\item The associated 1-endomorphism categories are given by
\begin{equation}
\text{1-End}_\C(1) \, = \, \text{1-End}_\C(V) \, = \,  \vect_{\bZ_2} \, , \qquad \text{End}_\C(D) = \vect \, .
\end{equation}
The first two are generated by a non-trivial topological line $\gamma$ which is the non-trivial Wilson line for the gauged $\mathbb{Z}_2$-symmetry. In particular, $\pi_1(\mathsf{C}) = \lbrace 1, \gamma \rbrace$. The restriction of $\gamma$ onto $D$ gives the identity line on $D$.
\end{itemize}

\subsubsection{Tube algebra}

The associated tube algebra is generated by equivalence classes of pairs $[B,\Psi]$ corresponding to the following combinations of an object $B$ and an intersection 2-morphism $\Psi$: 
\begin{itemize}
    \item For $\mu \in \pi_1(\mathsf{C})$, we denote by $\genfour[1]{\mu}{}$ the case $B = 1$ and $\Psi = \text{Id}_{\mu}$.
    
\item For $\mu \in \pi_1(\mathsf{C})$, we denote by $\genfour[V]{\mu}{}$ the case $B = V$ and $\Psi = \text{Id}_{\mu} \otimes \text{Id}_V^2$.
    
\item For $\mu, \nu \in \pi_1(\mathsf{C})$, we denote by $\genfour[D]{\mu}{\nu}$ the case $B = D$ and $\Psi = \text{Id}^2_D$. 
\end{itemize}
\vspace{4pt}
The algebra product of these generators can be computed using \eqref{eq:1-tube-composition} and knowledge of the $2$-associator. For example, we find 
\begin{equation}
\begin{aligned}
\genfour[V]{\mu}{} \, \circ \, \genfour[V]{\mu}{} \;\, &= \;\, \genfour[1]{\mu}{}\, , \\[4pt]
\genfour[D]{\nu}{\mu} \, \circ \, \genfour[D]{\mu}{\nu} \;\, &= \;\, \genfour[1]{\mu}{} \, +\, (-1)^{\delta_{\nu,\gamma}} \cdot \genfour[V]{\mu}{} \, .
\end{aligned}
\vspace{4pt}
\end{equation}
Note that these show similarities to the tube algebra of the Ising category in two dimensions and are compatible with the fusion rules on $S^2$.

\subsubsection{Tube representations}

We now describe the tube representations. On general grounds, irreducible tube representations are in 1:1-correspondence with irreducible representations of $D_8$. They may be determined by starting from genuine local operators in the theory with $D_8$ symmetry and gauging the non-normal subgroup $\bZ_2 \subset D_8$. The five irreducible representations of $D_8$ then translate into the following tube representations:
\begin{itemize}
\item There are two one-dimensional irreducible tube representations $\mathcal{F}_1^{\pm}$ with untwisted sector $\mathcal{F}_1^{\pm}(1) = \mathbb{C}$ and non-trivial generator actions
\begin{equation}
\mathcal{F}^{\pm}_1\big(\genfour[V]{1}{}\big) \;  = \; 1 \, , \qquad
\mathcal{F}^{\pm}_1\big(\genfour[D]{1}{1}\big) \;  = \; \pm \sqrt{2} \, .
\end{equation}
These correspond to genuine local operators that are not charged under the $\bZ_2$ symmetry generated by $V$ and that remain genuine after the action of $D$.

\item There are two one-dimensional irreducible tube representations $\mathcal{F}_{\gamma}^{\pm}$ with untwisted sector $\mathcal{F}_{\gamma}^{\pm}(\gamma) = \mathbb{C}$ and non-trivial generator actions
\begin{equation}
\mathcal{F}_{\gamma}^{\pm}\big(\genfour[V]{\gamma}{}\big) \;  = \; -1\, , \qquad
\mathcal{F}_{\gamma}^{\pm}\big(\genfour[D]{\gamma}{\gamma}\big) \;  = \; \pm \sqrt{2} \, .
\end{equation} 
These correspond to $\gamma$-twisted sector local operators that are charged under the $\bZ_2$ symmetry generated by $V$ and that remain twisted sector operators after the action of the non-invertible defect $D$.

\item There is one two-dimensional irreducible tube representation $\mathcal{F}$ with twisted sectors $\mathcal{F}(1) = \mathcal{F}(\gamma) = \mathbb{C}$ and non-trivial generator actions 
\begin{equation}
\begin{aligned}
& \mathcal{F}\big(\genfour[V]{1}{}\big) \; = \; \begin{pmatrix} -1 & 0 \\ 0 & 0 \end{pmatrix} \, , \qquad &&\mathcal{F}\big(\genfour[V]{\gamma}{}\big) \; = \; \begin{pmatrix} 0 & 0 \\ 0 & -1 \end{pmatrix} \, ,  \\[6pt]
& \mathcal{F}\big(\genfour[D]{1}{}\big) \; = \; \begin{pmatrix} 0 & 0 \\ \sqrt{2} & 0 \end{pmatrix} \, ,  &&\mathcal{F}\big(\genfour[D]{\gamma}{}\big) \; = \; \begin{pmatrix} 0 & \sqrt{2} \\ 0 & 0 \end{pmatrix} \, . 
\end{aligned}
\end{equation}
This corresponds to an untwisted and a $\gamma$-twisted sector local operator which are charged under the $\mathbb{Z}_2$ symmetry generated by $V$ and exchanged by the action of the non-invertible defect $D$.
\end{itemize}
Note that there are no one-dimensional tube representations with a genuine local operator charged under the $\bZ_2$ symmetry generated by $V$. Such a local operator must be transformed into a $\gamma$-twisted operator under the action of the non-invertible symmetry defect $D$.

\subsubsection{Gauge theory}

The above scenario is straightforward to realise in the context of gauge theory. Starting with a three-dimensional pure gauge theory with gauge group $\mathbb{G} = PSO(4N)$, we have an ordinary symmetry group $G = D_8 \cong D_4 \rtimes \bZ_2$ consisting of
\begin{itemize}
\item magnetic symmetry $\pi_1(\mathbb{G})^\vee = D_4$,
\item charge conjugation symmetry $\text{Out}(\mathbb{G}) = \bZ_2$.
\end{itemize}
This theory has genuine monopole operators labelled by representations of the Langlands dual group ${}^L \mathbb{G} = Spin(4N)$, which transform in irreducible representations of $D_8$. For example, consider the pair of genuine monopole operators $S^+$ and $S^-$ labelled by the spinor and conjugate spinor representations of $Spin(4N)$. They are charged under the two factors of the magnetic symmetry $D_4 = \bZ_2 \times \bZ_2$ respectively and exchanged under charge conjugation. This is the two-dimensional irreducible representation of $D_8$.

Now gauging the non-normal charge conjugation subgroup $\mathbb{Z}_2 \subset D_8$ results in a gauge theory with gauge group $\mathbb{G}' = PO(4N)$, whose symmetry category is of the aforementioned Ising-like non-invertible type. Concretely, the diagonal combination in $D_4 = \bZ_2 \times \bZ_2$ remains an invertible $\mathbb{Z}_2$ symmetry generated by $V$, while the broken combination becomes the non-invertible symmetry $D$.

The genuine monopole operators charged under charge conjugation now become fractional or twisted sector monopole operators and transform in irreducible tube representations as above. For example, we have that
\begin{itemize}
\item $S^++S^-$ remains an untwisted monopole operator,
\item $S^+-S^-$ becomes a $\gamma$-twisted monopole operator.
\end{itemize}
They are both charged under the $\bZ_2$ symmetry generated by $V$ and exchanged by the non-invertible symmetry $D$. This is precisely the two-dimensional irreducible tube representation described above.

\subsection{Example: braiding lines}
\label{subsec:3d-modB}

Let us now consider a three-dimensional theory with topological lines described by a braided fusion 1-category $\B$. This corresponds to symmetry 2-category
\be
\mathsf{C} = \text{Mod}(\mathsf{B}) \, .
\ee
Fusion 2-categories of this type are \textit{connected} in the sense that
$\pi_0(\mathsf{C}) = \lbrace \mathbf{1} \rbrace$. In other words, the only topological surfaces are condensations obtained by condensing topological lines in $\mathsf{B}$ on the identity surface. 

\begin{figure}[h]
	\centering
	\includegraphics[height=2.7cm]{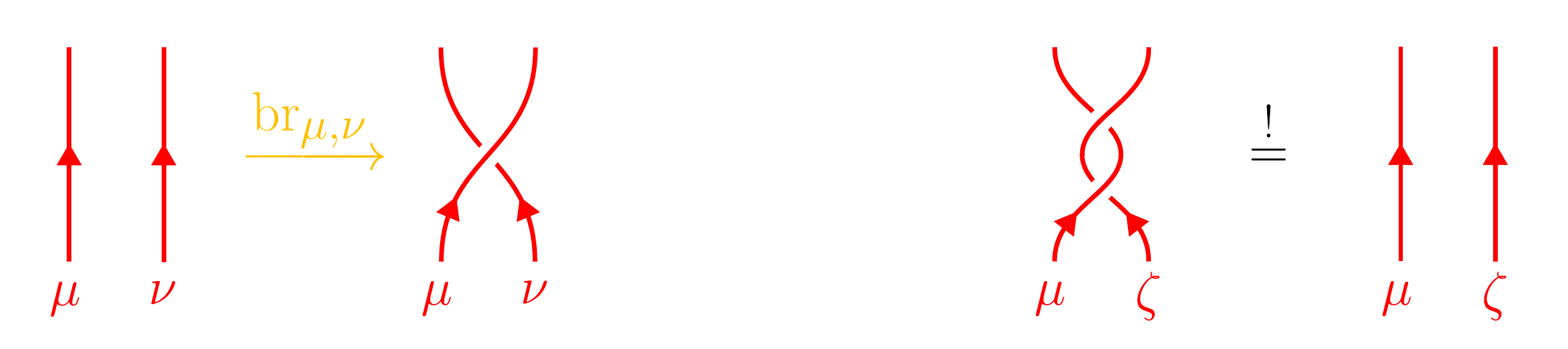}
	\vspace{-5pt}
	\caption{}
	\label{fig:3d-1-braiding}
\end{figure}

The braided fusion category $\mathsf{B}$ is equipped with isomorphisms
\vspace{-4pt}
\begin{equation}
\begin{gathered}
\includegraphics[height=1.1cm]{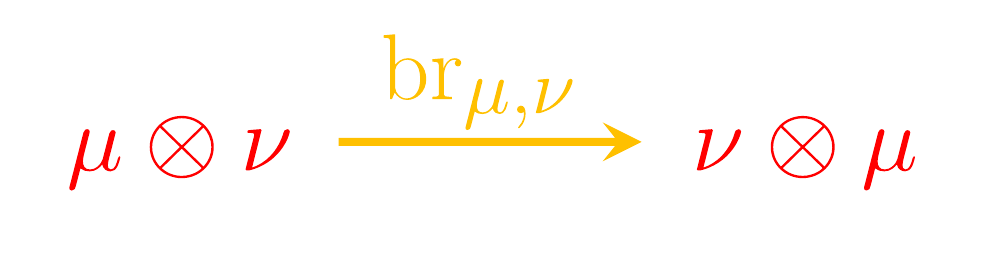}
\end{gathered}
\vspace{-8pt}
\end{equation}
for each $\mu,\nu \in \mathsf{B}$ that describe the braiding of topological lines, as illustrated on the left of figure \ref{fig:3d-1-braiding}. The \textit{M\"uger center} $\mathcal{Z}_2(\mathsf{B}) \subset \mathsf{B}$ is then defined as the sub-category of lines $\zeta \in \mathsf{B}$ that braid trivially in the sense that the diagram
\vspace{-4pt}
\begin{equation}
\begin{gathered}
\includegraphics[height=3cm]{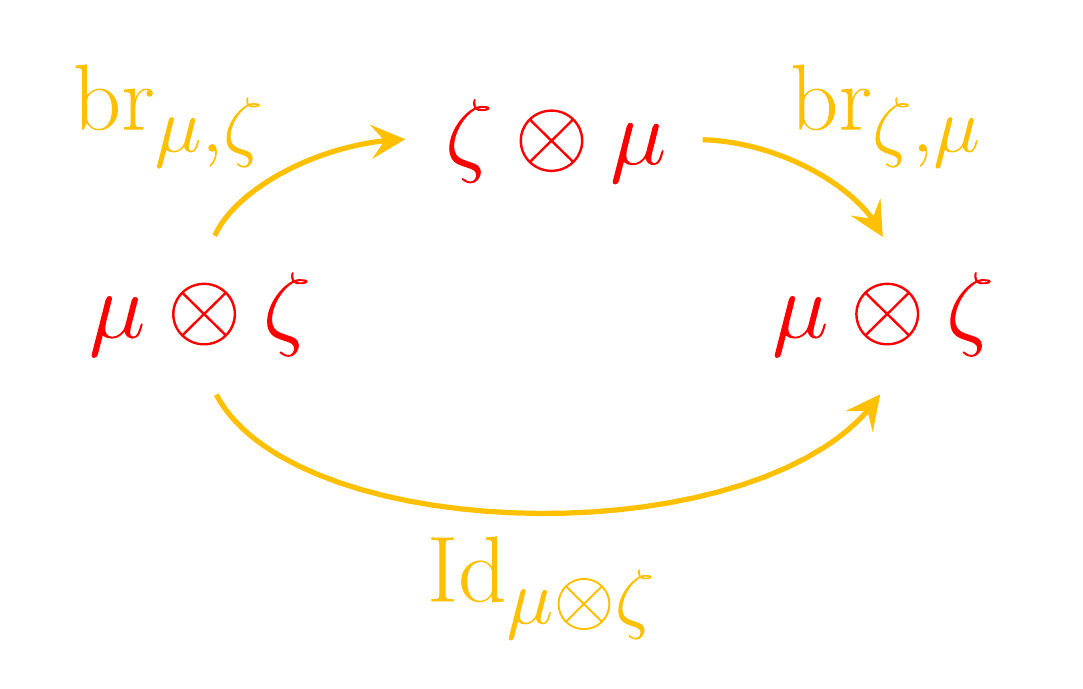}
\end{gathered}
\vspace{-6pt}
\end{equation}
commutes for all $\mu \in \mathsf{B}$, as illustrated on the right-hand side of figure \ref{fig:3d-1-braiding}. 
This is related to the Drinfeld centre of the symmetry category by the symmetric equivalence
\begin{equation}
\Omega \mathcal{Z}(\mathsf{C}) \; \cong \; \mathcal{Z}_2(\mathsf{B}) \, ,
\end{equation}
as discussed in~\cite{Johnson_Freyd_2023}. 

We therefore conclude that for a connected fusion 2-category $\C = \mathsf{Mod}(\B)$ there is an equivalence
\be
[T_{S^2}\C , \vect] \cong \cZ_2(\B)
\ee
between the representations of the tube category of $\C$ and the M\"uger center of $\B$. From a physical perspective, this equivalence simply captures the fact that topological lines with a non-trivial braiding cannot end, and are thus obstructed from having twisted sectors. The two extreme cases of this are as follows:
\begin{itemize}
\item $\B$ is \emph{non-degenerate} if $\cZ_2(\B) = \vect$. In this case, all topological lines braid and there are no twisted sector local operators. Tube representations simply correspond to the vector spaces of genuine local operators.
\item $\B$ is \emph{symmetric} if $\cZ_2(\B) = \B$. In this case, there are no constraints on twisted sector local operators and tube representations simply enumerate the vector spaces of local operators on which topological lines end.
\end{itemize}
In all cases, the tube representations or objects in $\cZ_2(\B)$ simply enumerate the possible non-trivial twisted sectors. 

A straightforward explicit example is an anomalous 1-form symmetry $A$, in which case all topological lines are invertible. This corresponds to choosing $\B = \vect^q_A$ for an abelian group $A$ and an abelian 3-cocyle 
\begin{equation}
q \, \in \, Z_{\text{ab}}^3(A,U(1)) \, .
\end{equation}
The associated fusion 2-category may be denoted $\C =  2\vect^q_{A[1]}$, where by an abuse of notation the 't Hooft anomaly $q$ is now regarded as a quadratic form for the 1-form symmetry $A$ via the chain of isomorphisms
\begin{equation}
H_{\text{ab}}^3(A,U(1)) \; \cong \; H^4(B^2A ,U(1)) \; \cong \; \text{Hom}(\Gamma(A),U(1)) \, .
\end{equation} 
The associated M\"uger center is simply given by
\begin{equation}
\mathcal{Z}_2(\mathsf{Vec}_A^q) \; = \; \mathsf{Vec}_{A^{\perp}} \, ,
\end{equation}
where $A^{\perp} \subset A$ denotes the orthogonal complement of $A$ with respect to the symmetric bilinear form
\begin{equation}
\braket{a,b}_q \; := \; \frac{q(a b)}{q(a)q(b)}
\end{equation}
induced by the quadratic form $q$.

The two extreme cases are as follows. If the 1-form symmetry $A$ is completely anomalous (meaning $\braket{.,.}_q$ is non-degenerate), there are no twisted sectors. If on the other hand the 1-form symmetry is anomaly-free (meaning $\braket{.,.}_q$ is trivial), there are no constraints on twisted sectors.

A non-invertible example is the symmetric braided fusion category $\B = \rep(G)$. The corresponding symmetry 2-category $\C = \mathsf{2Rep}(G)$ is obtained by gauging a finite non-abelian symmetry group $G$ in three dimensions~\cite{Bartsch:2022mpm,Bhardwaj:2022lsg}. In this case, $\cZ_2(\B) = \rep (G)$ and there are no constraints on twisted sectors: tube representations simply enumerate twisted sectors for topological Wilson lines.


\section{Three dimensions: line operators}
\label{sec:3d-lines}

In this section, we continue studying a theory with a spherical fusion 2-category symmetry $\C$ in three dimensions. However, we now consider the action on twisted sector line defects attached to topological surfaces. 

We will motivate and introduce the tube 2-category $\T_{S^1}\C$ and tube 2-algebra $\cA_{S^1}(\C)$ associated to the manifold $S^1$ linking a line in three dimensions. We show that twisted sector line defects transform in irreducible 2-representations thereof, which are in 1:1-correspondence with topological surfaces in the sandwich construction. We propose that this extends to an equivalence
\be
[\T_{S^1}\C,2\vect] \; \cong \; \text{TV}_\C(S^1)\;  := \; \int_{S^1} \C \; = \; \mathcal{Z}(\C) \, .
\ee
of braided fusion 2-categories.

We illustrate these ideas in two examples. The first is a finite 2-group symmetry with 't Hooft anomaly, in which case the tube 2-algebra is a higher analogue of a twisted Drinfeld double (this includes ordinary finite 0-form and 1-form symmetries). The second describes general non-invertible topological lines captured by a braided fusion category.

\subsection{Twisted sector lines}

\begin{figure}[h]
	\centering
	\includegraphics[height=3.2cm]{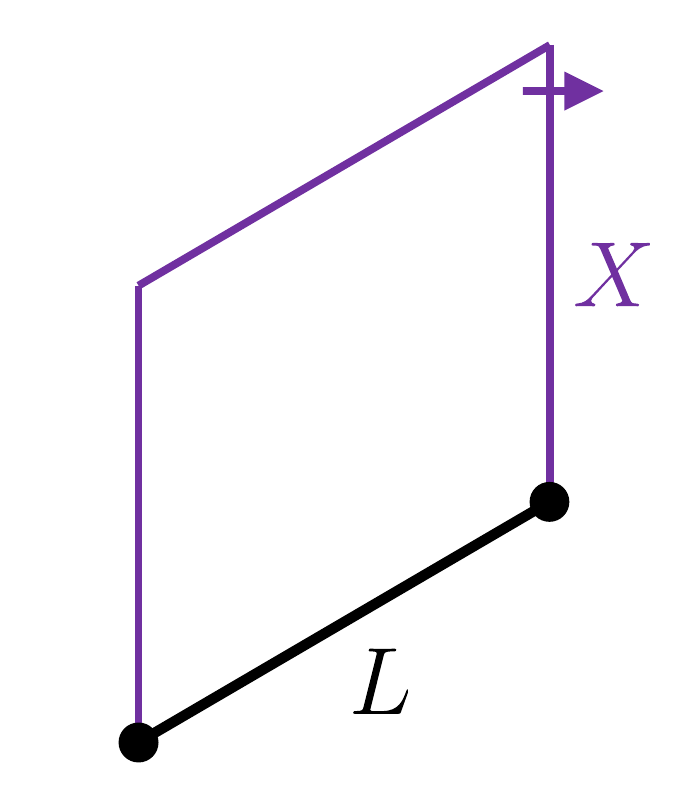}
	\vspace{-5pt}
	\caption{}
	\label{fig:3d-twisted-sectors}
\end{figure}

We consider \textit{2-twisted sector line defects}, which are line defects $L$ attached to an oriented topological surface $X \in \mathsf{C}$, as illustrated in figure \ref{fig:3d-twisted-sectors}. Line defects of this type are said to be in the \textit{$X$-twisted sector}. The line defects themselves need not be topological and correlation functions may depend on where they are inserted.

Following the perspective introduced in our previous work~\cite{Bartsch:2023pzl}, any finite collection of such \emph{simple} line operators $L$ may be regarded as a 2-vector space
\begin{equation}\label{eq:3d-F-objects}
    \mathcal{F}(X) \, \in \, \mathsf{2Vec} \, .
\end{equation}
It is convenient (but not necessary) to assume that the line operators are \emph{reduced}. Up to equivalence, the 2-vector space $\mathcal{F}(X)$ is then completely determined by the cardinality of the collection of line operators under consideration.

The action of a spherical fusion 2-category $\mathsf{C}$ on 2-twisted sector line operators now has two layers of structure, corresponding to wrapping with topological surfaces (i.e. objects of $\mathsf{C}$) and linking with topological lines (i.e. 1-morphisms in $\mathsf{C}$). We will consider the two actions in turn.

\subsubsection{Wrapping action}
\label{subsubsec:3d-wrapping-action}

\begin{figure}[h]
	\centering
	\includegraphics[height=4.5cm]{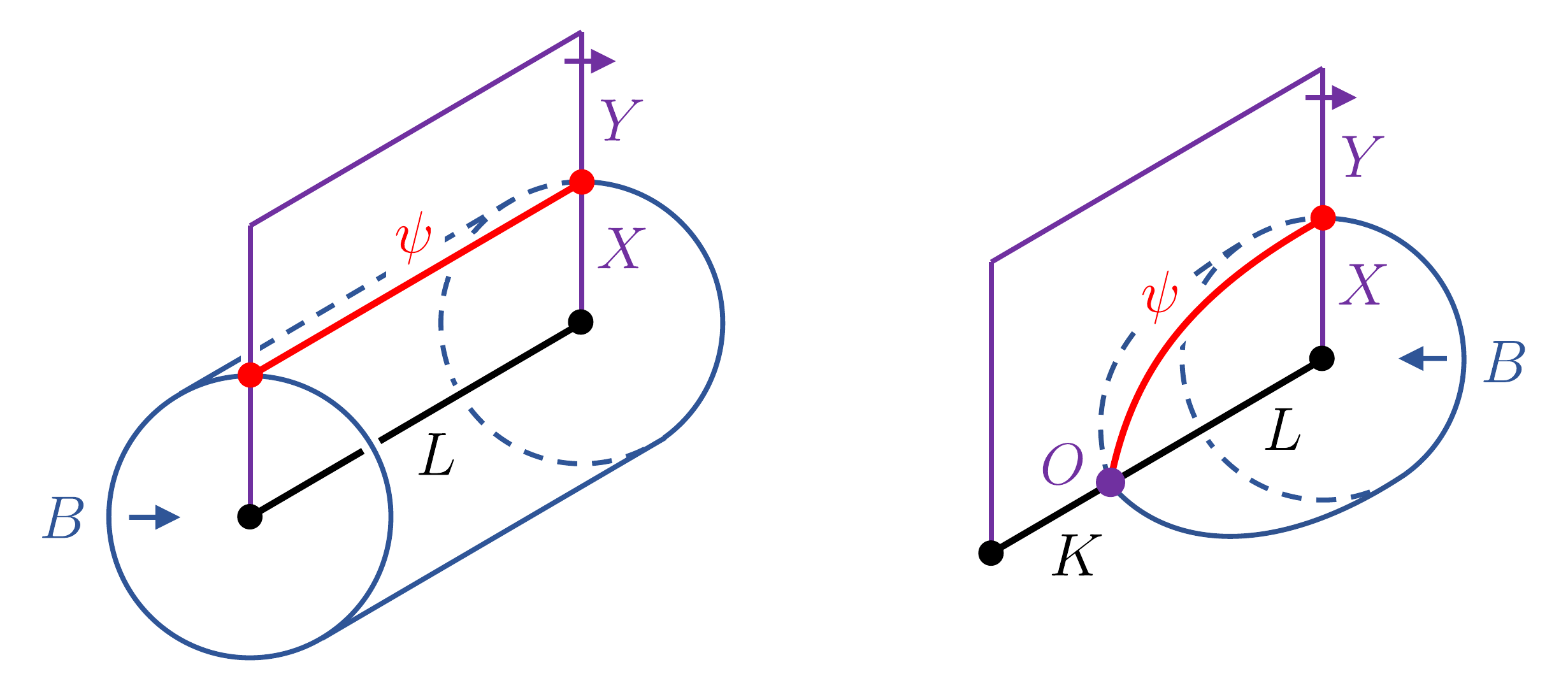}
	\vspace{-5pt}
	\caption{}
	\label{fig:3d-line-action}
\end{figure}

Topological surface defects $B \in \mathsf{C}$ act on $X$-twisted sector lines $L \in \mathcal{F}(X)$ by wrapping them with a cylinder $C^2 \cong \mathbb{R} \times S^1$, as shown on the left of figure \ref{fig:3d-line-action}. Due to the topological surface $X$ attached to $L$, this requires a choice of 1-morphism 
\vspace{-10pt}
\begin{equation}\label{eq:3d-tube-morphisms}
\begin{gathered}
\includegraphics[height=1.18cm]{2d-dia-morphism.pdf}
\end{gathered}
\vspace{-6pt}
\end{equation}
that specifies how $B$ intersects $X$ and transforms it into a new surface $Y$.

Upon shrinking the cylinder down onto $L$, we generate a point-like junction between $L$ and a line $K$ in the $Y$-twisted sector, as illustrated on the right of figure \ref{fig:3d-line-action}. The set of topological operators $O$ at this junction forms a finite-dimensional vector space, which we denote by 
\begin{equation}
\mathcal{F}(B,\psi)_{K,L} \; \in \; \mathsf{Vec} \; .
\end{equation} 
By scanning over $L$ and $K$, we then obtain a matrix $\mathcal{F}(B,\psi)$ of vector spaces determined by the pair $(B,\psi)$, which we interpret as a 1-morphism
\vspace{-8pt}
\begin{equation}\label{eq:3d-F-1-morphisms}
\begin{gathered}
\includegraphics[height=1.18cm]{2d-dia-F-morphism.pdf}
\end{gathered}
\vspace{-8pt}
\end{equation}
in the 2-category $\mathsf{2Vec}$ of 2-vector spaces. 

\begin{figure}[h]
	\centering
	\includegraphics[height=7.5cm]{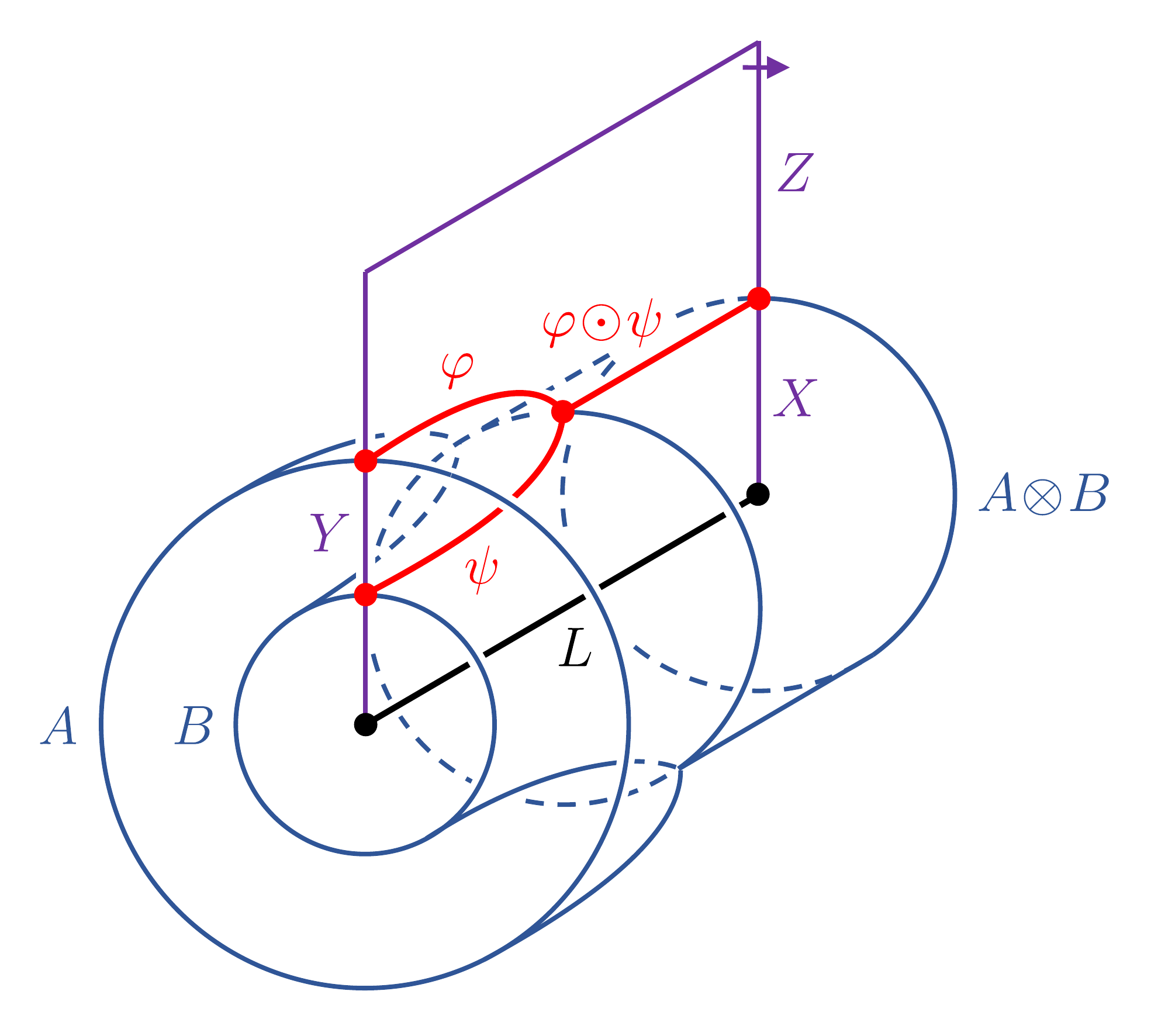}
	\vspace{-5pt}
	\caption{}
	\label{fig:3d-pseudo-2-morphism}
\end{figure}

The compatibility of this 1-morphism with the consecutive action of two symmetry defects $A$ and $B$ is now implemented by a 2-morphism
\vspace{3pt}
\begin{equation}\label{eq:3d-F-pseudo-morphisms}
\begin{gathered}
\includegraphics[height=3.3cm]{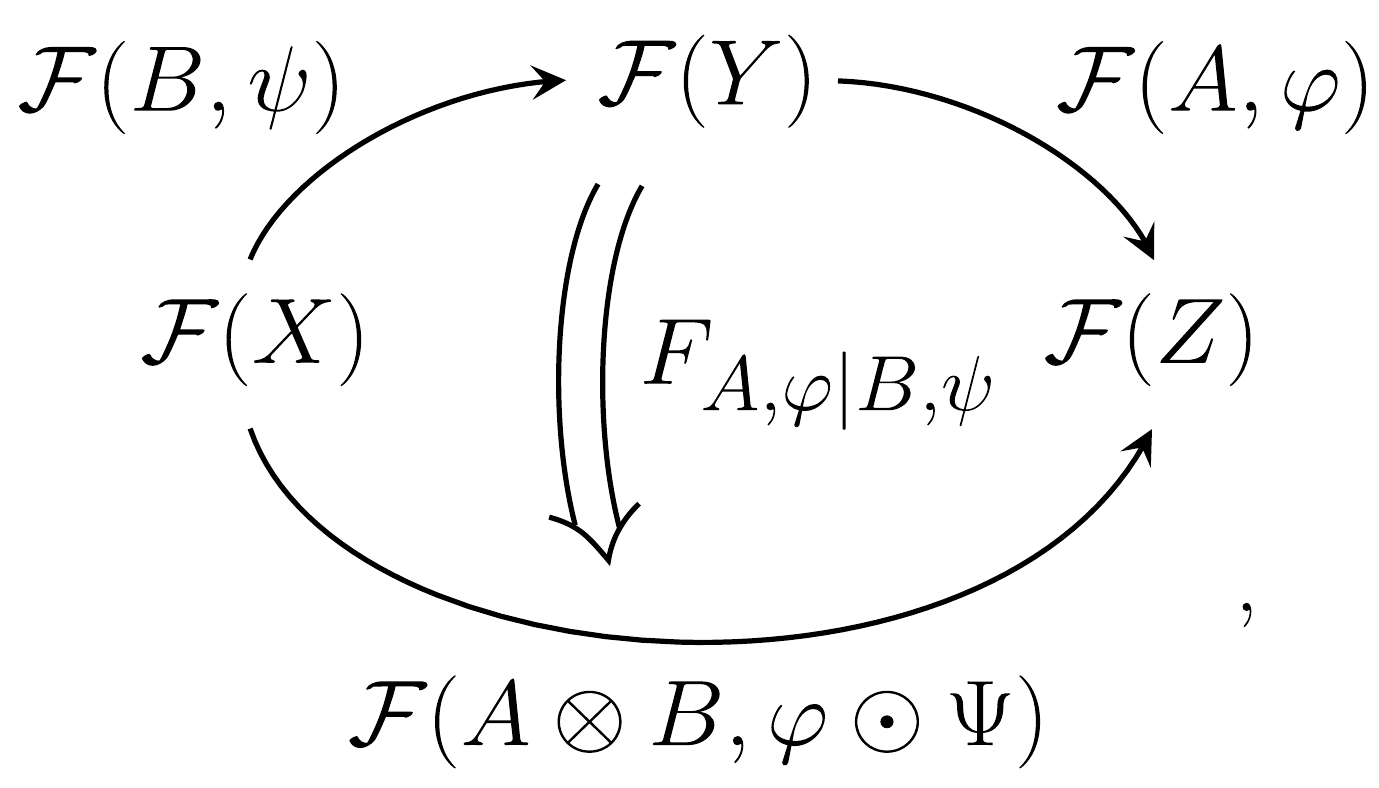}
\end{gathered}
\vspace{-2pt}
\end{equation}
where the 1-morphism $\varphi \odot \psi$ is defined analogously to equation (\ref{eq:tube-composition}). Physically, this is interpreted as the matrix of linear maps that relate the consecutive wrapping action of $A$ and $B$ to the wrapping action of $A\otimes B$, as illustrated in figure \ref{fig:3d-pseudo-2-morphism}.

\begin{figure}[h]
	\centering
	\includegraphics[height=14.8cm]{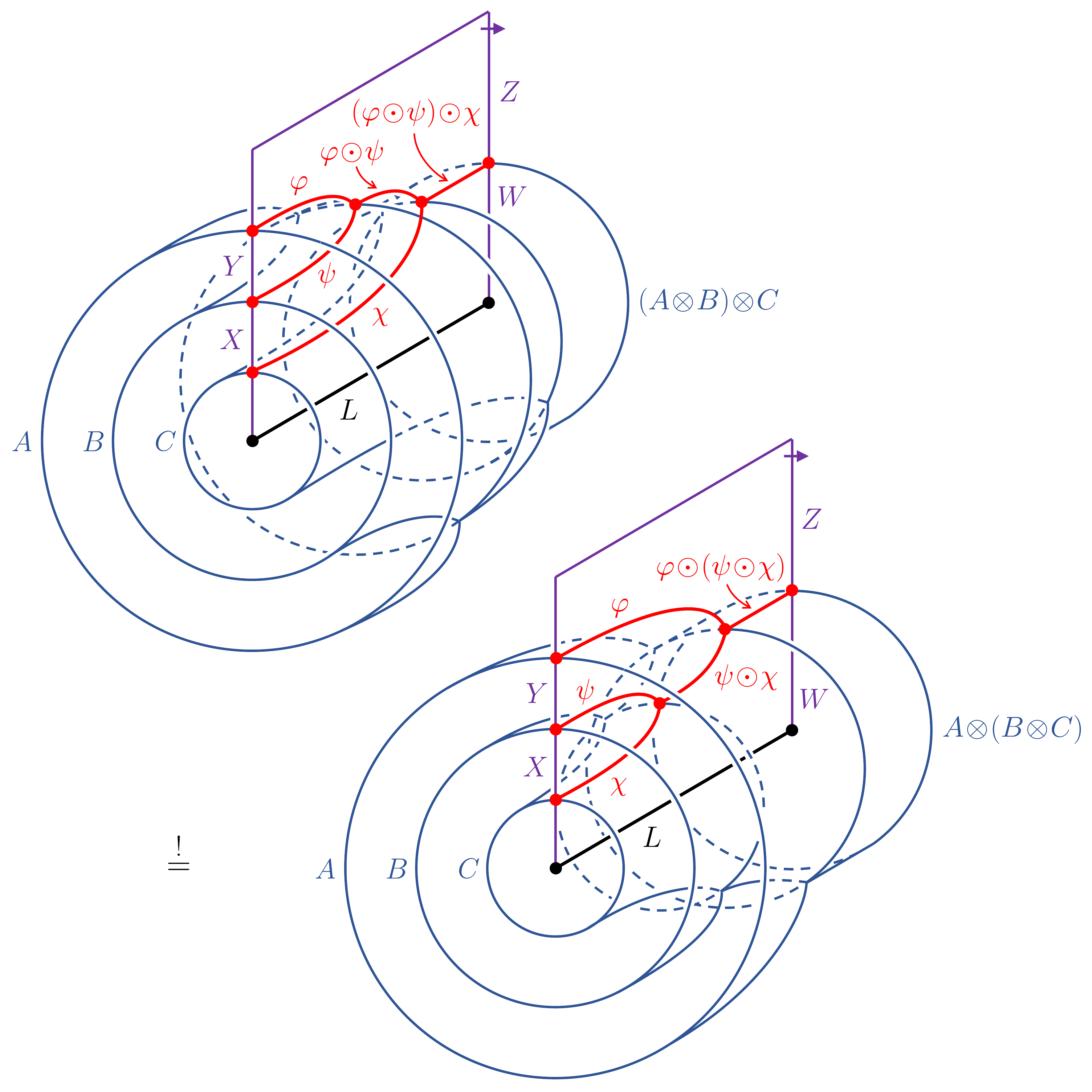}
	\vspace{-5pt}
	\caption{}
	\label{fig:3d-pseudo-2-morphism-compatibility}
\end{figure}

There is now further additional structure corresponding to the fact that the 2-morphisms $F_{A,\varphi\,|\,B,\psi}$ themselves must be compatible with the consecutive action of three symmetry defects $A$, $B$, $C$. Concretely, we demand that the diagram
\begin{equation}\label{eq:3d-pseudo-2-morphism-compatibility}
\begin{gathered}
\includegraphics[height=9.4cm]{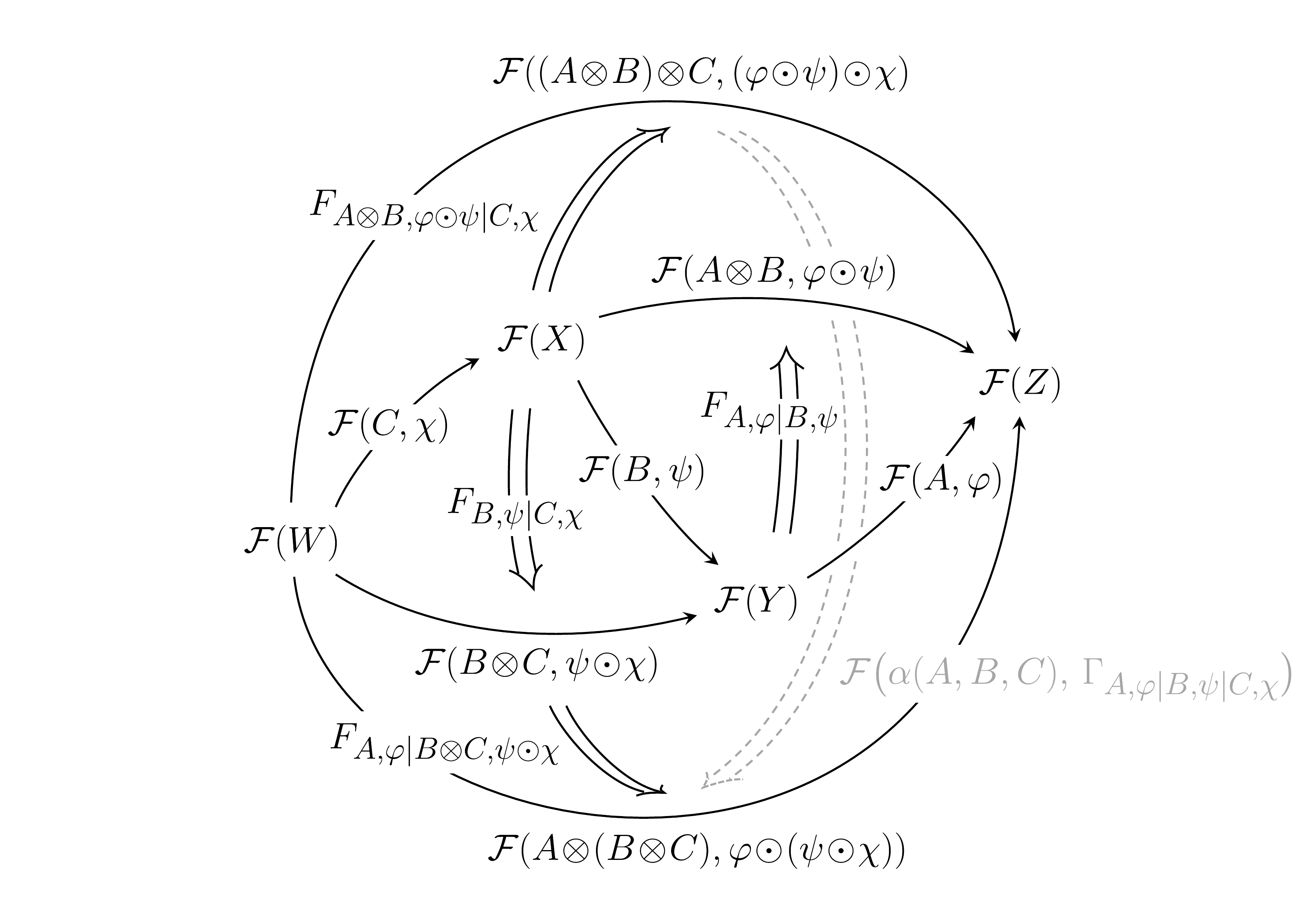}
\end{gathered}
\end{equation}
2-commutes, where we regard $\mathcal{F}\big(\alpha(A,B,C),\,\Gamma_{A,\varphi\,|\,B,\psi\,|\,C,\chi}\big)$ as a 2-morphism in $\mathsf{2Vec}$ that is induced by the associator $\alpha(A,B,C)$ and the 2-morphsim $\Gamma_{A,\varphi\,|\,B,\psi\,|\,C,\chi}$ defined through the diagram\footnote{In order to maintain readability of the diagram, we omitted the monoidal structure $\otimes$ in denoting the fusion of objects in $\mathsf{C}$.}
\begin{equation}\label{eq:3d-dia-composition-associator}
\begin{gathered}
\hspace{-8pt}
\includegraphics[height=14cm]{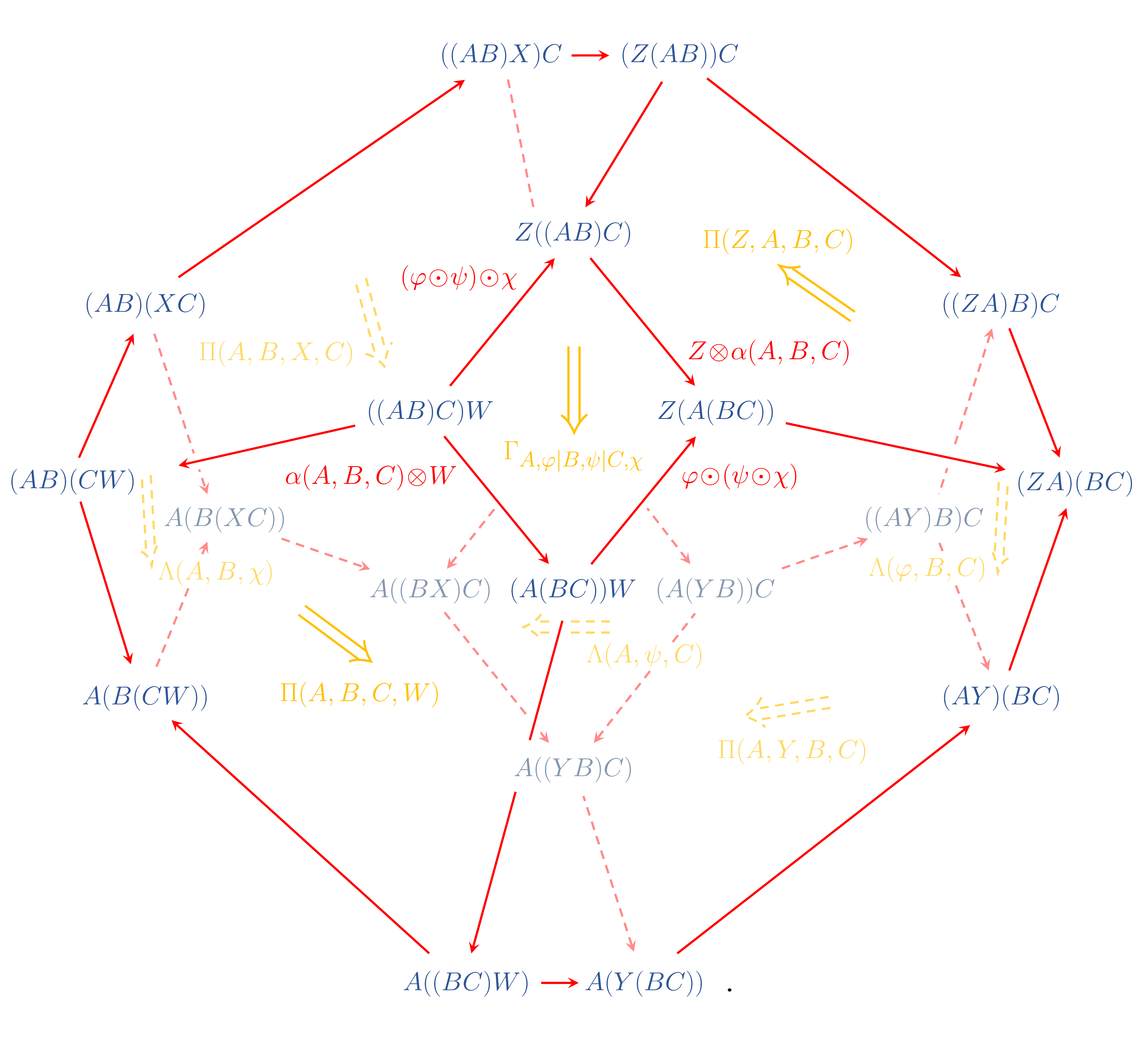}
\end{gathered}
\end{equation}
This rather formidable condition simply captures that fact that the two ways to parenthesise the wrapping action of three topological surfaces $A$, $B$ and $C$ are naturally equivalent, as illustrated graphically in figure \ref{fig:3d-pseudo-2-morphism-compatibility}.

\subsubsection{Linking action}

\begin{figure}[h]
	\centering
	\includegraphics[height=4.5cm]{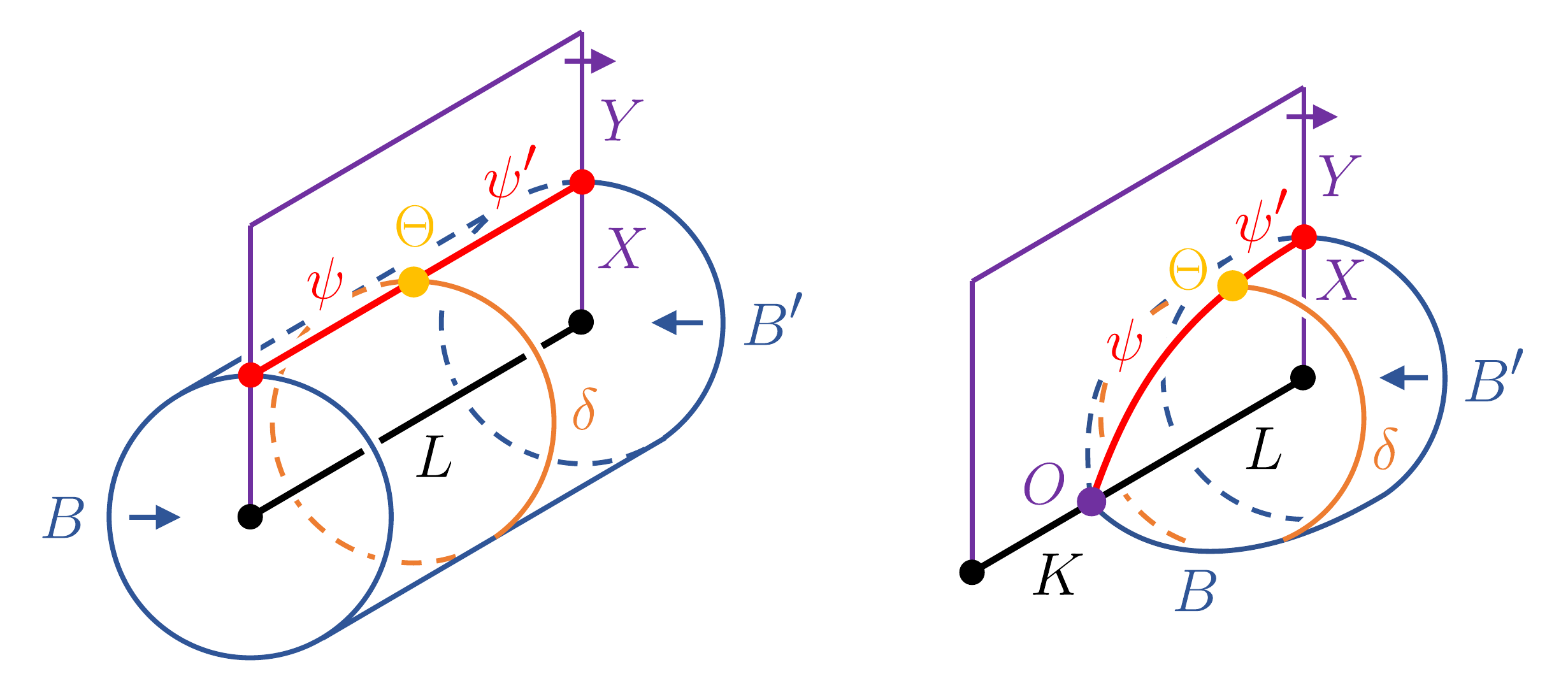}
	\vspace{-5pt}
	\caption{}
	\label{fig:3d-line-action-2}
\end{figure}

In addition to wrapping with a topological surface, we may link $L$ with a topological line $\delta : B \to B'$ forming an interface between topological surfaces $B$ and $B'$, as shown on the left-hand side of figure \ref{fig:3d-line-action-2}. 

Due to the 1-morphisms $\psi$ and $\psi'$ attached to the intersections of $B$, $B'$ and $X$, this requires an additional choice of 2-morphism
\begin{equation}\label{eq:3d-tube-2-morphism}
\begin{gathered}
\includegraphics[height=4.1cm]{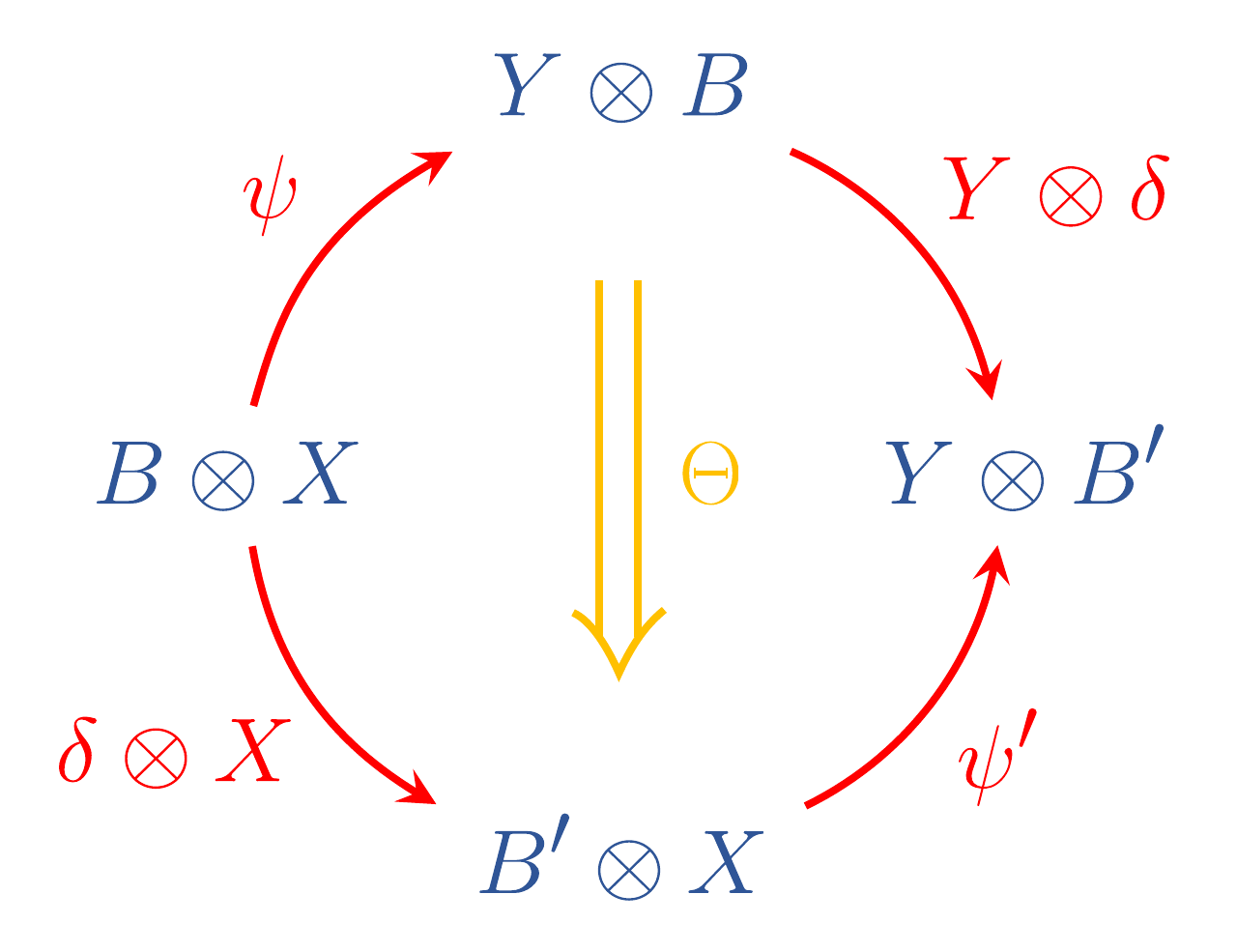}
\end{gathered}
\vspace{-4pt}
\end{equation}
that describes how $\delta$ intersects $\psi$ and $\psi'$.

Upon shrinking the cylinder down to $L$, we again obtain the vector space $\mathcal{F}(B,\psi)_{K,L}$ of topological local operators $O$ at the junction between $L$ and a new line $K$ in the $Y$-twisted sector, as illustrated on the right-hand side of figure \ref{fig:3d-line-action-2}. These operators can now be acted upon by sliding the 1-sphere formed by $\delta$ down onto $O$ from the right. The pair $(\delta,\Theta)$ thus induces a linear map
\begin{equation}
\mathcal{F}(\delta, \Theta)_{K,L} \; : \;\;\mathcal{F}(B,\psi)_{K,L} \,\; \to \;\, \mathcal{F}(B',\psi')_{K,L} \, .
\end{equation}
By scanning over $L$ and $K$, we then obtain a matrix $\mathcal{F}(\delta,\Theta)$ of linear maps determined by the pair $(\delta,\Theta)$, which we interpret as a 2-morphism
\begin{equation}\label{eq:3d-F-2-morphisms}
\begin{gathered}
\includegraphics[height=3.6cm]{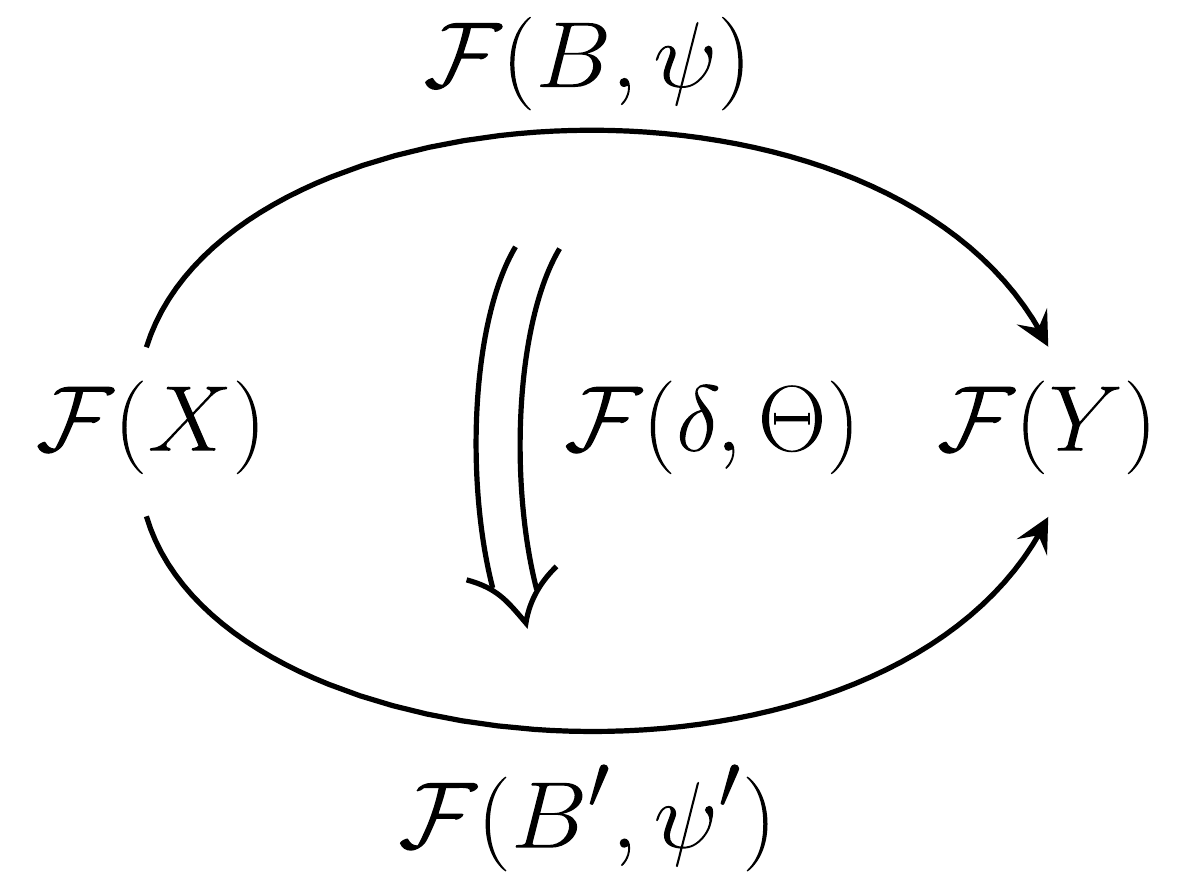}
\end{gathered}
\end{equation}
in the 2-category $\mathsf{2Vec}$ of 2-vector spaces. 

\begin{figure}[h]
	\centering
	\includegraphics[height=5cm]{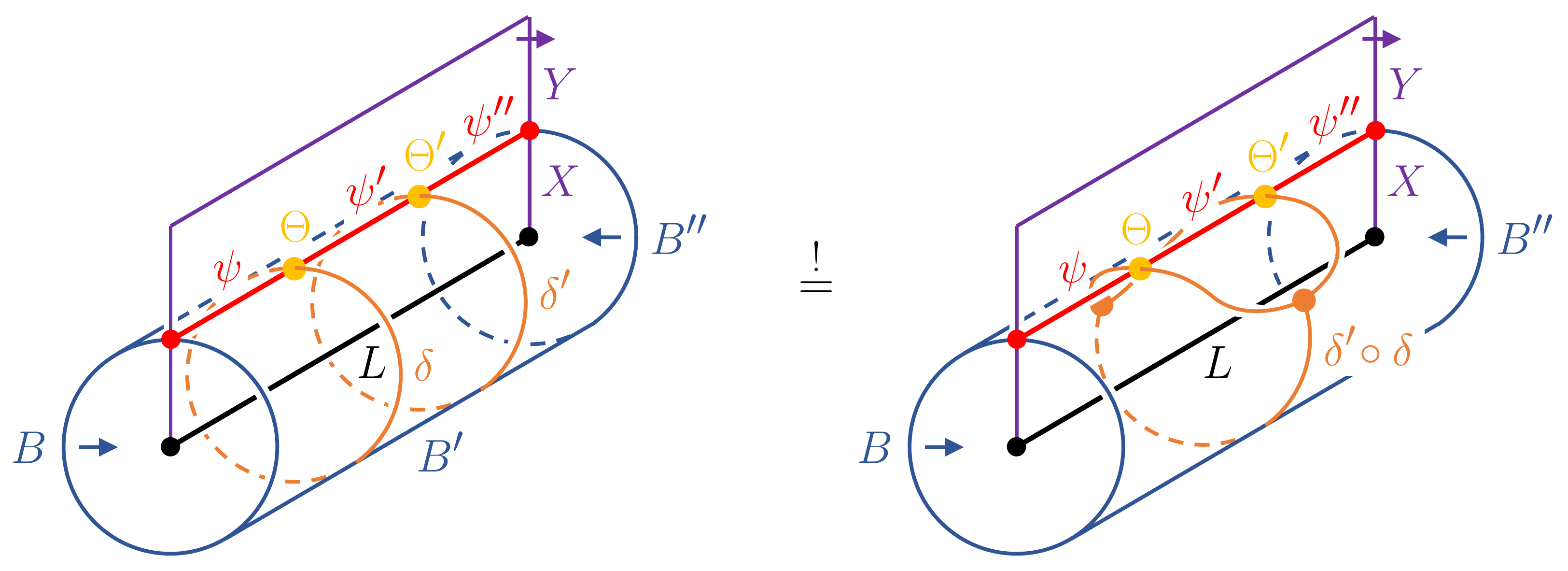}
	\vspace{-5pt}
	\caption{}
	\label{fig:3d-vertical-composition}
\end{figure}

These 2-morphisms must be compatible with the possibility to link $L$ with two parallel line defects $\delta$ and $\delta'$ as illustrated in figure \ref{fig:3d-vertical-composition}.
Mathematically, this is implemented by the 2-commutativity of the diagram
\begin{equation}\label{eq:3d-vertical-compatibility}
\begin{gathered}
\includegraphics[height=5.25cm]{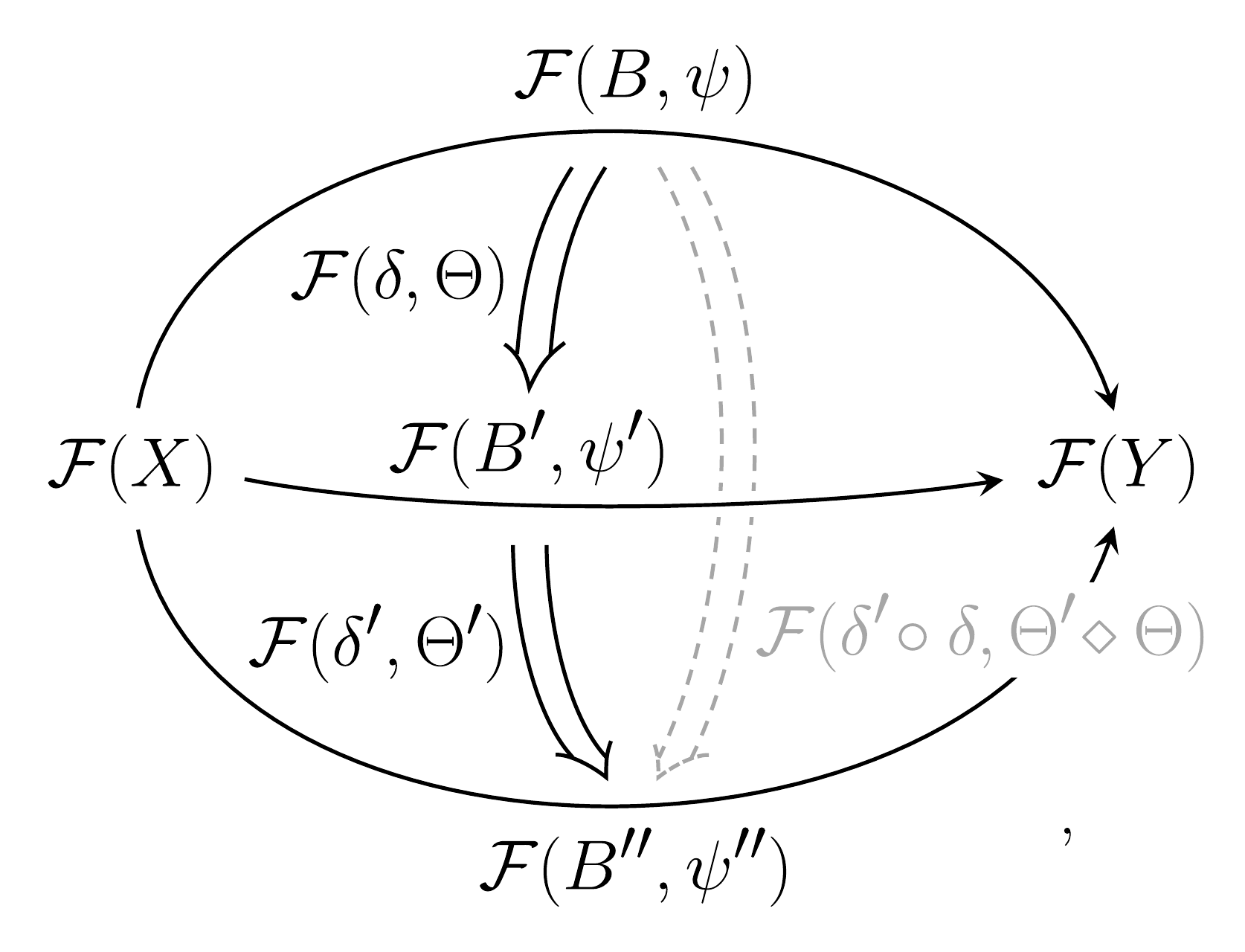}
\end{gathered}
\vspace{-6pt}
\end{equation}
where the 2-morphism $\Theta'\diamond \Theta$ is defined by the diagram
\begin{equation}\label{eq:3d-dia-vertical-composition}
\begin{gathered}
\includegraphics[height=6cm]{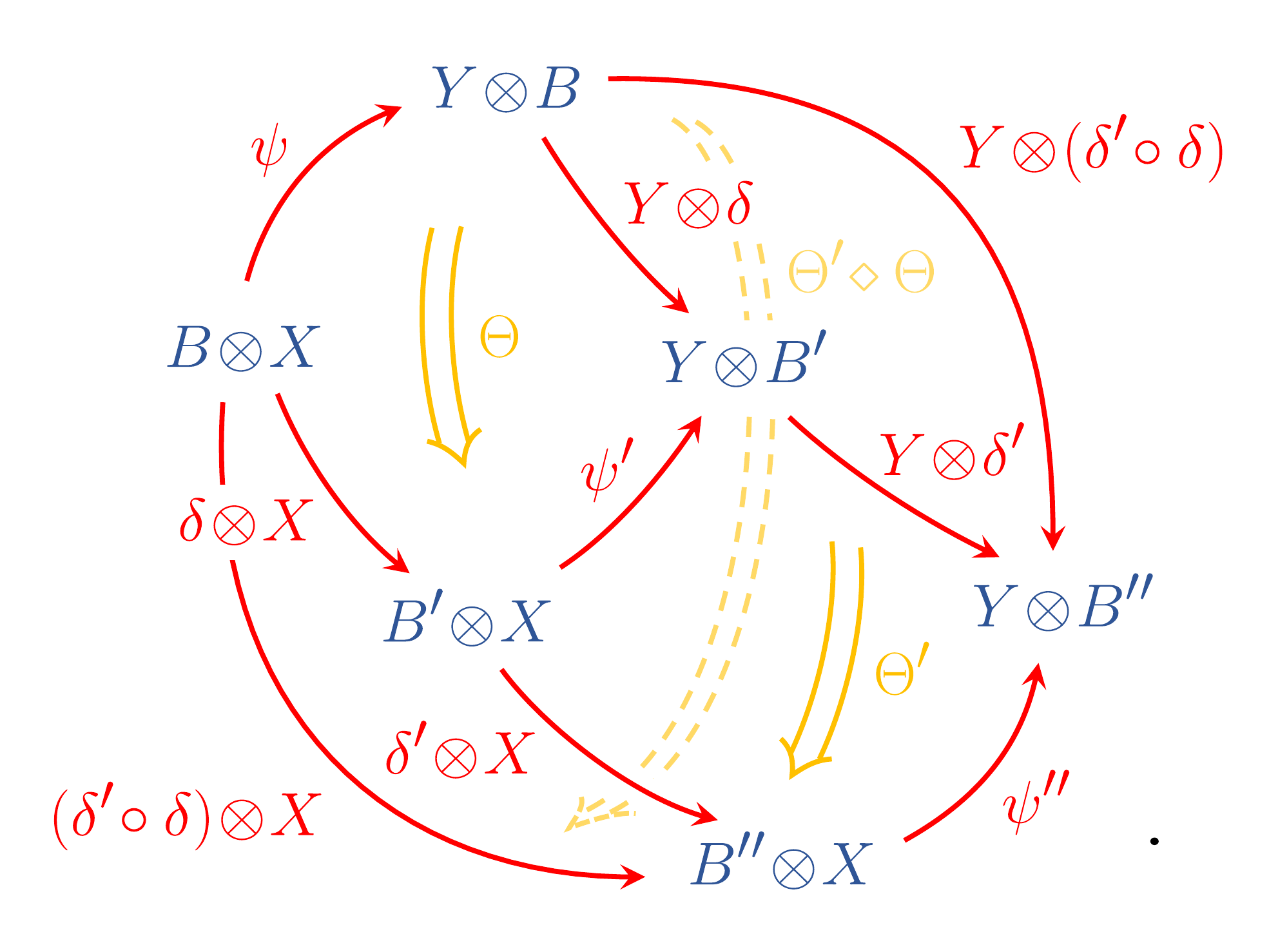}
\end{gathered}
\end{equation}
Physically, this condition simply states that linking $L$ with two parallel line defects $\delta$ and $\delta'$ is equivalent to linking with their composition $\delta \circ \delta'$, as illustrated in figure \ref{fig:3d-vertical-composition}.

\begin{figure}[h]
	\centering
	\includegraphics[height=6.8cm]{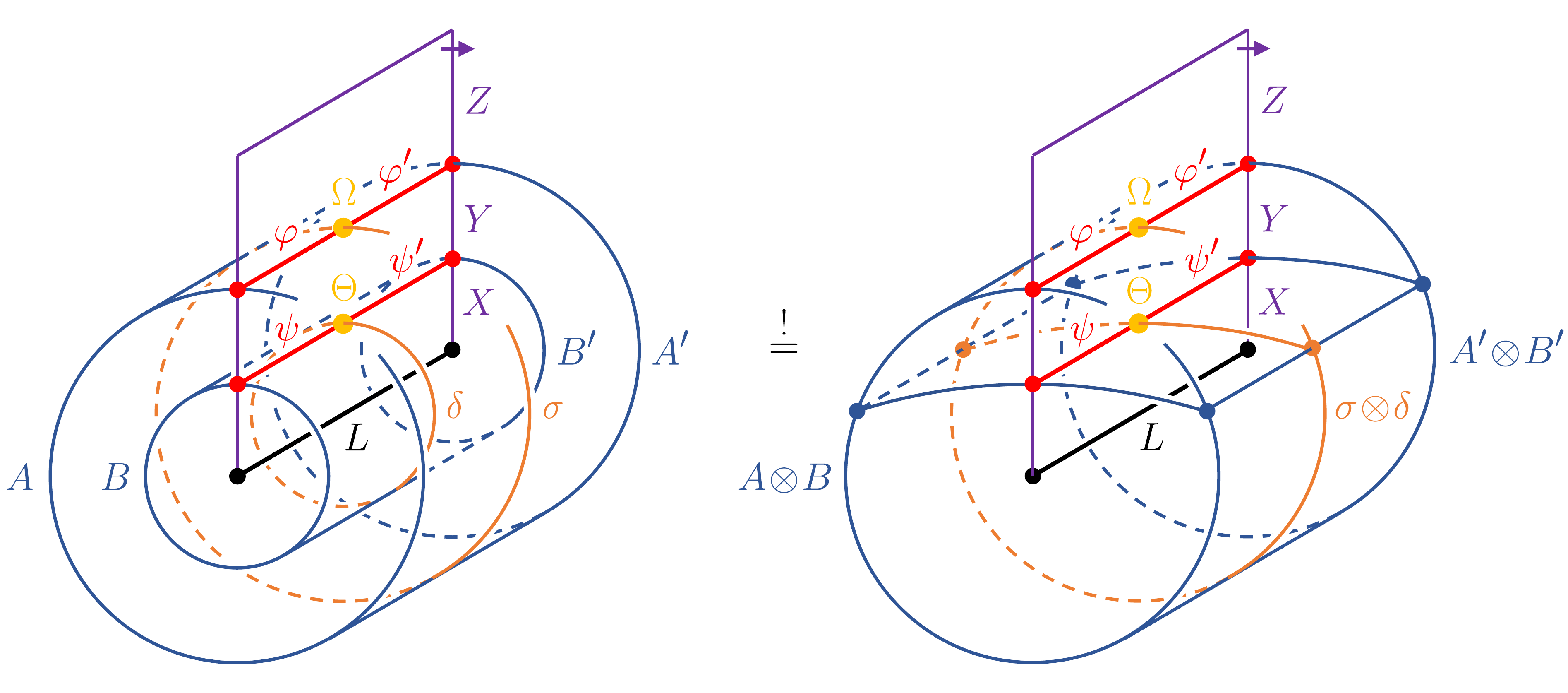}
	\vspace{-5pt}
	\caption{}
	\label{fig:3d-horizontal-composition}
\end{figure}

Moreover, the 2-morphisms $\mathcal{F}(\delta,\Theta)$ also need to be compatible with the possibility to link $L$ with two line defects $\sigma$ and $\delta$ consecutively as illustrated in figure \ref{fig:3d-horizontal-composition}. 
Mathematically, this is implemented by the 2-commutativity of the diagram 
\begin{equation}\label{eq:3d-horizontal-compatibility}
\begin{gathered}
\includegraphics[height=8.7cm]{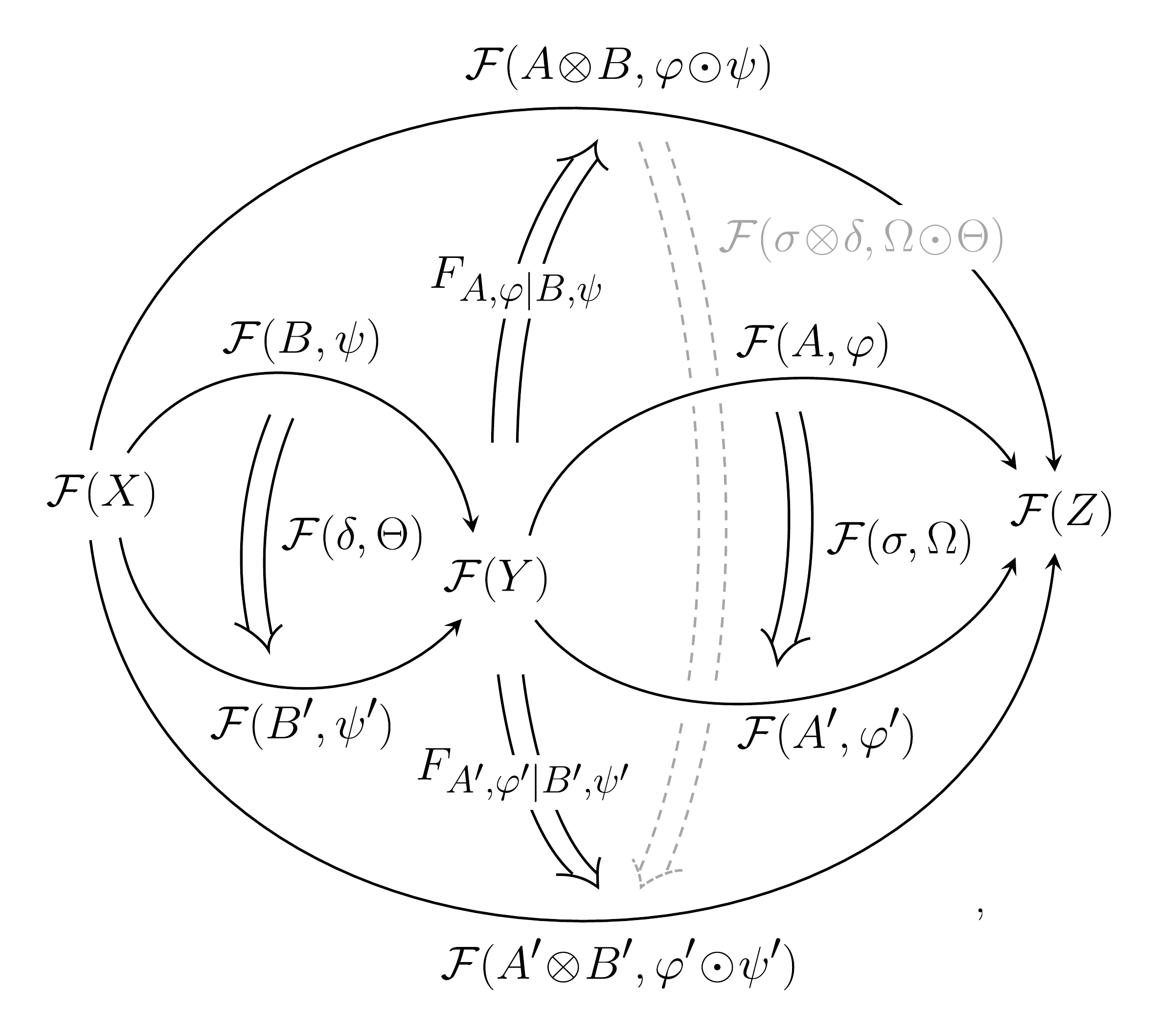}
\end{gathered}
\end{equation}
where the 2-morphism $\Omega \odot \Theta$ is defined by the diagram
\begin{equation}\label{eq:3d-dia-horizontal-composition}
\begin{gathered}
\includegraphics[height=12.75cm]{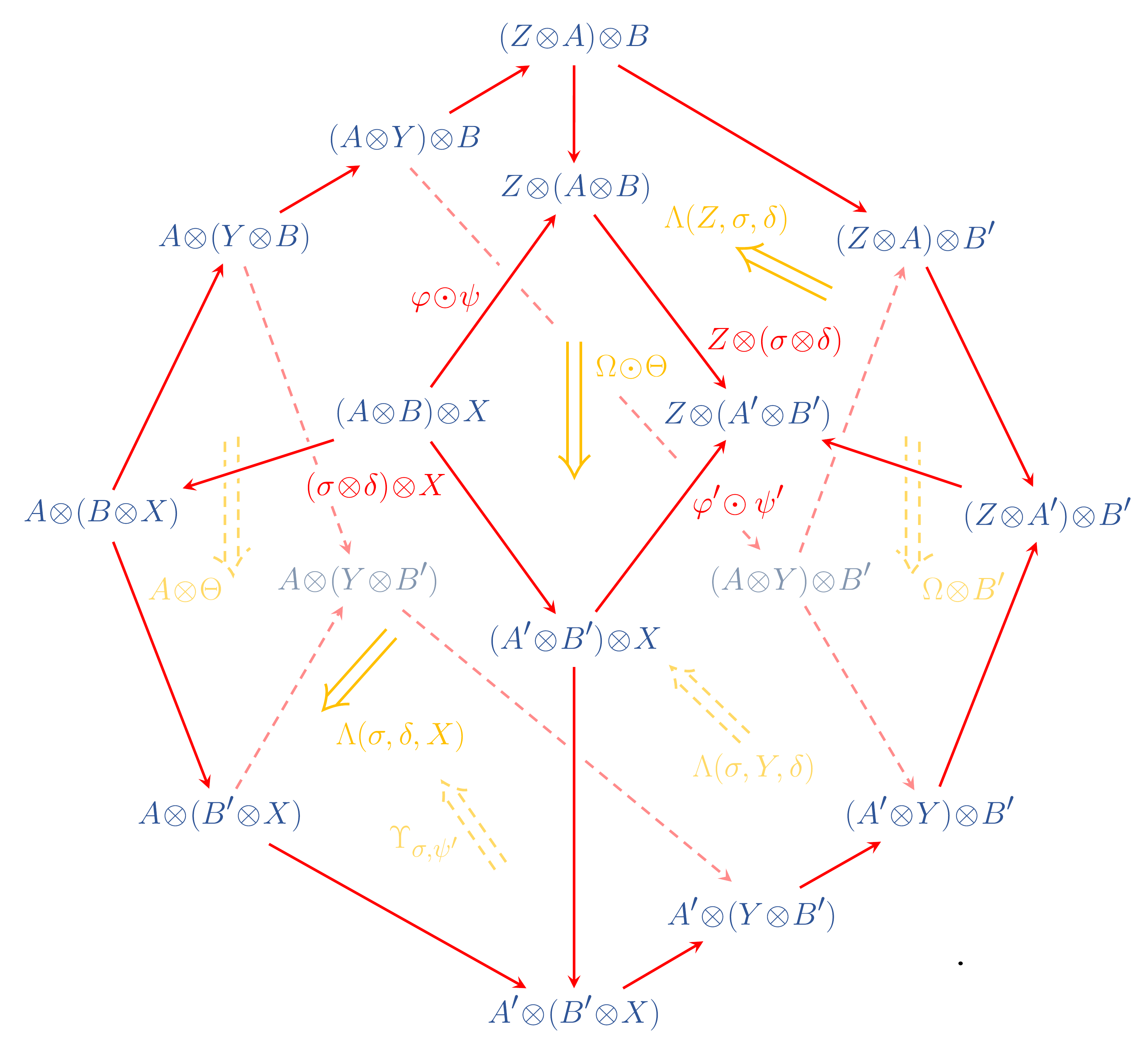}
\end{gathered}
\end{equation}
Physically, this condition simply states that linking $L$ with two line defects $\sigma$ and $\delta$ consecutively is equivalent to linking $L$ with their fusion $\sigma \otimes \delta$, as illustrated in figure \ref{fig:3d-horizontal-composition}.

\subsubsection{Equivalence relations}

The action of topological surfaces and lines on 2-twisted sector line defects must be compatible with the possibility to move topological lines and their junctions around the cylinder $C^2 \cong \mathbb{R} \times S^1$. This will generate equivalence relations that identify configurations acting in the same way on 2-twisted sector line defects. There are now two layers of structure, corresponding to the wrapping with surface and the linking with line defects, which we consider in turn.

\begin{figure}[h]
	\centering
	\includegraphics[height=5cm]{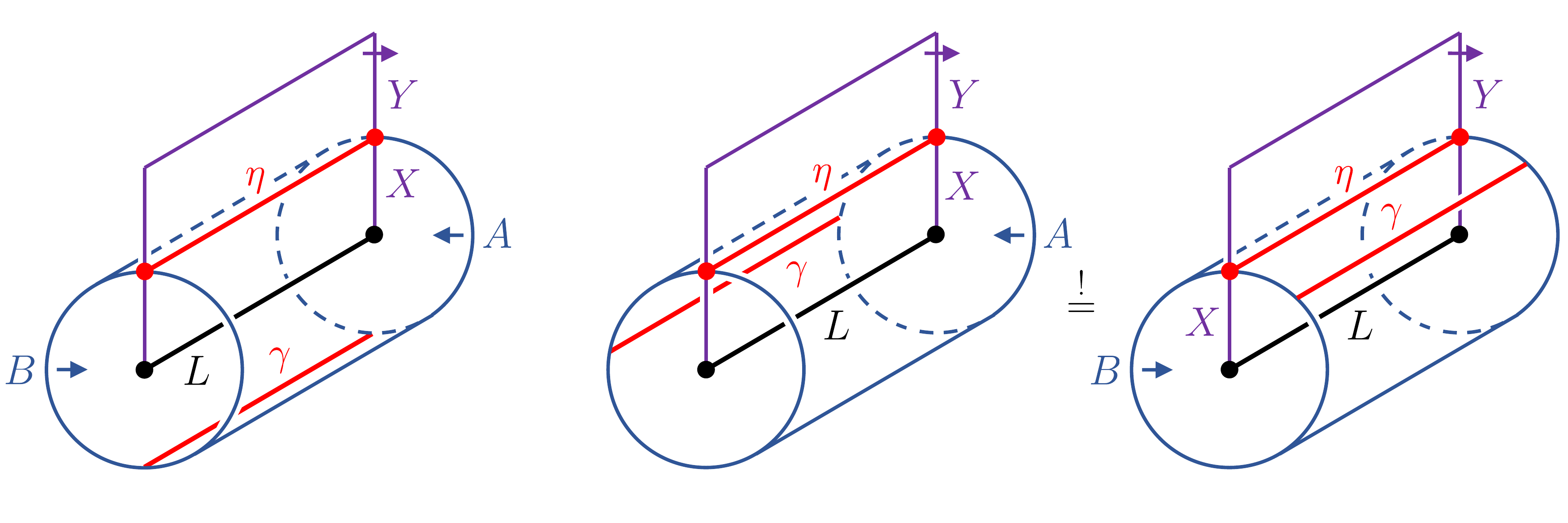}
	\vspace{-5pt}
	\caption{}
	\label{fig:3d-tube-equivalence-1}
\end{figure}

First, the wrapping action of topological surface defects must be compatible with the possibility to move parallel topological lines around the cylinder $C^2$. Concretely, consider a configuration as on the left-hand side of figure \ref{fig:3d-tube-equivalence-1}, where two topological surfaces $A$ and $B$ connected by a line interface $\gamma: A \to B$ wrap a line operator $L$ in the $X$-twisted sector via a specified 1-morphism
\vspace{-6pt}
\begin{equation}
\begin{gathered}
\includegraphics[height=1.19cm]{2d-dia-tube-equivalence-morphism.pdf}
\end{gathered}
\vspace{-8pt}
\end{equation}
By moving the topological junction $\gamma$ around the cylinder towards $\eta$ from the left and from the right, we can either regard this configuration as 
\begin{enumerate}
\item the defect $A$ wrapping $L$ via the intersection 1-morphism $\varphi = \eta \circ (\gamma \otimes X)$,
\item the defect $B$ wrapping $L$ via the intersection 1-morphism $\psi = (Y \otimes \gamma) \circ \eta$.
\end{enumerate}
Since both configurations are physically equivalent, the corresponding wrapping actions on $L$ must coincide in the sense that
\begin{equation}\label{eq:3d-F-invariance}
\mathcal{F}(A,\varphi) \; \stackrel{!}{=} \; \mathcal{F}(B,\psi) \, .
\end{equation}
This is illustrated on the right-hand side of figure \ref{fig:3d-tube-equivalence-1}.

\begin{figure}[h]
	\centering
	\includegraphics[height=5cm]{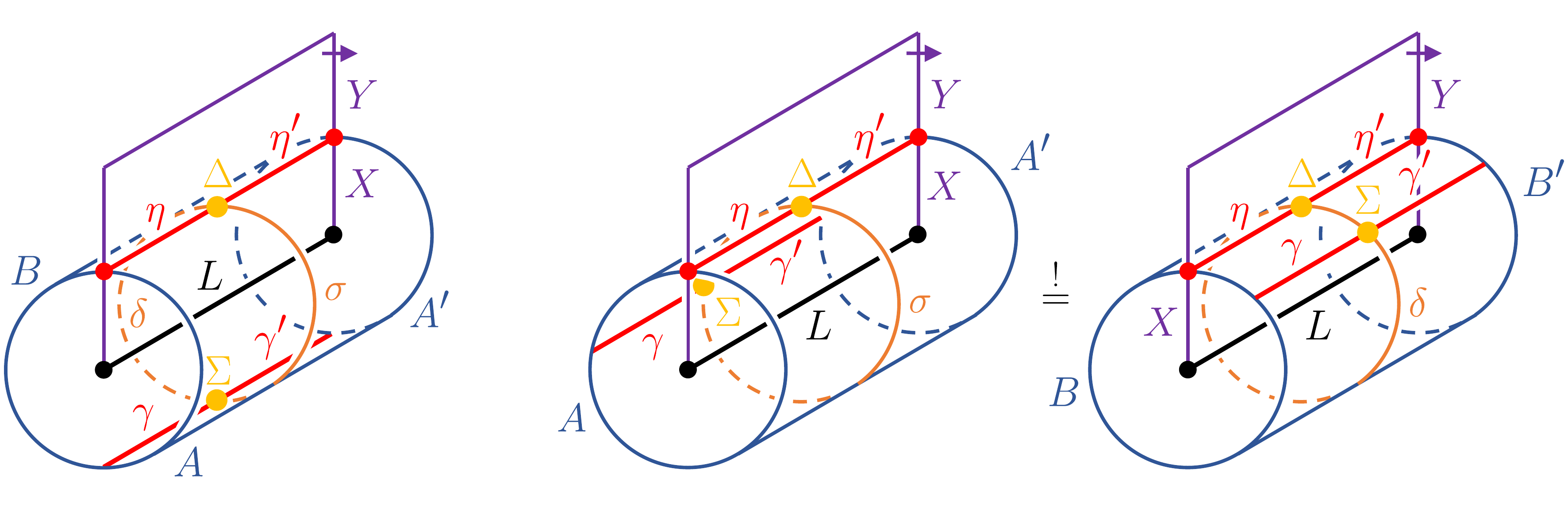}
	\vspace{-5pt}
	\caption{}
	\label{fig:3d-tube-equivalence-2}
\end{figure}

Second, the linking action of topological line defects must be compatible with the ability to move topological point-like junctions around the linking $S^1$. Consider the configuration on the left-hand side of figure \ref{fig:3d-tube-equivalence-2}, which can be described as follows:
\begin{itemize}
\item In the front, the line operator $L$ is wrapped by two surfaces $A$ and $B$ that are connected by a line interface $\gamma: A \to B$ and intersect the attached surface $X$ via a specified 1-morphism
\vspace{-6pt}
\begin{equation}
\begin{gathered}
\includegraphics[height=1.19cm]{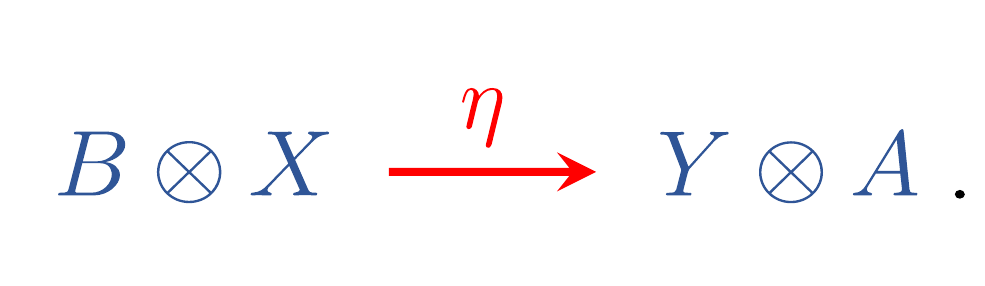}
\end{gathered}
\vspace{-8pt}
\end{equation}

\item In the back, the line operator $L$ is wrapped by two surfaces $A'$ and $B'$ that are connected by a line interface $\gamma': A' \to B'$ and intersect the attached surface $X$ via a specified 1-morphism
\vspace{-6pt}
\begin{equation}
\begin{gathered}
\includegraphics[height=1.19cm]{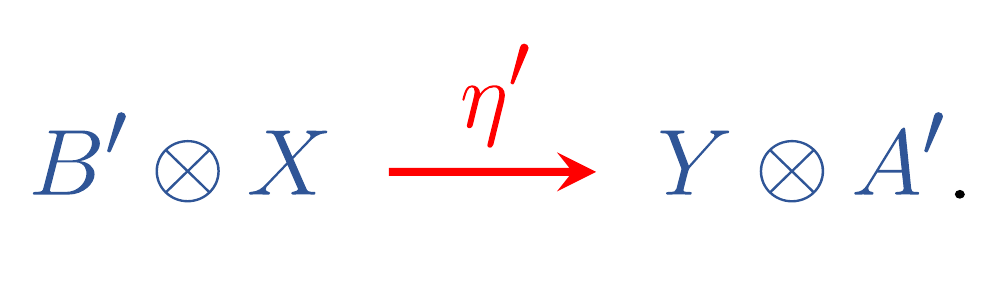}
\end{gathered}
\vspace{-8pt}
\end{equation}

\item In the middle, the surfaces $A$, $A'$ and $B$, $B'$ are connected by line interfaces $\sigma: A \to A'$ and $\delta: B \to B'$. At the bottom, these intersect the interfaces $\gamma$ and $\gamma'$ via a specified 2-morphism
\vspace{-0pt}
\begin{equation}
\begin{gathered}
\includegraphics[height=2.5cm]{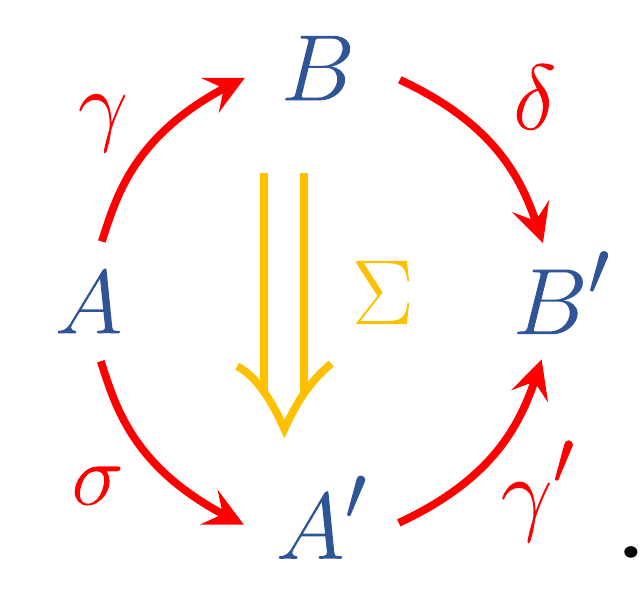}
\end{gathered}
\vspace{-4pt}
\end{equation}
At the top, they intersect $\eta$ and $\eta'$ via a specified 2-morphism
\vspace{-0pt}
\begin{equation}
\begin{gathered}
\includegraphics[height=3.4cm]{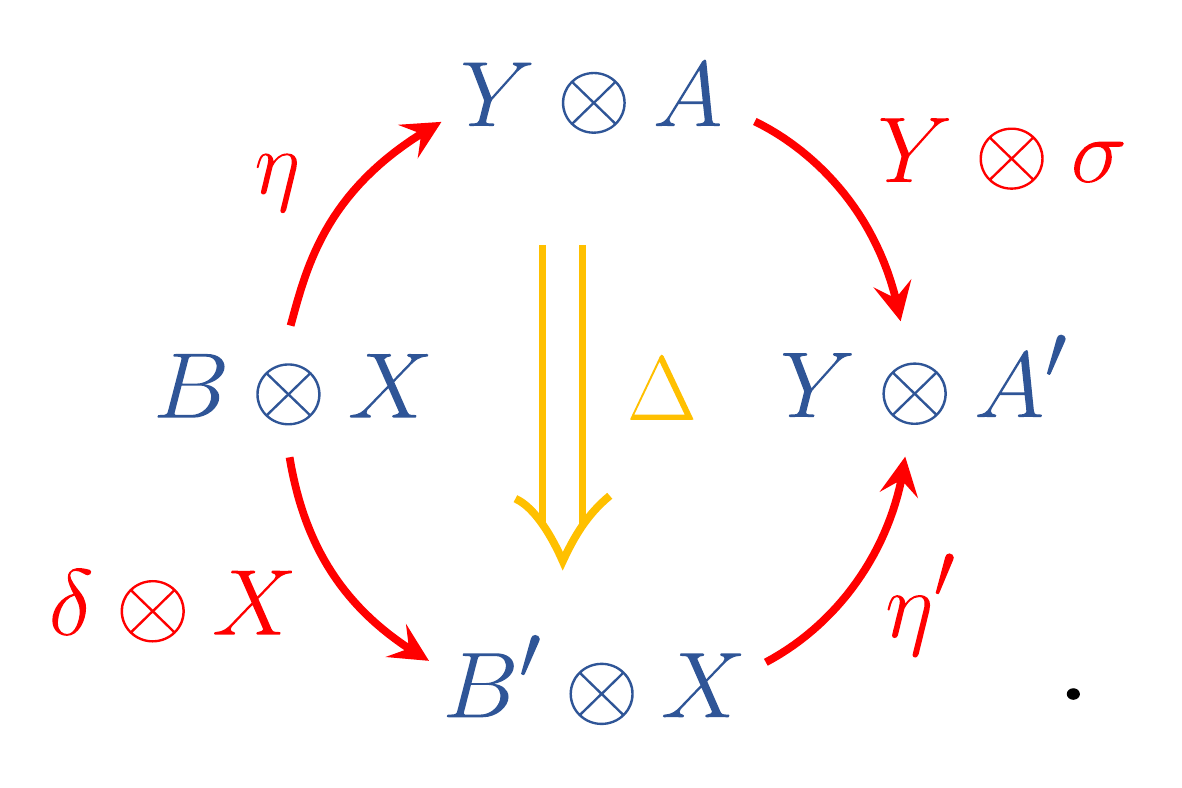}
\end{gathered}
\vspace{-4pt}
\end{equation}
\end{itemize}
By moving the parallel topological lines $\gamma$ and $\gamma'$ around the cylinder towards $\eta$ and $\eta'$ from the left and from the right respectively, we can interpret this configuration in the following two ways:
\begin{enumerate}
\item Two surfaces $A$ and $A'$ connected by a line $\sigma: A \to A'$ wrap $L$ via intersection 1-morphisms
\begin{equation}
\varphi \; = \; \eta \circ (\gamma \otimes X) \qquad \text{and} \qquad \varphi' \; = \; \eta' \circ (\gamma' \otimes X) \, .
\end{equation}
These are intersected by the topological line $\sigma$ via the intersection 2-morphism
\vspace{-6pt}
\begin{equation}\label{eq:3d-dia-tube-equivalence-1}
\begin{gathered}
\includegraphics[height=6.6cm]{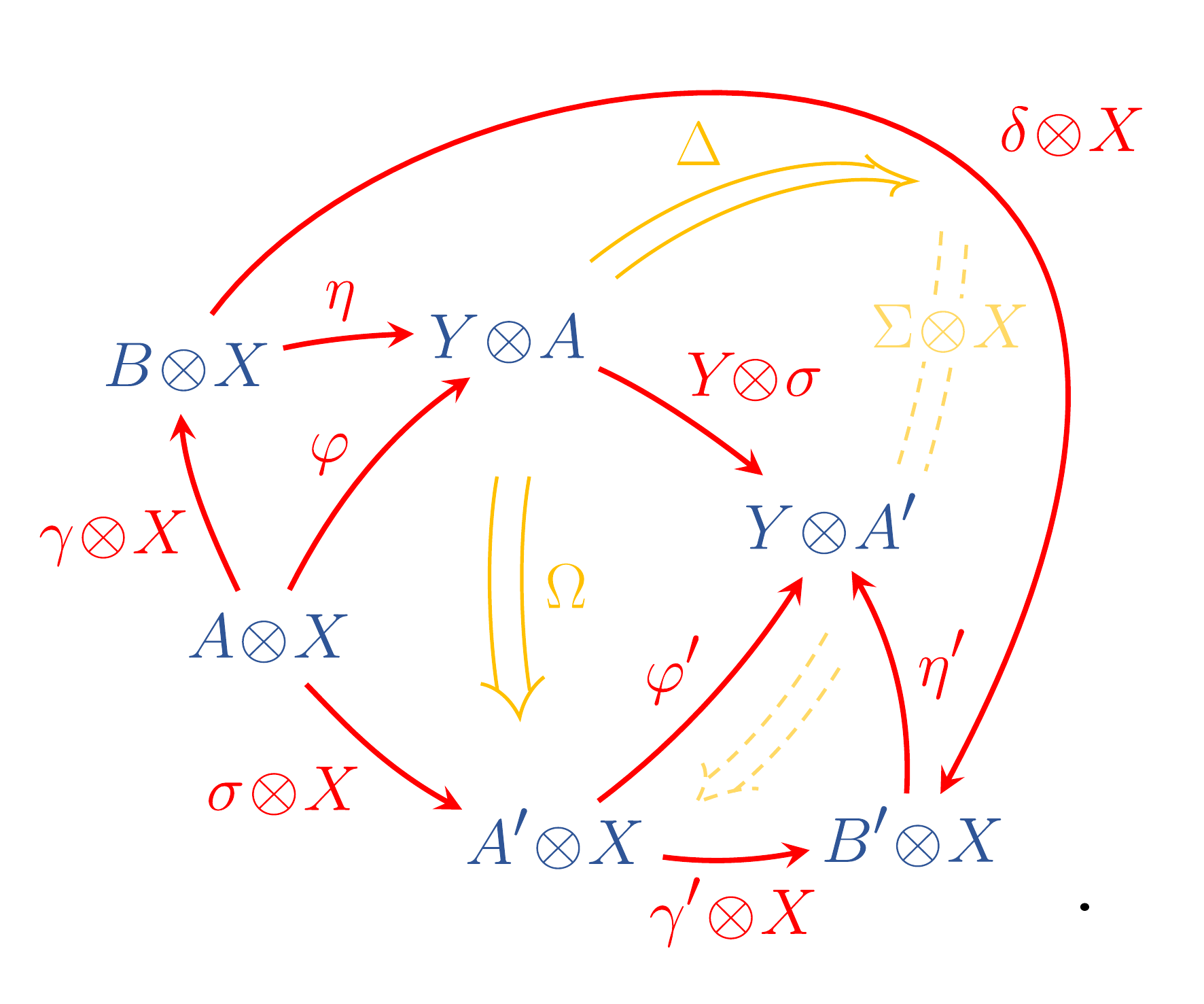}
\end{gathered}
\vspace{-4pt}
\end{equation}

\item Two surfaces $B$ and $B'$ connected by a line defect $\delta: B \to B'$ wrap $L$ via intersection 1-morphisms
\begin{equation}
\psi \; = \; (Y \otimes \gamma) \circ \eta \qquad \text{and} \qquad \psi' \; = \; (Y \otimes \gamma') \circ \eta' \, .
\end{equation}
These are intersected by the topological line $\delta$ via the intersection 2-morphism
\vspace{-6pt}
\begin{equation}\label{eq:3d-dia-tube-equivalence-2}
\begin{gathered}
\includegraphics[height=6.6cm]{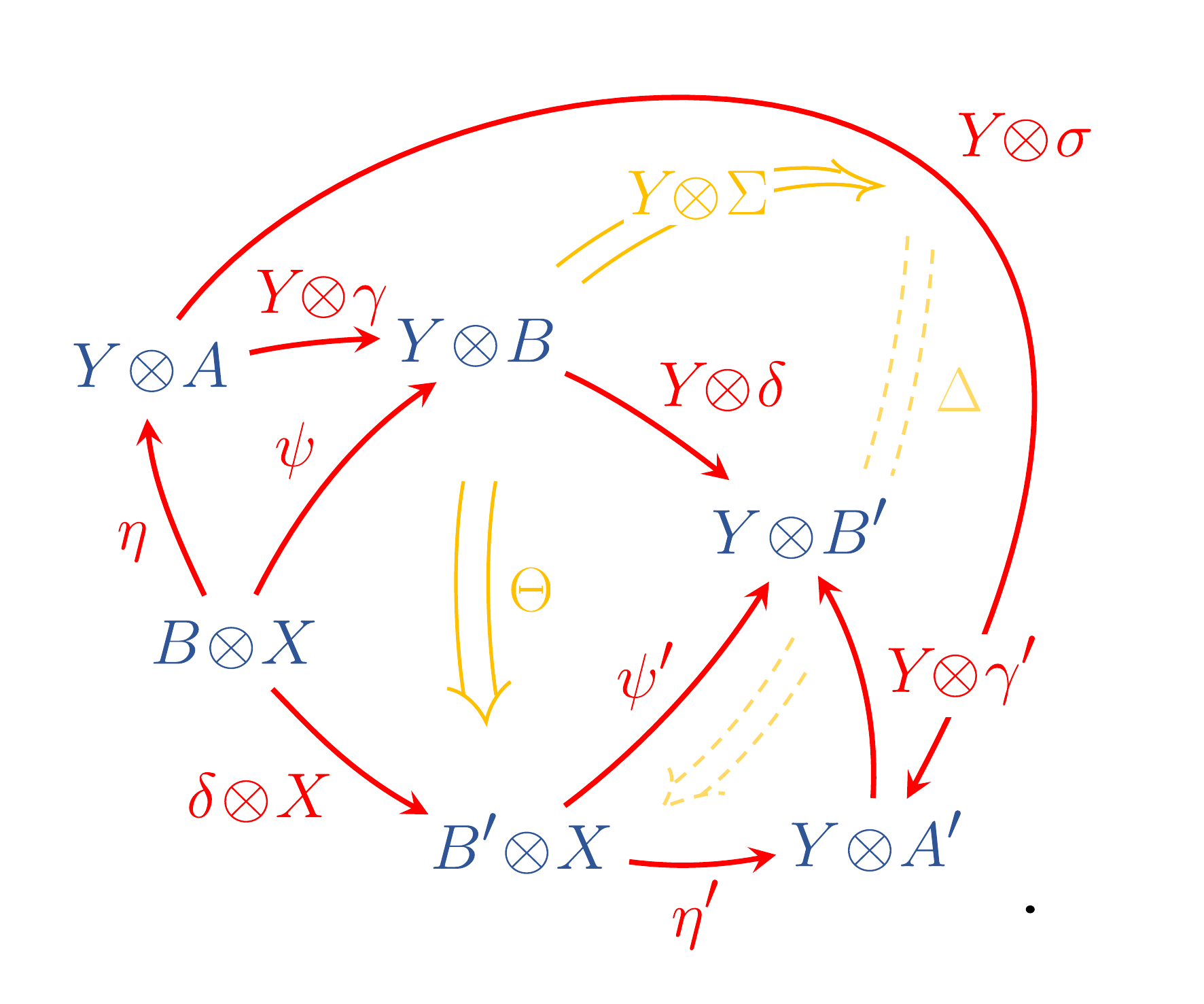}
\end{gathered}
\vspace{-4pt}
\end{equation}
\end{enumerate}
Since both configurations are physically equivalent, the corresponding linking actions must coincide in the sense that
\begin{equation}
\mathcal{F}(\sigma,\Omega) \; \stackrel{!}{=} \; \mathcal{F}(\delta,\Theta) \, .
\end{equation}
This is illustrated on the right-hand side of figure \ref{fig:3d-tube-equivalence-2}.

Similarly to before, for fixed $X$ and $Y$, we expect the above relations to generate equivalence relations on pairs $(B,\psi)$ and $(\delta,\Theta)$. Without providing further details, we will schematically denote their equivalence classes by $[B,\psi]$ and $[\delta,\Theta]$ in what follows.

\subsection{Tube 2-representations}

We now formulate the action of the symmetry 2-category $\mathsf{C}$ on 2-twisted sector line operators in terms of higher representation theory. We introduce the \textit{tube 2-category} $\T_{S^1}\C$ and \textit{tube 2-algebra} $\cA_{S^1}(\C)$, building on the results in sections~\ref{sec:2d} and~\ref{sec:3d-ops}, and show that their 2-representation theory captures the structures presented above. This incorporates and generalises the fact that genuine line defects in three dimensions transform in 2-representations of a finite invertible symmetry~\cite{Bartsch:2023pzl,Bhardwaj:2023wzd}.

\subsubsection{Tube 2-category}

We first introduce the tube 2-category $\T_{S^1}\C$. This is the finite semi-simple 2-category whose objects are twisted sectors and whose 1-morphisms and 2-morphisms are actions of topological surface and line defects on twisted sectors via wrapping and linking respectively. 

It has the following explicit description:
\begin{itemize}
\item Objects are objects of $\mathsf{C}$, i.e. $\text{Ob}(\T_{S^1}\C) = \text{Ob}(\mathsf{C})$,

\item 1-morphisms between objects $X,Y \in \T_{S^1}\C$ are given by direct sums of equivalence classes of pairs 
\begin{equation}
\begin{gathered}
\includegraphics[height=1.18cm]{2d-dia-tube-morphism.pdf}
\end{gathered}
\vspace{-4pt}
\end{equation}
consisting of an object $B \in \mathsf{C}$ and a 1-morphism
\begin{equation}
\psi \; \in \; \text{Hom}_{\mathsf{C}}(B \otimes X, Y \otimes B) \, .
\end{equation}
The composition of 1-morphisms is defined analogously to (\ref{eq:2d-tube-composition}). 

\item 2-morphisms between 1-morphisms $[B,\psi], [B',\psi'] \in \text{Hom}_{\mathsf{T_{S^1}C}}(X,Y)$ are linear combinations of equivalence classes of pairs  
\begin{equation}
\begin{gathered}
\includegraphics[height=3.25cm]{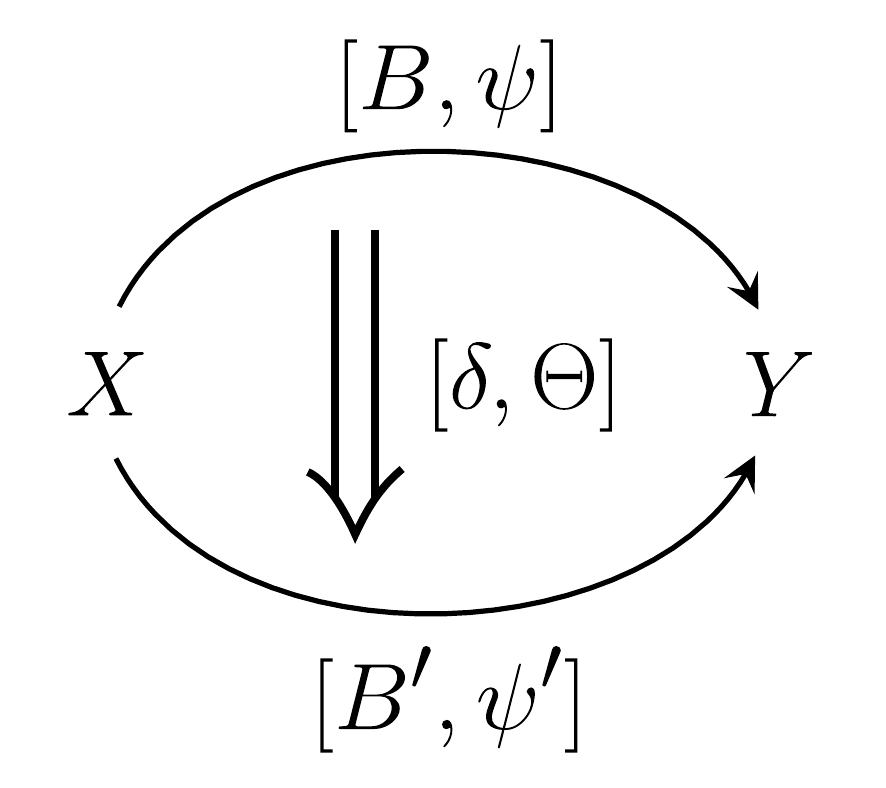}
\end{gathered}
\vspace{-4pt}
\end{equation}
consisting of a 1-morphism $\delta \in \text{1-Hom}_{\mathsf{C}}(B,B')$ and a 2-morphism
\begin{equation}
\Theta \; \in \; \text{2-Hom}_{\mathsf{C}}\big( (Y \otimes \delta) \circ \psi, \, \psi' \circ (\delta \otimes X) \big) \, .
\end{equation}

\item The vertical composition of two 2-morphisms
\begin{equation}
\begin{gathered}
\includegraphics[height=4cm]{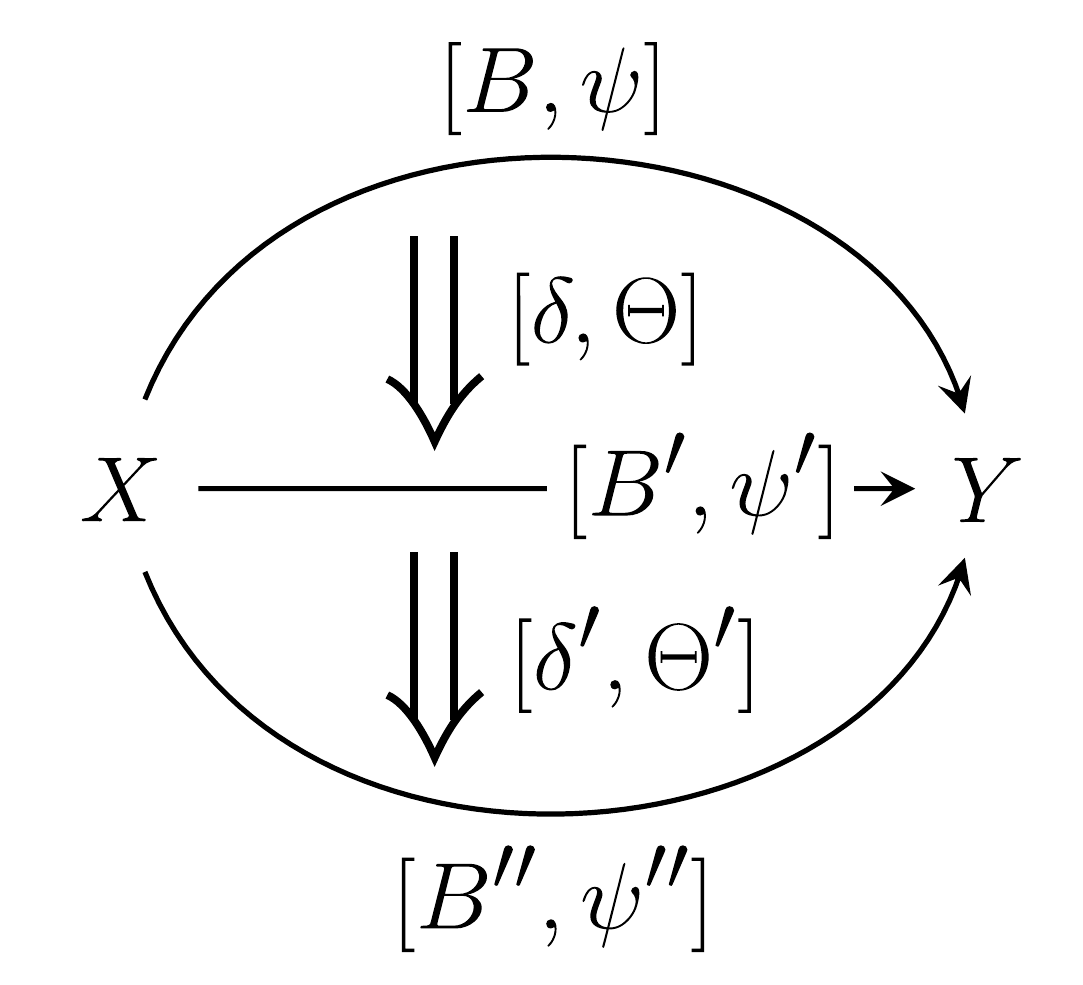}
\end{gathered}
\vspace{-4pt}
\end{equation}
is defined by the composition law
\begin{equation}\label{eq:3d-tube-vertical-composition}
[\delta',\Theta'] \, \circ \, [\delta, \Theta] \; := \; [\delta' \circ \delta, \, \Theta' \diamond \Theta]
\end{equation}
with the 2-morphism $\Theta' \diamond \Theta$ defined as in (\ref{eq:3d-dia-vertical-composition}). The horizontal composition of two 2-morphisms
\begin{equation}
\begin{gathered}
\includegraphics[height=3.15cm]{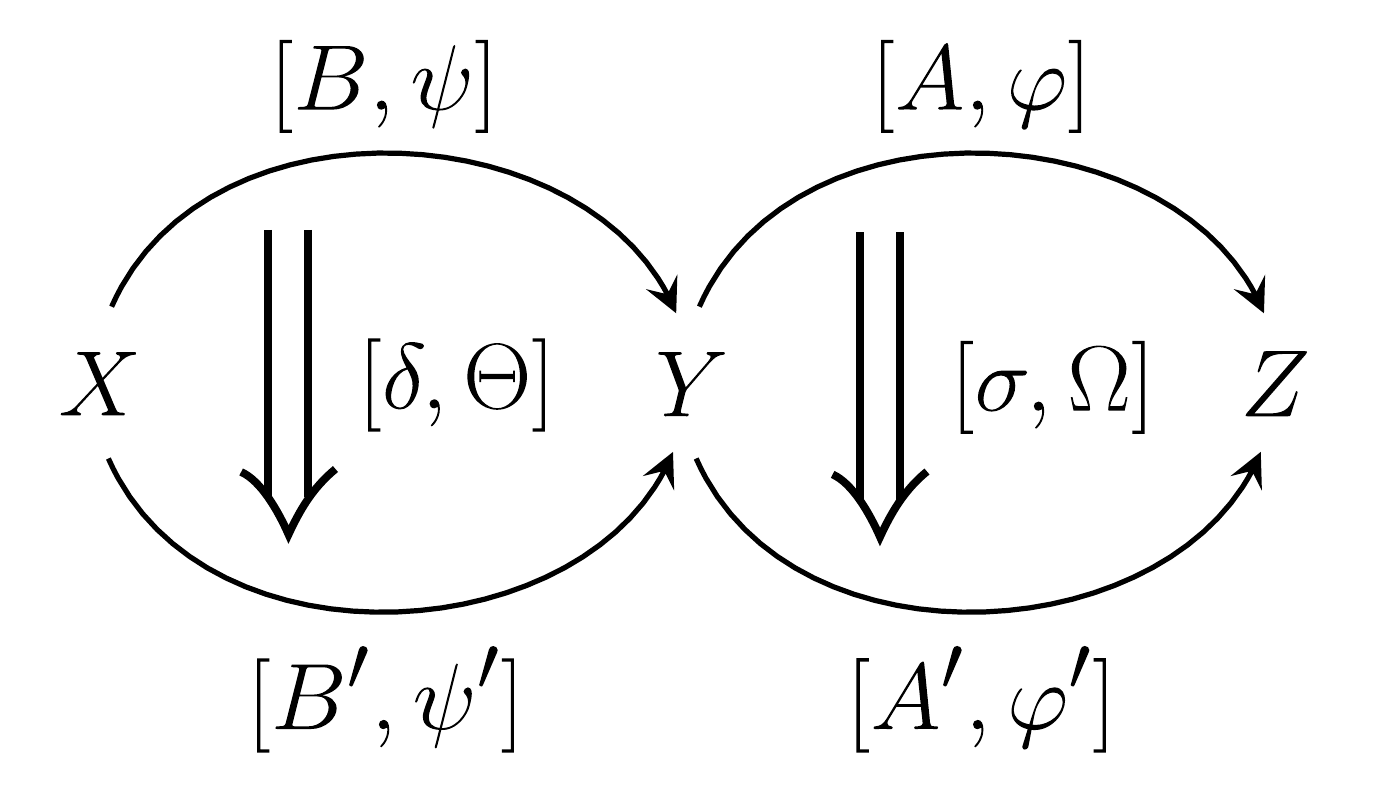}
\end{gathered}
\vspace{-4pt}
\end{equation}
is defined by the composition law
\begin{equation}\label{eq:3d-tube-horizontal-composition}
[\sigma,\Omega] \, \ast \, [\delta, \Theta] \; := \; [\sigma \otimes \delta, \, \Omega \odot \Theta]
\end{equation}
with the 2-morphism $\Omega \odot \Theta$ defined as in (\ref{eq:3d-dia-horizontal-composition}).

\item The 2-associator for the composition of 1-morphisms is given by
\begin{equation}
\begin{gathered}
\includegraphics[height=4.3cm]{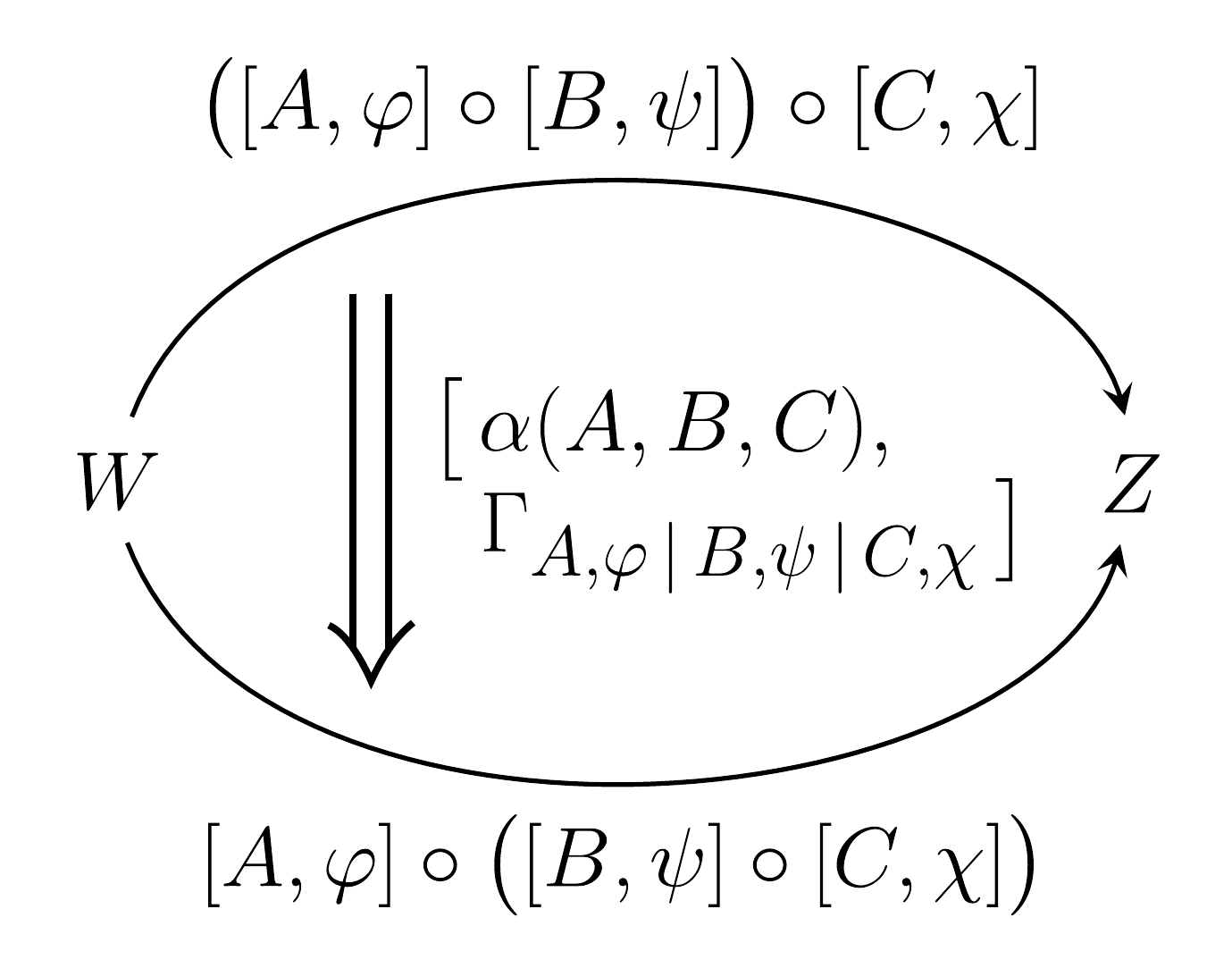}
\end{gathered}
\vspace{-4pt}
\end{equation}
with the 2-morphism $\Gamma_{A,\varphi\,|\,B,\psi\,|\,C,\chi}$ defined as in (\ref{eq:3d-dia-composition-associator}).
\end{itemize}

\begin{figure}[h]
	\centering
	\includegraphics[height=5cm]{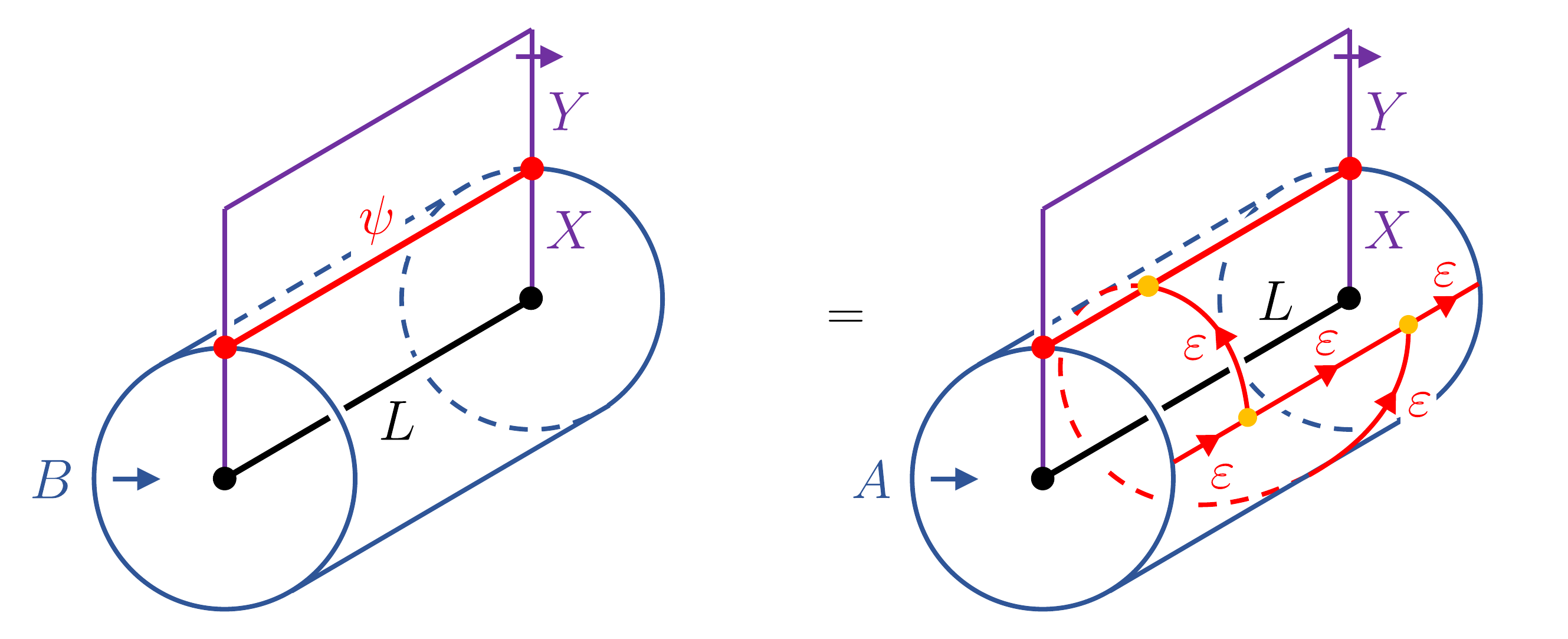}
	\vspace{-5pt}
	\caption{}
	\label{fig:3d-condensation-action}
\end{figure}

We may now ask whether the 1-morphisms and 2-morphisms admit a simpler but less canonical description in terms of representatives of homotopy groups of $\mathsf{C}$. Unlike the analysis in sections~\ref{sec:2d} and~\ref{sec:3d-ops}, the topology of the cylinder $C^2 \cong \R \times S^2$ does not immediately identify the wrapping action of topological surfaces related by condensation. Nevertheless, the actions of condensations are determined as follows:

Consider a 2-condensation $A \cond B$ from $A$ onto $B$ as discussed in equation (\ref{eq:2-condensation-1-2-morphisms}). Physically, wrapping with $B$ can be viewed as wrapping with $A$ together with a network of topological lines associated to the condensation monad $\varepsilon \in\text{1-End}_\C(A)$. As a consequence, the wrapping action of $B$ is determined by the wrapping action of $A$ together with the parallel and the linking action of $\varepsilon$. This is illustrated in figure \ref{fig:3d-condensation-action}.

This motivates the proposal that
\begin{equation}\label{eq:3d-tube-hom-isomorphism}
\pi_0\big(\text{1-Hom}_{\mathsf{\T_{S^1}\C}}(X,Y)\big) \;\; \cong \bigsqcup_{[S] \, \in \, \pi_0(\mathsf{C})} \pi_0\big( \text{1-Hom}_{\mathsf{C}}(S \otimes X, Y \otimes X) \big) \, ,
\end{equation}
which states that, up to isomorphism, the wrapping action of generic topological surfaces is determined by the wrapping action of their simple constituents. Furthermore, up to isomorphism, the latter only need to be considered up to condensation. We provide further justification for this proposal in appendix \ref{app:2-condensations-cylinder}.

With the above definition of the tube 2-category $\T_{S^1}\C$ at hand, the collection of data in (\ref{eq:3d-F-objects}), (\ref{eq:3d-F-1-morphisms}), (\ref{eq:3d-F-pseudo-morphisms}) and (\ref{eq:3d-F-2-morphisms}) together with the compatibility conditions (\ref{eq:3d-pseudo-2-morphism-compatibility}), (\ref{eq:3d-vertical-compatibility}) and (\ref{eq:3d-horizontal-compatibility}) can now be summarised conveniently as a pseudo-2-functor
\begin{equation}
\mathcal{F}: \; \T_{S^1}\C \, \to \, \mathsf{2Vec} \, .
\end{equation}
In other words, line operators in the 2-twisted sector transform in 2-representations of the tube 2-category $\T_{S^1}\C$. 

We denote the 2-category of all such linear 2-representations by 
\begin{equation}
[\T_{S^1}\C,\mathsf{2Vec}] \, .
\end{equation}
We expect this is the Karoubi or idempotent completion of the tube category $\T_{S^1}\C$ and has the structure of a braided fusion 2-category. The fusion structure captures the action of $\C$ on products of parallel twisted sector line defects supported at separate points in the transverse plane. The braided fusion structure arises because the associated configuration space is homotopic to $S^1$.

\subsubsection{Tube 2-algebra}

There is an equivalent formulation of the above 2-representation theory in terms of the \textit{tube 2-algebra} $\mathcal{A}_{S^1}(\mathsf{C})$. This is the multi-fusion category
\begin{equation}
\mathcal{A}_{S^1}(\mathsf{C}) \; := \; \text{1-End}_{\T_{S^1}\C}\Big( \bigoplus_{[S] \, \in \, \pi_0(\C)} \! S \; \Big) \, ,
\end{equation} 
where $S$ runs over a complete set of representatives of connected components of $\mathsf{C}$ and the monoidal structure is given by (horizontal) composition. 

\begin{figure}[h]
	\centering
	\includegraphics[height=3.6cm]{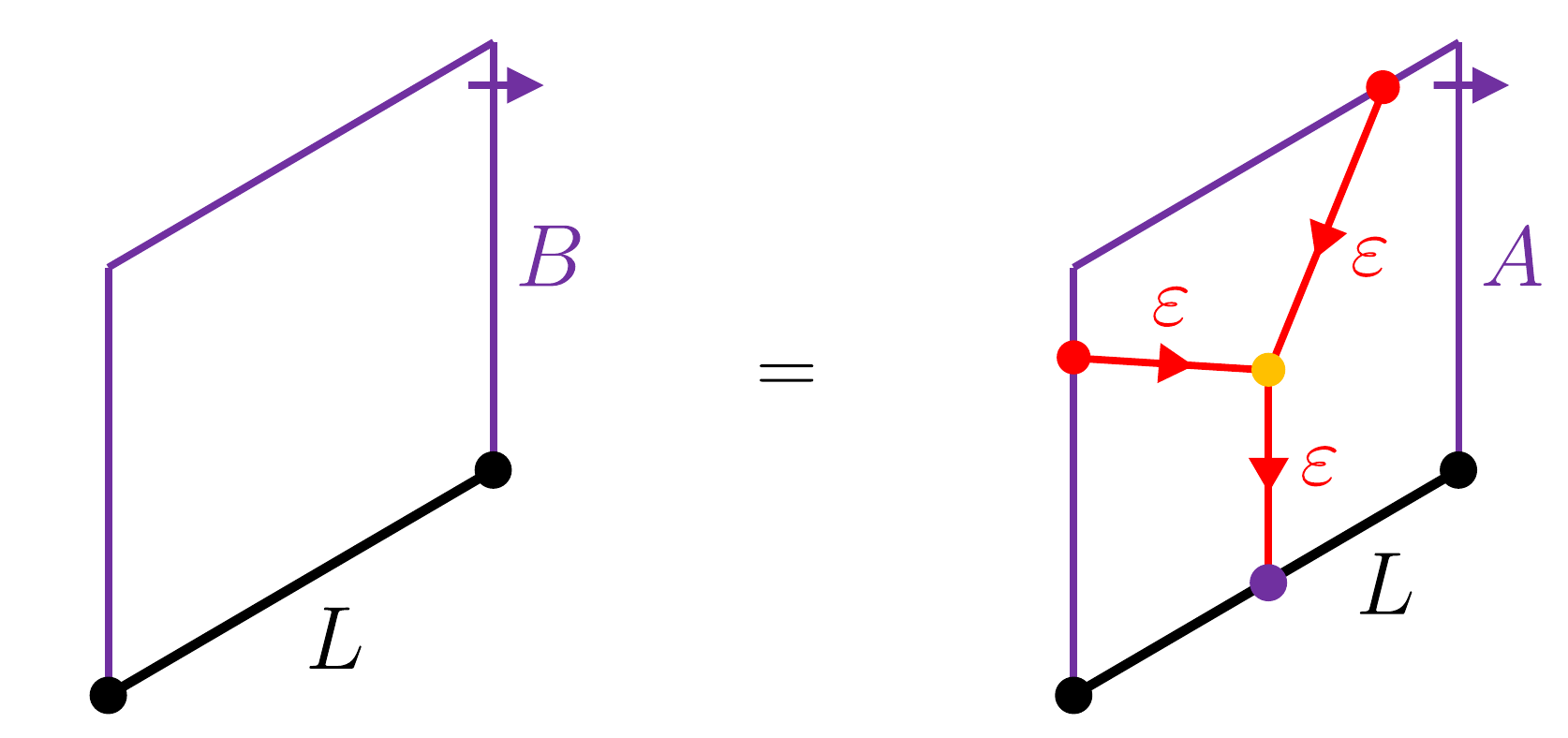}
	\vspace{-5pt}
	\caption{}
	\label{fig:3d-condensation-twisted-sectors}
\end{figure}

From a physical perspective, the restriction to simple objects up to condensation can be motivated by the fact that, given a condensation $A \cond B$ from $A$ onto $B$, we can view 2-twisted sector lines at the end of $B$ as 2-twisted sector lines at the end of $A$ that are acted upon by the associated condensation monad $\varepsilon \in \text{1-End}_{\mathsf{C}}(A)$. This is illustrated in figure \ref{fig:3d-condensation-twisted-sectors}. We will give a more detailed discussion of twisted sectors for condensations in the context invertible higher group-like symmetries later.

Any 2-representation $\mathcal{F} : \T_{S^1}\C \to \mathsf{2Vec}$ of the tube 2-category then determines a 2-representation of the tube 2-algebra $\mathcal{A}_{S^1}(\mathsf{C})$ on the 2-vector space 
\begin{equation}
\mathcal{S} \; := \; \mathcal{F}\Big(\bigoplus_{[S]}\, S \Big)
\end{equation}
by sending 2-algebra elements
\begin{gather}
\begin{aligned}
a \in \mathcal{A}_2(\mathsf{C}) \;\; &\mapsto \;\; \mathcal{F}(a) \in \text{1-End}(\mathcal{S}) \, , \\
\phi \in \text{Hom}(a,b) \;\; &\mapsto \;\; \mathcal{F}(\phi) \in \text{2-Hom}(\mathcal{F}(a),\mathcal{F}(b)) \, .
\end{aligned}
\end{gather}
This sets up an equivalence of braided fusion 2-categories
\begin{equation}
[\mathsf{T_{S^1}C},\mathsf{2Vec}] \; \cong \; \mathsf{2Rep}(\mathcal{A}_{S^1}(\mathsf{C})) \, .
\end{equation}
From either perspective, this provides a complete description of the 2-representation theory of a fusion 2-category symmetry $\C$ on 2-twisted sector line operators in dimension $D = 3$. We will speak uniformly of tube 2-algebra as \textit{tube 2-representations} in what follows.

\subsection{Sandwich construction}

We now rephrase the above construction of tube representations in the context of the sandwich construction. This is realised by an equivalence between the 2-category of tube 2-representations and the 2-category of topological surfaces in the associated four-dimensional Turaev-Viro theory $\text{TV}_\C$.

The latter is the Drinfeld center
\be
\text{TV}_\C(S^1) \; :=\; \int_{S^1} \C \; = \; \mathcal{Z}(\C)
\ee
of the fusion 2-category $\C$. We begin with a brief summary of the Drinfeld center $\cZ(\C)$ following~\cite{BAEZ1996196,davydov2021braided,Kong:2019brm}, before giving an explicit description of the equivalence with tube 2-representations in one direction. 

An object in $\cZ(\C)$ is a triple $z = (U,\tau,\Sigma)$ consisting of 
\begin{enumerate}
\item an object $U \in \mathsf{C}$,

\item a half-braiding $(.) \otimes U \xRightarrow{\, \tau \;} U \otimes (.)$ with component 1-morphisms as in (\ref{eq:2d-half-braiding}) and component 2-morphisms
\vspace{-5pt}
\begin{equation}\label{eq:3d-half-braiding}
\begin{gathered}
\includegraphics[height=3.3cm]{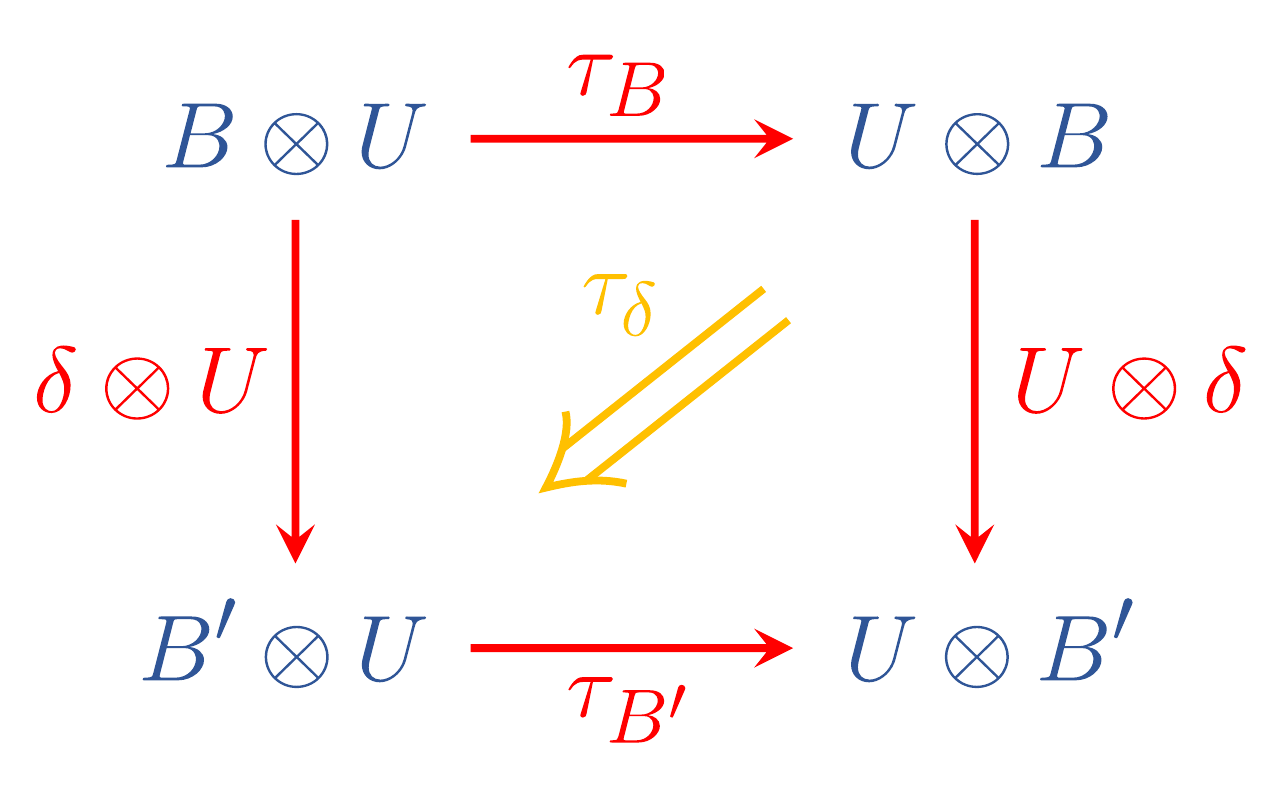}
\end{gathered}
\vspace{-2pt}
\end{equation}
for each 1-morphism $\delta \in \text{Hom}_{\mathsf{C}}(B,B')$,

\item a modification $\Sigma$ with component 2-morphisms
\begin{equation}
\begin{gathered}
\includegraphics[height=7.2cm]{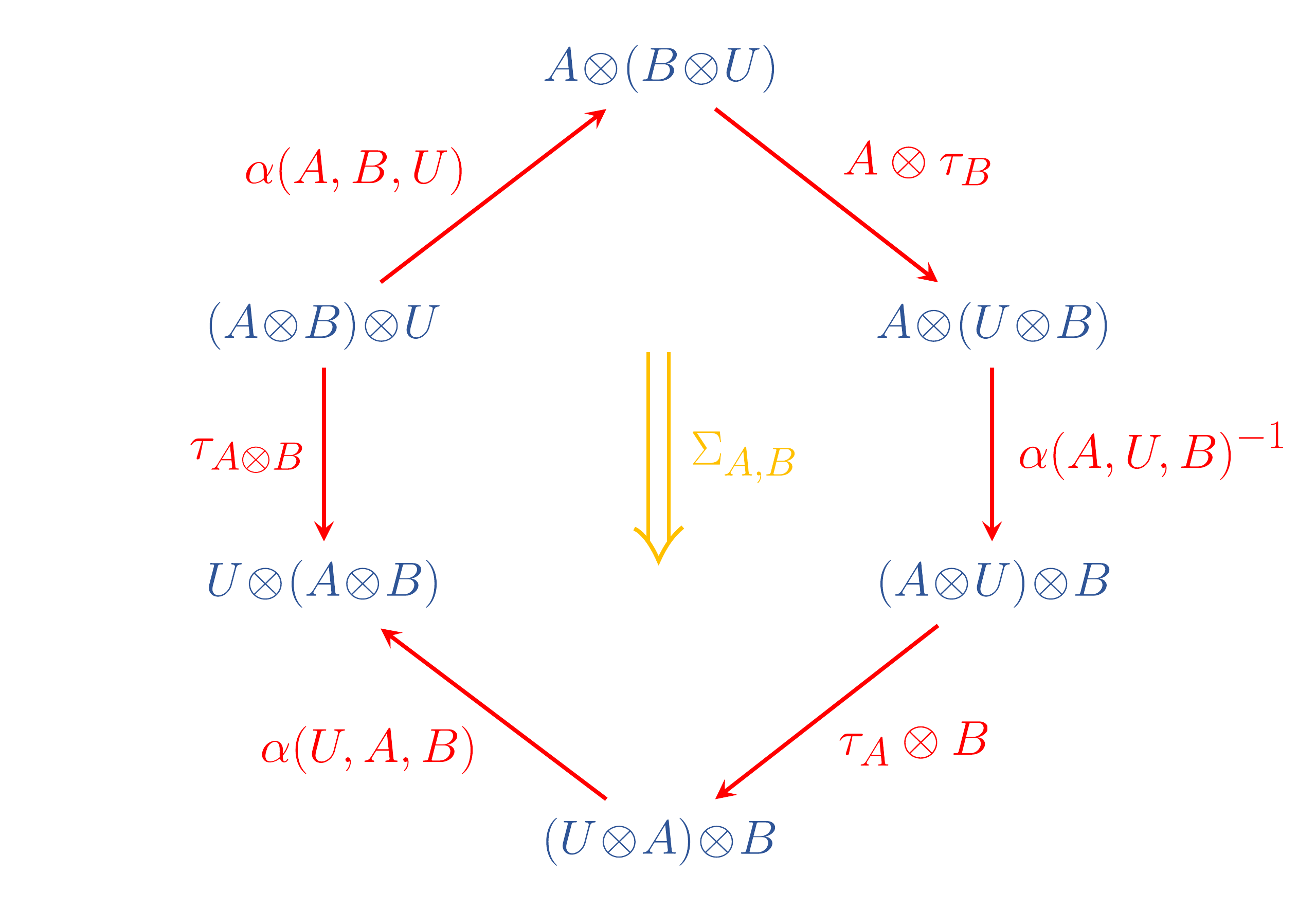}
\end{gathered}
\end{equation}
for each pair of objects $A,B \in \mathsf{C}$, subject to suitable coherence conditions \cite{Kong:2019brm}.
\end{enumerate}
Note that there is a forgetful 2-functor $F: \mathcal{Z}(\C) \to \C$ that discards the information of $\tau$ and $\Sigma$ and sends $z \mapsto F(z) = U$. This forgetful functor will become the bulk-to-boundary map in the sandwich construction discussed below.

To each object $z \in \mathcal{Z}(\mathsf{C})$ of the Drinfeld centre we can associate a tube 2-representation $\mathcal{F}_z \in [\T_{S^1}\C,\mathsf{2Vec}]$ as follows:
\begin{itemize}
\item To an object $X \in \T_{S^1}\C$ it assigns the 2-vector space $\mathcal{F}_z(X)$ given by the set of simple objects in $\text{1-Hom}_{\mathsf{C}}(U,X)$. 

\item To a 1-morphisms $[B,\psi] \in \text{1-Hom}_{\T_{S^1}\C}(X,Y)$ it assigns the matrix of vector spaces
\begin{equation}
\begin{gathered}
\includegraphics[height=1.08cm]{2d-dia-drinfeld-morphism-1.pdf}
\end{gathered}
\end{equation}
whose entry at $K \in \mathcal{F}_z(Y)$ and $L \in \mathcal{F}_z(X)$ is given by the vector space of 2-morphisms
\begin{equation}
\begin{gathered}
\includegraphics[height=4.2cm]{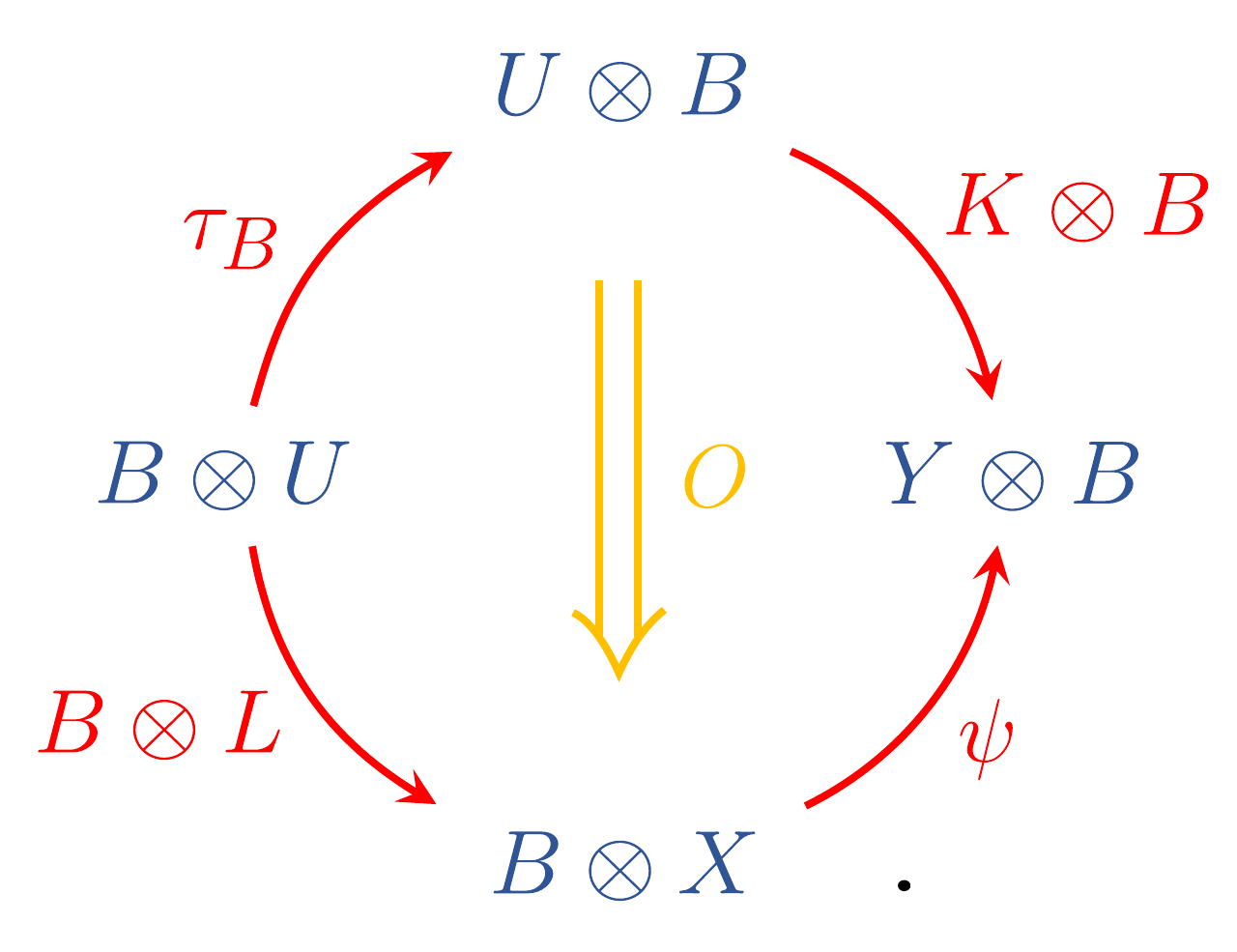}
\end{gathered}
\end{equation}

\item The composition of two matrices $\mathcal{F}_z(A,\varphi)$ and $\mathcal{F}_z(B,\psi)$ is controlled by the matrix of linear maps
\vspace{3pt}
\begin{equation}\label{eq:3d-drinfeld-pseudo-morphisms}
\begin{gathered}
\includegraphics[height=3.3cm]{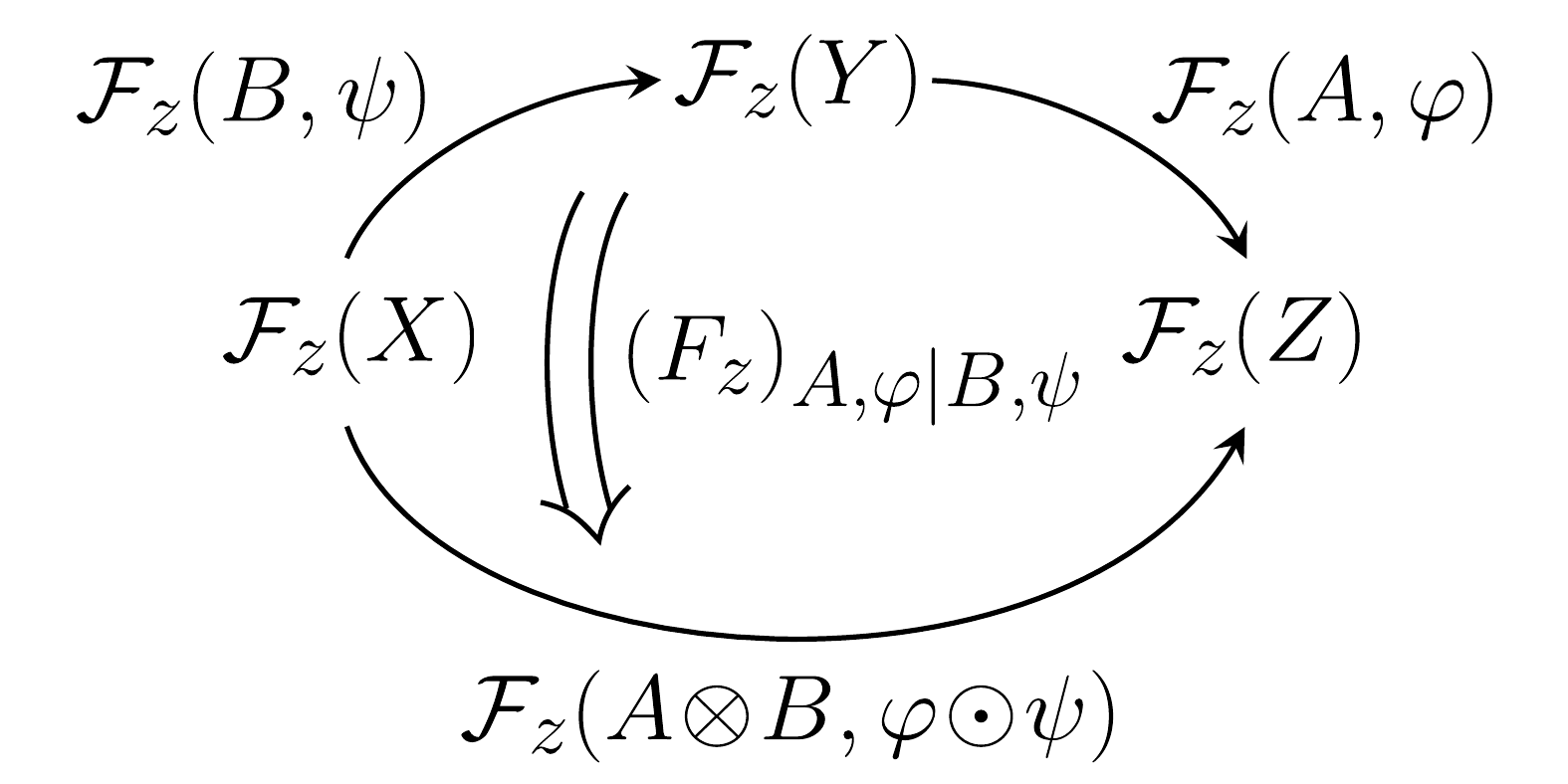}
\end{gathered}
\vspace{-2pt}
\end{equation}
whose entry at $J \in \mathcal{F}_z(Z)$ and $L \in \mathcal{F}_z(X)$ is given by the linear map that sends the tensor product of two vectors 
\begin{equation}
O_1 \, \in \, \mathcal{F}_z(A,\varphi)_{J,K} \qquad \text{and} \qquad O_2 \, \in \, \mathcal{F}_z(B,\psi)_{K,L}
\end{equation}
to the vector $O \in \mathcal{F}_z(A \otimes B, \varphi \odot \psi)_{J,L}$ defined by the diagram
\begin{equation}
\begin{gathered}
\includegraphics[height=9.1cm]{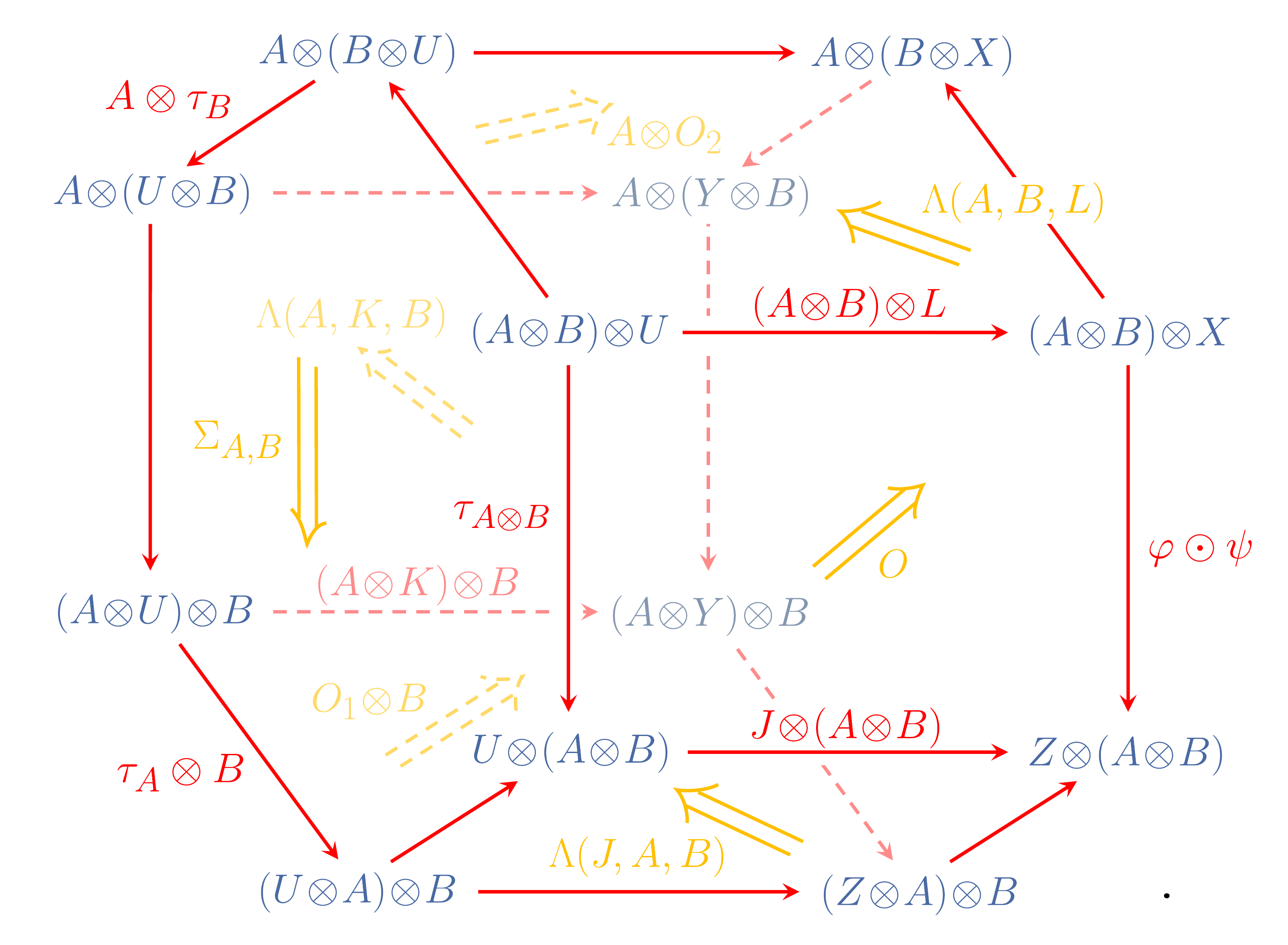}
\end{gathered}
\end{equation}

\item To a 2-morphism $[\delta,\Theta] \in \text{2-Hom}_{\T_{S^1}\C}([B,\psi],[B',\psi'])$ it assigns the matrix of linear maps
\begin{equation}
\begin{gathered}
\includegraphics[height=3.7cm]{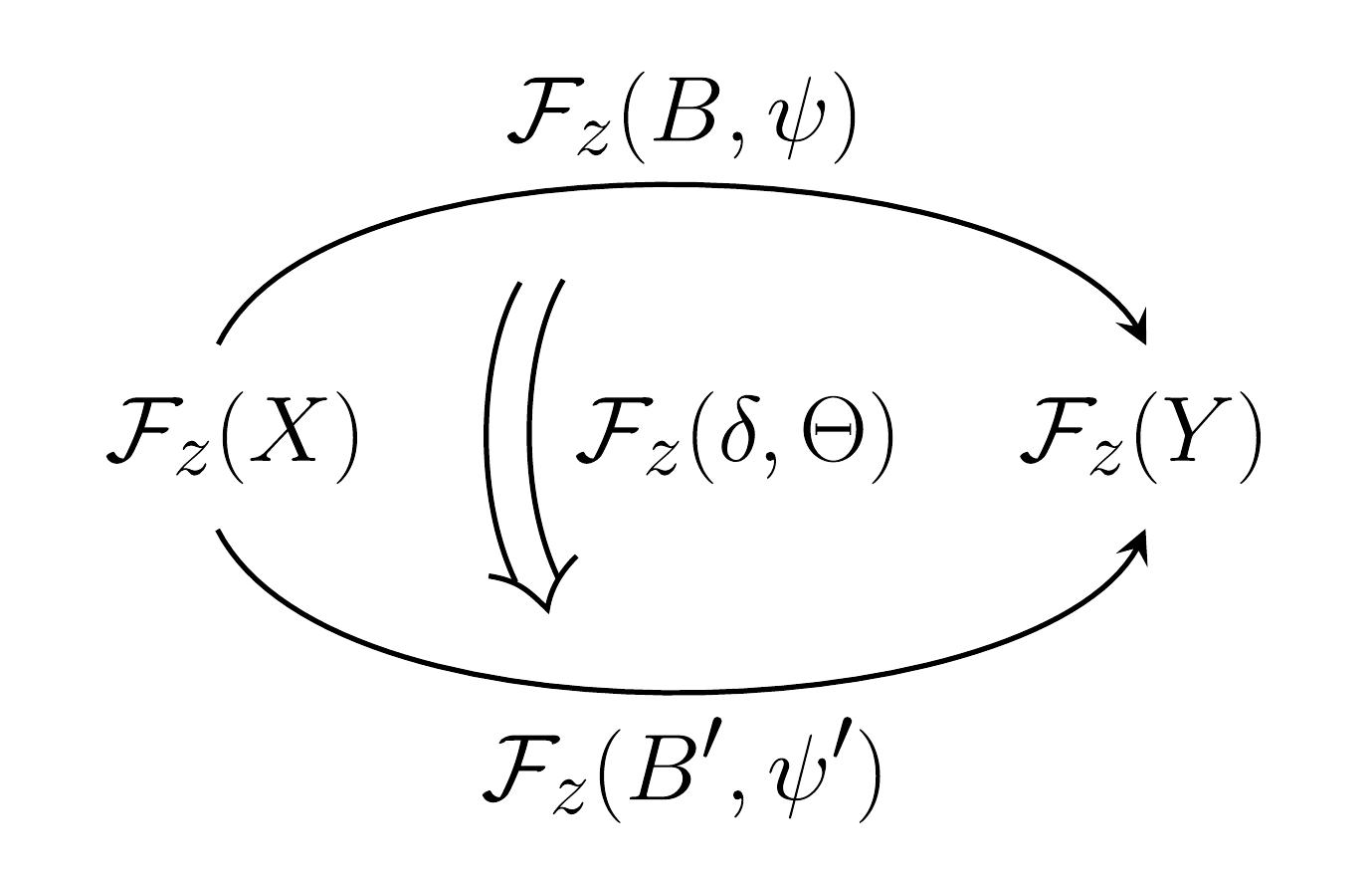}
\end{gathered}
\end{equation}
whose entry at $K \in \mathcal{F}_z(Y)$ and $L \in \mathcal{F}_z(X)$ is given by the linear map that sends a vector $O \in \mathcal{F}_z(B,\psi)_{K,L}$ to the vector $O' \in \mathcal{F}_z(B',\psi')_{K,L}$ defined by the diagram
\begin{equation}
\begin{gathered}
\includegraphics[height=8cm]{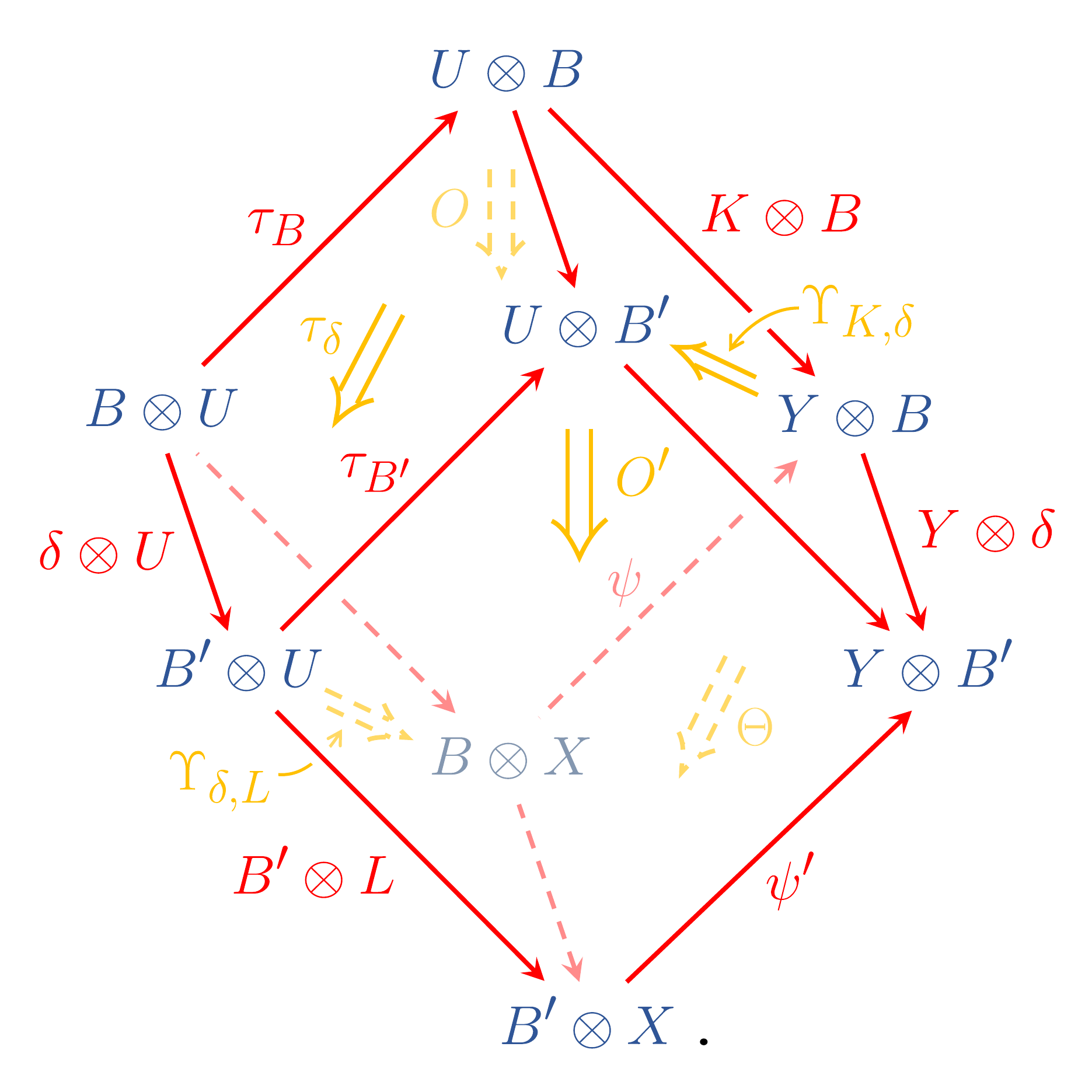}
\end{gathered}
\end{equation}
\end{itemize}

We propose that, up to equivalence, every tube 2-representation is of the form $\mathcal{F} = \mathcal{F}_z$ for some $z \in \mathcal{Z}(\mathsf{C})$, and the mapping $z \mapsto \mathcal{F}_z$ extends to an equivalence 
\begin{equation}\label{eq:3d-equivalence-drinfeld-tube-2-reps}
\mathcal{Z}(\mathsf{C}) \; \cong \; [\T_{S^1}\C,\mathsf{2Vec}] \, 
\end{equation}
of braided fusion 2-categories.

Let us explain the physical picture underpinning this equivalence. We first put forward an intrinsically three-dimensional perspective that extends the categorical perspective in our previous work~\cite{Bartsch:2023pzl} to a fusion 2-category symmetry acting on 2-twisted sector line operators, before discussing its interpretation in the four-dimensional sandwich construction.

\begin{figure}[h]
	\centering
	\includegraphics[height=6.2cm]{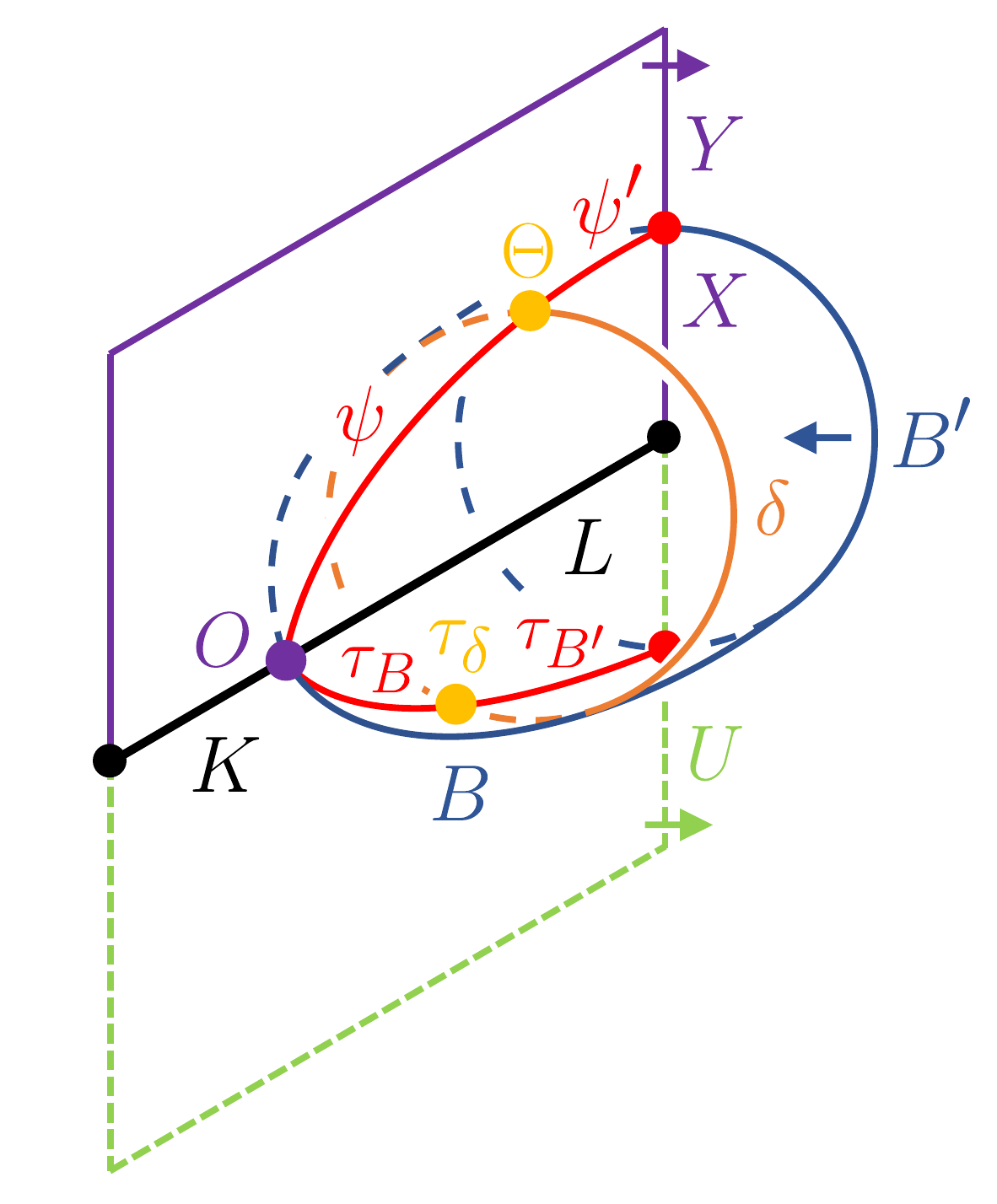}
	\vspace{-5pt}
	\caption{}
	\label{fig:3d-drinfeld-isomorphism}
\end{figure}

From an intrinsically three-dimensional perspective, line operators $L$ in the $X$-twisted sector are viewed as simple interfaces between an auxiliary topological surface defect $U$ and $X$, as illustrated in figure \ref{fig:3d-drinfeld-isomorphism}. We then have the following interpretations:
\begin{itemize}
\item $\mathcal{F}_z(X)$ is identified with the set of simple objects of $\text{1-Hom}_{\mathsf{C}}(U,X)$. 
\item The action of 1-morphisms $[B,\psi]$ is obtained by wrapping with a cylindrical surface $B$ that intersects $U$ via the associated component 1-morphism of the half-braiding $\tau$. Shrinking the cylinder then produces the matrix of vector spaces $\mathcal{F}_z(B,\psi)$ whose elements are junctions $O$. 
\item The action of 2-morphisms $[\delta, \Theta]$ is obtained by linking with a circular line $\delta$ that intersects $U$ via the associated component 2-morphism of the half-braiding $\tau$. They act on junctions by sliding down the corresponding circle towards $O$ from the right, which produces the matrix $\mathcal{F}_z(\delta,\Theta)$ of linear maps. 
\end{itemize}

\begin{figure}[h]
	\centering
	\includegraphics[height=4.7cm]{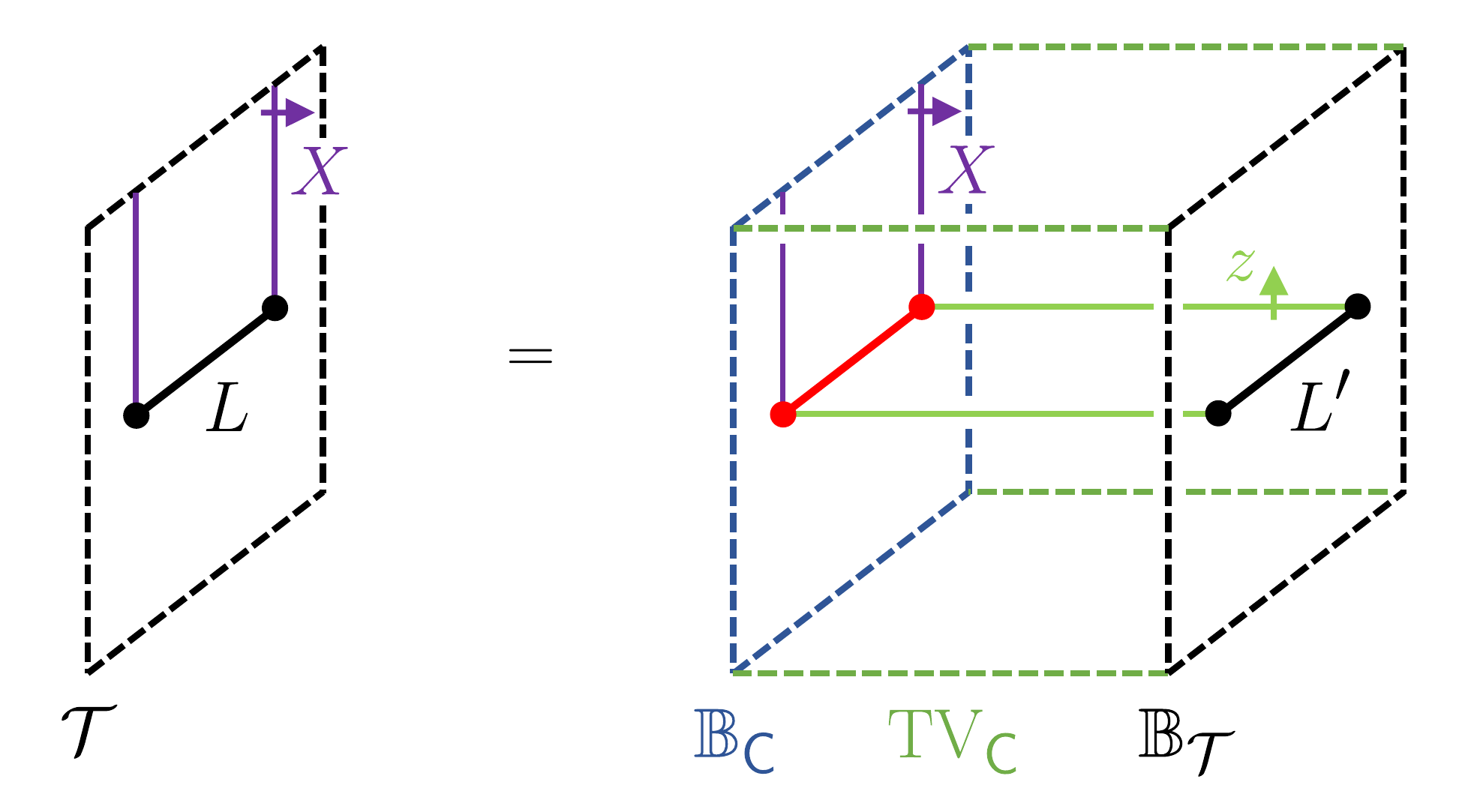}
	\vspace{-5pt}
	\caption{}
	\label{fig:3d-sandwich-construction}
\end{figure}

In the sandwich construction, we instead view a three-dimensional theory $\cT$ with spherical fusion 2-category symmetry $\C$ as an interval compactification of the associated four-dimensional topological theory $\text{TV}_{\mathsf{C}}$. The Drinfeld center is identified with the braided fusion 2-category of topological surfaces
\be
\text{TV}_\C(S^1) \; :=\; \mathcal{Z}(\C) \; = \; \int_{S^1} \C \, ,
\ee
which is what the four-dimensional topological theory $\text{TV}_\C$ assigns to $S^1$. The left boundary condition is the canonical gapped boundary condition $\mathbb{B}_\C$ with associated bulk-to-boundary 2-functor $F : \cZ(\C) \to \C$, while the right boundary condition $\mathbb{B}_\cT$ contains informations about the theory $\cT$. This setup is illustrated in figure~\ref{fig:3d-sandwich-construction}.

A tube 2-representation $\cF_z \in [\T_{S^1}\C,\vect]$ with underlying 2-vector space
$\mathcal{F}_z(X)$ given by the simple objects of $\text{1-Hom}_{\mathsf{C}}(U,X)$ now corresponds to line operators on the boundary $\mathbb{B}_\C$ that sit at the junction between a bulk surface $z = (U,\tau,\Sigma) \in \mathcal{Z}(\C)$ and a boundary surface $X \in \C$, as illustrated in figure \ref{fig:3d-sandwich-construction}.

As a special case of the above construction, we define the \textit{adjoint 2-representation} as the tube 2-representation that is induced by the monoidal unit $\mathbf{1} \in \mathcal{Z}(\mathsf{C})$ of the Drinfeld center, i.e.
\begin{equation}\label{eq:adjoint-2-rep}
\text{Ad} \; := \; \mathcal{F}_{\mathbf{1}} \, .
\end{equation}
From the discussion above, it is clear that this tube 2-representation describes the action of the symmetry 2-category $\mathsf{C}$ on its own topological line defects, which explains our choice of notation.

\subsection{Junction operators}
\label{subsec:junction-operators}

\begin{figure}[h]
	\centering
	\includegraphics[height=3.4cm]{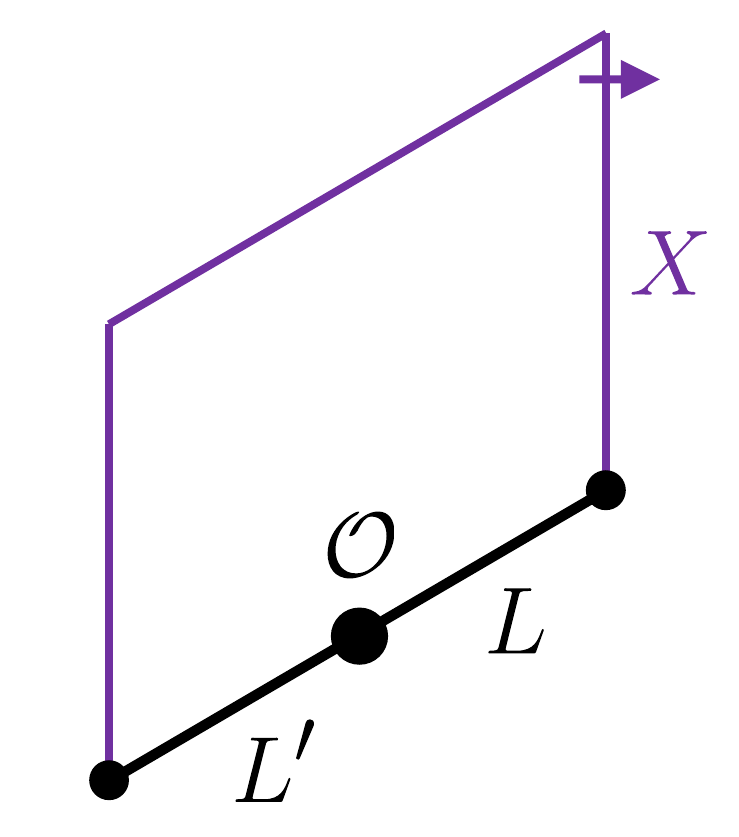}
	\vspace{-5pt}
	\caption{}
	\label{fig:3d-twisted-sectors-2}
\end{figure}

So far we considered 2-twisted sector line operators that are attached to topological surface defects $X \in \mathsf{C}$. We now consider local operators $\mathcal{O}$ that sit at the junction between 2-twisted sector lines as illustrated in figure \ref{fig:3d-twisted-sectors-2}. 

Concretely, given two line operators $L$ and $L'$ in the $X$-twisted sector transforming in tube 2-representations $\mathcal{F},\mathcal{F}'$, respectively, we denote by
\begin{equation}
(\Xi_X)_{L',L} \, \in \, \mathsf{Vec}
\end{equation}
the vector space of local operators $\mathcal{O}$ that can sit at the junction of $L$ and $L'$, as illustrated in figure \ref{fig:3d-twisted-sectors-2}. By scanning over $L$ and $L'$, we obtain a matrix $\Xi_X$ of vector spaces, which we interpret as a 1-morphism
\vspace{-4pt}
\begin{equation}\label{eq:3d-xi-objects}
\begin{gathered}
\includegraphics[height=1.1cm]{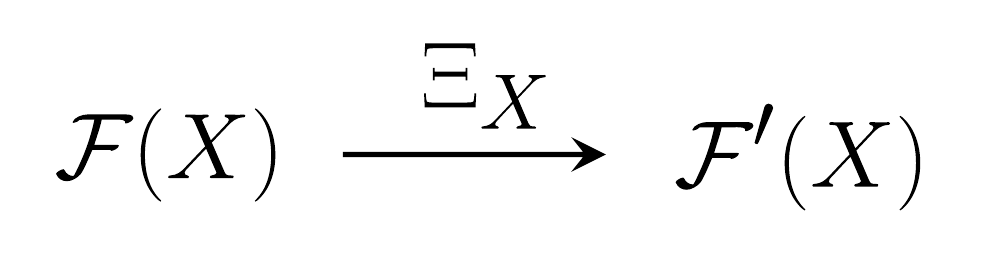}
\end{gathered}
\vspace{-4pt}
\end{equation}
in the 2-category $\mathsf{2Vec}$ of 2-vector spaces.

\begin{figure}[h]
	\centering
	\includegraphics[height=4.5cm]{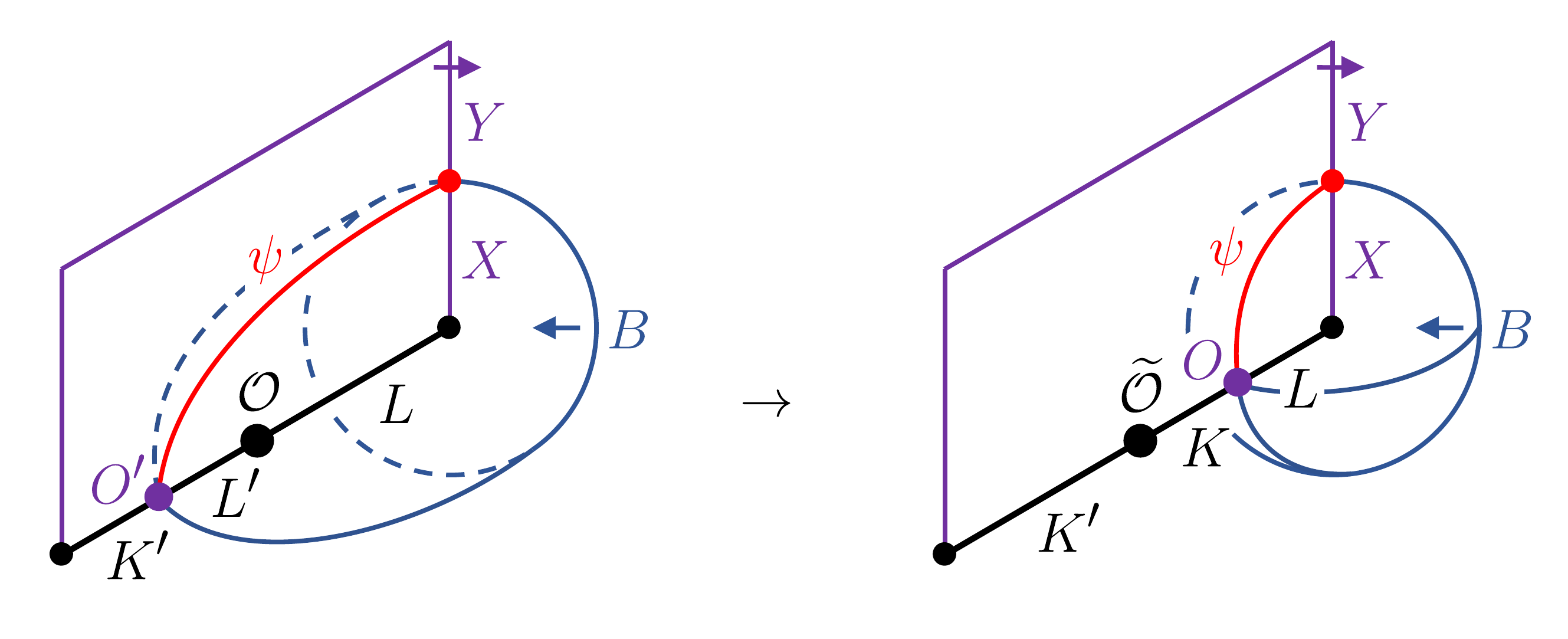}
	\vspace{-5pt}
	\caption{}
	\label{fig:3d-junction-action}
\end{figure}

Symmetry defects $B \in \mathsf{C}$ can act on elements $\mathcal{O} \in (\Xi_X)_{L',L}$ in the manner illustrated in figure \ref{fig:3d-junction-action}. 
Concretely, consider wrapping the lines $L$ and $L'$ with a cylindrical surface $B$ via a specified intersection 1-morphism
\vspace{-8pt}
\begin{equation}
\begin{gathered}
\includegraphics[height=1.18cm]{2d-dia-morphism.pdf}
\end{gathered}
\vspace{-6pt}
\end{equation}
as before. Shrinking the cylinder down towards $L'$ gives a choice of topological operator
\begin{equation}
O' \; \in \; \mathcal{F}'(B,\psi)_{K',L'}
\end{equation}
sitting in-between the wrapped line $L'$ and a new line $K'$ in the $Y$-twisted sector. Upon dragging the topological operator $O'$ through $\mathcal{O}$ from left to right, the latter is transformed into a new junction operator 
\begin{equation}
\widetilde{\mathcal{O}} \; \in \; (\Xi_Y)_{K',K}
\end{equation}
between lines $K$ and $K'$ in the $Y$-twisted sector, whereas the former becomes a new topological operator
\begin{equation}
O \; \in \; \mathcal{F}(B,\psi)_{K,L}
\end{equation}
to the right of $\widetilde{\mathcal{O}}$. For fixed $L$ and $K'$, the pair $(B,\psi)$ thus determines a linear map
\begin{equation}
\begin{tikzcd}[row sep=3pc]
\displaystyle \bigoplus_{L' \, \in \, \mathcal{F}'(X)} 
\mathcal{F}'(B,\psi)_{K',L'} \; \otimes \; (\Xi_X)_{L',L} \arrow[d, "\displaystyle (\Xi_{(B,\psi)})_{K',L}", pos=0.4, outer sep = 3pt, shorten >= 1pt, shorten <= -6pt] \\
\displaystyle \bigoplus_{K \, \in \, \mathcal{F}(Y)} 
(\Xi_Y)_{K',K} \; \otimes \; \mathcal{F}(B,\psi)_{K,L} \;\, .
\end{tikzcd}
\end{equation}
By scanning over $L$ and $K'$, we then obtain a matrix $\Xi_{(B,\psi)}$ of linear maps, which we interpret as a 2-morphism
\vspace{-4pt}
\begin{equation}\label{eq:3d-xi-morphisms}
\begin{gathered}
\includegraphics[height=3.3cm]{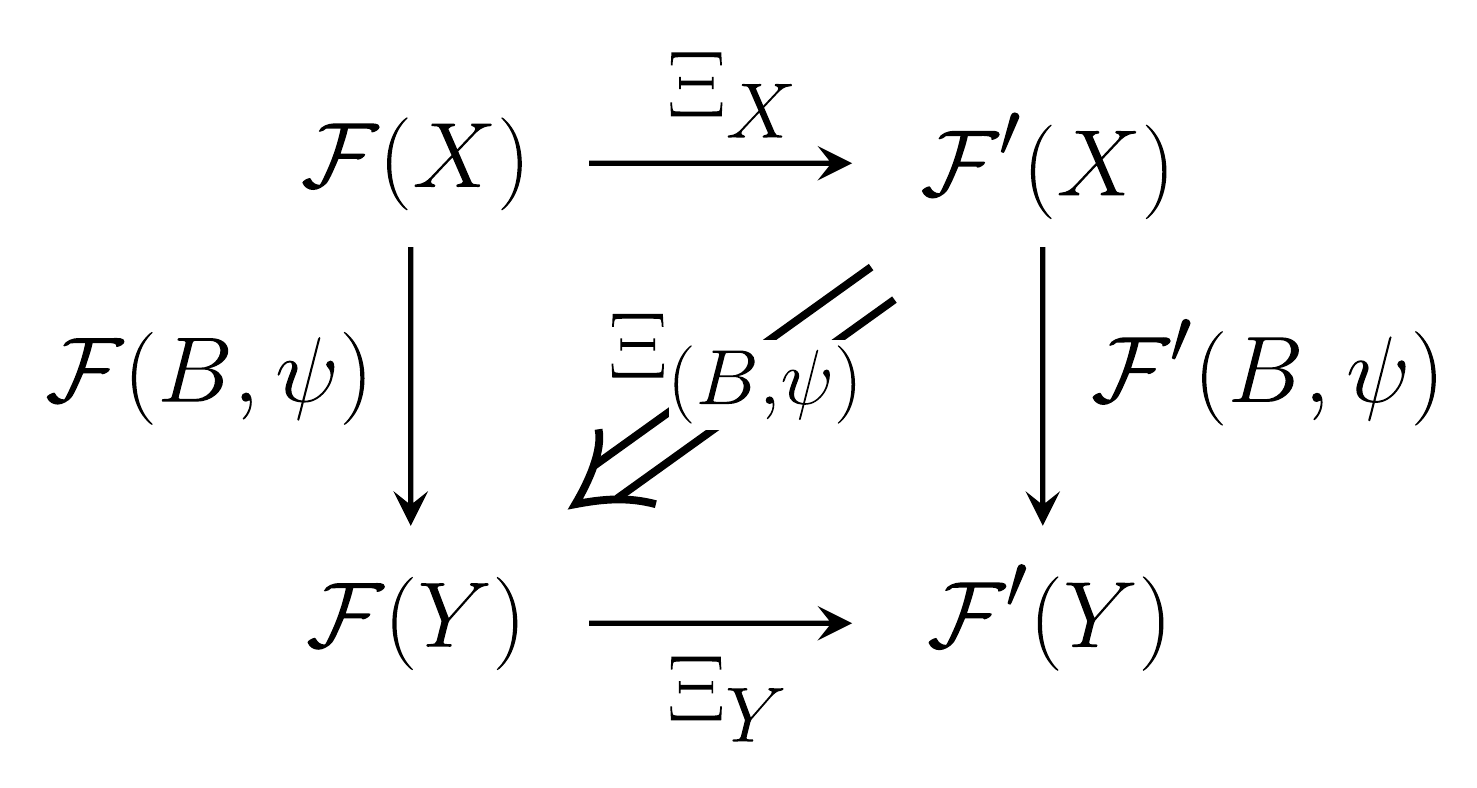}
\end{gathered}
\vspace{-4pt}
\end{equation}
in the 2-category $\mathsf{2Vec}$. By similar arguments to before, this 2-morphism only depends on the equivalence class of the pair $(B,\psi)$ under the relations discussed at the end of sub-section \ref{subsubsec:3d-wrapping-action}. 

\begin{figure}[h]
	\centering
	\includegraphics[height=5.1cm]{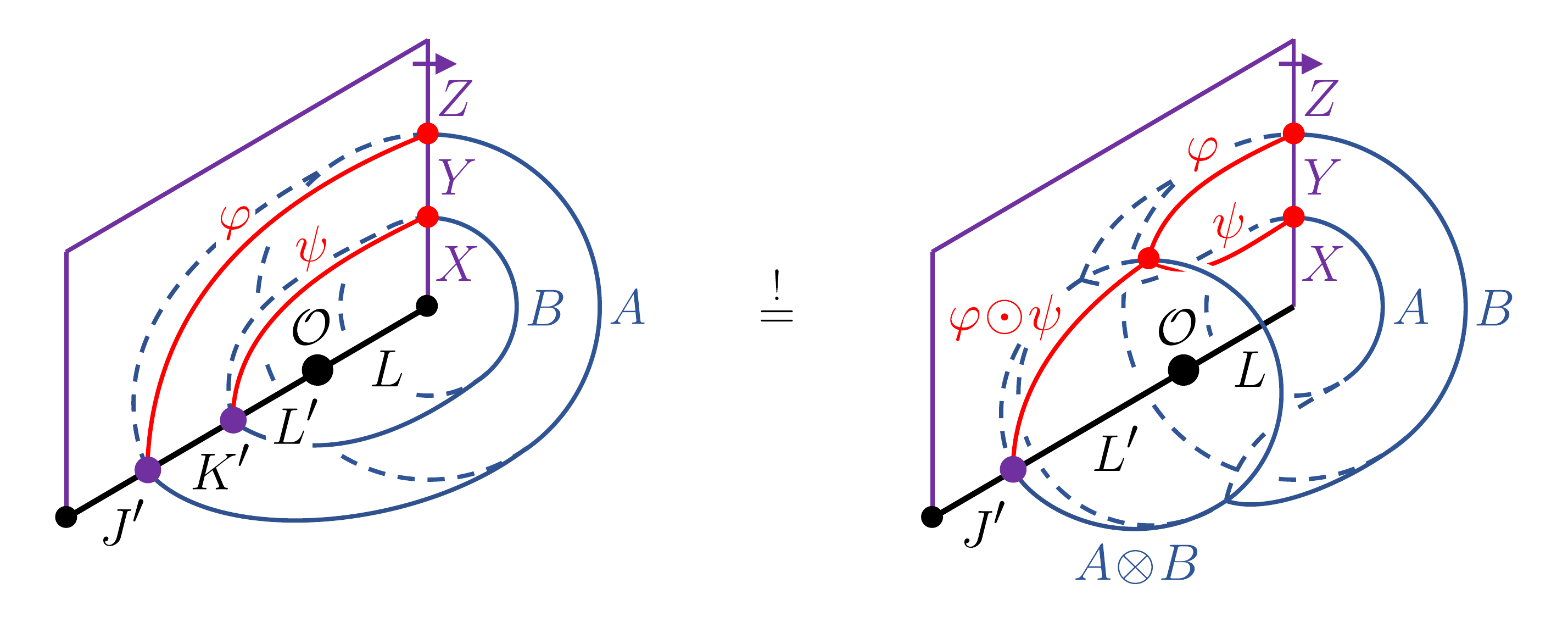}
	\vspace{-5pt}
	\caption{}
	\label{fig:3d-junction-action-2}
\end{figure}

Furthermore, the 2-morphisms $\Xi_{(B,\psi)}$ need to be compatible with the consecutive action of two symmetry defects $A$ and $B$ as illustrated in figure \ref{fig:3d-junction-action-2}. 
Mathematically, this is implemented by the 2-commutativity of the diagram
\vspace{-4pt}
\begin{equation}\label{eq:3d-xi-compatibility}
\begin{gathered}
\includegraphics[height=5.1cm]{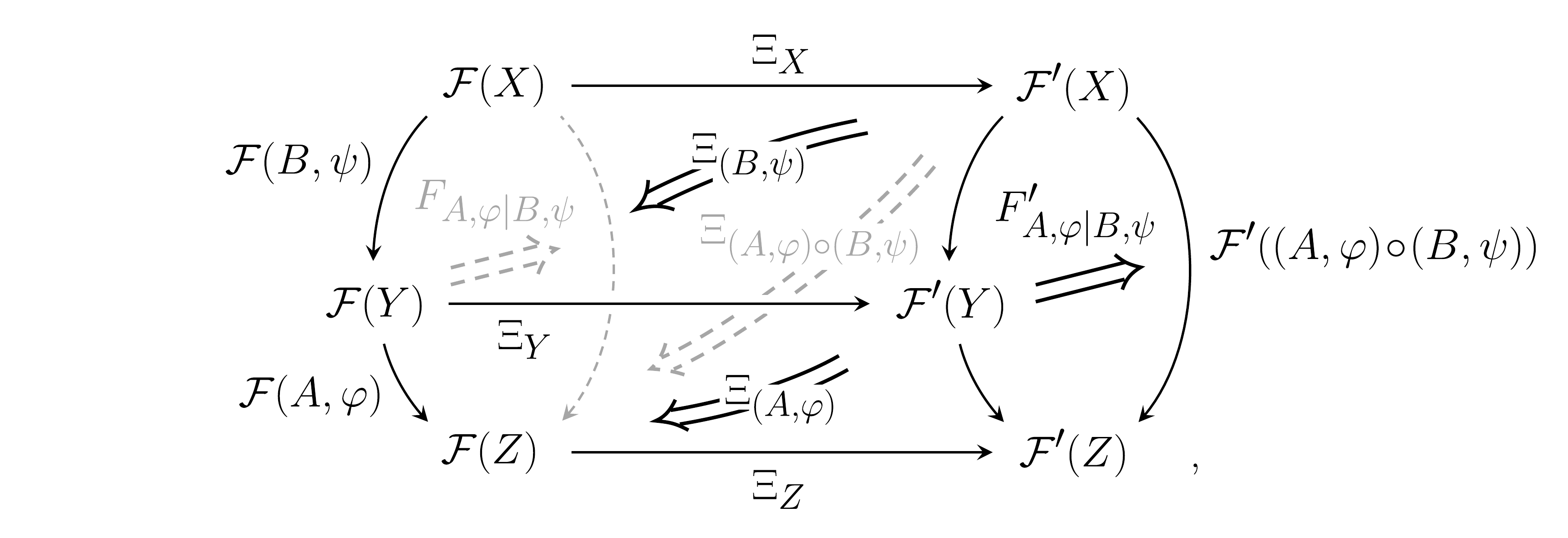}
\end{gathered}
\vspace{4pt}
\end{equation}
which states that acting on a junction operator $\mathcal{O}$ with two symmetry defects $A$ and $B$ consecutively is equivalent to acting on $\mathcal{O}$ with their fusion $A \otimes B$.

The collection of data in (\ref{eq:3d-xi-objects}) and (\ref{eq:3d-xi-morphisms}) together with the compatibility condition (\ref{eq:3d-xi-compatibility}) can now be summarised conveniently as a natural transformation
\begin{equation}
\Xi\, : \; \mathcal{F} \, \Rightarrow \, \mathcal{F}' \, .
\end{equation}
Local junction operators between lines $L$ and $L'$ thus transform in natural transformations between the corresponding tube 2-representations. These correspond to 1-morphisms in the 2-category of tube 2-representations $[\T_{S^1}\C,\mathsf{2Vec}]$.

As a special case, let us consider the category of 1-endomorphisms of the adjoint 2-representation defined in (\ref{eq:adjoint-2-rep}). Using the equivalence (\ref{eq:3d-equivalence-drinfeld-tube-2-reps}), this can be identified with
\begin{equation}
\text{1-End}(\text{Ad}) \; \cong \; \Omega \mathcal{Z}(\mathsf{C}) \, ,
\end{equation}
which using the equivalence (\ref{eq:3d-1-tube-reps-and-drinfeld}) allows us to recover the category of tube representations for 1-twisted local operators discussed in section~\ref{sec:3d-ops}. In this sense, tube 2-representations for 2-twisted sector line defects incorporate tube representations for 1-twisted sector local operators.

\subsection{Example: 2-group symmetry}

Let us now consider a finite 2-group symmetry $\mathcal{G} = (G,A,\rho, \alpha)$ consisting of 
\begin{enumerate}
\item a 0-form symmetry group $G$,
\item an abelian 1-form symmetry group $A$,
\item a group action $\rho: G \to \text{Aut}(A)$ of $G$ on $A$,
\item a representative $\alpha$ of the Postnikov class $[\alpha] \in H^3_{\rho}(G,A)$.
\end{enumerate}
The possible 't Hooft anomalies for a finite 2-group symmetry were classified in~\cite{Kapustin:2013uxa,Benini:2018reh}. For simplicity, we restrict attention to a pure 't Hooft anomaly for the 0-form symmetry specified by a normalised 4-cocycle $\pi \in Z^4(G,U(1))$.

\subsubsection{Symmetry category}

The associated symmetry category is the spherical fusion 2-category $\C = 2\vect_{\cG}^\pi$ of finite-dimensional $\cG$-graded 2-vector spaces with pentagonator twisted by the 't Hooft anomaly $\pi$. It has the following explicit description:
\begin{itemize}
\item The simple objects up to condensation are one-dimensional G-graded 2-vector spaces $1_g$ with graded components $(1_g)_h = \delta_{g,h}$ and fusion $1_g \otimes 1_h \; = \; 1_{gh}$.

\item The 1-morphism spaces between simple objects are the fusion categories
\begin{equation}
\text{1-Hom}(1_g,1_h) \; = \; \delta_{g,h} \cdot \mathsf{Vec}_A
\end{equation}
whose simple objects are one-dimensional vector spaces $\mathbb{C}_a$ with graded components $(\mathbb{C}_a)_b = \delta_{a,b} \cdot \mathbb{C}$. The fusion of 1-morphisms $\mathbb{C}_a \in \text{1-End}(1_g)$ and $\mathbb{C}_b \in \text{1-End}(1_h)$ is
\begin{equation}
\mathbb{C}_a \, \otimes \, \mathbb{C}_b \; = \; \mathbb{C}_{a \, \cdot \, {}^gb} \; \in \; \text{1-End}(1_{gh}) \, .
\end{equation}
\item The associator on simple objects is given by the $A$-graded vector space
\vspace{-5pt}
\begin{equation}
\begin{gathered}
\includegraphics[height=1.18cm]{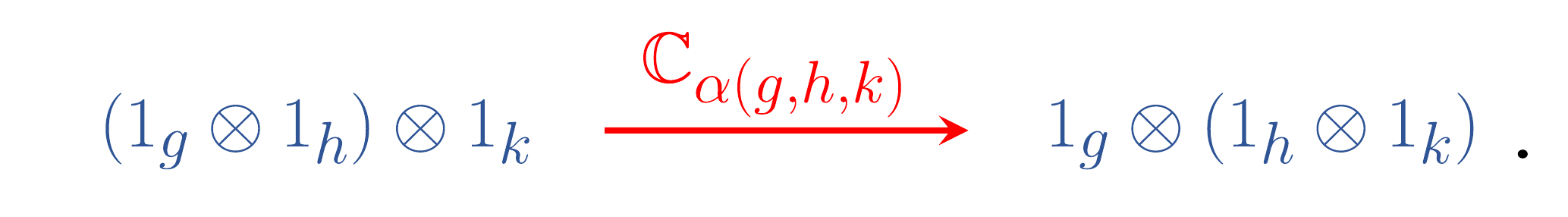}
\end{gathered}
\vspace{-4pt}
\end{equation}

\item The pentagonator on simple objects is given by
\begin{equation}
\begin{gathered}
\includegraphics[height=7.2cm]{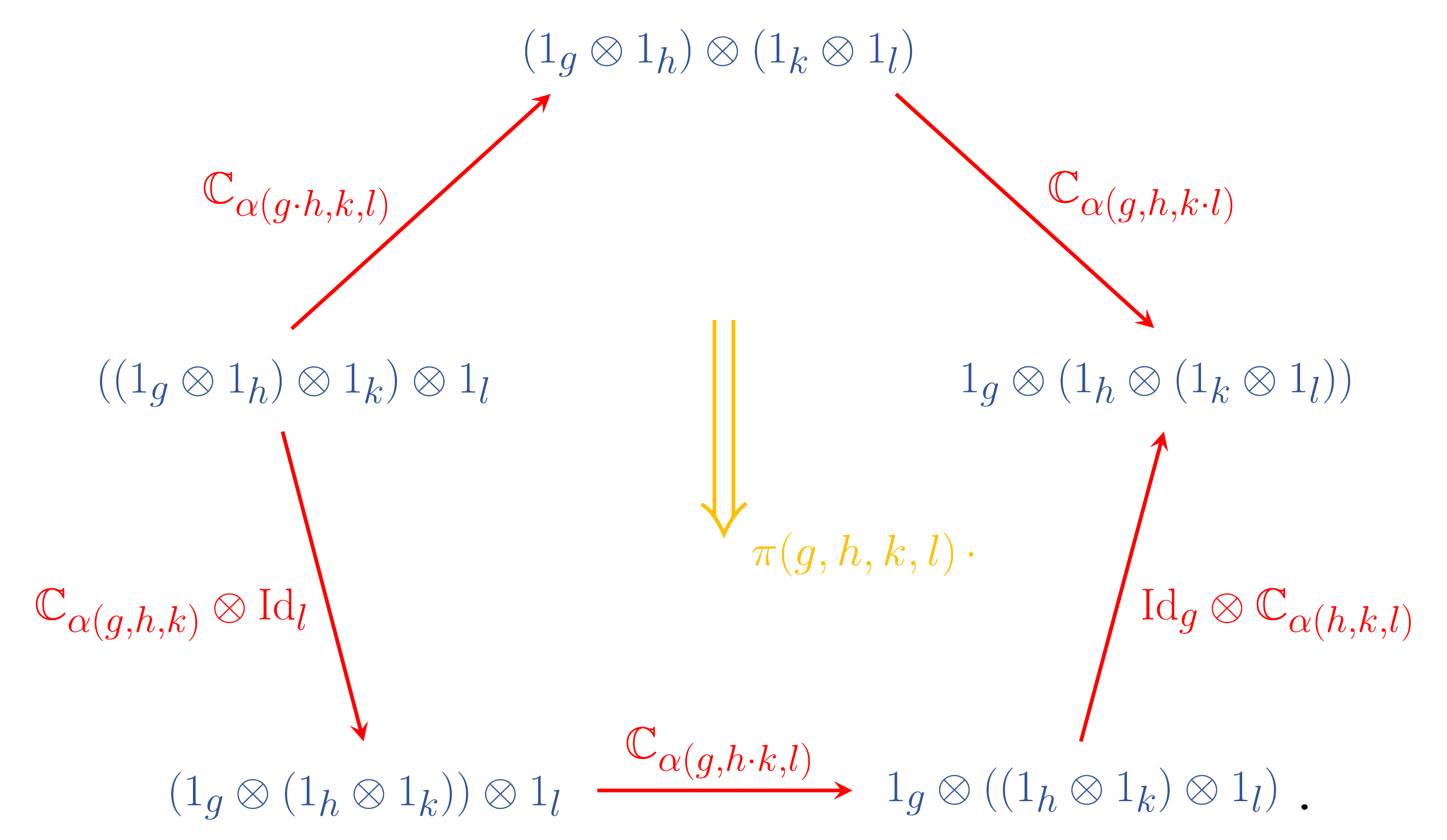}
\end{gathered}
\end{equation}
\end{itemize}

The simple objects up to condensation are the topological surface defects generating the 0-form symmetry $G$, while simple 1-endomorphisms are topological lines generating the 1-form symmetry $A$. They correspond to the homotopy groups $\pi_0(\C)$ and $\pi_1(\C)$ respectively. As before, passing a line labelled by $a \in A$ through a surface labelled by $g \in G$ transforms it into a new line $\rho_g(a) \in A$. 

\begin{figure}[h]
	\centering
	\includegraphics[height=7.8cm]{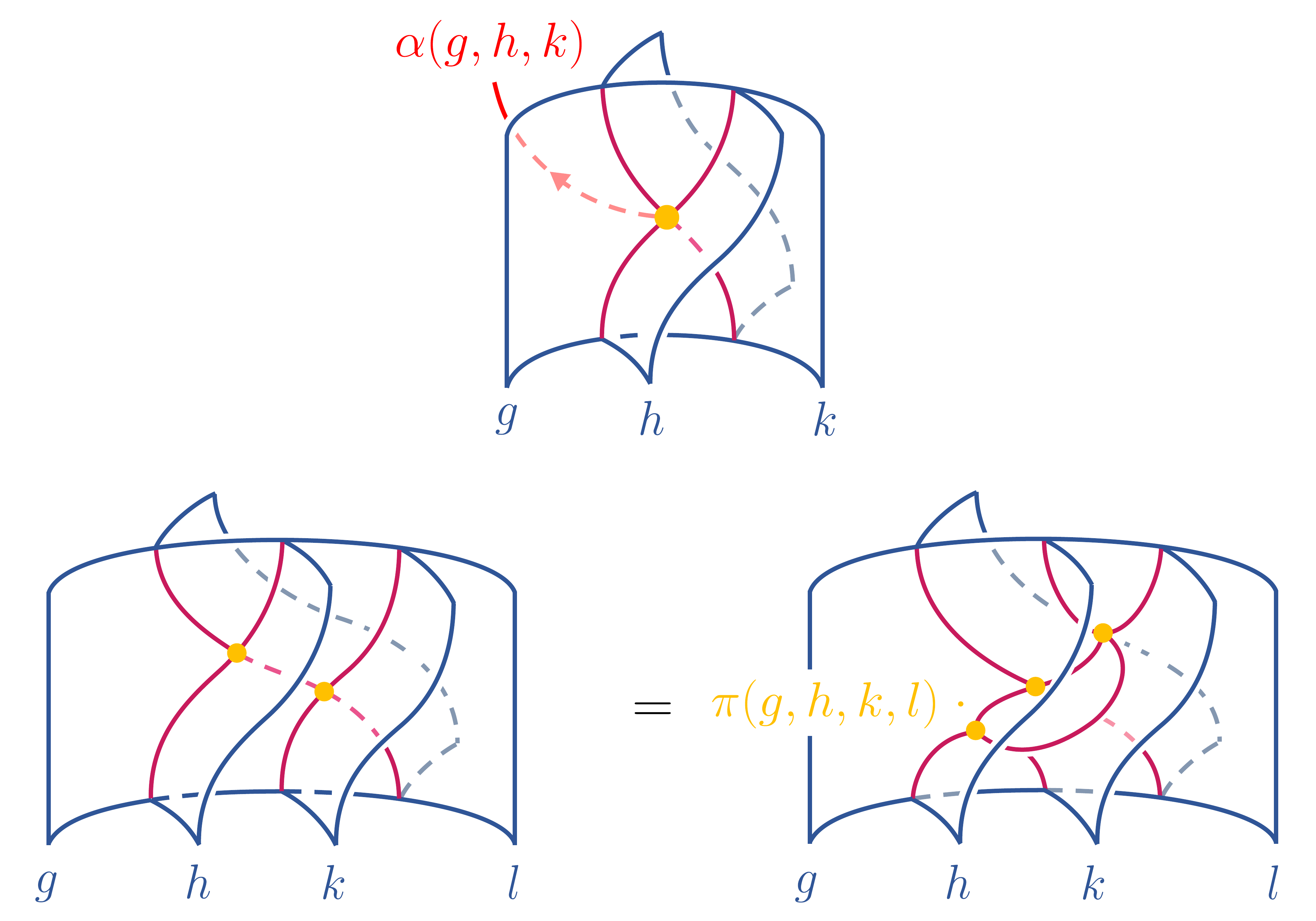}
	\vspace{-5pt}
	\caption{}
	\label{fig:3d-grpex-fusion+associator+pentagonator}
\end{figure}

However, the fusion of three surfaces labelled by $g$, $h$, $k$ is now only associative up to an emitted line  $\alpha(g,h,k) \in A$. The fusion of four topological surfaces labelled by $g$, $h$, $k$, $l$ is invariant under the pentagon move up to a multiplicative phase $\pi(g,h,k,l)$. These properties are illustrated in figure \ref{fig:3d-grpex-fusion+associator+pentagonator}.

\subsubsection{Tube 2-algebra}

The associated tube 2-algebra $\cA_{S^1}(\C)$ has the following description:
\begin{itemize}
\item The simple objects up to ismorphism are given by equivalence classes of pairs $[B,\psi]$ where $B = 1_g$ and $\psi$ has a single graded component
\vspace{-5pt}
\begin{equation}
\begin{gathered}
\includegraphics[height=1.18cm]{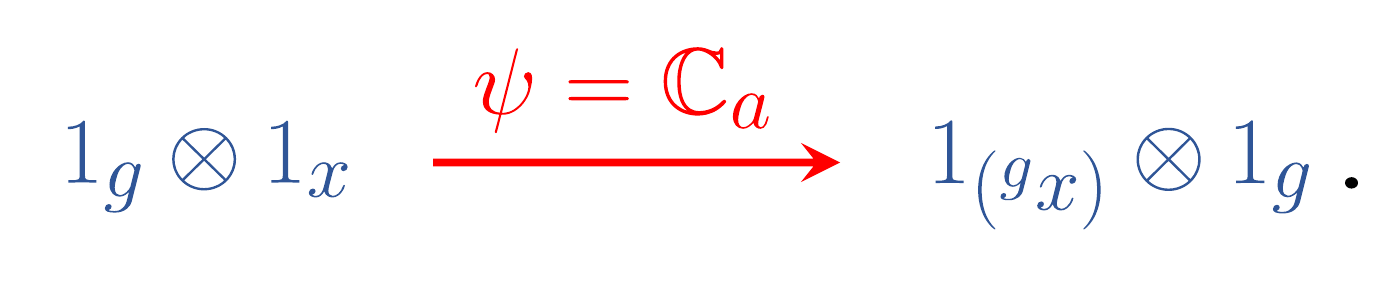}
\end{gathered}
\vspace{-5pt}
\end{equation}
Emulating our previous notation, we denote these objects by $\big\langle \! \yleftarrow[\;a\,]{\;g\,}\! x \big\rangle$. Using the definition (\ref{eq:2d-tube-composition}), their 2-algebra product can be determined to be
\begin{equation}\label{eq:3d-tube-2-alg-product}
\big\langle \! \yleftarrow[\;a\,]{\;g\,}\! x \big\rangle \, \otimes \, \big\langle \! \yleftarrow[\;b\,]{\;h\,}\! y \big\rangle \;\; = \;\; \delta_{x, {}^{h}y} \, \cdot \, \Big\langle \! \yleftarrow[\;a \,\cdot\, {}^{g}b \;\cdot\; \tau_y(\alpha)(g,h)\;]{\;g \,\cdot\, h\;}\! y \Big\rangle \, ,
\end{equation}
where $\tau(\alpha) \in Z^2_{\rho}(G/\! /_{\!\,} G,A)$ again denotes the transgression of $\alpha$ as in (\ref{eq:2d-transgression}). 

\item The morphisms between simple objects $\big\langle \! \yleftarrow[\;a\,]{\;g\,}\! x \big\rangle$ and $\big\langle \! \yleftarrow[\;b\,]{\;h\,}\! y \big\rangle$ are spanned by equivalence classes of pairs $[\delta,\Theta]$ consisting of a 1-morphism $\delta: 1_g \to 1_h$ and a 2-morphism
\begin{equation}\label{eq:3d-dia-grpex-tube-2-alg-morphism}
\begin{gathered}
\includegraphics[height=4.1cm]{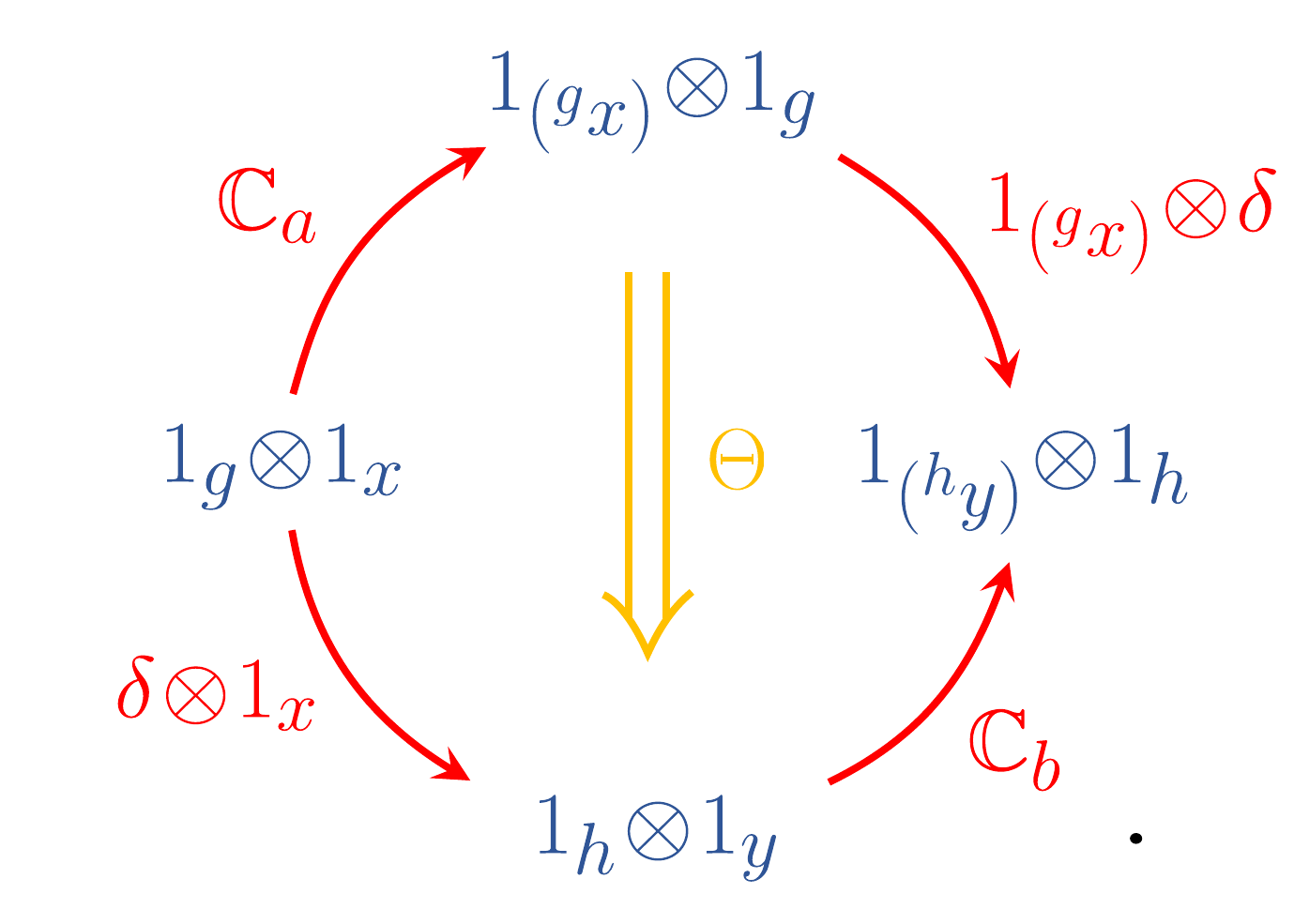}
\end{gathered}
\vspace{-4pt}
\end{equation}
The former can only be non-trivial when $g=h$, whereas the latter can only be non-trivial when $x=y$. Upon choosing $\delta = \mathbb{C}_c$, the diagram in (\ref{eq:3d-dia-grpex-tube-2-alg-morphism}) enforces
\begin{equation}
a \cdot {}^{({}^gx)}c \; \stackrel{!}{=} \; c \cdot b \, .
\end{equation}
The morphism spaces are therefore
\begin{equation}
\text{Hom}\big(\big\langle \! \yleftarrow[\;a\,]{\;g\,}\! x \big\rangle \, , \big\langle \! \yleftarrow[\;b\,]{\;h\,}\! y \big\rangle \big) \,\; = \;\, \delta^{\mspace{1.5mu} g,h}_{\mspace{1.5mu} x,y} \, \cdot \, \mathbb{C}\big[ \mspace{1mu} C_{b / a}^{\mspace{2mu} {}^{g\mspace{-1mu}}x}(A) \mspace{1mu} \big] \, ,
\end{equation}
where for $u \in G$ and $p \in A$ we defined $C^{\mspace{1mu} u}_{p}(A) := \lbrace c \in A \, | \, {}^uc = p \cdot c \rbrace$. 

\item Using the definition (\ref{eq:3d-tube-vertical-composition}), the composition of morphisms is induced by the natural map
\begin{equation}\label{eq:3d-grpex-tube-alg-composition}
C^{\mspace{1mu} u}_{p}(A) \, \times \, C^{\mspace{1mu} u}_{q}(A) \; \to \; C^{\mspace{1mu} u}_{p \cdot q}(A)
\end{equation}
given by multiplication in $A$. Using the definition (\ref{eq:3d-tube-horizontal-composition}), the 2-algebra product of two morphisms
\begin{gather}
\begin{aligned}
n \; &\in \; \text{Hom}\big(\big\langle \! \yleftarrow[\;a\,]{\;g\,}\! x \big\rangle \, , \big\langle \! \yleftarrow[\;b\,]{\;g\,}\! x \big\rangle \big) \,\; = \;\,  \mathbb{C}\big[ \mspace{1mu} C_{b / a}^{\mspace{2mu} {}^{g\mspace{-1mu}}x}(A) \mspace{1mu} \big] \\
m \; &\in \; \text{Hom}\big(\big\langle \! \yleftarrow[\;c\,]{\;h\,}\! y \big\rangle\, , \big\langle \! \yleftarrow[\;d\,]{\;h\,}\! y \big\rangle \big) \,\; = \;\,  \mathbb{C}\big[ \mspace{1mu} C_{d / c}^{\mspace{2mu} {}^{h\mspace{-1mu}}y}(A) \mspace{1mu} \big]
\end{aligned}
\end{gather}
is given by the element of
\begin{equation}
\text{Hom}\Big(\big\langle \! \yleftarrow[\;a\,]{\;g\,}\! x \big\rangle \otimes \big\langle \! \yleftarrow[\;c\,]{\;h\,}\! y \big\rangle \, , \, \big\langle \! \yleftarrow[\;b\,]{\;g\,}\! x \big\rangle \otimes \big\langle \! \yleftarrow[\;d\,]{\;h\,}\! y \big\rangle \Big) \,\; = \;\,  \delta_{x, {}^{h}y} \cdot \mathbb{C}\big[ \mspace{1mu} C_{(b/a) \, \cdot \, {}^{g\mspace{-1.5mu}}(d/c)}^{\, {}^{gh\mspace{-1mu}}y}(A) \mspace{1mu} \big]
\end{equation}
that is defined by
\begin{equation}\label{eq:3d-grpex-tube-alg-2-product}
n \otimes m \,\; = \;\, \delta_{x, {}^{h}y} \cdot n \cdot {}^gm \, .
\end{equation}

\item The associator on simple objects is given by
\vspace{-4pt}
\begin{equation}\label{eq:3d-grpex-tube-2-alg-associator}
\begin{gathered}
\includegraphics[height= 1.55cm]{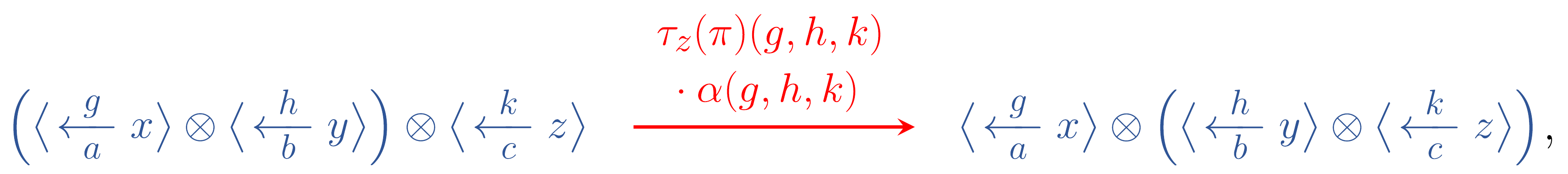}
\end{gathered}
\vspace{-4pt}
\end{equation}
where the multiplicative phase is given by
\begin{equation}
\tau_z(\pi)(g,h,k) \; := \; \frac{\pi(g,h,k,z) \cdot \pi(g,{}^{hk}z,h,k)}{\pi(g,h,{}^kz,k) \cdot \pi({}^{ghk}z,g,h,k)} \, .
\end{equation}
Similarly to before, this defines a groupoid 3-cocycle 
\begin{equation}
\tau(\pi) \, \in \, Z^3\big(G/\! / G,U(1)\big)
\end{equation}
which is the \textit{transgression} of the 't Hooft anomaly $\pi \in Z^4(G,U(1))$.
\end{itemize}

In summary, the tube 2-algebra for an anomalous finite 2-group symmetry may be viewed as a higher categorical analogue of the twisted Drinfeld double. We will motivate this interpretation further though special cases below.

\subsubsection{Tube 2-representations}

A general tube 2-representation $\mathcal{F}$ is given by a collection of finite-dimensional 2-vector spaces $n_x := \mathcal{F}(1_x)$ together with 1-morphisms
\vspace{-3pt}
\begin{equation}\label{eq:3d-dia-grpex-tube-rep-1-morphism}
\begin{gathered}
\includegraphics[height=1.1cm]{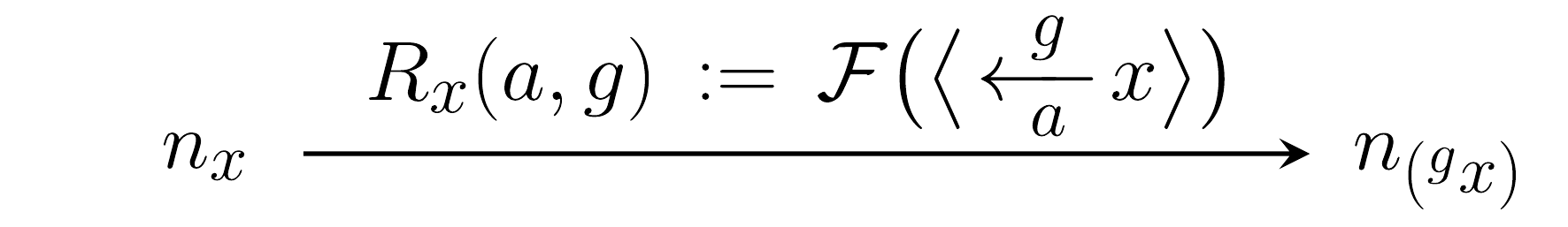}
\end{gathered}
\vspace{-3pt}
\end{equation}
in $\mathsf{2Vec}$, whose composition is controlled by 2-isomorphisms
\vspace{1pt}
\begin{equation}\label{eq:3d-dia-grpex-tube-rep-pseudo-2-morphism}
\begin{gathered}
\includegraphics[height=3.3cm]{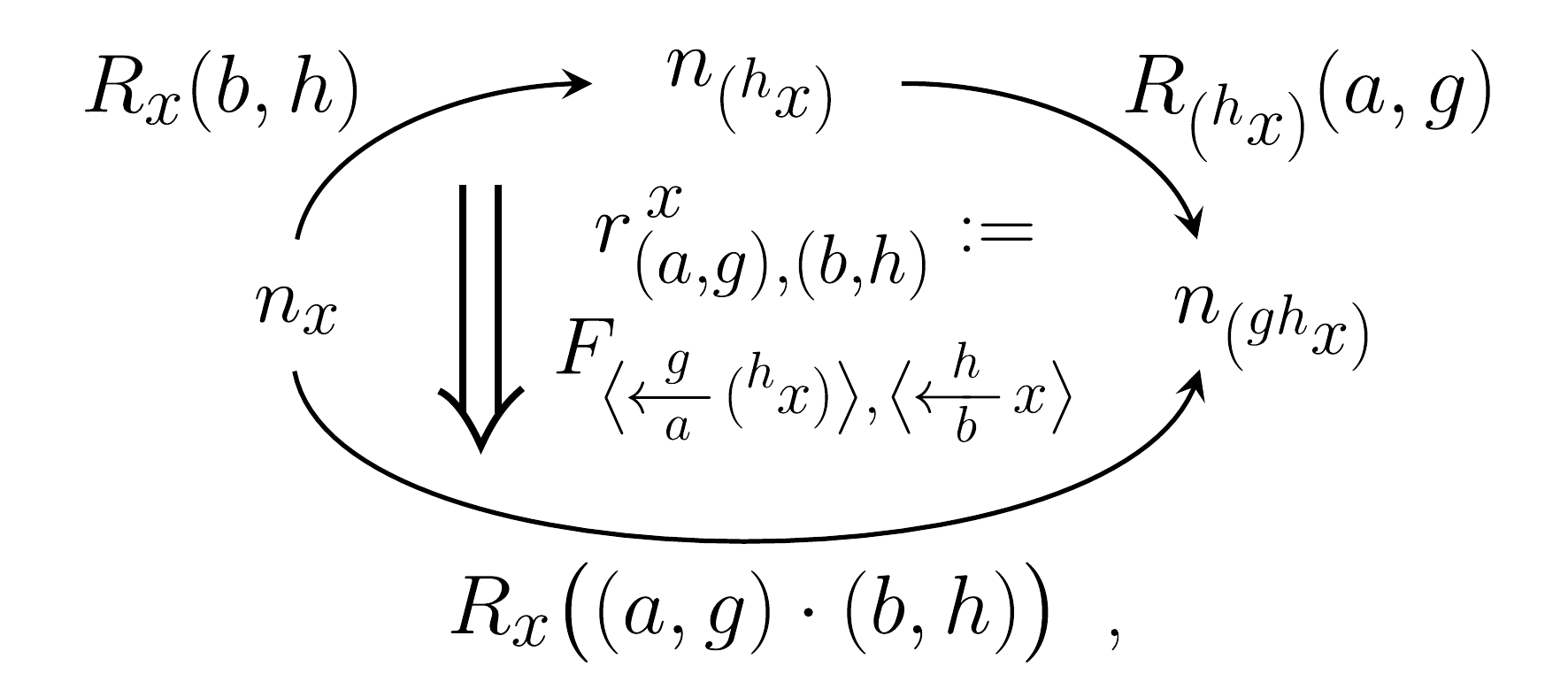}
\end{gathered}
\vspace{-1pt}
\end{equation}
where we used the abbreviation
\begin{equation}
(a,g) \cdot (b,h) \; := \; \big( a \cdot {}^{g}b \cdot \tau_x(\alpha)(g,h), \, g \cdot h \big) \, .
\end{equation}
Furthermore, for each $d \in C_{b / a}^{\mspace{2mu} {}^{g\mspace{-1mu}}x}(A)$ there exists a 2-morphism
\vspace{-3pt}
\begin{equation}\label{eq:3d-dia-grpex-tube-rep-2-morphism}
\begin{gathered}
\includegraphics[height=3.3cm]{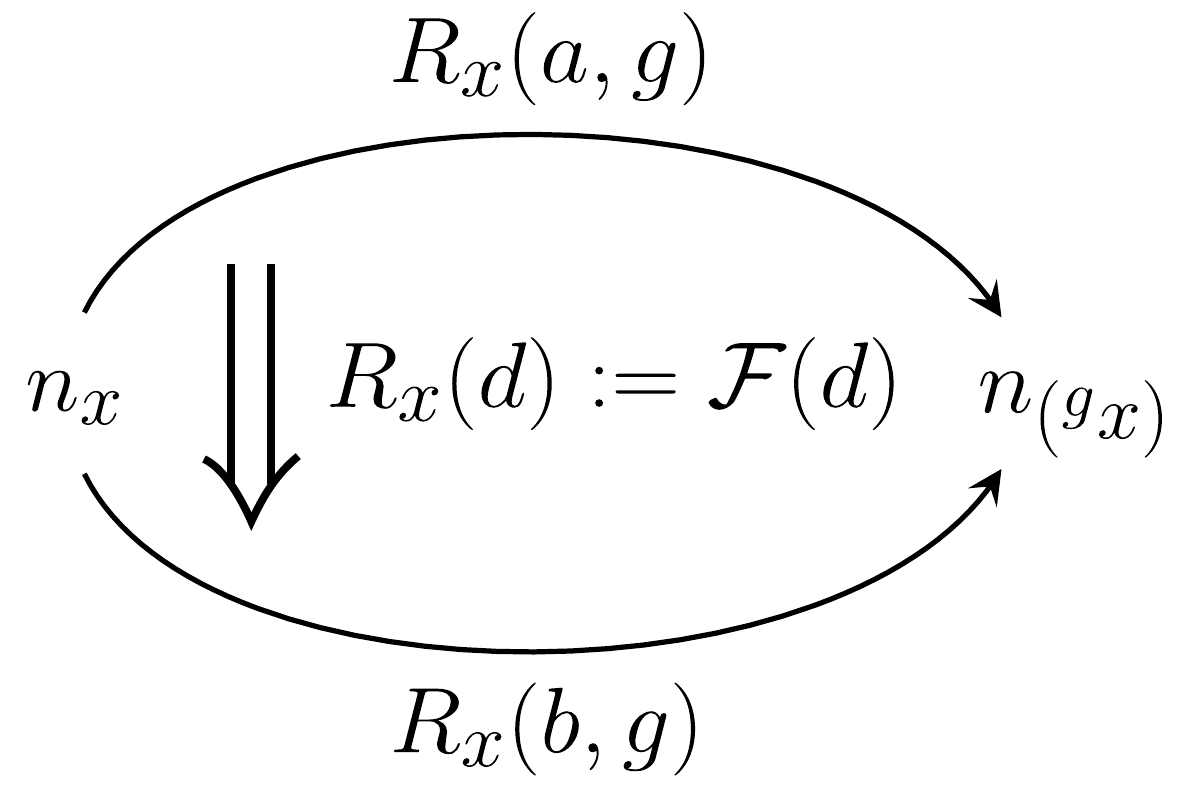}
\end{gathered}
\vspace{-3pt}
\end{equation}
that preserves the composition (\ref{eq:3d-grpex-tube-alg-composition}) and the 2-algebra product (\ref{eq:3d-grpex-tube-alg-2-product}). This data needs to satisfy the compatibility condition
\begin{gather}\label{eq:3d-grpex-tube-rep-compatibility}
\begin{aligned}
\tau_x(\pi)(g,h,k) \, \cdot \, R_x(\alpha(g,h,k)) \; \circ \; r^{\,x}_{(a,g) \, \cdot \, (b,h), \, (c,k)} \, \circ \, &\big[ \, r^{\,({}^k x)}_{(a,g),\,(b,h)} \, \ast \, \text{Id}_{(c,k)} \, \big]   \\[2pt]
= \;\; r^{\, x}_{(a,g), \, (b,h) \, \cdot \, (c,k)} \, \circ \, &\big[ \, \text{Id}_{(a,g)} \, \ast \, r^{\, x}_{(b,h),\,(c,k)} \, \big] \, ,
\end{aligned}
\end{gather}
where we denoted by $\text{Id}_{(c,k)}$ the identity 2-morphism of $R_x(c,k)$ in $\mathsf{2Vec}$.

Similarly to two dimensions, it is clear that the tube 2-representation $\mathcal{F}$ decomposes as a direct sum of 2-representations supported on conjugacy classes of $G$. Let us therefore fix a conjugacy class $[x] \in G/G$ with representative $x \in G$. If we restrict attention to
\begin{enumerate}
\item elements $g,h,k \in C_x(G)$ in the centraliser of $x$ in $G$,
\item elements $d \in C_1^x(A) =: C_x(A)$ in the group of $x$-invariants in $A$, 
\end{enumerate}
the data (\ref{eq:3d-dia-grpex-tube-rep-1-morphism}), (\ref{eq:3d-dia-grpex-tube-rep-pseudo-2-morphism}) and (\ref{eq:3d-dia-grpex-tube-rep-2-morphism}) together with the compatibility condition (\ref{eq:3d-grpex-tube-rep-compatibility}) determines a projective 2-representation $R_x$ of the finite 2-group $C_x(\mathcal{G})$ defined as follows:
\begin{itemize}
\item Its 0-form symmetry group is the extension $A \rtimes_{\tau_x(\alpha)} C_x(G)$, where $C_x(G)$ is the centraliser of $x$ in $G$ and the extension class is represented by
\begin{equation}
\tau_x(\alpha) \, \in \, Z^2_{\rho}(C_x(G),A) \, .
\end{equation}
\item Its 1-form symmetry group is $C_x(A) = \lbrace d \in A \, | \, {}^xd = d \rbrace$.
\item Its action of the 0-form on the 1-form symmetry group is given by ${}^{(a,g)}d = {}^gd$.
\item Its representative $\widetilde{\alpha}$ of the Postnikov class is given by
\begin{equation}
\widetilde{\alpha}\big( \,(a,g), (b,h), (c,k) \, \big) \; = \; \alpha(g,h,k) \, .
\end{equation}
\end{itemize}
We will call the 2-group $C_x(\mathcal{G})$ the \textit{centraliser of $x$ in $\mathcal{G}$}. The 3-cocycle associated to the projective 2-representation $R_x$ is then given by
\begin{equation}
\tau_x(\pi) \, \in \, Z^3(C_x(G),U(1)) \, .
\end{equation}
A tube 2-representation supported on $[x] \in G/G$ is irreducible if the associated projective 2-representation $R_x$ of $C_x(\mathcal{G})$ is irreducible. Conversely, any irreducible projective 2-representation of $C_x(\mathcal{G})$ determines an irreducible tube 2-representation by induction.

In summary, the irreducible tube 2-representations are determined by
\begin{enumerate}
\item a group element $x \in G$,
\item a projective 2-representation of $C_x(\mathcal{G})$ with 3-cocycle
\begin{equation}
\tau_z(\pi)(g,h,k) \; := \; \frac{\pi(g,h,k,x) \cdot \pi(g,x,h,k)}{\pi(g,h,x,k) \cdot \pi(x,g,h,k)} \, .
\end{equation}
\end{enumerate}
Up to equivalence, this data depends only on the conjugacy class $[x] \in G/G$ and the group cohomology class $[\tau_x(\pi)] \in H^3(C_x(G),U(1))$. 

From a physical perspective, the 2-group $C_x(\mathcal{G})$ describes configurations that leave the surface defect $x \in G$ invariant under the intersection with other 0-form defects and the parallel collision and linking with 1-form lines. A 2-representation of $C_x(\mathcal{G})$ then describes the action of $C_x(\mathcal{G})$ on a set of 2-twisted sector lines at the end of the surface $x$. The projectivity is induced by the anomalous phases that arise when intersecting the surface $x$ with multiple 0-form defects wrapping $x$-twisted sector lines.

The above provides an explicit description of simple objects in the braided fusion 2-category $\mathcal{Z}(\mathsf{2Vec}^{\pi}_{\mathcal{G}})$ describing topological surfaces in the four-dimensional Dikgraaf-Witten theory associated to the data $\cG$ and $\pi$. The latter is the relevant topological theory $\text{TV}_\C$ for the symmetry category $\C = \mathsf{2Vec}_{\cG}^\pi$ in the sandwich construction. This is a higher categorical analogue of the fact that representations of the twisted Drinfeld double of a finite anomalous group provide a description of topological lines in the associated three-dimensional Dijkgraaf-Witten theory.

\subsubsection{Example: genuine lines}

As a special case of the above considerations, let us consider tube 2-representations that are supported on the trivial conjugacy class $[e] \in G/G$. They describe the transformation behaviour of genuine line operators attached to the trivial surface defect $e \in G$.

Concretely, such a tube 2-representation is given by a 2-representation of the centraliser 2-group $C_e(\mathcal{G})$ whose
\begin{itemize}
\item 0-form symmetry is $A \rtimes G$,
\item 1-form symmetry is $A$,
\item group action is the pullback of $\rho : G \to \text{Aut}(A)$,
\item Postnikov representative is the pullback of $\alpha \in Z^3_{\rho}(G,A)$.
\end{itemize}
In particular, this implies that genuine line operators transform in 2-representations of the original 2-group $\mathcal{G} \subset C_e(\mathcal{G})$~\cite{Bartsch:2023pzl,Bhardwaj:2023wzd}. However, there is additional structure that arises from the parallel action of topological line defects in $A$ on genuine line operators. This parallel action will be discussed further below.

\subsubsection{Example: 0-form symmetry}
\label{sssec:0-form-symmetry}

Let us now consider the special case of an ordinary finite 0-form symmetry group $G$ with 't Hooft anomaly $\pi \in Z^4(G,U(1))$. The corresponding symmetry category is $\C = \mathsf{2Vec}_{G}^{\pi}$. The associated tube 2-algebra is the multi-fusion category
\begin{equation}
\cA_{S^1}(\C) \; = \; \vect^{\tau(\pi)}_{G/\! /_{\!\,} G}
\end{equation}
of finite-dimensional vector spaces graded by the inertia groupoid ${G/\! /_{\!\,} G}$ with associator twisted by the transgression of the 't Hooft anomaly $\pi \in Z^4(G,U(1))$. Concretely, simple objects correspond to one-dimensional vector spaces $\bC_{g,h}$ that fuse according to
\begin{equation}
\bC_{g,x} \, \otimes \, \bC_{h,y} \; = \; \delta_{x, {}^{h}y} \cdot \bC_{gh,y} \, .
\end{equation}
This can be seen as a higher analogue of the twisted Drinfeld double of a finite group.

The irreducible tube 2-representations are labelled by pairs consisting of
\begin{enumerate}
\item a group element $x \in G$,
\item an irreducible projective 2-representation of $C_x(G)$ with 3-cocycle $\tau_x(\pi)$,
\end{enumerate}
and depend up to equivalence only on the conjugacy class of $x$ and the group cohomology class of $\tau_x(\pi)$. This reproduces the classification of simple objects in the Drinfeld center $\mathcal{Z}(\mathsf{2Vec}^{\pi}_G)$, or equivalently of simple topological surfaces in the associated four-dimensional Dijkgraaf-Witten theory~\cite{Kong:2019brm}. 

\begin{figure}[h]
	\centering
	\includegraphics[height=3.9cm]{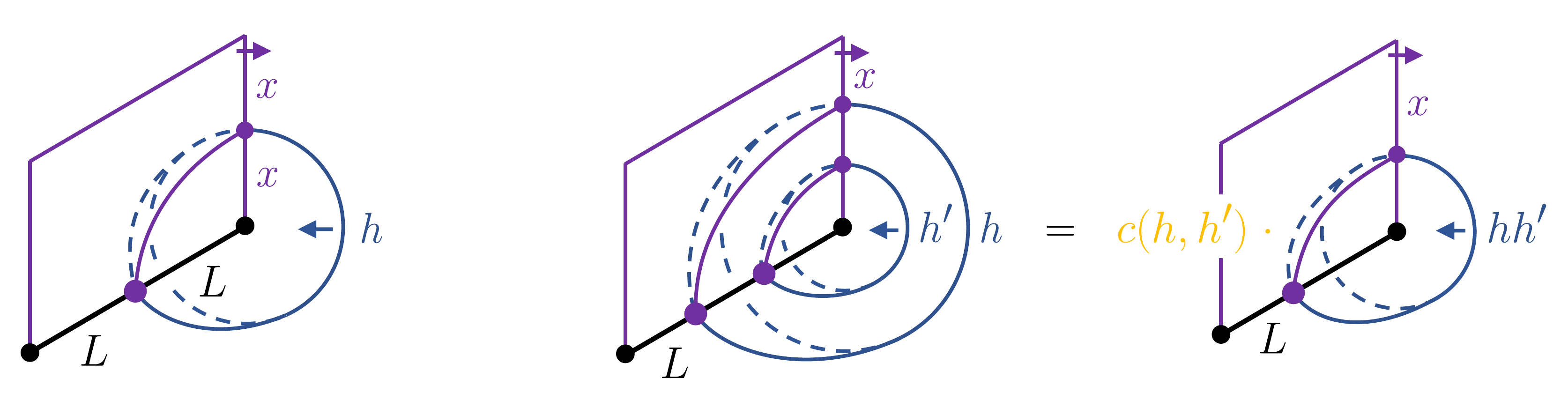}
	\vspace{-5pt}
	\caption{}
	\label{fig:3d-grpex-intersection-action}
\end{figure}

From a physical perspective, this corresponds to a finite collection of line operators in the $x$-twisted sector that transform in an irreducible projective 2-representation under the wrapping action of the centraliser $C_x(G)$. Concretely, such a 2-representation is labelled by a pair consisting of
\begin{enumerate}
\item a subgroup $H \subset C_x(G)$,
\item a 2-cochain $c \in C^2(H,U(1))$ satisfying $\delta c = \tau_x(\pi)|_H$.
\end{enumerate}
The former corresponds to the subgroup of $C_x(G)$ that leaves a given $x$-twisted sector line operator $L$ invariant under the wrapping action, whereas the latter captures projective phases that relate the consecutive wrapping action of two surface defects $h,h' \in H$ to the wrapping action of their fusion $hh' \in H$. This is illustrated in figure \ref{fig:3d-grpex-intersection-action}.

The above provides a generalisation of our previous work~\cite{Bartsch:2023pzl} to twisted sector line operators. The latter are crucial for the detection of the 't Hooft anomaly: Genuine line defects transform in tube 2-representations supported on the trivial conjugacy class $[e] \in G/G$, which are simply 2-representations of the group $G$. Since the projective 3-cocycle of the latter is trivial, they do not detect the 't Hooft anomaly.

\subsubsection{Example: 1-form symmetry}
\label{sssec:1-form-symmetry}

Next, we consider the case of an anomaly-free 1-form symmetry $A$. The corresponding symmetry category is $\C = \mathsf{2Vec}_{A[1]}$. The associated tube 2-algebra is the multi-fusion category whose simple objects are labelled by group elements $a \in A$ and whose morphism spaces are given by
\begin{equation}
\text{Hom}(a, b) \; = \delta_{a,b} \; \cdot \bC[A] \, .
\end{equation}
The irreducible tube 2-representations are labelled by pairs consisting of
\begin{enumerate}
\item an irreducible 2-representations of $A$,
\item a character $\chi \in \widehat{A} := \text{Hom}(A,U(1))$.
\end{enumerate}

\begin{figure}[h]
	\centering
	\includegraphics[height=2.2cm]{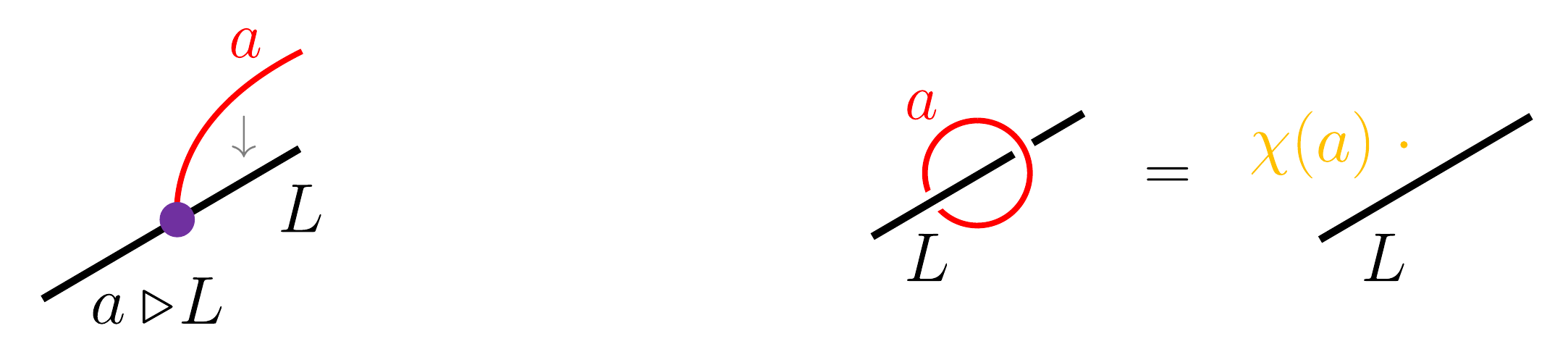}
	\vspace{-5pt}
	\caption{}
	\label{fig:3d-grpex-collision-action}
\end{figure}

From a physical perspective, the former describes a finite collection of line operators $L$ that are acted upon by parallel collision with 1-form symmetry defects $a \in A$. The latter describes a collection of phases that arise when linking the line operators $L$ with 1-form defects $a \in A$. This is illustrated in figure \ref{fig:3d-grpex-collision-action}.

From a mathematical point of view, the line operators $L$ may be viewed as simple objects of a finite semi-simple 1-category that is acted upon by the symmetric braided fusion 1-category $\mathsf{Vec}_A$. This is due to the equivalence
\begin{equation}\label{eq:2-reps-as-modules}
\mathsf{2Rep}(A) \; \cong \; \text{Mod}(\mathsf{Vec}_A)
\end{equation}
of 2-representations of $A$ and $\mathsf{Vec}_A$-module categories. The additional character $\chi$ then equips such a $\mathsf{Vec}_A$-module category with the structure of a \textit{module braiding}, which sets up an equivalence
\begin{equation}
\mathsf{2Rep}(\mathcal{A}_{S^1}(\mathsf{C})) \; \cong \; \text{BrMod}(\mathsf{Vec}_A)
\end{equation}
between the 2-category of tube 2-representations and the 2-category of braided module categories over $\mathsf{Vec}_A$. This agrees with the description of the Drinfeld center $\cZ(\C)$ of the symmetry category $\C = \mathsf{2Vec}_{A[1]}$ \cite{Johnson_Freyd_2023}.

\subsubsection{Compatibility with gauging}

As a consistency check, let us describe how the above two special cases of a pure 0-form and a pure 1-form symmetry are compatible with gauging. Concretely, it is known that gauging a finite abelian 0-form symmetry $A$ of a three-dimensional theory $\cT$ with symmetry category $\mathsf{C} = \mathsf{2Vec}_A$ results in theory $\mathcal{T}' = \cT / A$ with symmetry category
\begin{equation}
\mathsf{C}' \; = \; \mathsf{2Rep}(A) \; \cong \; \mathsf{2Vec}_{\widehat{A}[1]} \, ,
\end{equation}
where $\widehat A = \text{Hom}(A,U(1))$ denotes the dual 1-form symmetry. Although the associated tube 2-algebras of $\mathcal{T}$ and $\mathcal{T}'$ are distinct, their 2-representation theories are expected to be equivalent as the symmetry categories $\C$, $\C'$ have the same Drinfeld center.

It is illuminating to check this non-trivial equivalence explicitly, as it is intimately connected to how twisted sectors for condensation defects are encoded in tube 2-representations. Concretely, the two sides of the equivalence can be described as follows:
\begin{itemize}
\item In the theory $\mathcal{T}$ with symmetry category $\mathsf{C} = \mathsf{2Vec}_A$, tube 2-representations are labelled by pairs consisting of
\begin{enumerate}
\item a group element $a \in A$,
\item an irreducible 2-representation $R$ of $A$,
\end{enumerate}
following the analysis of subsection \ref{sssec:0-form-symmetry}. This data corresponds to a collection of line operators in the $a$-twisted sector transforming in a 2-representation $R$ of the abelian 0-form symmetry $A$.

\item In the theory $\mathcal{T}'$ with symmetry category $\mathsf{C}' = \mathsf{2Vec}_{\widehat{A}[1]}$, tube 2-representations are labelled by pairs consisting of
\begin{enumerate}
\item an irreducible 2-representation $\widehat{R}$ of $\widehat{A}$,
\item a character $a \in \widehat{\widehat{A\,}} = A$, 
\end{enumerate}
following the analysis of subsection \ref{sssec:1-form-symmetry}. This data corresponds to a collection of line operators transforming in a 2-representation $\widehat{R}$ under the parallel action of topological Wilson lines in $\widehat{A}[1]$ with 1-form charge $a \in A$.
\end{itemize}

At first glance, this may seem like a contradiction. Its resolution requires a non-trivial correspondence between 2-representations $R$ of $A$ and 2-representations $\widehat R$ of $\widehat A$. Such a correspondence was discussed from a physical perspective in~\cite{Bhardwaj:2022lsg} in terms of a relationship between different constructions of condensation defects in the gauged theory $\mathcal{T}'$. From a mathematical perspective, it is induced by the non-trivial Morita equivalence
\begin{equation}
\text{Mod}(\mathsf{Vec}_A) \; \cong \; \text{Mod}(\mathsf{Rep}(A)) \, ,
\end{equation}
which using (\ref{eq:2-reps-as-modules}) and the canonical identification $\mathsf{Rep}(A) = \mathsf{Vec}_{\widehat{A}}$ implies that
\begin{equation}\label{eq:2-rep-morita}
\mathsf{2Rep}(A) \; \cong \; \mathsf{2Rep}(\widehat{A}) \, .
\end{equation}

Let us give a physical intuition behind this equivalence. We start with a collection of line operators $L$ in the $a$-twisted sector of $\mathcal{T}$ transforming in an irreducible 2-representation $R$ of the 0-form symmetry $A$. Upon gauging $A$, these lines become 2-twisted sector lines for the condensation defect in $\mathcal{T}'$ labelled by $R$, whose charge w.r.t. the dual 1-form symmetry $\widehat{A}$ is given by $a \in A$. In order to think of this as a module for $\mathsf{Vec}_{\widehat{A}}$, we employ the fact that condensation defects always admit gapped boundaries, which can be used to transform a twisted sector for a condensation to a genuine line defect.

\begin{figure}[h]
	\centering
	\includegraphics[height=7.8cm]{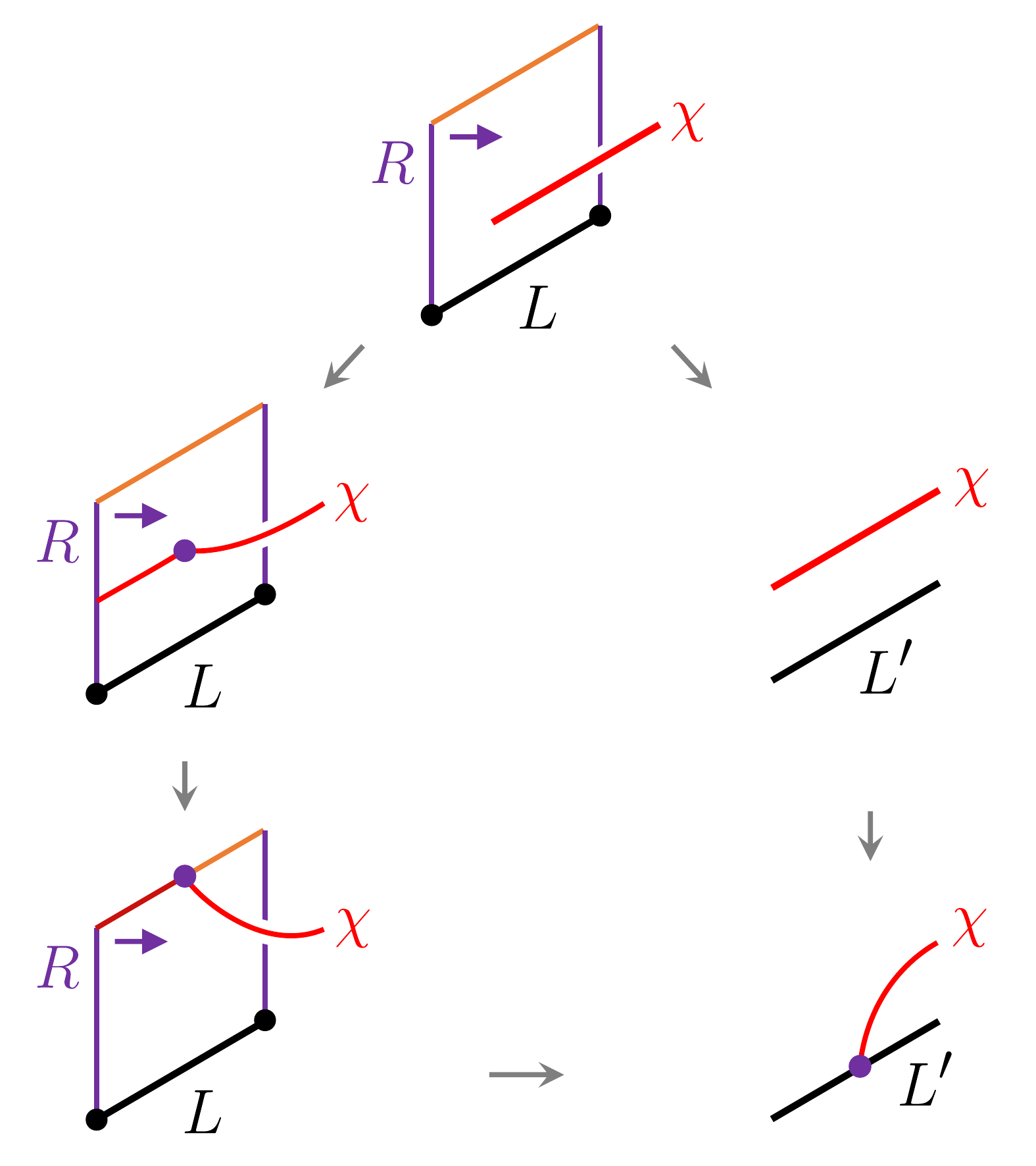}
	\vspace{-5pt}
	\caption{}
	\label{fig:3d-compatibility-with-gauging}
\end{figure}

Concretely, upon thinking of $R$ as a pair $(B,c)$ consisting of a subgroup $B \subset A$ and a 2-cocycle $c \in Z^2(B,U(1))$, the condensation defect $R$ in $\mathcal{T}'$ admits a gapped boundary condition that supports badly quantised Wilson lines in projective representations of $B$ with Schur-multiplier $c$. The $\mathsf{Vec}_{\widehat{A}\,}$-module structure then comes from transporting a bulk Wilson line $\chi \in \rep(A)$ onto the condensation defect $R$, which restricts it to the subgroup $B \subset A$, and subsequently moving it onto the gapped boundary, which is the tensor product of (projective) representations. This is illustrated on the left-hand side of figure \ref{fig:3d-compatibility-with-gauging}.

Alternatively, we may perform an interval compactification and collapse the gapped boundary onto the 2-twisted sector lines $L$. The resulting genuine lines operators $L'$  in $\mathcal{T}'$ then form a $\mathsf{Vec}_{\widehat{A}\,}$-module (or equivalently a 2-representation $\widehat{R}$ of $\widehat{A}$) w.r.t. the parallel action of topological Wilson lines. This is illustrated on the right-hand side of figure \ref{fig:3d-compatibility-with-gauging}.

\subsubsection{Gauge theory}

Now consider a three-dimensional gauge theory with connected simple gauge group $\mathbb{G}$. This has an anomaly free split 2-group symmetry given by the automorphism 2-group of $\mathbb{G}$. This can be expressed $\cG = Z(\mathbb{G})[1] \rtimes \text{Out}(\mathbb{G})$, where $\text{Out}(\mathbb{G})$ is charge conjugation symmetry and $Z(\mathbb{G})$ is the electric 1-form symmetry.

The irreducible tube 2-representations are labelled as follows:
\begin{itemize}
\item An outer automorphism $[x] \in \text{Out}(\mathbb{G})$ with representative $x$.
\item A irreducible 2-representation of the centraliser 2-subgroup $C_{x}(\cG)$.
\end{itemize}
The twisted sectors are labelled outer automorphisms $\text{Out}(G)$. The centraliser 2-group has 0-form component $Z(\mathbb{G}) \rtimes C_{x}(\text{Out}(\mathbb{G}))$, 1-form component $\text{Stab}_x(Z(\mathbb{G}))$ and natural action.  The additional factor of $Z(\mathbb{G})$ in the 0-form component arises from the parallel action of 1-form symmetry generators, which are topological Gukov-Witten defects.

Let us consider $\mathbb{G} = Spin(4N)$. In this case the 2-group symmetry is $\cG = (\bZ_2 \times \bZ_2)[1] \rtimes \bZ_2$. There are two twisted sectors corresponding to genuine line defects and $s$-twisted line defects where $s$ is the generator of charge conjugation symmetry. We consider each in turn.
\begin{itemize}
\item The genuine line defects transform in 2-representations of $(\bZ_2 \times \bZ_2)[1] \rtimes D_8$. The additional $\bZ_2 \times \bZ_2$ that gives rise to $D_8$ is generated by the parallel action of topological Gukov-Witten lines. 
\item The $s$-twisted line defects transform in $2$-representations of $\bZ_2[1] \times D_8$, where $\bZ_2[1]$ is the diagonal 1-form symmetry that is invariant under charge conjugation. The additional $\bZ_2 \times \bZ_2$ that gives rise to $D_8$ is again generated by the parallel action of topological Gukov-Witten lines. 
\end{itemize}
An example is the pair of Wilson lines $W_{S^+}$, $W_{S^-}$ transforming in the spinor and conjugate spinor representations, which transform together with topological Gukov-Witten lines in an eight-dimensional irreducible 2-representation of $(\bZ_2 \times \bZ_2)[1] \rtimes D_8$.

Let us now consider the $\mathbb{G}=PSO(4N)$ gauge theory, which may be obtained by gauging $(\bZ_2\times \bZ_2)[1]$ in the example above. This has symmetry group $G=D_8$ combining the $\bZ_2\times \bZ_2$ magnetic symmetry and charge conjugation symmetry. There are now eight twisted sectors forming five conjugacy classes and with centralizers:
\begin{equation}
\begin{aligned}
&[e]=\{e\}, &\quad & C_{e}(D_8)=D_8,\\
&[a]=\{a,b\}, &\quad & C_{a}(D_8)=\bZ_2\times\bZ_2,\\
&[s]=\{s,abs\},&\quad & C_{s}(D_8)=\bZ_2\times\bZ_2,\\
&[ab]=\{ab\},&\quad & C_{ab}(D_8)=D_8,\\
&[as]=\{as,bs\},&\quad & C_{as}(D_8)=\bZ_4.
\end{aligned}
\end{equation}
The irreducible tube 2-representations are classified by choosing 2-representations of the centralizers.

The Wilson lines $W_{S^+}$ and $W_{S^-}$ now become $a$-twisted and $b$-twisted in $\mathbb{G} = PSO(4N)$. They are exchanged by charge conjugation and transform in a tube 2-representation associated to the conjugacy class $\{a,b\}$. This 2-representation corresponds to the wrapping action of the $\bZ_2 \times \bZ_2$ magnetic symmetry. The relevant 2-representation for the Wilson lines $W_{S^+}$, $W_{S^-}$ is the trivial 2-representation.

\subsection{Example: braiding lines}

Let us now again consider a three-dimensional theory with connected symmetry 2-category
\be
\mathsf{C} \; = \; \text{Mod}(\mathsf{B})
\ee
determined by a braided fusion 1-category $\mathsf{B}$ of topological lines and their condensations, as discussed previously in section~\ref{subsec:3d-modB}.

The Drinfeld center in this case is the braided fusion 2-category
\begin{equation}
\mathcal{Z}(\mathsf{C}) \; = \; \text{BrMod}(\mathsf{B})
\end{equation}
of braided (left) module categories over $\mathsf{B}$ \cite{Johnson_Freyd_2023}. We do not provide a complete description here but settle for a brief discussion that will allow a graphical interpretation in terms of tube 2-representations. Namely, an object of in the Drinfeld center is given by a finite semi-simple 1-category $\mathsf{M}$ together with a $\mathsf{B}$-action
\begin{equation}
\triangleright : \; \mathsf{B} \, \boxtimes \, \mathsf{M} \; \to \; \mathsf{M}
\end{equation}
and a collection of 1-morphisms
\vspace{-4pt}
\begin{equation}
\begin{gathered}
\includegraphics[height=1.1cm]{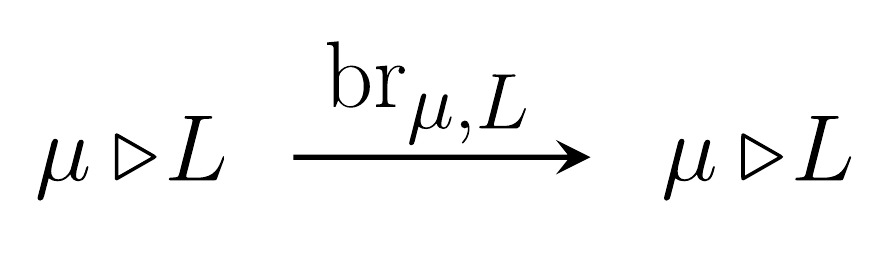}
\end{gathered}
\vspace{-3pt}
\end{equation}
for all $\mu \in \mathsf{B}$ and $L \in \mathsf{M}$. These 1-morphisms must satisfy coherence conditions that may be found in~\cite{Johnson_Freyd_2023} and that we will illustrate graphically below.

\begin{figure}[h]
	\centering
	\includegraphics[height=2.15cm]{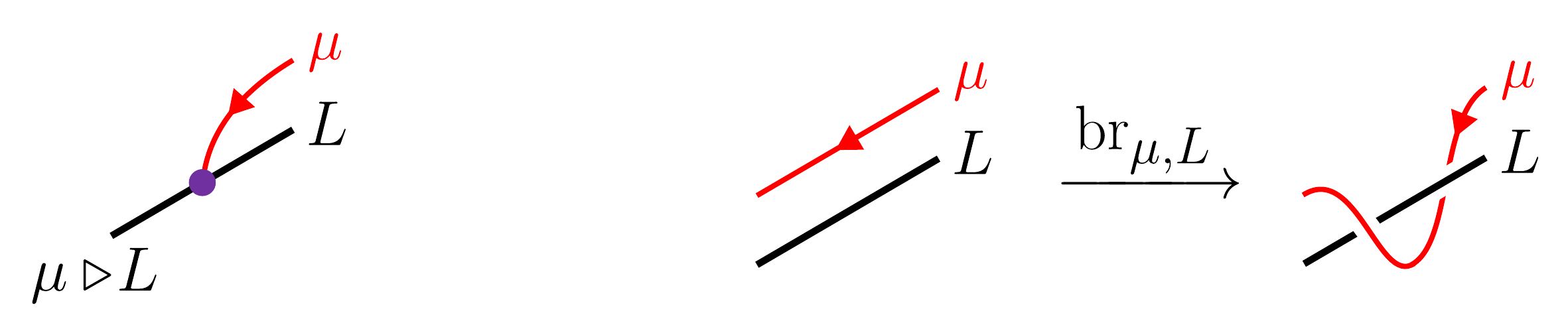}
	\vspace{-5pt}
	\caption{}
	\label{fig:3d-braided-module-1}
\end{figure}

From a physical perspective, all topological surfaces are condensations whose wrapping action on line defects is completely determined by the parallel and linking of the topological lines. Concretely, a topological line $\mu \in \mathsf{B}$ can act on line defects $L \in \mathsf{M}$ by parallel collision or linking. This provides the data of the $\B$-action and 1-morphisms $\text{br}_{\mu,L}$ respectively, as illustrated in figure \ref{fig:3d-braided-module-1}. 

\begin{figure}[h]
	\centering
	\includegraphics[height=8.3cm]{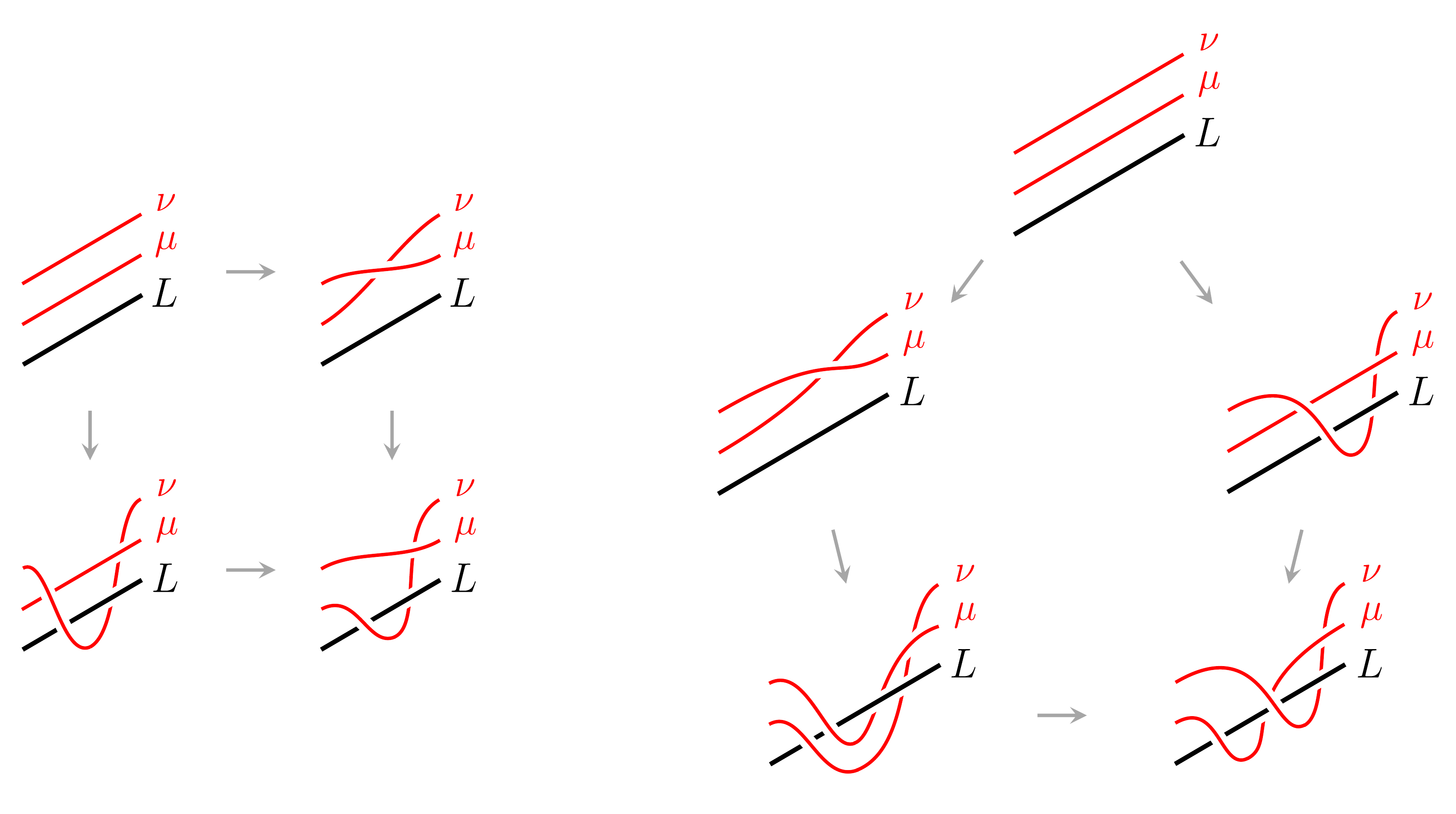}
	\vspace{-5pt}
	\caption{}
	\label{fig:3d-braided-module-3}
\end{figure}

This data needs to be compatible with the braiding of topological lines in the ambient three-dimensional spacetime in the sense that the sequences of topological moves illustrated in figure \ref{fig:3d-braided-module-3} commute. This reproduce the coherence conditions presented in~\cite{Johnson_Freyd_2023}.

\acknowledgments

It is a please to thank Alex Bullivant, Andrea Ferrari, David Jordan, Theo Johnson-Freyd, Lukas M\"uller and Jamie Pearson for discussions and feedback. The work of MB is supported by the EPSRC Early Career Fellowship EP/T004746/1 and the STFC Research Grant ST/T000708/1. The work of MB and AG is supported by the Simons Collaboration on Global Categorical Symmetry.


\appendix

\section{Condensation defects}

In this appendix, we review the notion of condensation defects in higher categories introduced in \cite{Gaiotto:2019xmp}. We then use these results to give a schematic proof of equations (\ref{eq:2d-tube-hom-isomorphism}) and (\ref{eq:3d-1-tube-hom-isomorphism}), which state that the action of generic symmetry defects $B \in \mathsf{C}$ on twisted sector local operators can be reduced to the action of representatives of the connected components of the (higher) fusion category $\mathsf{C}$.

Following \cite{Gaiotto:2019xmp}, we take an inductive approach to define the notion of an \textit{$n$-condensation} between objects $A$ and $B$ in an $n$-category $\mathsf{C}$. For $n=0$, a $0$-condensation $A \cond B$ is an equality $A = B$ of objects of a $0$-category (i.e. elements of a set). For generic $n$, an $n$-condensation $A \cond B$ in an $n$-category $\mathsf{C}$ is a pair of 1-morphisms
\vspace{-4pt}
\begin{equation}\label{eq:app-condensation}
\begin{gathered}
\includegraphics[height=2.2cm]{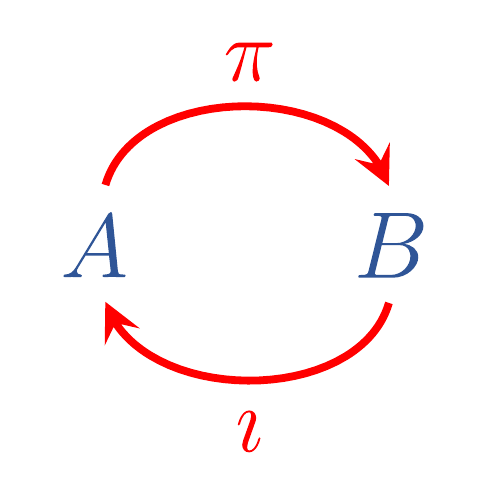}
\end{gathered}
\vspace{-4pt}
\end{equation}
together with an $(n-1)$-condensation $\pi \circ \imath \cond \text{Id}_B$ in the $(n-1)$-category $\text{1-End}_{\mathsf{C}}(B)$. We will unravel this definition in more detail for the cases $n=1,2$ below.

\subsection{1-condensations}
\label{app:1-condensations}

A 1-condensation $A \cond B$ between objects of a 1-category is a pair of morphisms $\imath$ and $\pi$ as in (\ref{eq:app-condensation}) such that $\pi \circ \imath = \text{Id}_B$. If we restrict attention to simple objects $S$ and $T$ of a fusion category $\mathsf{C}$, a 1-condensation $S \cond T$ is equivalent to an isomorphism $S \cong T$. The set of connected components of $\mathsf{C}$ can thus be identified with
\begin{equation}
\pi_0(\mathsf{C}) \; = \; \lbrace \text{simple objects} \; S \in \mathsf{C} \rbrace \,/ \, \text{1-condensation} \, .
\end{equation}

\begin{figure}[h]
	\centering
	\includegraphics[height=4.2cm]{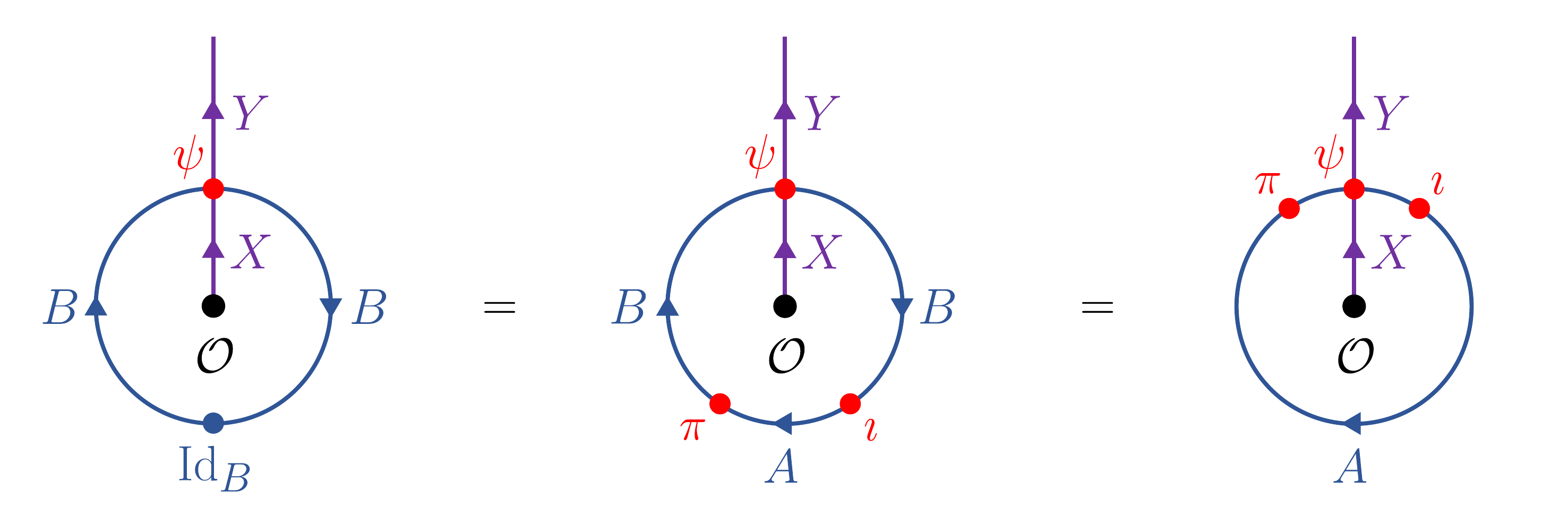}
	\vspace{-5pt}
	\caption{}
	\label{fig:2d-tube-hom-isomorphism-2}
\end{figure}

From a physical perspective, a 1-condensation $A \cond B$ implies that the linking action of $B$ on twisted sectors is completely determined by the linking action of $A$. To see this, consider linking an $X$-twisted sector local operator $\mathcal{O}$ with the symmetry defect $B$ via a specified intersection morphism $\psi$. Using the splitting $\text{Id}_B = \pi \circ \imath$, we can then insert a small $A$-interval into the line $B$, which can be blown up to give a linking of $\mathcal{O}$ by $A$ as illustrated in figure \ref{fig:2d-tube-hom-isomorphism-2}. This shows that the linking action of $B$ can be deduced from the linking action of $A$ on $\mathcal{O}$.

\begin{figure}[h]
	\centering
	\includegraphics[height=4.2cm]{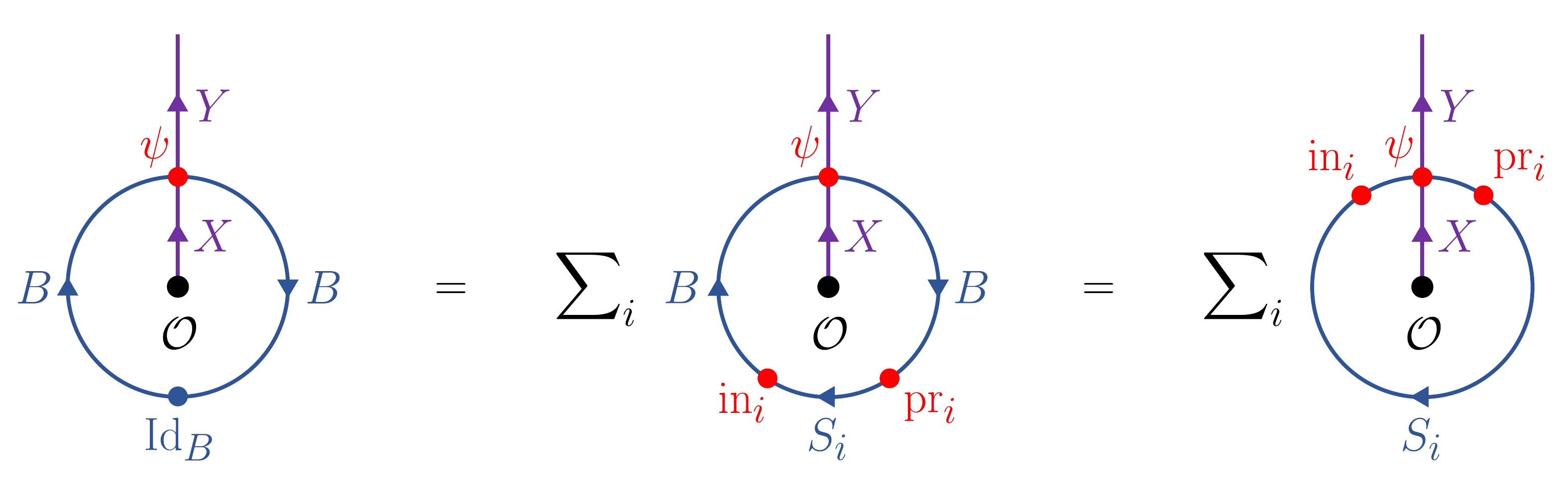}
	\vspace{-5pt}
	\caption{}
	\label{fig:2d-tube-hom-isomorphism}
\end{figure}

Let us now give a schematic proof of equation (\ref{eq:2d-tube-hom-isomorphism}), which states that the linking action of a generic symmetry defect $B \in \mathsf{C}$ on twisted sectors in two dimensions is completely determined by the action of its simple constituents. To see this, consider a decomposition
\begin{equation}
B \; = \; \bigoplus\nolimits_i S_i
\end{equation}
of $B$ into simple objects $S_i \in \mathsf{C}$ together with inclusion and projection morphisms
\begin{equation}
\text{in}_i : \, S_i \, \xhookrightarrow \, B  \qquad \text{and} \qquad \text{pr}_i: \, B \, \twoheadrightarrow \, S_i \, .
\end{equation}
Using the completeness relation
\begin{equation}
\text{Id}_B \; = \; \sum\nolimits_i \, \text{in}_i \, \circ \, \text{pr}_i \, ,
\end{equation}
we can then insert a sum of small $S_i$-intervals into the line $B$, which can be blown up to give a sum of linkings of $\mathcal{O}$ by the $S_i$'s as illustrated in figure \ref{fig:2d-tube-hom-isomorphism}. This shows that the linking action of $B$ is completely determined by the linking action of its simple constituents $S_i$ on $\mathcal{O}$. Furthermore, as discussed above, we can replace each of the $S_i$'s by an arbitrary representative of its connected component $[S_i] \in \pi_0(\mathsf{C})$, which proves equation (\ref{eq:2d-tube-hom-isomorphism}).

\subsection{2-condensations on the sphere}
\label{app:2-condensations-sphere}

A 2-condensation $A \cond B$ between objects of a 2-category is a pair of morphisms $\imath$ and $\pi$ as in (\ref{eq:app-condensation}) together with a pair of 2-morphisms
\vspace{-4pt}
\begin{equation}
\begin{gathered}
\includegraphics[height=2.8cm]{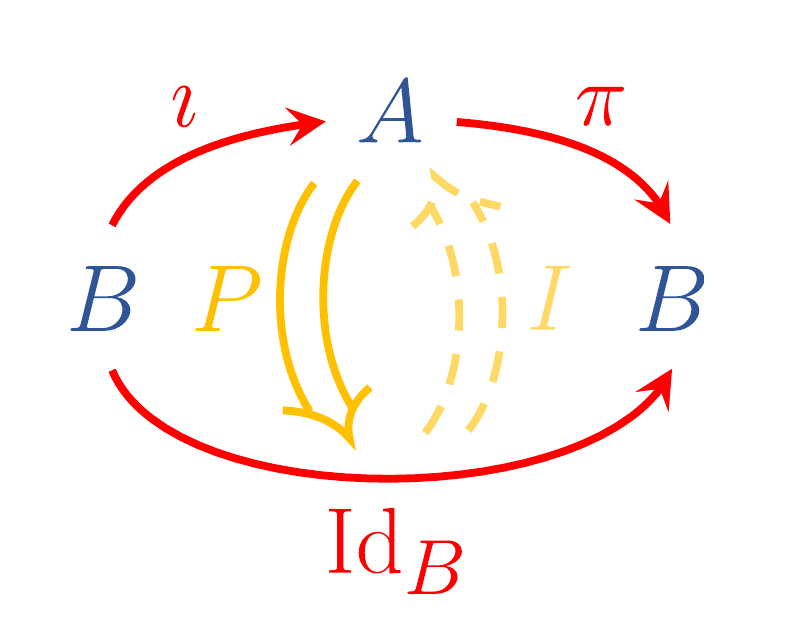}
\end{gathered}
\vspace{-4pt}
\end{equation}
such that $P \circ I = \text{Id}^2_B$. Unlike in the case of 1-condensations, a 2-condensation $S \cond T$ between simple objects of a fusion 2-category $\mathsf{C}$ does not induce an isomorphism of $S$ and $T$ in general. Nevertheless, it induces an equivalence relation on the set of simple objects, which can be used to identify the set of connected components of $\mathsf{C}$ as
\begin{equation}
\pi_0(\mathsf{C}) \; = \; \lbrace \text{simple objects} \; S \in \mathsf{C} \rbrace \,/ \, \text{2-condensation} \, .
\end{equation}

\begin{figure}[h]
	\centering
	\includegraphics[height=4.4cm]{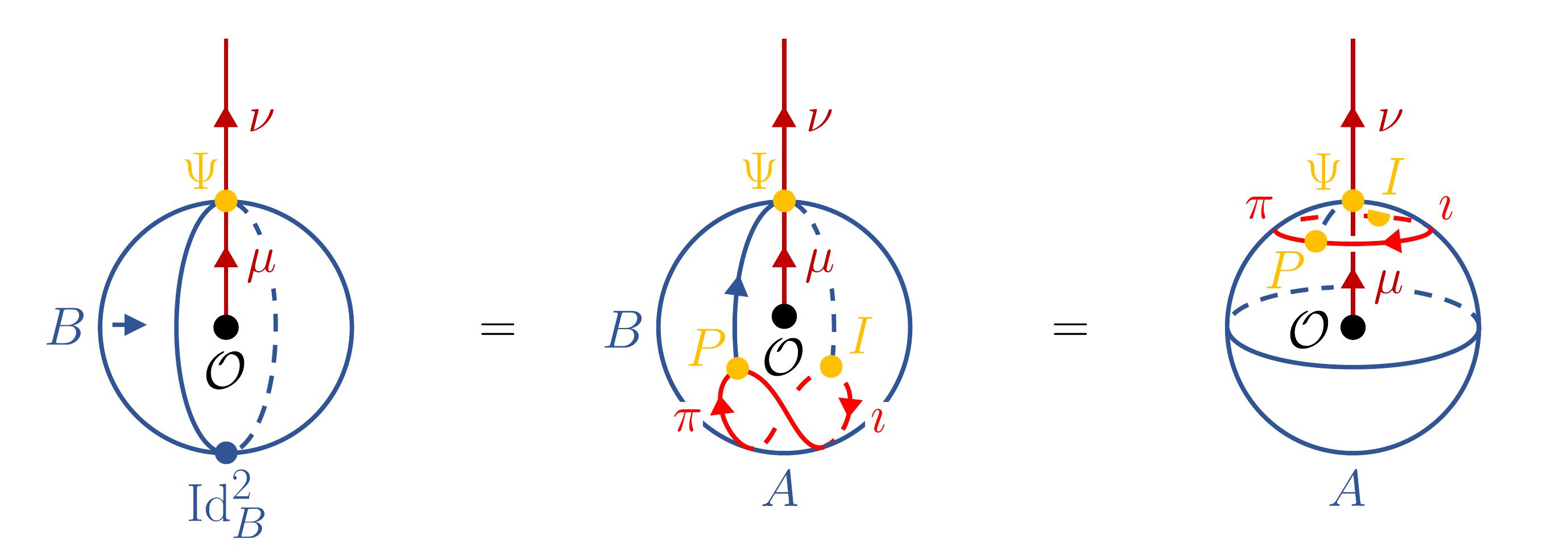}
	\vspace{-5pt}
	\caption{}
	\label{fig:3d-1-tube-hom-isomorphism-2}
\end{figure}

From a physical perspective, a 2-condensation $A \cond B$ implies that the linking action of $B$ on 1-twisted sectors is completely determined by the linking action of $A$. To see this, consider linking a $\mu$-twisted sector local operator $\mathcal{O}$ with the symmetry defect $B$ via a specified intersection 2-morphism $\Psi$. Using the splitting $\text{Id}^2_B = P \circ I$, we can then insert a small $A$-bubble into the surface $B$, which can be blown up to give a linking of $\mathcal{O}$ by $A$ as illustrated in figure \ref{fig:3d-1-tube-hom-isomorphism-2}. This shows that the linking action of $B$ can be deduced from the linking action of $A$ on $\mathcal{O}$.

\begin{figure}[h]
	\centering
	\includegraphics[height=4.2cm]{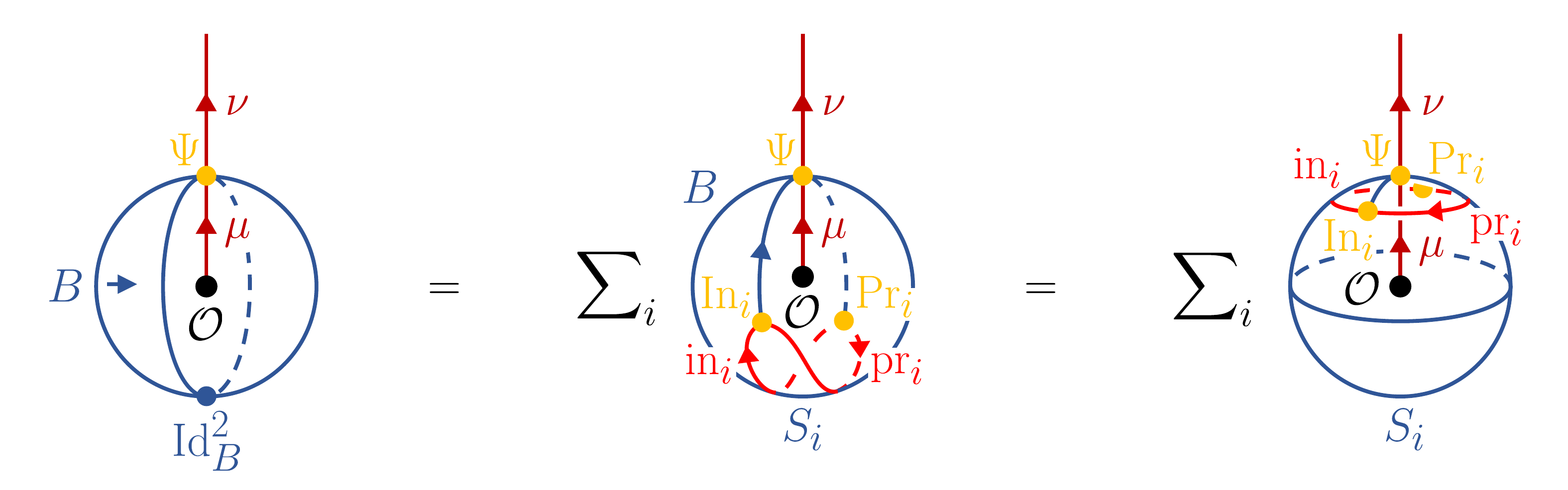}
	\vspace{-5pt}
	\caption{}
	\label{fig:3d-1-tube-hom-isomorphism}
\end{figure}

Let us now give a schematic proof of equation (\ref{eq:3d-1-tube-hom-isomorphism}), which states that the linking action of a generic symmetry defect $B \in \mathsf{C}$ on 1-twisted sectors in three dimensions is completely determined by the action of its simple constituents. To see this, consider a decomposition
\begin{equation}
B \; = \; \bigoplus\nolimits_i S_i
\end{equation}
of $B$ into simple objects $S_i \in \mathsf{C}$ together with inclusion and projection 1-morphisms
\begin{equation}
\text{in}_i : \, S_i \, \xhookrightarrow \, B  \qquad \text{and} \qquad \text{pr}_i: \, B \, \twoheadrightarrow \, S_i
\end{equation}
and inclusion and projection 2-morphisms
\begin{equation}
\text{In}_i : \,\; \text{in}_i \,\circ \, \text{pr}_i \; \Rightarrow \; \text{Id}_B \qquad \text{and} \qquad \text{Pr}_i : \,\; \text{Id}_B \; \Rightarrow \; \text{in}_i \,\circ \, \text{pr}_i \, .
\end{equation}
Using the completeness relation
\begin{equation}
\text{Id}^2_B \; = \; \sum\nolimits_i \, \text{In}_i \, \circ \, \text{Pr}_i \, ,
\end{equation}
we can then insert a sum of small $S_i$-bubbles into the surface $B$, which can be blown up to give a sum of linkings of $\mathcal{O}$ by the $S_i$'s as illustrated in figure \ref{fig:3d-1-tube-hom-isomorphism}. This shows that the linking action of $B$ is completely determined by the linking action of its simple constituents $S_i$ on $\mathcal{O}$. Furthermore, as discussed above, we can replace each of the $S_i$'s by an arbitrary representative of its connected component $[S_i] \in \pi_0(\mathsf{C})$, which proves equation (\ref{eq:3d-1-tube-hom-isomorphism}).
 
\subsection{2-condensations on the cylinder}
\label{app:2-condensations-cylinder}

\begin{figure}[h]
	\centering
	\includegraphics[height=8.5cm]{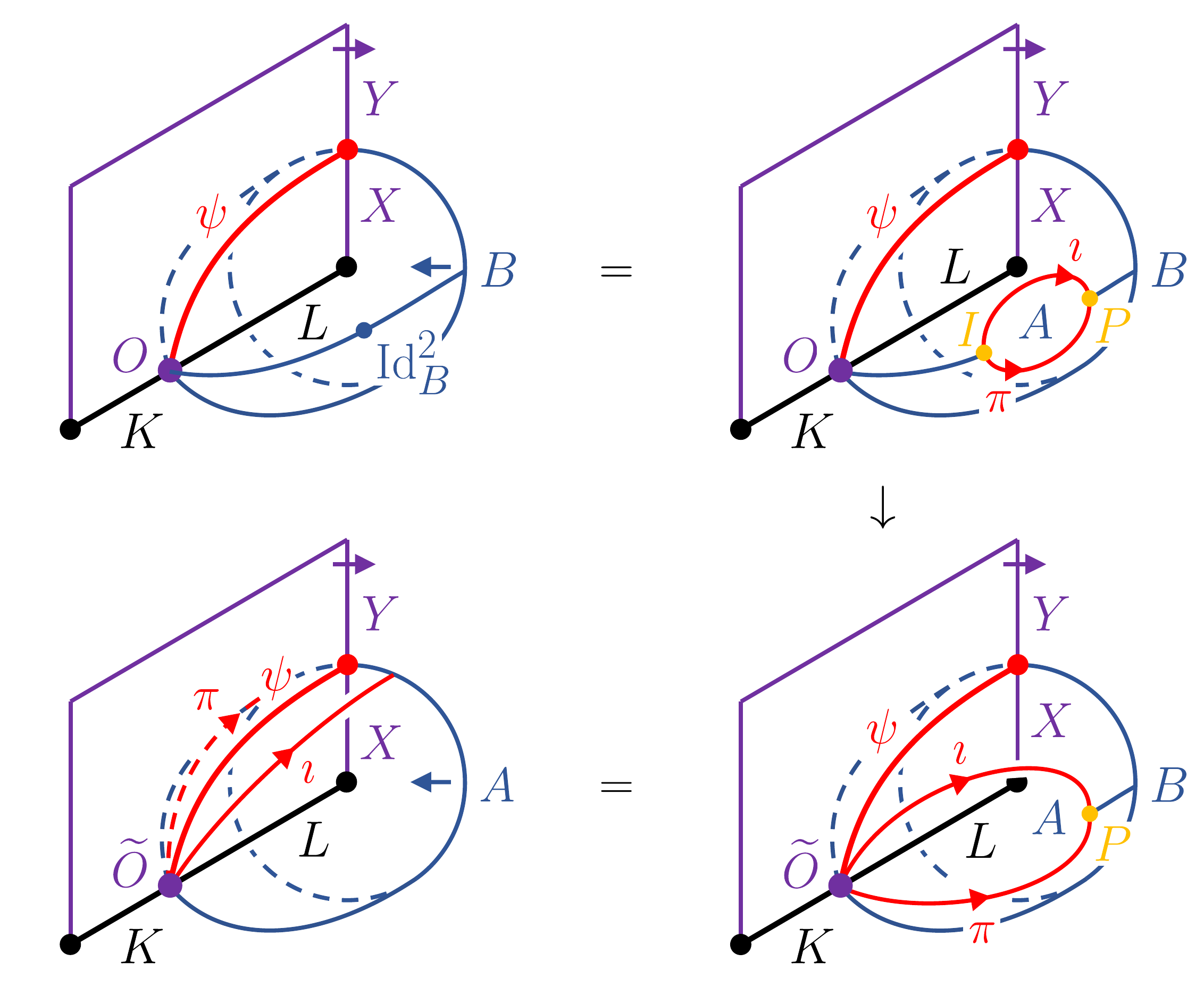}
	\vspace{-2pt}
	\caption{}
	\label{fig:3d-condensation-action-2}
\end{figure}

In the case of 2-twisted sector line operators, the topology of the wrapping cylinder obstructs a literal identification of the wrapping actions of symmetry defects related by condensation. Nevertheless, we expect them to be related as follows:

Given a 2-condensation $A \cond B$, the wrapping action of $B$ on 2-twisted sectors is determined the wrapping action of $A$ up to a 1-condensation. To see this, consider wrapping a line operator $L$ in the $X$-twisted sector with the symmetry defect $B$ via a specified intersection 1-morphism $\psi$. Using the splitting $\text{Id}^2_B = P \circ I$, we can then insert a small $A$-bubble into the surface $B$ as illustrated in the top right corner of figure \ref{fig:3d-condensation-action-2}. Topological local operators $O$ sitting at the junction between the wrapped line $L$ and another line $K$ can then be collided with the topological junction $I$ to give new topological operators $\widetilde{O}$ as illustrated in the bottom right corner of figure \ref{fig:3d-condensation-action-2}. The map $O \mapsto \widetilde{O}$ is induced by the 2-morphism
\begin{equation}
\begin{gathered}
\includegraphics[height=3.7cm]{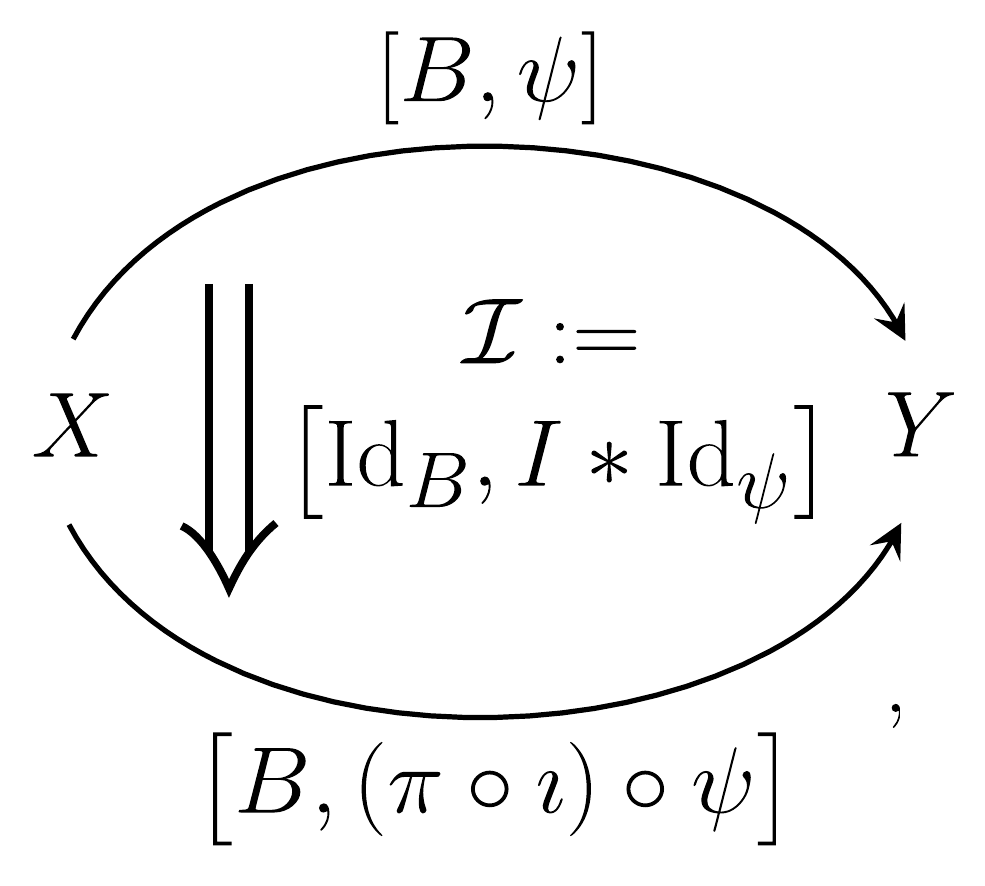}
\end{gathered}
\vspace{-2pt}
\end{equation}
where $(\pi \circ \imath)$ should be read as $\text{Id}_Y \otimes (\pi \circ \imath)$ and $I$ should be read as $\text{Id}_Y^2 \otimes I$. Similarly, there exists a 2-morphism 
\begin{equation}
\begin{gathered}
\includegraphics[height=3.7cm]{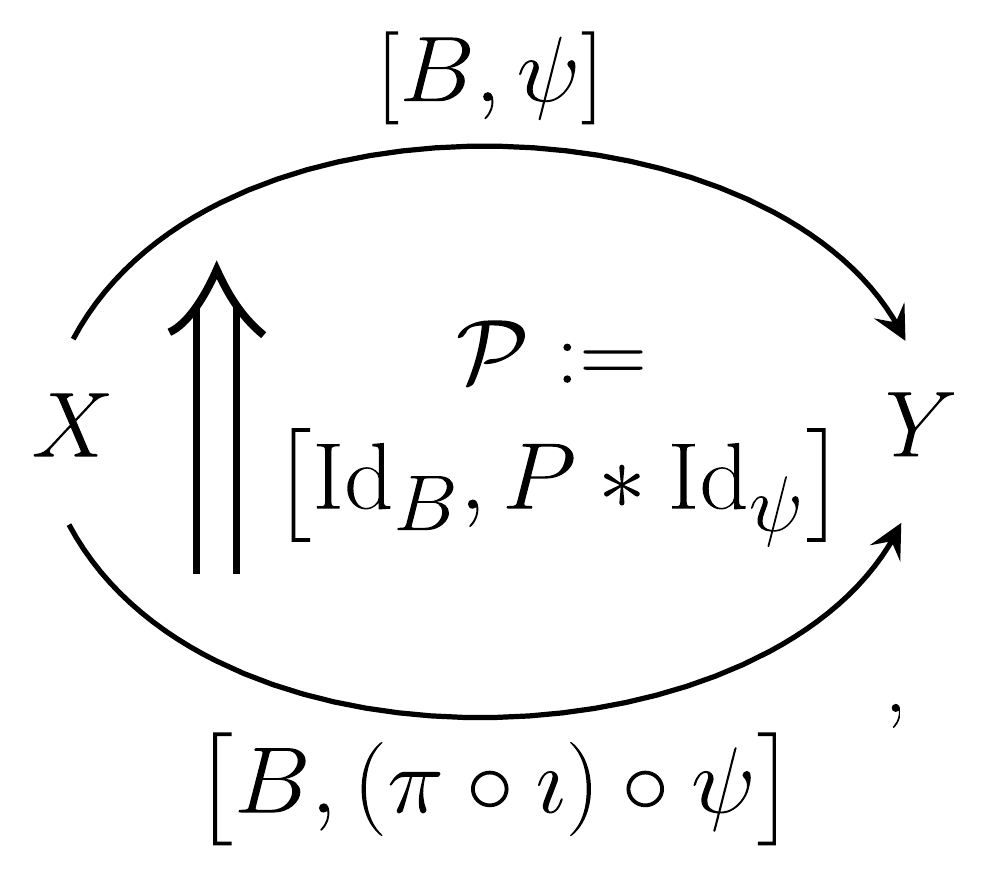}
\end{gathered}
\vspace{-2pt}
\end{equation}
where $P$ should be read as $ \text{Id}_Y^2 \otimes P$. As a consequence of $P \circ I = \text{Id}^2_B$, the 2-morphisms $\mathcal{I}$ and $\mathcal{P}$ satisfy the relation
\begin{equation}
\mathcal{P} \, \circ \, \mathcal{I} \; = \; \text{Id}_{[B,\psi]} \, .
\end{equation}
Furthermore, by moving the topological line interface $\pi$ around the wrapping cylinder as illustrated in the bottom left corner of figure \ref{fig:3d-condensation-action-2}, we may identify
\begin{equation}
\big[B,\, (\pi \circ \imath) \circ \psi\big] \; = \; \big[A,\,\imath \circ \psi \circ \pi\big] \, ,
\end{equation}
where $\imath \circ \psi \circ \pi$ should be read as $( Y \otimes \imath ) \circ \psi \circ  ( \pi \otimes X )$. Consequently, the pair $\mathcal{I}$, $\mathcal{P}$ defines a 1-condensation 
\begin{equation}
[A,\,\imath \circ \psi \circ \pi ] \,\; \cond \,\; [B,\psi]
\end{equation}
in the 1-category $\text{1-Hom}_{\mathsf{T_2C}}(X,Y)$. This shows that the wrapping action of $B$ on 2-twisted sectors is determined the wrapping action of $A$ up to 1-condensation.

\begin{figure}[h]
	\centering
	\includegraphics[height=8.5cm]{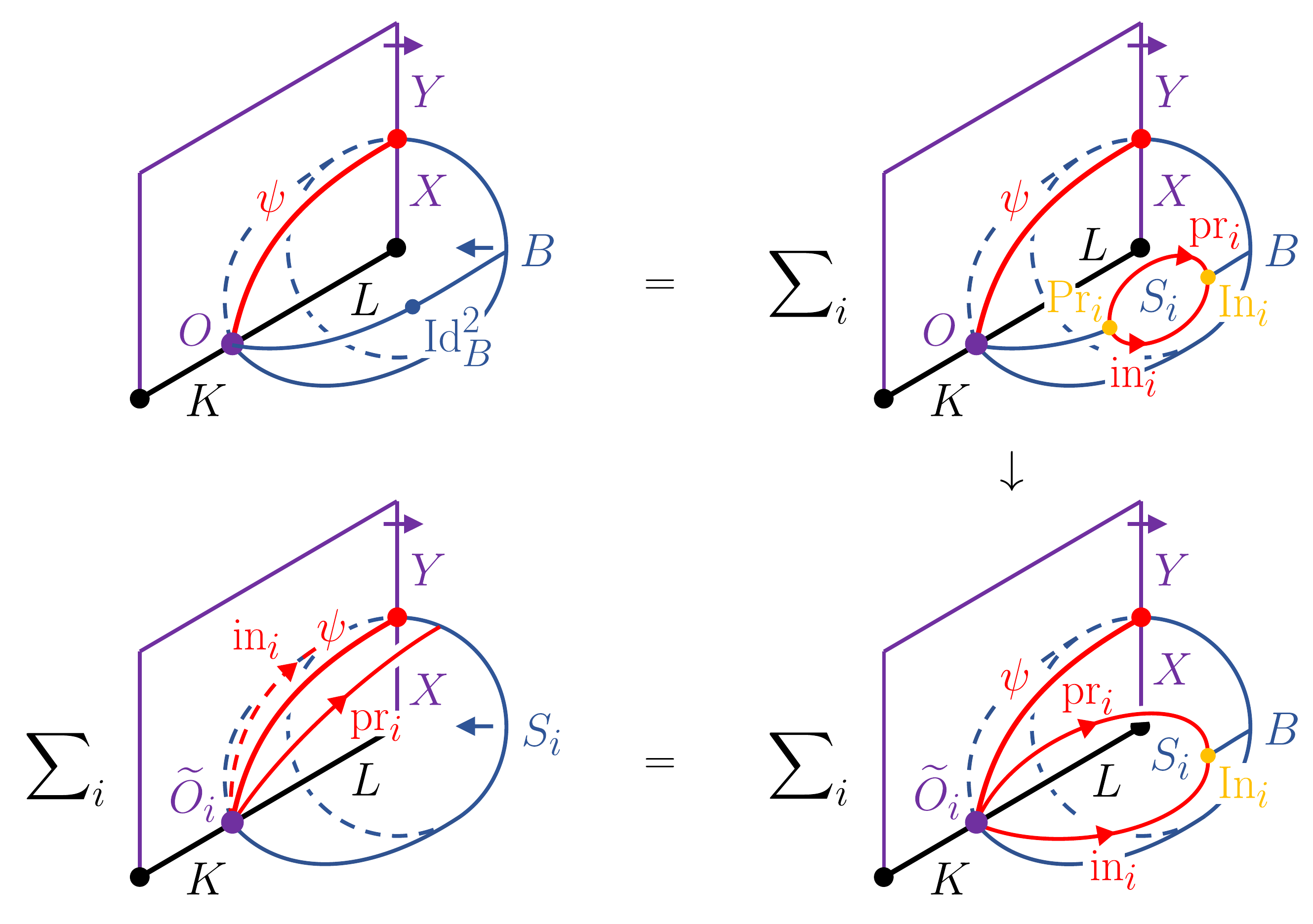}
	\vspace{-2pt}
	\caption{}
	\label{fig:3d-tube-hom-isomorphism}
\end{figure}

Let us now give a schematic justification for equation (\ref{eq:3d-tube-hom-isomorphism}), which states that, up to isomorphism, the wrapping action of a generic symmetry defect $B \in \mathsf{C}$ on 2-twisted sectors is determined the wrapping action of its simple constituents. To see this, consider a decomposition
\begin{equation}
B \; = \; \bigoplus\nolimits_i S_i
\end{equation}
of $B$ into simple objects $S_i \in \mathsf{C}$ together with inclusion and projection 1-morphisms
\begin{equation}
\text{in}_i : \, S_i \, \xhookrightarrow \, B  \qquad \text{and} \qquad \text{pr}_i: \, B \, \twoheadrightarrow \, S_i
\end{equation}
and inclusion and projection 2-morphisms
\begin{equation}
\text{In}_i : \,\; \text{in}_i \,\circ \, \text{pr}_i \; \Rightarrow \; \text{Id}_B \qquad \text{and} \qquad \text{Pr}_i : \,\; \text{Id}_B \; \Rightarrow \; \text{in}_i \,\circ \, \text{pr}_i
\end{equation}
as before. Using the completeness relation
\begin{equation}\label{eq:3d-completeness-relation}
\text{Id}^2_B \; = \; \sum\nolimits_i \, \text{In}_i \, \circ \, \text{Pr}_i \, ,
\end{equation}
we can then insert a sum of small $S_i$-bubbles into the surface $B$ as illustrated in the top right corner of figure \ref{fig:3d-tube-hom-isomorphism}. Topological local operators $O$ sitting at the junction between the wrapped line $L$ and a line $K$ can then be collided with the topological junctions $\text{Pr}_i$ to give new topological operators $\widetilde{O}_i$ as illustrated in the bottom right corner of figure \ref{fig:3d-tube-hom-isomorphism}. The maps $O \mapsto \widetilde{O}_i$ are induced by 2-morphisms
\begin{equation}
\begin{gathered}
\includegraphics[height=3.7cm]{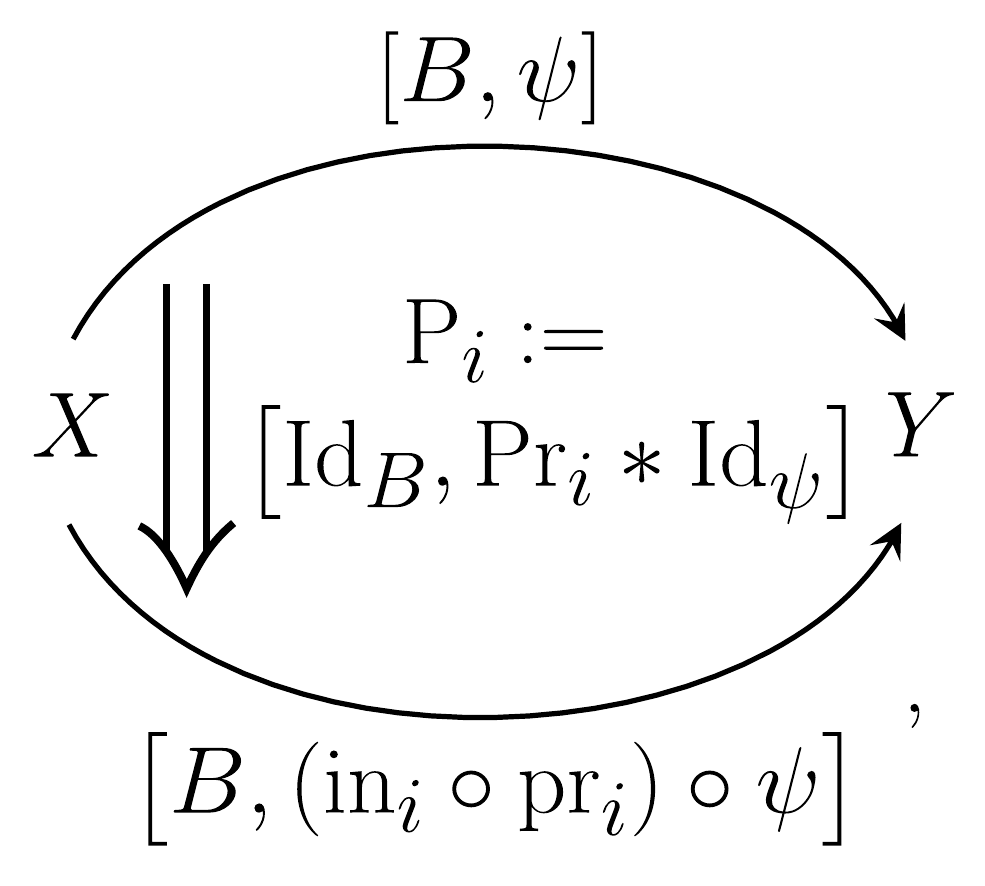}
\end{gathered}
\vspace{-2pt}
\end{equation}
where $(\text{in}_i \circ \text{pr}_i)$ should be read as $\text{Id}_Y \otimes (\text{in}_i \circ \text{pr}_i)$ and $\text{Pr}_i$ should be read as $\text{Id}_Y^2 \otimes \text{Pr}_i$. Similarly, there exist 2-morphisms
\begin{equation}
\begin{gathered}
\includegraphics[height=3.7cm]{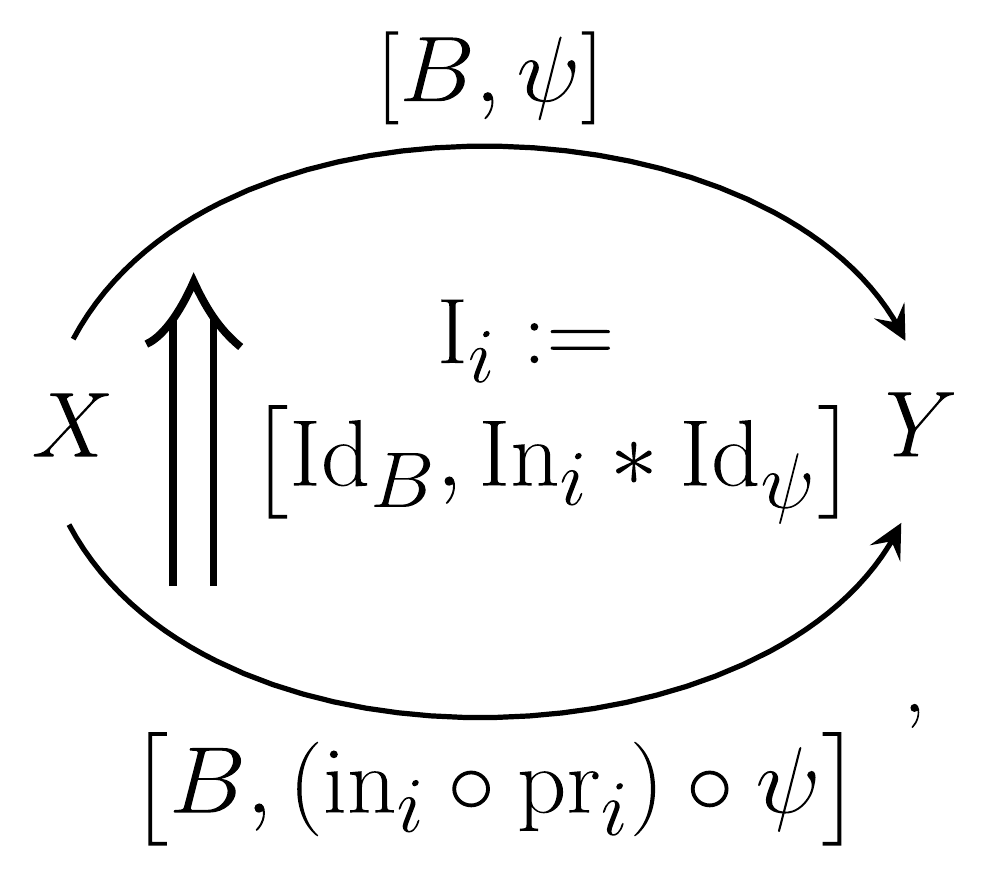}
\end{gathered}
\vspace{-2pt}
\end{equation}
where $\text{In}_i$ should be read as $ \text{Id}_Y^2 \otimes \text{In}_i$. As a consequence of the completeness relation (\ref{eq:3d-completeness-relation}) and the fact that $\text{Pr}_i \circ \text{In}_i = \text{Id}_{(\text{in}_i \, \circ \, \text{pr}_i)}$, the 2-morphisms $\text{P} := \sum_i \text{P}_i$ and $\text{I} := \sum_i \text{I}_i$ then satisfy the relations
\begin{equation}
\text{I} \, \circ \, \text{P} \; = \; \text{Id}_{[B,\psi]} \qquad \text{and} \qquad \text{P} \, \circ \, \text{I} \; = \; \text{Id}_{\, \oplus_i \,[B, \,(\text{in}_i \, \circ \, \text{pr}_i)\, \circ \, \psi]} \, .
\end{equation}
Furthermore, by moving the topological line interfaces $\text{in}_i$ around the wrapping cylinder as illustrated in the bottom left corner of figure \ref{fig:3d-tube-hom-isomorphism}, we may identify
\begin{equation}
\big[B,\, (\text{in}_i \, \circ \, \text{pr}_i) \circ \psi\big] \; = \; \big[S_i,\,\text{pr}_i \circ \psi \circ \text{in}_i \big] \, ,
\end{equation}
where $\text{pr}_i \circ \psi \circ \text{in}_i$ should be read as $( Y \otimes \text{pr}_i ) \circ \psi \circ  ( \text{in}_i \otimes X )$. Consequently, the pair $\text{P}$, $\text{I}$ defines an isomorphism
\begin{equation}
[B,\psi] \;\, \cong \;\, \bigoplus\nolimits_i \; \big[S_i, \, \text{pr}_i \circ \psi \circ \text{in}_i\big]
\end{equation}
of objects in the 1-category $\text{1-Hom}_{\mathsf{T_2C}}(X,Y)$. This shows that, up to isomorphism, the wrapping action of $B$ on 2-twisted sectors is determined the wrapping action of its simple constituents. Furthermore, as discussed above, up to 1-condensations we can replace each of the $S_i$'s by an arbitrary representative of its connected component $[S_i] \in \pi_0(\mathsf{C})$, which justifies equation (\ref{eq:3d-tube-hom-isomorphism}).

\bibliographystyle{JHEP}
\bibliography{gauging}

\providecommand{\href}[2]{#2}\begingroup\raggedright\begin{thebibliography}{100}

\bibitem{2018arXiv181211933D}
C.~L. {Douglas} and D.~J. {Reutter}, {\it {Fusion 2-categories and a state-sum
  invariant for 4-manifolds}},  {\em arXiv e-prints} (Dec., 2018)
  arXiv:1812.11933, [\href{http://arxiv.org/abs/1812.11933}{{\tt
  arXiv:1812.11933}}].

\bibitem{Gaiotto:2019xmp}
D.~Gaiotto and T.~Johnson-Freyd, {\it {Condensations in higher categories}},
  \href{http://arxiv.org/abs/1905.09566}{{\tt arXiv:1905.09566}}.

\bibitem{Johnson-Freyd:2022wm}
T.~Johnson-Freyd, {\it On the classification of topological orders},  {\em
  Communications in Mathematical Physics} {\bf 393} (2022), no.~2 989--1033.

\bibitem{Verlinde:1988sn}
E.~P. Verlinde, {\it {Fusion Rules and Modular Transformations in 2D Conformal
  Field Theory}},  {\em Nucl. Phys. B} {\bf 300} (1988) 360--376.

\bibitem{Petkova:2000ip}
V.~B. Petkova and J.~B. Zuber, {\it {Generalized twisted partition functions}},
   {\em Phys. Lett. B} {\bf 504} (2001) 157--164,
  [\href{http://arxiv.org/abs/hep-th/0011021}{{\tt hep-th/0011021}}].

\bibitem{Frohlich:2004ef}
J.~Frohlich, J.~Fuchs, I.~Runkel, and C.~Schweigert, {\it {Kramers-Wannier
  duality from conformal defects}},  {\em Phys. Rev. Lett.} {\bf 93} (2004)
  070601, [\href{http://arxiv.org/abs/cond-mat/0404051}{{\tt
  cond-mat/0404051}}].

\bibitem{Frohlich:2006ch}
J.~Frohlich, J.~Fuchs, I.~Runkel, and C.~Schweigert, {\it {Duality and defects
  in rational conformal field theory}},  {\em Nucl. Phys. B} {\bf 763} (2007)
  354--430, [\href{http://arxiv.org/abs/hep-th/0607247}{{\tt hep-th/0607247}}].

\bibitem{Frohlich:2009gb}
J.~Frohlich, J.~Fuchs, I.~Runkel, and C.~Schweigert, {\it {Defect lines,
  dualities, and generalised orbifolds}},  in {\em {16th International Congress
  on Mathematical Physics}}, 9, 2009.
\newblock \href{http://arxiv.org/abs/0909.5013}{{\tt arXiv:0909.5013}}.

\bibitem{Carqueville:2012dk}
N.~Carqueville and I.~Runkel, {\it {Orbifold completion of defect
  bicategories}},  {\em Quantum Topol.} {\bf 7} (2016) 203,
  [\href{http://arxiv.org/abs/1210.6363}{{\tt arXiv:1210.6363}}].

\bibitem{Brunner:2013ota}
I.~Brunner, N.~Carqueville, and D.~Plencner, {\it {Orbifolds and topological
  defects}},  {\em Commun. Math. Phys.} {\bf 332} (2014) 669--712,
  [\href{http://arxiv.org/abs/1307.3141}{{\tt arXiv:1307.3141}}].

\bibitem{Brunner:2013xna}
I.~Brunner, N.~Carqueville, and D.~Plencner, {\it {A quick guide to defect
  orbifolds}},  {\em Proc. Symp. Pure Math.} {\bf 88} (2014) 231--242,
  [\href{http://arxiv.org/abs/1310.0062}{{\tt arXiv:1310.0062}}].

\bibitem{Bhardwaj:2017xup}
L.~Bhardwaj and Y.~Tachikawa, {\it {On finite symmetries and their gauging in
  two dimensions}},  {\em JHEP} {\bf 03} (2018) 189,
  [\href{http://arxiv.org/abs/1704.02330}{{\tt arXiv:1704.02330}}].

\bibitem{Chang:2018iay}
C.-M. Chang, Y.-H. Lin, S.-H. Shao, Y.~Wang, and X.~Yin, {\it {Topological
  Defect Lines and Renormalization Group Flows in Two Dimensions}},  {\em JHEP}
  {\bf 01} (2019) 026, [\href{http://arxiv.org/abs/1802.04445}{{\tt
  arXiv:1802.04445}}].

\bibitem{Thorngren:2019iar}
R.~Thorngren and Y.~Wang, {\it {Fusion Category Symmetry I: Anomaly In-Flow and
  Gapped Phases}},  \href{http://arxiv.org/abs/1912.02817}{{\tt
  arXiv:1912.02817}}.

\bibitem{Ji:2019ugf}
W.~Ji, S.-H. Shao, and X.-G. Wen, {\it {Topological Transition on the Conformal
  Manifold}},  {\em Phys. Rev. Res.} {\bf 2} (2020), no.~3 033317,
  [\href{http://arxiv.org/abs/1909.01425}{{\tt arXiv:1909.01425}}].

\bibitem{Lin:2019hks}
Y.-H. Lin and S.-H. Shao, {\it {Duality Defect of the Monster CFT}},  {\em J.
  Phys. A} {\bf 54} (2021), no.~6 065201,
  [\href{http://arxiv.org/abs/1911.00042}{{\tt arXiv:1911.00042}}].

\bibitem{Komargodski:2020mxz}
Z.~Komargodski, K.~Ohmori, K.~Roumpedakis, and S.~Seifnashri, {\it {Symmetries
  and strings of adjoint QCD$_{2}$}},  {\em JHEP} {\bf 03} (2021) 103,
  [\href{http://arxiv.org/abs/2008.07567}{{\tt arXiv:2008.07567}}].

\bibitem{Chang:2020imq}
C.-M. Chang and Y.-H. Lin, {\it {Lorentzian dynamics and factorization beyond
  rationality}},  {\em JHEP} {\bf 10} (2021) 125,
  [\href{http://arxiv.org/abs/2012.01429}{{\tt arXiv:2012.01429}}].

\bibitem{Nguyen:2021naa}
M.~Nguyen, Y.~Tanizaki, and M.~\"Unsal, {\it {Noninvertible 1-form symmetry and
  Casimir scaling in 2D Yang-Mills theory}},  {\em Phys. Rev. D} {\bf 104}
  (2021), no.~6 065003, [\href{http://arxiv.org/abs/2104.01824}{{\tt
  arXiv:2104.01824}}].

\bibitem{Burbano:2021loy}
I.~M. Burbano, J.~Kulp, and J.~Neuser, {\it {Duality defects in E$_{8}$}},
  {\em JHEP} {\bf 10} (2022) 186, [\href{http://arxiv.org/abs/2112.14323}{{\tt
  arXiv:2112.14323}}].

\bibitem{Huang:2021nvb}
T.-C. Huang, Y.-H. Lin, K.~Ohmori, Y.~Tachikawa, and M.~Tezuka, {\it {Numerical
  Evidence for a Haagerup Conformal Field Theory}},  {\em Phys. Rev. Lett.}
  {\bf 128} (2022), no.~23 231603, [\href{http://arxiv.org/abs/2110.03008}{{\tt
  arXiv:2110.03008}}].

\bibitem{Thorngren:2021yso}
R.~Thorngren and Y.~Wang, {\it {Fusion Category Symmetry II: Categoriosities at
  $c$ = 1 and Beyond}},  \href{http://arxiv.org/abs/2106.12577}{{\tt
  arXiv:2106.12577}}.

\bibitem{Sharpe:2021srf}
E.~Sharpe, {\it {Topological operators, noninvertible symmetries and
  decomposition}},  \href{http://arxiv.org/abs/2108.13423}{{\tt
  arXiv:2108.13423}}.

\bibitem{Lin:2022dhv}
Y.-H. Lin, M.~Okada, S.~Seifnashri, and Y.~Tachikawa, {\it {Asymptotic density
  of states in 2d CFTs with non-invertible symmetries}},
  \href{http://arxiv.org/abs/2208.05495}{{\tt arXiv:2208.05495}}.

\bibitem{Chang:2022hud}
C.-M. Chang, J.~Chen, and F.~Xu, {\it {Topological Defect Lines in Two
  Dimensional Fermionic CFTs}},  \href{http://arxiv.org/abs/2208.02757}{{\tt
  arXiv:2208.02757}}.

\bibitem{Lin:2023uvm}
Y.-H. Lin and S.-H. Shao, {\it {Bootstrapping Non-invertible Symmetries}},
  \href{http://arxiv.org/abs/2302.13900}{{\tt arXiv:2302.13900}}.

\bibitem{Rudelius:2020orz}
T.~Rudelius and S.-H. Shao, {\it {Topological Operators and Completeness of
  Spectrum in Discrete Gauge Theories}},  {\em JHEP} {\bf 12} (2020) 172,
  [\href{http://arxiv.org/abs/2006.10052}{{\tt arXiv:2006.10052}}].

\bibitem{Heidenreich:2021xpr}
B.~Heidenreich, J.~McNamara, M.~Montero, M.~Reece, T.~Rudelius, and
  I.~Valenzuela, {\it {Non-invertible global symmetries and completeness of the
  spectrum}},  {\em JHEP} {\bf 09} (2021) 203,
  [\href{http://arxiv.org/abs/2104.07036}{{\tt arXiv:2104.07036}}].

\bibitem{Koide:2021zxj}
M.~Koide, Y.~Nagoya, and S.~Yamaguchi, {\it {Non-invertible topological defects
  in 4-dimensional $\mathbb {Z}_2$ pure lattice gauge theory}},  {\em PTEP}
  {\bf 2022} (2022), no.~1 013B03, [\href{http://arxiv.org/abs/2109.05992}{{\tt
  arXiv:2109.05992}}].

\bibitem{Choi:2021kmx}
Y.~Choi, C.~Cordova, P.-S. Hsin, H.~T. Lam, and S.-H. Shao, {\it
  {Non-Invertible Duality Defects in 3+1 Dimensions}},
  \href{http://arxiv.org/abs/2111.01139}{{\tt arXiv:2111.01139}}.

\bibitem{Kaidi:2021xfk}
J.~Kaidi, K.~Ohmori, and Y.~Zheng, {\it {Kramers-Wannier-like Duality Defects
  in (3+1)D Gauge Theories}},  {\em Phys. Rev. Lett.} {\bf 128} (2022), no.~11
  111601, [\href{http://arxiv.org/abs/2111.01141}{{\tt arXiv:2111.01141}}].

\bibitem{Wang:2021vki}
J.~Wang and Y.-Z. You, {\it {Gauge Enhanced Quantum Criticality Between Grand
  Unifications: Categorical Higher Symmetry Retraction}},
  \href{http://arxiv.org/abs/2111.10369}{{\tt arXiv:2111.10369}}.

\bibitem{Chen:2021xuc}
X.~Chen, A.~Dua, P.-S. Hsin, C.-M. Jian, W.~Shirley, and C.~Xu, {\it {Loops in
  4+1d Topological Phases}},  \href{http://arxiv.org/abs/2112.02137}{{\tt
  arXiv:2112.02137}}.

\bibitem{Cordova:2022rer}
C.~Cordova, K.~Ohmori, and T.~Rudelius, {\it {Generalized symmetry breaking
  scales and weak gravity conjectures}},  {\em JHEP} {\bf 11} (2022) 154,
  [\href{http://arxiv.org/abs/2202.05866}{{\tt arXiv:2202.05866}}].

\bibitem{Benini:2022hzx}
F.~Benini, C.~Copetti, and L.~Di~Pietro, {\it {Factorization and global
  symmetries in holography}},  \href{http://arxiv.org/abs/2203.09537}{{\tt
  arXiv:2203.09537}}.

\bibitem{Roumpedakis:2022aik}
K.~Roumpedakis, S.~Seifnashri, and S.-H. Shao, {\it {Higher Gauging and
  Non-invertible Condensation Defects}},
  \href{http://arxiv.org/abs/2204.02407}{{\tt arXiv:2204.02407}}.

\bibitem{DelZotto:2022ras}
M.~Del~Zotto and I.~n. Garc\'\i{}a~Etxebarria, {\it {Global Structures from the
  Infrared}},  \href{http://arxiv.org/abs/2204.06495}{{\tt arXiv:2204.06495}}.

\bibitem{Bhardwaj:2022yxj}
L.~Bhardwaj, L.~Bottini, S.~Schafer-Nameki, and A.~Tiwari, {\it {Non-Invertible
  Higher-Categorical Symmetries}},  \href{http://arxiv.org/abs/2204.06564}{{\tt
  arXiv:2204.06564}}.

\bibitem{Hayashi:2022fkw}
Y.~Hayashi and Y.~Tanizaki, {\it {Non-invertible self-duality defects of
  Cardy-Rabinovici model and mixed gravitational anomaly}},  {\em JHEP} {\bf
  08} (2022) 036, [\href{http://arxiv.org/abs/2204.07440}{{\tt
  arXiv:2204.07440}}].

\bibitem{Arias-Tamargo:2022nlf}
G.~Arias-Tamargo and D.~Rodriguez-Gomez, {\it {Non-Invertible Symmetries from
  Discrete Gauging and Completeness of the Spectrum}},
  \href{http://arxiv.org/abs/2204.07523}{{\tt arXiv:2204.07523}}.

\bibitem{Choi:2022zal}
Y.~Choi, C.~Cordova, P.-S. Hsin, H.~T. Lam, and S.-H. Shao, {\it
  {Non-invertible Condensation, Duality, and Triality Defects in 3+1
  Dimensions}},  \href{http://arxiv.org/abs/2204.09025}{{\tt
  arXiv:2204.09025}}.

\bibitem{Kaidi:2022uux}
J.~Kaidi, G.~Zafrir, and Y.~Zheng, {\it {Non-Invertible Symmetries of
  $\mathcal{N}=4$ SYM and Twisted Compactification}},
  \href{http://arxiv.org/abs/2205.01104}{{\tt arXiv:2205.01104}}.

\bibitem{Choi:2022jqy}
Y.~Choi, H.~T. Lam, and S.-H. Shao, {\it {Non-invertible Global Symmetries in
  the Standard Model}},  \href{http://arxiv.org/abs/2205.05086}{{\tt
  arXiv:2205.05086}}.

\bibitem{Cordova:2022ieu}
C.~Cordova and K.~Ohmori, {\it {Non-Invertible Chiral Symmetry and Exponential
  Hierarchies}},  \href{http://arxiv.org/abs/2205.06243}{{\tt
  arXiv:2205.06243}}.

\bibitem{Antinucci:2022eat}
A.~Antinucci, G.~Galati, and G.~Rizi, {\it {On Continuous 2-Category Symmetries
  and Yang-Mills Theory}},  \href{http://arxiv.org/abs/2206.05646}{{\tt
  arXiv:2206.05646}}.

\bibitem{Bashmakov:2022jtl}
V.~Bashmakov, M.~Del~Zotto, and A.~Hasan, {\it {On the 6d Origin of
  Non-invertible Symmetries in 4d}},
  \href{http://arxiv.org/abs/2206.07073}{{\tt arXiv:2206.07073}}.

\bibitem{Damia:2022rxw}
J.~A. Damia, R.~Argurio, and L.~Tizzano, {\it {Continuous Generalized
  Symmetries in Three Dimensions}},
  \href{http://arxiv.org/abs/2206.14093}{{\tt arXiv:2206.14093}}.

\bibitem{Damia:2022bcd}
J.~A. Damia, R.~Argurio, and E.~Garcia-Valdecasas, {\it {Non-Invertible Defects
  in 5d, Boundaries and Holography}},
  \href{http://arxiv.org/abs/2207.02831}{{\tt arXiv:2207.02831}}.

\bibitem{Moradi:2022lqp}
H.~Moradi, S.~F. Moosavian, and A.~Tiwari, {\it {Topological Holography:
  Towards a Unification of Landau and Beyond-Landau Physics}},
  \href{http://arxiv.org/abs/2207.10712}{{\tt arXiv:2207.10712}}.

\bibitem{Choi:2022rfe}
Y.~Choi, H.~T. Lam, and S.-H. Shao, {\it {Non-invertible Time-reversal
  Symmetry}},  \href{http://arxiv.org/abs/2208.04331}{{\tt arXiv:2208.04331}}.

\bibitem{Bhardwaj:2022lsg}
L.~Bhardwaj, S.~Schafer-Nameki, and J.~Wu, {\it {Universal Non-Invertible
  Symmetries}},  \href{http://arxiv.org/abs/2208.05973}{{\tt
  arXiv:2208.05973}}.

\bibitem{Bartsch:2022mpm}
T.~Bartsch, M.~Bullimore, A.~E.~V. Ferrari, and J.~Pearson, {\it
  {Non-invertible Symmetries and Higher Representation Theory I}},
  \href{http://arxiv.org/abs/2208.05993}{{\tt arXiv:2208.05993}}.

\bibitem{Lin:2022xod}
L.~Lin, D.~G. Robbins, and E.~Sharpe, {\it {Decomposition, condensation
  defects, and fusion}},  {\em Fortsch. Phys.} {\bf 70} (2022) 2200130,
  [\href{http://arxiv.org/abs/2208.05982}{{\tt arXiv:2208.05982}}].

\bibitem{Lu:2022ver}
D.-C. Lu and Z.~Sun, {\it {On Triality Defects in 2d CFT}},
  \href{http://arxiv.org/abs/2208.06077}{{\tt arXiv:2208.06077}}.

\bibitem{GarciaEtxebarria:2022vzq}
I.~n. Garc\'\i{}a~Etxebarria, {\it {Branes and Non-Invertible Symmetries}},
  \href{http://arxiv.org/abs/2208.07508}{{\tt arXiv:2208.07508}}.

\bibitem{Heckman:2022muc}
J.~J. Heckman, M.~H\"ubner, E.~Torres, and H.~Y. Zhang, {\it {The Branes Behind
  Generalized Symmetry Operators}},
  \href{http://arxiv.org/abs/2209.03343}{{\tt arXiv:2209.03343}}.

\bibitem{Kaidi:2022cpf}
J.~Kaidi, K.~Ohmori, and Y.~Zheng, {\it {Symmetry TFTs for Non-Invertible
  Defects}},  \href{http://arxiv.org/abs/2209.11062}{{\tt arXiv:2209.11062}}.

\bibitem{Niro:2022ctq}
P.~Niro, K.~Roumpedakis, and O.~Sela, {\it {Exploring Non-Invertible Symmetries
  in Free Theories}},  \href{http://arxiv.org/abs/2209.11166}{{\tt
  arXiv:2209.11166}}.

\bibitem{Mekareeya:2022spm}
N.~Mekareeya and M.~Sacchi, {\it {Mixed Anomalies, Two-groups, Non-Invertible
  Symmetries, and 3d Superconformal Indices}},
  \href{http://arxiv.org/abs/2210.02466}{{\tt arXiv:2210.02466}}.

\bibitem{Antinucci:2022vyk}
A.~Antinucci, F.~Benini, C.~Copetti, G.~Galati, and G.~Rizi, {\it {The
  holography of non-invertible self-duality symmetries}},
  \href{http://arxiv.org/abs/2210.09146}{{\tt arXiv:2210.09146}}.

\bibitem{Chen:2022cyw}
S.~Chen and Y.~Tanizaki, {\it {Solitonic symmetry beyond homotopy:
  invertibility from bordism and non-invertibility from TQFT}},
  \href{http://arxiv.org/abs/2210.13780}{{\tt arXiv:2210.13780}}.

\bibitem{Bashmakov:2022uek}
V.~Bashmakov, M.~Del~Zotto, A.~Hasan, and J.~Kaidi, {\it {Non-invertible
  Symmetries of Class $\mathcal{S}$ Theories}},
  \href{http://arxiv.org/abs/2211.05138}{{\tt arXiv:2211.05138}}.

\bibitem{Karasik:2022kkq}
A.~Karasik, {\it {On anomalies and gauging of U(1) non-invertible symmetries in
  4d QED}},  \href{http://arxiv.org/abs/2211.05802}{{\tt arXiv:2211.05802}}.

\bibitem{Cordova:2022fhg}
C.~Cordova, S.~Hong, S.~Koren, and K.~Ohmori, {\it {Neutrino Masses from
  Generalized Symmetry Breaking}},  \href{http://arxiv.org/abs/2211.07639}{{\tt
  arXiv:2211.07639}}.

\bibitem{Decoppet:2022dnz}
T.~D. D\'ecoppet and M.~Yu, {\it {Gauging Noninvertible Defects: A
  2-Categorical Perspective}},  \href{http://arxiv.org/abs/2211.08436}{{\tt
  arXiv:2211.08436}}.

\bibitem{GarciaEtxebarria:2022jky}
I.~n. Garc\'\i{}a~Etxebarria and N.~Iqbal, {\it {A Goldstone theorem for
  continuous non-invertible symmetries}},
  \href{http://arxiv.org/abs/2211.09570}{{\tt arXiv:2211.09570}}.

\bibitem{Choi:2022fgx}
Y.~Choi, H.~T. Lam, and S.-H. Shao, {\it {Non-invertible Gauss Law and
  Axions}},  \href{http://arxiv.org/abs/2212.04499}{{\tt arXiv:2212.04499}}.

\bibitem{Bhardwaj:2022kot}
L.~Bhardwaj, S.~Schafer-Nameki, and A.~Tiwari, {\it {Unifying Constructions of
  Non-Invertible Symmetries}},  \href{http://arxiv.org/abs/2212.06159}{{\tt
  arXiv:2212.06159}}.

\bibitem{Bartsch:2022ytj}
T.~Bartsch, M.~Bullimore, A.~E.~V. Ferrari, and J.~Pearson, {\it
  {Non-invertible Symmetries and Higher Representation Theory II}},
  \href{http://arxiv.org/abs/2212.07393}{{\tt arXiv:2212.07393}}.

\bibitem{Bhardwaj:2022maz}
L.~Bhardwaj, L.~E. Bottini, S.~Schafer-Nameki, and A.~Tiwari, {\it
  {Non-Invertible Symmetry Webs}},  \href{http://arxiv.org/abs/2212.06842}{{\tt
  arXiv:2212.06842}}.

\bibitem{Antinucci:2022cdi}
A.~Antinucci, C.~Copetti, G.~Galati, and G.~Rizi, {\it {''Zoology'' of
  non-invertible duality defects: the view from class $\mathcal{S}$}},
  \href{http://arxiv.org/abs/2212.09549}{{\tt arXiv:2212.09549}}.

\bibitem{Apte:2022xtu}
A.~Apte, C.~Cordova, and H.~T. Lam, {\it {Obstructions to Gapped Phases from
  Non-Invertible Symmetries}},  \href{http://arxiv.org/abs/2212.14605}{{\tt
  arXiv:2212.14605}}.

\bibitem{Delcamp:2023kew}
C.~Delcamp and A.~Tiwari, {\it {Higher categorical symmetries and gauging in
  two-dimensional spin systems}},  \href{http://arxiv.org/abs/2301.01259}{{\tt
  arXiv:2301.01259}}.

\bibitem{Kaidi:2023maf}
J.~Kaidi, E.~Nardoni, G.~Zafrir, and Y.~Zheng, {\it {Symmetry TFTs and
  Anomalies of Non-Invertible Symmetries}},
  \href{http://arxiv.org/abs/2301.07112}{{\tt arXiv:2301.07112}}.

\bibitem{Brennan:2023kpw}
T.~D. Brennan, S.~Hong, and L.-T. Wang, {\it {Coupling a Cosmic String to a
  TQFT}},  \href{http://arxiv.org/abs/2302.00777}{{\tt arXiv:2302.00777}}.

\bibitem{Radhakrishnan:2023zcq}
R.~Radhakrishnan, {\it {On Reconstructing Finite Gauge Group from Fusion
  Rules}},  \href{http://arxiv.org/abs/2302.08419}{{\tt arXiv:2302.08419}}.

\bibitem{Putrov:2023jqi}
P.~Putrov and J.~Wang, {\it {Categorical Symmetry of the Standard Model from
  Gravitational Anomaly}},  \href{http://arxiv.org/abs/2302.14862}{{\tt
  arXiv:2302.14862}}.

\bibitem{Carta:2023bqn}
F.~Carta, S.~Giacomelli, N.~Mekareeya, and A.~Mininno, {\it {Comments on
  Non-invertible Symmetries in Argyres-Douglas Theories}},
  \href{http://arxiv.org/abs/2303.16216}{{\tt arXiv:2303.16216}}.

\bibitem{Bhardwaj:2023wzd}
L.~Bhardwaj and S.~Schafer-Nameki, {\it {Generalized Charges, Part I:
  Invertible Symmetries and Higher Representations}},
  \href{http://arxiv.org/abs/2304.02660}{{\tt arXiv:2304.02660}}.

\bibitem{Bartsch:2023pzl}
T.~Bartsch, M.~Bullimore, and A.~Grigoletto, {\it {Higher representations for
  extended operators}},  \href{http://arxiv.org/abs/2304.03789}{{\tt
  arXiv:2304.03789}}.

\bibitem{Cao:2023doz}
W.~Cao, L.~Li, M.~Yamazaki, and Y.~Zheng, {\it {Subsystem Non-Invertible
  Symmetry Operators and Defects}},
  \href{http://arxiv.org/abs/2304.09886}{{\tt arXiv:2304.09886}}.

\bibitem{Inamura:2023qzl}
K.~Inamura and K.~Ohmori, {\it {Fusion Surface Models: 2+1d Lattice Models from
  Fusion 2-Categories}},  \href{http://arxiv.org/abs/2305.05774}{{\tt
  arXiv:2305.05774}}.

\bibitem{Freed:2012bs}
D.~S. Freed and C.~Teleman, {\it {Relative quantum field theory}},  {\em
  Commun. Math. Phys.} {\bf 326} (2014) 459--476,
  [\href{http://arxiv.org/abs/1212.1692}{{\tt arXiv:1212.1692}}].

\bibitem{Gaiotto:2020iye}
D.~Gaiotto and J.~Kulp, {\it {Orbifold groupoids}},  {\em JHEP} {\bf 02} (2021)
  132, [\href{http://arxiv.org/abs/2008.05960}{{\tt arXiv:2008.05960}}].

\bibitem{Ji:2019jhk}
W.~Ji and X.-G. Wen, {\it {Categorical symmetry and noninvertible anomaly in
  symmetry-breaking and topological phase transitions}},  {\em Phys. Rev. Res.}
  {\bf 2} (2020), no.~3 033417, [\href{http://arxiv.org/abs/1912.13492}{{\tt
  arXiv:1912.13492}}].

\bibitem{Kong:2020cie}
L.~Kong, T.~Lan, X.-G. Wen, Z.-H. Zhang, and H.~Zheng, {\it {Algebraic higher
  symmetry and categorical symmetry -- a holographic and entanglement view of
  symmetry}},  {\em Phys. Rev. Res.} {\bf 2} (2020), no.~4 043086,
  [\href{http://arxiv.org/abs/2005.14178}{{\tt arXiv:2005.14178}}].

\bibitem{Freed:2022qnc}
D.~S. Freed, G.~W. Moore, and C.~Teleman, {\it {Topological symmetry in quantum
  field theory}},  \href{http://arxiv.org/abs/2209.07471}{{\tt
  arXiv:2209.07471}}.

\bibitem{Freed:2022iao}
D.~S. Freed, {\it {Introduction to topological symmetry in QFT}},
  \href{http://arxiv.org/abs/2212.00195}{{\tt arXiv:2212.00195}}.

\bibitem{Apruzzi:2021nmk}
F.~Apruzzi, F.~Bonetti, I.~n.~G. Etxebarria, S.~S. Hosseini, and
  S.~Schafer-Nameki, {\it {Symmetry TFTs from String Theory}},
  \href{http://arxiv.org/abs/2112.02092}{{\tt arXiv:2112.02092}}.

\bibitem{Apruzzi:2022dlm}
F.~Apruzzi, {\it {Higher form symmetries TFT in 6d}},  {\em JHEP} {\bf 11}
  (2022) 050, [\href{http://arxiv.org/abs/2203.10063}{{\tt arXiv:2203.10063}}].

\bibitem{Apruzzi:2022rei}
F.~Apruzzi, I.~Bah, F.~Bonetti, and S.~Schafer-Nameki, {\it {Noninvertible
  Symmetries from Holography and Branes}},  {\em Phys. Rev. Lett.} {\bf 130}
  (2023), no.~12 121601, [\href{http://arxiv.org/abs/2208.07373}{{\tt
  arXiv:2208.07373}}].

\bibitem{vanBeest:2022fss}
M.~van Beest, D.~S.~W. Gould, S.~Schafer-Nameki, and Y.-N. Wang, {\it {Symmetry
  TFTs for 3d QFTs from M-theory}},  {\em JHEP} {\bf 02} (2023) 226,
  [\href{http://arxiv.org/abs/2210.03703}{{\tt arXiv:2210.03703}}].

\bibitem{Chen:2023qnv}
J.~Chen, W.~Cui, B.~Haghighat, and Y.-N. Wang, {\it {SymTFTs and Duality
  Defects from 6d SCFTs on 4-manifolds}},
  \href{http://arxiv.org/abs/2305.09734}{{\tt arXiv:2305.09734}}.

\bibitem{TURAEV1992865}
V.~Turaev and O.~Viro, {\it State sum invariants of 3-manifolds and quantum
  6j-symbols},  {\em Topology} {\bf 31} (1992), no.~4 865--902.

\bibitem{Barrett:1993ab}
J.~W. Barrett and B.~W. Westbury, {\it {Invariants of piecewise linear three
  manifolds}},  {\em Trans. Am. Math. Soc.} {\bf 348} (1996) 3997--4022,
  [\href{http://arxiv.org/abs/hep-th/9311155}{{\tt hep-th/9311155}}].

\bibitem{Evans1995ONOT}
D.~E. Evans and Y.~Kawahigashi, {\it On ocneanu's theory of asymptotic
  inclusions for subfactors, topological quantum field theories and quantum
  doubles},  {\em International Journal of Mathematics} {\bf 06} (1995)
  205--228.

\bibitem{Izumi:2000aa}
M.~Izumi, {\it The structure of sectors associated with longo--rehren
  inclusions i. general theory},  {\em Communications in Mathematical Physics}
  {\bf 213} (2000), no.~1 127--179.

\bibitem{MUGER2003159}
M.~M{\"u}ger, {\it From subfactors to categories and topology ii: The quantum
  double of tensor categories and subfactors},  {\em Journal of Pure and
  Applied Algebra} {\bf 180} (2003), no.~1 159--219.

\bibitem{Kong:2019brm}
L.~Kong, Y.~Tian, and S.~Zhou, {\it {The center of monoidal 2-categories in
  3+1D Dijkgraaf-Witten theory}},  {\em Adv. Math.} {\bf 360} (2020) 106928,
  [\href{http://arxiv.org/abs/1905.04644}{{\tt arXiv:1905.04644}}].

\bibitem{Bullivant:2020xhy}
A.~Bullivant and C.~Delcamp, {\it {Gapped boundaries and string-like
  excitations in (3+1)d gauge models of topological phases}},  {\em JHEP} {\bf
  07} (2021) 025, [\href{http://arxiv.org/abs/2006.06536}{{\tt
  arXiv:2006.06536}}].

\bibitem{Bullivant:2021pkd}
A.~Bullivant and C.~Delcamp, {\it {Crossing with the circle in
  Dijkgraaf\textendash{}Witten theory and applications to topological phases of
  matter}},  {\em J. Math. Phys.} {\bf 63} (2022), no.~8 081901,
  [\href{http://arxiv.org/abs/2103.12717}{{\tt arXiv:2103.12717}}].

\bibitem{kirillov2011stringnet}
A.~K.~J. au2, {\it String-net model of turaev-viro invariants},  2011.

\bibitem{hardiman2020graphical}
L.~Hardiman, {\it A graphical approach to the drinfeld centre},  2020.

\bibitem{hardiman_king_2020}
L.~Hardiman and A.~King, {\it Decomposing the tube category},  {\em Glasgow
  Mathematical Journal} {\bf 62} (2020), no.~2 441--458.

\bibitem{Mousaaid_2021}
Y.~Mousaaid and A.~Savage, {\it Affinization of monoidal categories},  {\em
  Journal de l'{\'{E}}cole polytechnique {\textemdash} Math{\'{e}}matiques}
  {\bf 8} (mar, 2021) 791--829.

\bibitem{Kirillov_Jr__2022}
J.~Alexander~Kirillov and Y.~H. Tham, {\it Factorization homology and 4d
  {TQFT}},  {\em Quantum Topology} {\bf 13} (apr, 2022) 1--54.

\bibitem{bhowmick2018tube}
J.~Bhowmick, S.~Ghosh, N.~Rakshit, and M.~Yamashita, {\it Tube representations
  and twisting of graded categories},  2018.

\bibitem{DIJKGRAAF199160}
R.~Dijkgraaf, V.~Pasquier, and P.~Roche, {\it Quasi hope algebras, group
  cohomology and orbifold models},  {\em Nuclear Physics B - Proceedings
  Supplements} {\bf 18} (1991), no.~2 60--72.

\bibitem{Freed_1994}
D.~S. Freed, {\it Higher algebraic structures and quantization},  {\em
  Communications in Mathematical Physics} {\bf 159} (jan, 1994) 343--398.

\bibitem{Willerton_2008}
S.~Willerton, {\it The twisted drinfeld double of a finite group via gerbes and
  finite groupoids},  {\em Algebraic and Geometric Topology} {\bf 8} (sep,
  2008) 1419--1457.

\bibitem{tambara1998tensor}
D.~Tambara and S.~Yamagami, {\it Tensor categories with fusion rules of
  self-duality for finite abelian groups},  {\em Journal of Algebra} {\bf 209}
  (1998), no.~2 692--707.

\bibitem{gelaki2009centers}
S.~Gelaki, D.~Naidu, and D.~Nikshych, {\it Centers of graded fusion
  categories},  2009.

\bibitem{BAEZ1996196}
J.~C. Baez and M.~Neuchl, {\it Higher dimensional algebra: I. braided monoidal
  2-categories},  {\em Advances in Mathematics} {\bf 121} (1996), no.~2
  196--244.

\bibitem{davydov2021braided}
A.~Davydov and D.~Nikshych, {\it Braided picard groups and graded extensions of
  braided tensor categories},  2021.

\bibitem{Kapustin:2013uxa}
A.~Kapustin and R.~Thorngren, {\it {Higher symmetry and gapped phases of gauge
  theories}},  \href{http://arxiv.org/abs/1309.4721}{{\tt arXiv:1309.4721}}.

\bibitem{Benini:2018reh}
F.~Benini, C.~C\'ordova, and P.-S. Hsin, {\it {On 2-Group Global Symmetries and
  their Anomalies}},  {\em JHEP} {\bf 03} (2019) 118,
  [\href{http://arxiv.org/abs/1803.09336}{{\tt arXiv:1803.09336}}].

\bibitem{Johnson_Freyd_2023}
T.~Johnson-Freyd and D.~Reutter, {\it Minimal nondegenerate extensions}, .

\end{thebibliography}\endgroup

\end{document}